\documentclass{ieeeaccess}
\usepackage{cite}
\usepackage{amsmath,amssymb,amsfonts}
\usepackage{algorithmic}
\usepackage{graphicx}
\usepackage{textcomp}
\usepackage{hyperref}
\usepackage{enumitem}
\usepackage{siunitx}
\usepackage{multirow}
\usepackage{wrapfig}
\usepackage{float}
\usepackage{pifont}
\newcommand{\cmark}{\ding{51}}%
\newcommand{\xmark}{\ding{53}}%
\usepackage[none]{hyphenat}

\newcommand{\orcidicon}{\includegraphics[width=0.32cm]{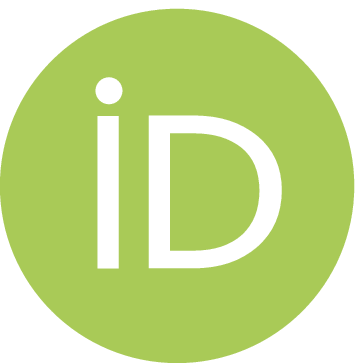}}
\expandafter\xdef\csname orcidA\endcsname{\noexpand\href{https://orcid.org/\csname orcidauthorA\endcsname}{\noexpand\orcidicon}}
\expandafter\xdef\csname orcidB\endcsname{\noexpand\href{https://orcid.org/\csname orcidauthorB\endcsname}{\noexpand\orcidicon}}
\expandafter\xdef\csname orcidC\endcsname{\noexpand\href{https://orcid.org/\csname orcidauthorC\endcsname}{\noexpand\orcidicon}}
\expandafter\xdef\csname orcidD\endcsname{\noexpand\href{https://orcid.org/\csname orcidauthorD\endcsname}{\noexpand\orcidicon}}
\expandafter\xdef\csname orcidE\endcsname{\noexpand\href{https://orcid.org/\csname orcidauthorE\endcsname}{\noexpand\orcidicon}}
\expandafter\xdef\csname orcidF\endcsname{\noexpand\href{https://orcid.org/\csname orcidauthorF\endcsname}{\noexpand\orcidicon}}

\def\BibTeX{{\rm B\kern-.05em{\sc i\kern-.025em b}\kern-.08em
    T\kern-.1667em\lower.7ex\hbox{E}\kern-.125emX}}

\begin{document}
\history{Date of publication xxxx 00, 0000, date of current version xxxx 00, 0000.}
\doi{10.1109/ACCESS.2020.3039858}

\title{Hardware and Software Optimizations for Accelerating Deep Neural Networks: Survey of Current Trends, Challenges, and the Road Ahead}

\author{\uppercase{Maurizio Capra}\authorrefmark{1,$ \dagger$}\orcidA{}, \IEEEmembership{Student Member, IEEE},
\uppercase{Beatrice Bussolino}\authorrefmark{1, $\dagger$}\orcidB{}, \IEEEmembership{Student Member, IEEE},
\uppercase{Alberto Marchisio}\authorrefmark{2}\orcidC{},\IEEEmembership{Student Member, IEEE},
\uppercase{Guido Masera}\authorrefmark{1}\orcidF{}, \IEEEmembership{Senior Member, IEEE},
\uppercase{Maurizio Martina}\authorrefmark{1}\orcidD{}, \IEEEmembership{Senior Member, IEEE},
\uppercase{and\\Muhammad Shafique}\authorrefmark{3}\orcidE{}, \IEEEmembership{Senior Member, IEEE}
}
\address[1]{Department of Electrical, Electronics and Telecommunication Engineering, Politecnico di Torino, 10129 Torino TO, Italy; name.surname@polito.it}
\address[2]{Institute of Computer Engineering, Technische Universit\"{a}t Wien (TU Wien), 1040 Vienna, Austria; alberto.marchisio@tuwien.ac.at}
\address[3]{Division of Engineering, New York University Abu Dhabi, UAE; muhammad.shafique@nyu.edu}

\tfootnote{$\dagger$ These authors contributed equally to the work herein presented.\\
This work has been partially supported by the Doctoral College Resilient Embedded Systems which is run jointly by TU Wien's Faculty of Informatics and FH-Technikum Wien.}

\markboth
{Capra \headeretal: Hardware and Software Optimizations for Accelerating Deep Neural Networks}
{Capra \headeretal: Hardware and Software Optimizations for Accelerating Deep Neural Networks}

\corresp{Corresponding author: Maurizio Capra (e-mail: maurizio.capra@polito.it).}

\begin{abstract}
Currently, Machine Learning (ML) is becoming ubiquitous in everyday life. Deep Learning (DL) is already present in many applications ranging from computer vision for medicine to autonomous driving of modern cars as well as other sectors in security, healthcare, and finance. However, to achieve impressive performance, these algorithms employ very deep networks, requiring a significant computational power, both during the training and inference time. A single inference of a DL model may require billions of multiply-and-accumulated operations, making the DL extremely compute- and energy-hungry. In a scenario where several sophisticated algorithms need to be executed with limited energy and low latency, the need for cost-effective hardware platforms capable of implementing energy-efficient DL execution arises.
This paper first introduces the key properties of two brain-inspired models like Deep Neural Network (DNN), and Spiking Neural Network (SNN), and then analyzes techniques to produce efficient and high-performance designs. This work summarizes and compares the works for four leading platforms for the execution of algorithms such as CPU, GPU, FPGA and ASIC describing the main solutions of the state-of-the-art, giving much prominence to the last two solutions since they offer greater design flexibility and bear the potential of high energy-efficiency, especially for the inference process.
In addition to hardware solutions, this paper discusses some of the important security issues that these DNN and SNN models may have during their execution, and offers a comprehensive section on benchmarking, explaining how to assess the quality of different networks and hardware systems designed for them.
\end{abstract}


\titlepgskip=-15pt

\maketitle

\section{Introduction}\label{sec:introduction}
\begin{figure*}[ht]
    \centering
    \includegraphics [width=1\textwidth] {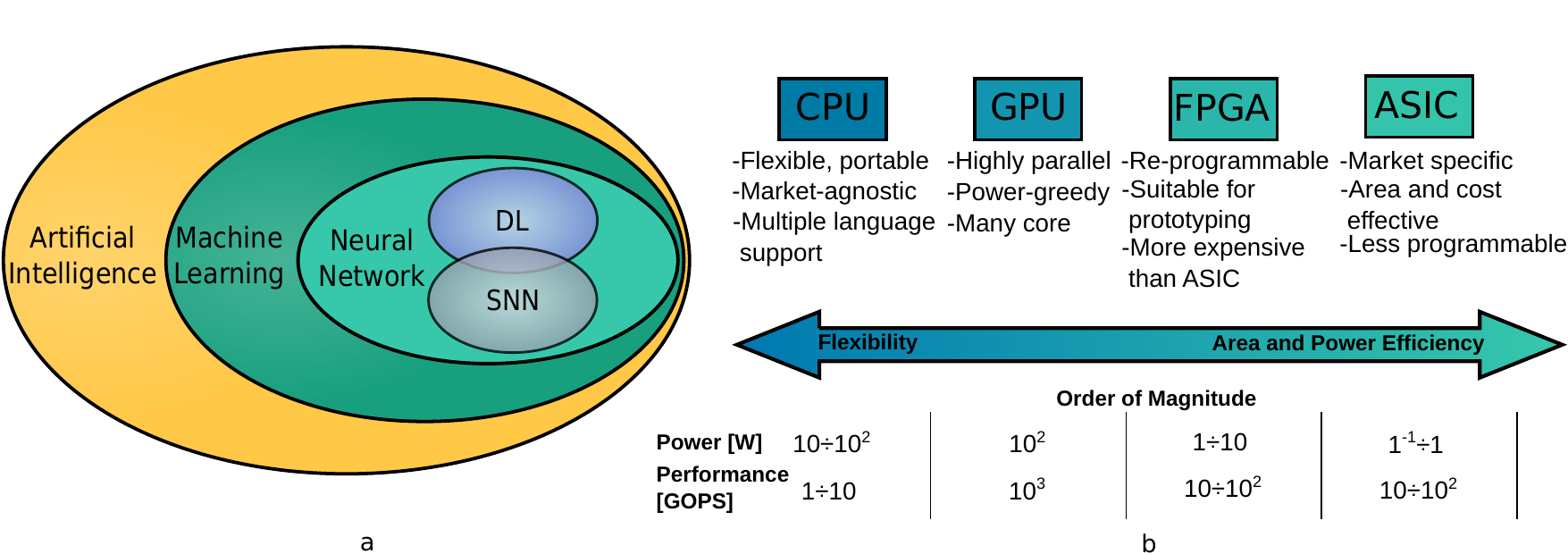}
    \caption{a) Artificial intelligence overview. b) Hardware platforms comparison\cite{energy_eff}.}
    \label{fig:intro}
\end{figure*}
Artificial intelligence (AI) has become a fundamental pillar in many applications and systems in recent years. It is transforming the way we interact with technology, to the point that, very often, we use it without even realizing it. Many techniques fall under the domain of AI, while one in particular raised among all, the Machine Learning (ML). In the last two decades, ML has been extensively employed in various application domains, thanks to the wide range of flexible and easy to learn statistical patterns. ML further consists of several sub-topics, as shown in Figure~\ref{fig:intro}. The most popular ones are the brain-inspired models such as the Neural Networks (NNs), including the Spiking Neural Networks (SNNs) and the Deep Learning (DL) with Deep Neural Networks (DNNs).

DL shows superior accuracy, even claimed to exceed the human one  in certain cases, e.g., in image classification and other problems of computer vision~\cite{SurpassingHuman-Level}. 
This is mainly enabled because of the two factors: (1) the computational power of the latest generation processors, and (2) the enormous amount of available data for training, from which DL can learn different patterns and can effectively derive certain predictions, using deeper and complex models. The larger the training dataset, the more and better the DL algorithms can learn and cover corner cases, achieving the performance never seen before. Since the training is a time-consuming task, effective hardware solutions are required to provide ready-to-use models within a reasonable time. This article mainly focuses on the hardware solutions related to those Deep Neural Networks (DNNs) that have captured much of the attention in the recent years, discussed in Section \ref{sec:background}. This article will also provide a brief overview of work on SNNs, which are becoming increasingly popular due to their similarities to the human brain and their energy-efficient computations.
The applications that are already DL-based are numerous, and cover many key areas:
\begin{itemize}[leftmargin=*]
\item \textbf{Computer Vision}: It is fundamental to extract meaningful features from video and pictures. Such tasks include object localization~\cite{8633218}, image classification~\cite{8793320}, and image segmentation~\cite{minaee2020image}. Their use is valuable for controlling web traffic~\cite{8840035} or for example, video surveillance~\cite{8876949}.
\item \textbf{Business and Finance}: Financial techs deploy such models to forecast market behavior~\cite{8959715}, including insurance~\cite{8258131} and lending~\cite{8701943}.
\item \textbf{Healthcare}: DL is widely used in cancer detection such as lung cancer~\cite{8875896}, brain cancer~\cite{brain}, skin cancer~\cite{8759561}, and many others are continuously rising.  Moreover, there is also a wide applicability of DL techniques in the IoT-Healthcare use cases and Wearables, for instance, to derive short-term and long-term health predictions.
\item \textbf{Robotics}: In robotics, DNNs served in a wide range of use cases like autonomous vehicles~\cite{Grigorescu_2019}, humanoid robots~\cite{humanoid}, assistive robots~\cite{assistive}, swarms~\cite{swarm}, and drone control system~\cite{drone}.
\item \textbf{Smart Energy Management}: DL can also be used to preserve valuable resources such as electricity. Indeed both managing~\cite{8792636} and forecasting~\cite{8403442} the required amount of energy consumption can lead to significant savings. 
\end{itemize}
DNNs learn intelligent activities without the explicit hand-crafted guidelines of experts.
Although DNNs, particularly CNNs and RNNs, represent the state-of-the-art in a wide range of applications, their increasing complexity demands for powerful hardware. Indeed, both inference and training processes require tens of billions of multiply-and-accumulate (MAC) operations that make these models extremely compute-intensive. Moreover, for each MAC, at least two input elements must be fetched from memory. As a result, performing these algorithms with minimal latency entails an additional critical constraint over the memory bandwidth.

For the reasons stated above, in many cases CPUs are not enough, therefore GPUs are one of the most appealing alternative to execute such complex models.
However, today's trend is driven by the Internet-of-Things (IoT)~\cite{IoT_shafique} applications that require more computation capability near the sensors. This process of moving resources towards the IoT nodes is also known as edge computing~\cite{edge_shafique}. This has become possible for two main reasons. Firstly the cost per silicon area has fallen to such an extent that the production of large scale devices to embed in IoT nodes is not an impediment anymore. Secondly, by performing on-site operations, it is no longer necessary to transmit the data to a central server, thus distributing the computing capacity reduces both latency and the large amounts of energy required for transmission, as well as preserving the privacy of data of edge nodes. 
The mesh of these nodes is subjected to strict power constraints, indeed, many of them are battery-powered or rely on energy harvesting systems~\cite{edgecomp}. Therefore, the integration of a high-end GPU into such a system is unfeasible since the required power would go far beyond the power envelope of the IoT-edge platforms.

In this scenario, DL algorithms need to be accelerated with alternative technologies such as low-power FPGAs, that are flexible and can be reprogrammed, or specialized accelerators in form of ASIC-IPs that are highly optimized and tailored for the application use case. This is also justified by the recent trend of integrated systems to move towards heterogeneous multicore systems (or heterogeneous multi-processor system on chip, MPSoCs)~\cite{diannao}, which embed a mix of low-power general-purpose cores and specialized hardware accelerators. The flexibility of FPGA and ASIC designs (Figure \ref{fig:intro}b) opens up a whole series of possible hardware optimizations, analyzed in the following, that are required for energy-efficient acceleration of DL models.
This work analyzes several hardware aspects that different platforms (CPU, GPU, FPGA, and ASIC) provide for the acceleration of DNN models with a comprehensive focus on dedicated accelerators. The latter, as explained before, gained much attention in recent years, thanks to their low-power and cost-effectiveness processing profile. Having a broad overview of the latest state-of-the-art concepts and methodologies can be very valuable for designers.

Table~\ref{tab:acronym} lists the acronyms used in this paper for a better understanding.

\begin{figure*}[ht]
    \centering
    \includegraphics [width=1\textwidth] {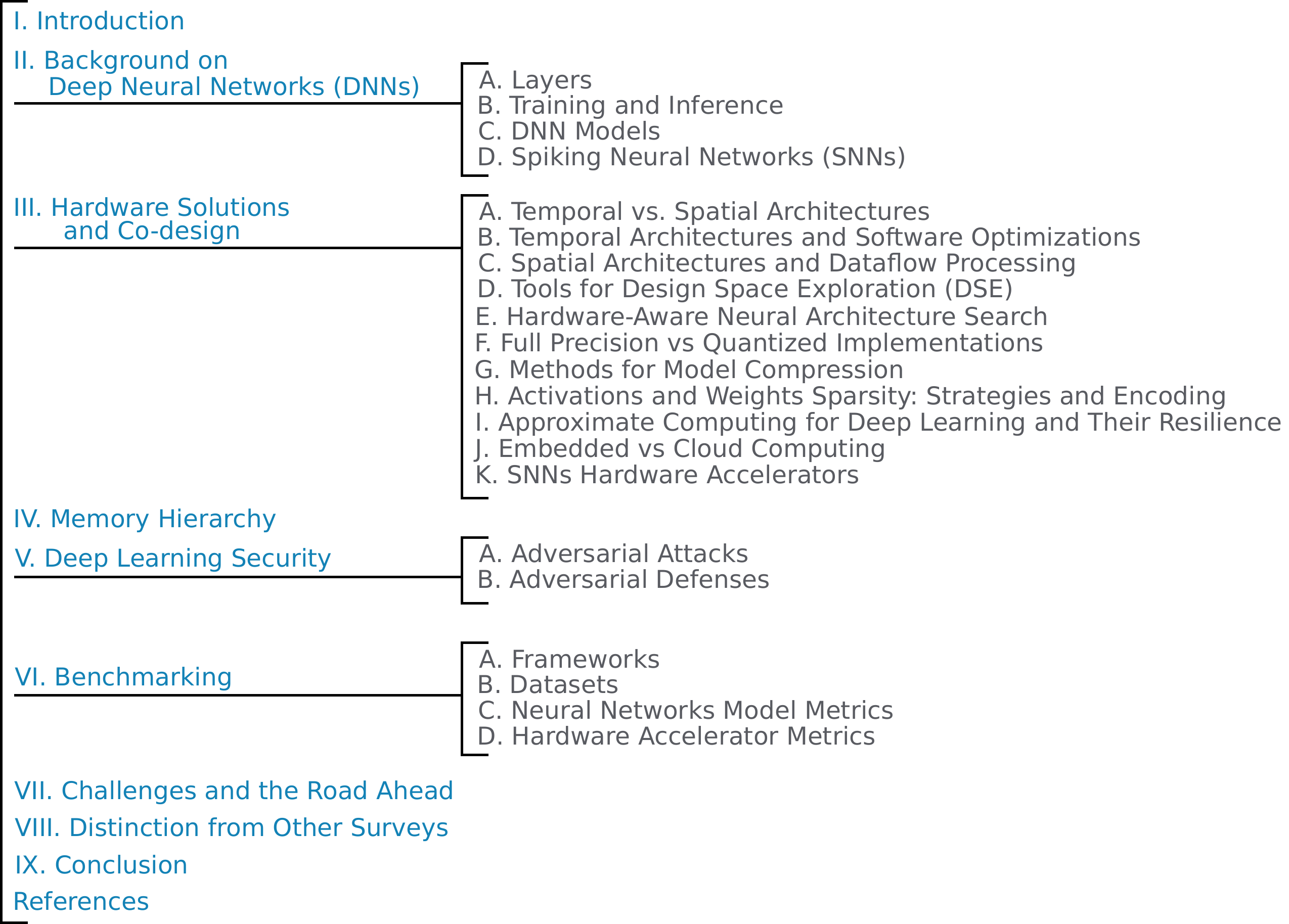}
    \caption{Paper outline.}
    \label{fig:outline}
\end{figure*}

Paper organization: this survey paper is organized systematically in different sections and sub-sections, as depicted in Figure~\ref{fig:outline}. Section~\ref{sec:background} describes the background of DNNs and SNNs, describing the evolution of networks over the years and providing examples of DNN architectures considered the milestones of the DL. Section~\ref{sec:hwsolutions} analyses different co-design techniques to translate and map an efficient dataflow onto the hardware. Section~\ref{memory_hier} outlines the characteristics of the memory hierarchy, being this an extremely power-greedy element. Section~\ref{security} presents the security issues related to ML models, providing examples on how to handle them. Section~\ref{sec:benchmarking} identifies the most important DL frameworks besides the datasets and the essential metrics to characterize both models and hardware devices. Section~\ref{sec:challenges} provides some hints about the research trends and future directions of ML and DL. Section~\ref{sec:relatedworks} provides a description of related survey works, and our distinction. Finally, Section~\ref{sec:conclusion} is reserved for the conclusion and summary.


\begin{table}[h]
\caption{List of Acronyms.}
\label{tab:acronym}
\begin{tabular}{ll}
\hline
A    & Activation                               \\
AI   & Artificial Intelligence                 \\
ASIC & Application Specific Integrated Circuit \\
BC   & Binary Connect                         \\
BLAS & Basic Linear Algebra Subroutines        \\
BN   & Batch Normalization                     \\
BNN  & Binarized Neural Network                  \\
BWN  & Binary Weight Net                     \\
CIS  & Compressed Image Size                   \\
CNN  & Convolutional Neural Network            \\
Conv & Convolutional                           \\
CP   & Canonical Polyadic                       \\
CPU  & Central Processing Unit                 \\
CSC  & Compressed Sparse Column                \\
CSR  & Compressed Sparse Row                   \\
DL   & Deep Learning                           \\
DNN  & Deep Neural Network                     \\
DRAM & Dynamic Random Access Memory             \\
FC   & Fully Connected                         \\
FFT  & Fast Fourier Transform                  \\
FM   & Feature Map                             \\
FPGA & Field Programmable Gate Array           \\
GAN  & Generative Adversarial Network           \\
GD   & Gradient Descent                         \\
GLB  & Global Buffer                           \\
GeMM & General Matrix Multiplication           \\
GPU  & Graphic Processing Unit                 \\
IFM  & Input Feature Map                       \\
IoT  & Internet-of-Things                      \\
L    & Loss                                    \\
LIF  & Leaky-Integrate-and-Fire                \\
LIM  & Logic-in-Memory                         \\
MAC  & Multiply-and-Accumulate                 \\
ML   & Machine Learning                        \\
MSE  & Mean Squared Error                       \\
NAS  & Neural Architecture Search              \\
NN   & Neural Network                          \\
NoC  & Network-on-Chip                          \\
NPU  & Neural Processing Unit                   \\
OFM  & Output Feature Map                      \\
OS   & Output Stationary                       \\
PE   & Processing Element                      \\
QWN  & Quatized Neural Network                  \\
ReLU & Rectified Linear Unit                   \\
RF   & Register File                           \\
RLC  & Run Length Coding                       \\
RS   & Row Stationary                          \\
SIMD & Single-Instruction Multiple-Data        \\
SIMT & Single-Instruction Multiple-Threads     \\
SNN  & Spiking Neural Network                  \\
SRAM & Static Random Access Memory              \\
STDP & Spike Time Dependent Plasticity          \\
TPU  & Tensor Processing Unit                   \\
TTFS & Time To First Spike                      \\
TWN  & Ternary Weight Net                       \\
VLSI & Very Large Scale Integration            \\
W    & Weight                                   \\
WS   & Weight Stationary                       \\ \hline
\end{tabular}
\end{table}

\section{Background on Deep Neural Networks (DNNs)}\label{sec:background}
\begin{figure}[h]
    \centering
    \includegraphics [width=\linewidth] {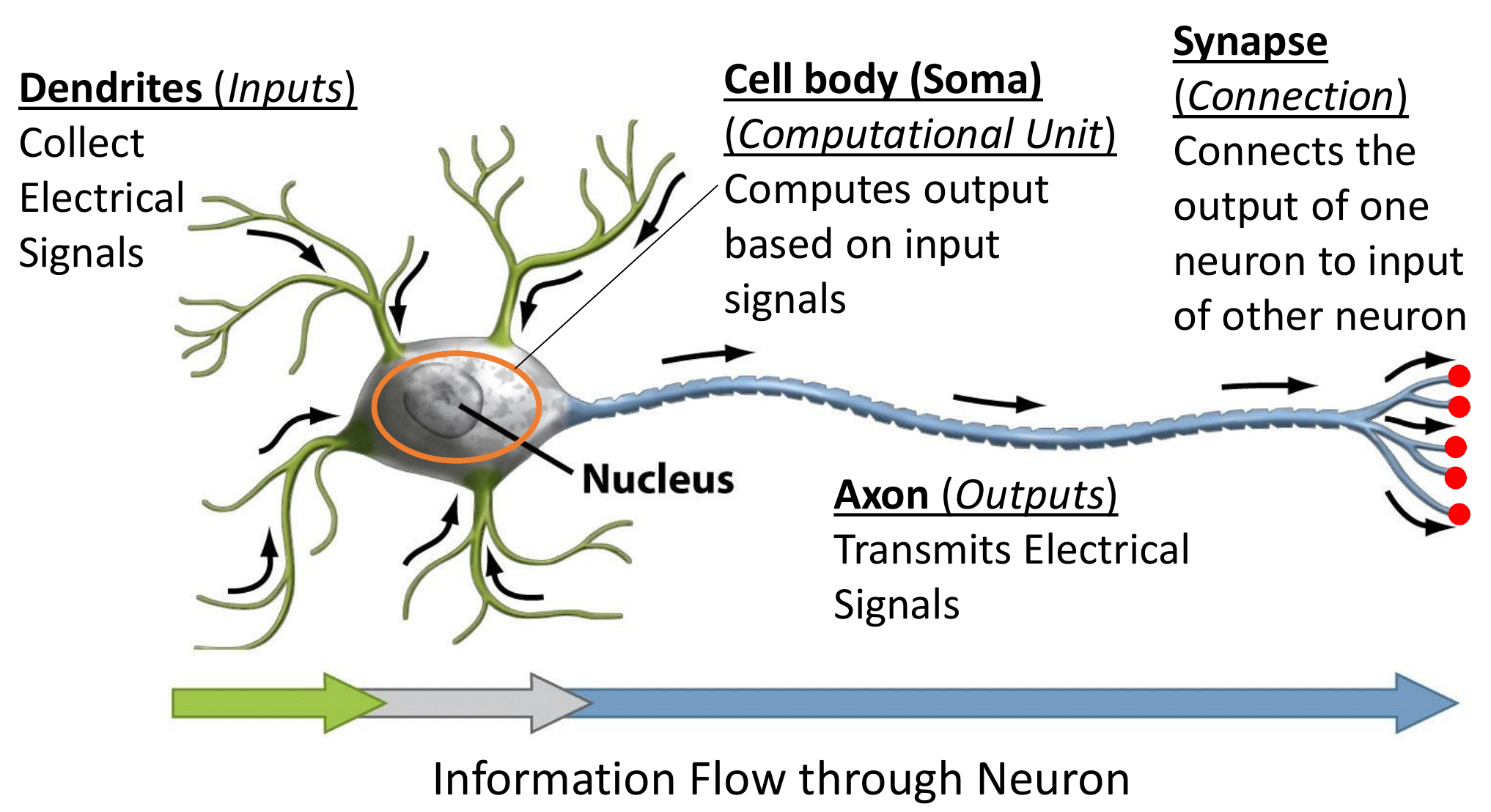}
    \caption{Model of a biological neuron, adapted from~\cite{Freeman2005biological}.}
    \label{fig:biological_neuron}
\end{figure}

The constituent element of a neural network is the \textit{neuron}, also called \textit{perceptron}, a computational block that attempts to model the behavior of a biological neuron, which is shown in Figure~\ref{fig:biological_neuron}. 

\begin{figure}[!h]
    \centering
    \begin{minipage}[c]{.49\linewidth}
    \centering
        \includegraphics[width=\linewidth]{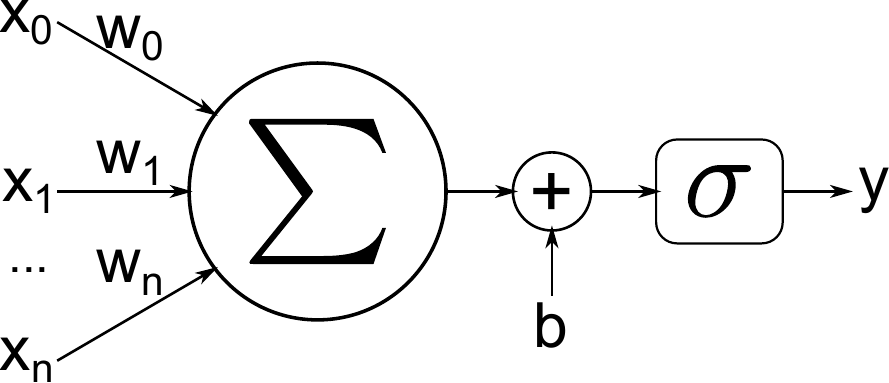}
        \caption{Model of an artificial neuron.}
        \label{fig:neuron}
    \end{minipage}
\hfill
    \begin{minipage}[c]{.49\linewidth}
    \begin{center}
        \begin{equation}
    \label{eq:sum_neuron}
    g(\mathbf{x}) = \sum_{n=0}^{N-1} \mathbf{x}[n]\mathbf{w}[n]
\end{equation}

\begin{equation}
    \label{eq:nonlin_neuron}
    y = \sigma \left( g(\mathbf{x}) + b \right)
\end{equation} 
\end{center}
    \end{minipage}
\end{figure}

A biological neuron consists of the cell body (soma), the dendrites and an axon~\cite{Freeman2005biological}. The dendrites and the axon are filaments; the former receive stimuli, that are then processed by the soma, while the latter takes the neuron output signal to other neurons. Neurons are electrically excitable; when the input voltage exceeds a certain threshold, a pulse, called \textit{action potential}, is generated on the axon. The neuron's response is \textit{all-or-none}, i.e., the neuron can only have no response or full response depending on the input voltage value. The computational model adopted in artificial neural networks has been modified in time~\cite{mcculloch1943}\cite{rosenblatt} until reaching the configuration now adopted (Figure~\ref{fig:neuron}). In essence, it performs a weighted sum of all its inputs (Eq. \ref{eq:sum_neuron}), to which a bias term $b$ is added to include a possible offset. The output of the neuron is then obtained applying a non-linear function $\sigma(\cdot)$ (Eq. \ref{eq:nonlin_neuron}). 

Artificial neural networks are constructed as directed graphs whose nodes represent the neurons. If the graph is acyclic, the network is a \textit{feedforward} NN. If the graph is cyclic, the network is \textit{recurrent} and has a temporal dynamic behavior.

\begin{figure}[t]
    \centering
    \includegraphics[width=0.9\linewidth]{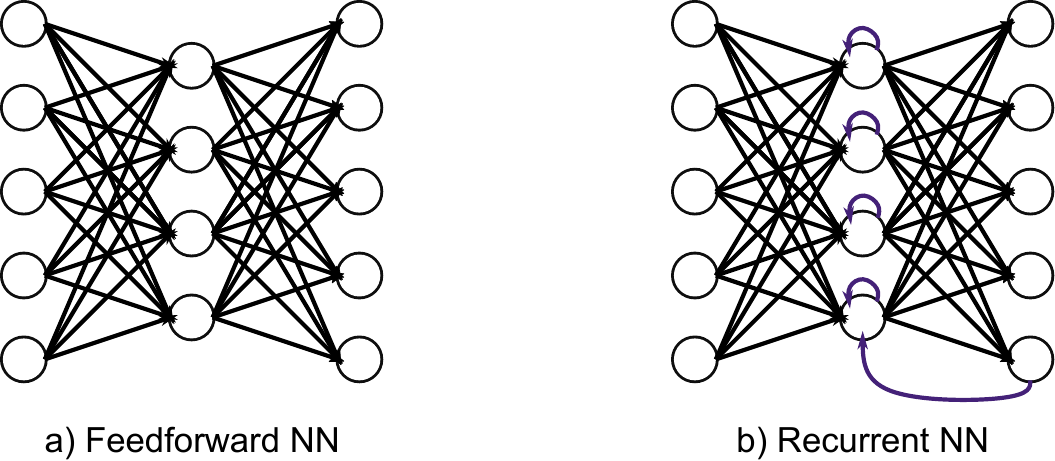}
    \caption{Examples of \textbf{(a)} a feedforward NN and \textbf{(b)} a recurrent NN.}
    \label{fig:fwd_rec}
\end{figure}

As shown in Figure~\ref{fig:fwd_rec}, the nodes are organized in layers: in a feedforward NN, each neuron of layer $l$ receives its inputs from layer $l-1$ and sends its activation to the neurons of layer $l+1$. The inputs to the network form the \textit{input layer}, and there is at least one layer that processes the input, which is called \textit{output layer}. All the layers inserted between the input and output layers are defined as \textit{hidden layers}. The number of hidden layers determines the \textit{depth} of a neural network. If there are more than three hidden layers, the neural network is typically called a \textit{Deep Neural Network} (DNN)~\cite{Bengio2009}. An NN learns how to solve different problems by finding the optimal values for the weights and the biases of its neurons, that can be organized and connected in different ways, as discussed in the following section. 

\subsection{Layers}
\label{sec:dnnlayers}
\noindent \textbf{Fully Connected (FC) layers.} In a Fully Connected layer, each neuron of layer $l$ receives as inputs all the activations of layer $l-1$, therefore, each output neuron performs a weighted sum of all the input neurons: 

\begin{equation*}
    \begin{gathered}
    \mathbf{O}[c_o] = \sum_{c_i=0}^{C_i-1} \mathbf{W}[c_o,c_i] \mathbf{I}[c_i] + \mathbf{b}[c_o] \\
    0 \leq c_o < C_o, \hspace{0.2cm} 0 \leq c_i < C_i
    \end{gathered}
\end{equation*}

\begin{figure}[h]
    \centering
    \includegraphics[width=0.45\linewidth]{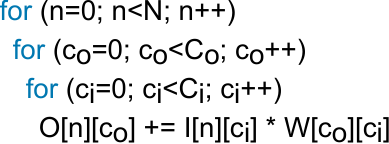}
    \caption{Pseudocode of an FC layer.}
    \label{fig:fc_pseudocode}
\end{figure}

Where $C_i$ and $C_o$ are the number of neurons of layers $l-1$ and $l$ respectively. Figure~\ref{fig:fc_pseudocode} shows the pseudocode that implements an FC layer. In Figure~\ref{fig:fc_pseudocode} N is the \textit{batch size}, where a \textit{batch} is a collection of inputs that can be processed in parallel.

\begin{figure}[h]
    \centering
    \includegraphics[width=0.8\linewidth]{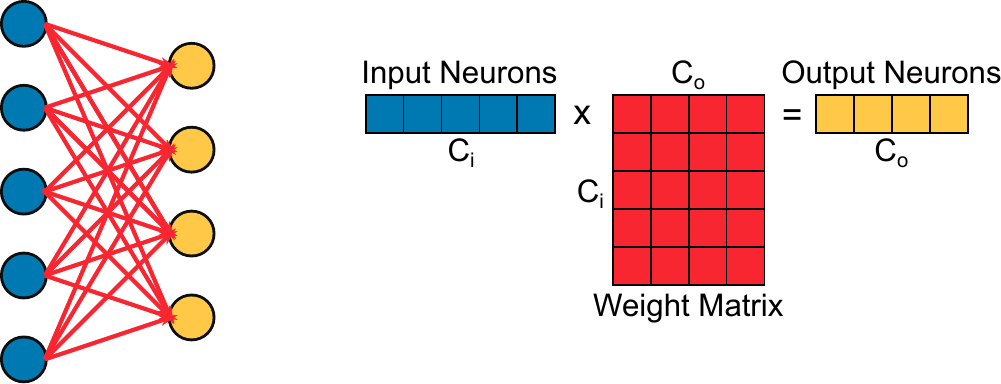}
    \caption{Example of an FC layer (left) and of how it can be modeled by a vector-matrix multiplication (right).}
    \label{fig:fc_layer}
\end{figure}

From the equation and the pseudocode that describes it, it is possible to see that an FC layer is a vector-matrix multiplication with the weights arranged in a $C_i\times C_o$ matrix (see Figure~\ref{fig:fc_layer}). 

Since $C_i$ and $C_o$ can assume high values, the number of parameters of an FC layer is potentially huge. However, it is not always necessary for an output neuron to receive information from all the input neurons. For this reason, Convolutional layers have been introduced. 

\medskip 
\noindent \textbf{Convolutional (Conv) layers.} 
FC layers are not well suited for tasks like object detection and recognition since their high degree of connectivity leads to an explosion of the number of parameters required to deal with high-resolution images. Moreover, FC layers treat inputs that are close together or far apart equivalently, ignoring the spatial structure present in images. To overcome these two problems, in 1998 a new architecture was proposed~\cite{MNIST}, known as Convolutional Neural Network (CNN), that includes Conv layers and exploits the ideas of \textit{local receptive fields} and \textit{shared weights}. The idea of local receptive fields has its biological counterpart in the study of David H. Hubel and Torsten Wiesel~\cite{hubel1959} on the visual cortex of a cat. They demonstrate that some neurons are activated when the cat is visually exposed to vertical lines, while different neurons respond to lines oriented along different angles. There are thus \textit{locally sensitive} neurons that are sensitive to a small portion of the visual field and higher-level neurons that are sensitive to larger portions and therefore analyze more complex patterns. Adapting the same idea to a neural network, the neurons are organized in a 2D grid, i.e., a \textit{feature map}, and a neuron of layer $l$ does not receive all the activations of the layer $l-1$, but it is instead connected to a small receptive field of dimension $[H_k \times W_k]$. The size of the receptive field and consequently of the weight matrix is commonly referred to as \textit{kernel size} and the distance between adjacent receptive fields is defined by a stride parameter $S$. Applying the idea of shared weights, all the neurons of layer $l$ have the same matrix of weights, detecting the same feature in different locations of layer $l-1$. To detect multiple features, a Conv layer has multiple channels, i.e., there are multiple feature maps. 

The computations performed in a Conv layer involve an \textit{input feature map} \textbf{Ifm} of size $[C_i \times H_i \times W_i]$, the \textit{weights} \textbf{W} of size $[C_i \times C_o \times H_k \times W_k]$, and a \textit{bias term} \textbf{b} of size $[C_o]$. The result of the computation is an \textit{output feature map} \textbf{Ofm} of size $[C_o \times H_o \times W_o]$,  computed as follows:
\begin{gather*}
    \begin{align*}
     & \mathbf{Ofm}[c_o,h_o,w_o] = \sum_{c_i=0}^{C_i-1} \sum_{h_k=0}^{H_k-1} \sum_{w_k=0}^{W_k-1} \bigg( \\ 
    & \mathbf{W}[c_i,c_o,h_k,w_k] \mathbf{Ifm}[c_i,Sh_o+h_k,Sw_o+w_k] + \mathbf{b}[c_o] \bigg)
    \\
    \end{align*}
    \\ \vspace{1cm} 0 \leq c_o < C_o, \hspace{0.2cm} 0 \leq h_o < H_o, \hspace{0.2cm} 0 \leq w_o < W_o \\
    0 \leq h_k < H_k, \hspace{0.2cm} 0 \leq w_k < W_k 
\end{gather*}

\begin{figure}[h]
    \centering
    \includegraphics{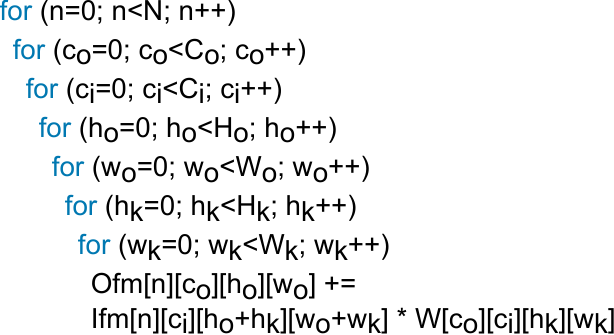}
    \caption{Pseudocode of a Conv layer.}
    \label{fig:conv_pseudocode}
\end{figure}

\begin{figure}[h]
    \centering
    \includegraphics[width=0.8\linewidth]{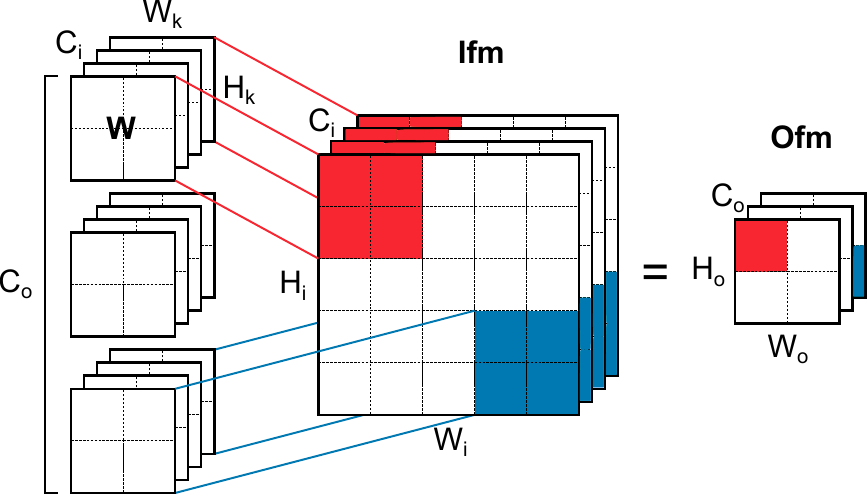}
    \caption{Graphical representation of the convolution operation in a Conv layer. A sub-tensor of \textbf{Ifm} (red) is multiplied by a sub-tensor of \textbf{W} and the results are accumulated to produce a single  value (red) of the \textbf{Ofm}.}
    \label{fig:conv_layer}
\end{figure}

Figure~\ref{fig:conv_pseudocode} shows the pseudocode of a Conv layer, and Figure~\ref{fig:conv_layer} gives a graphical representation.


\begin{figure}[h]
    \centering
    \includegraphics[width=\linewidth]{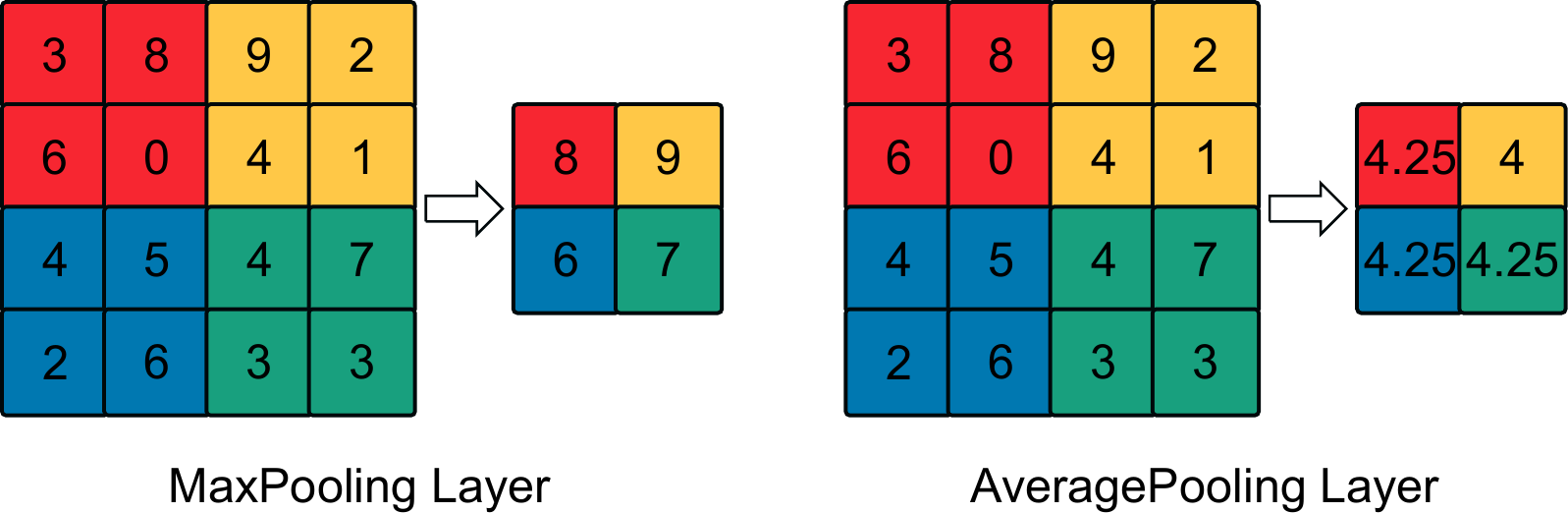}
    \caption{Examples of MaxPooling (left) and AveragePooling (right) layers.}
    \label{fig:pooling}
\end{figure}

\medskip
\noindent \textbf{Pooling layers.} Pooling layers are commonly placed after a Conv layer. Pooling layers have receptive fields, similarly to Conv layers. For the group of neurons in each receptive field, they return a single value that contains a statistic of the group, e.g., the maximum or the average value, as shown in Figure~\ref{fig:pooling}. The stride parameter is usually set equal to the dimension of the receptive field to have non-overlapping windows. 

Pooling layers reduce the number of activations of a layer, and consequently decrease the memory requirements and the number of computations to be performed after. Moreover, pooling layers achieve invariance to small local translations. The outputs of Conv layers depend heavily on the position of the input, so even for minor variations of the inputs, there are significant variations of the outputs. Pooling layers down-sample the outputs, making them more robust to small input variations. 

\medskip
\noindent \textbf{Normalization layers.} The inputs to neural networks are usually preprocessed to have a normal distribution, i.e., zero mean and unit variance. Normalization is beneficial because it keeps different inputs in the same range of values, making them easier to analyze by the same model. Also, as will be seen in the following paragraph, layers sometimes use saturating non-linear functions, such as Sigmoid or Softmax. So having values centred on zero avoids early-saturation of activations. To apply the same normalization constraint that applies to the inputs to internal activations, Normalization layers are inserted between Conv and FC layers. It must also be noted that activations normalization speeds up the training, as the layers do not need to adapt to different distributions at each training step. 

The commonly adopted normalization method is Batch Normalization (BatchNorm)~\cite{batchnorm} (Eq. \ref{eq:batchnorm}). The operation performed by the BatchNorm layer is standardization:

\begin{equation}
\label{eq:batchnorm}
    y = \frac{x - E[x]}{Var[x]+\epsilon} \cdot \gamma + \beta
\end{equation}

Where $E[x]$ and $Var[x]$ are the mean and standard deviation of the input tensor $x$, respectively. $\epsilon$ is a value necessary for numerical stability, and $\gamma$ and $\beta$ are two trainable parameters for the integration of the BatchNorm layer in the training process.

\medskip
\noindent \textbf{Non-linear activation functions.} Without a non-linear activation function, the NN would be a simple cascade of linear algebra operations, unable to solve complex non-linear problems. For this reason, different non-linear functions are applied to the weighted sum of the inputs of a neuron. Some of the most popular functions are: 
\begin{itemize}[leftmargin=*]
    \item \textit{Rectified Linear Unit (ReLU)} function forces the activations to be greater than or equal to zero. It is prevalent as it is computationally efficient since it requires a simple comparison between $x$ and 0.
    \vspace*{-0.4cm}
    \begin{figure}[h]
    \centering
        \begin{minipage}[c]{0.49\linewidth}
        \begin{equation*}
        y = \begin{cases}
             0 & \mbox{if } x<0 \\
            x & \mbox{otherwise}
        \end{cases}
    \end{equation*}
        \end{minipage}
        \begin{minipage}[c]{0.49\linewidth}
        \centering
        \includegraphics[width=0.8\linewidth]{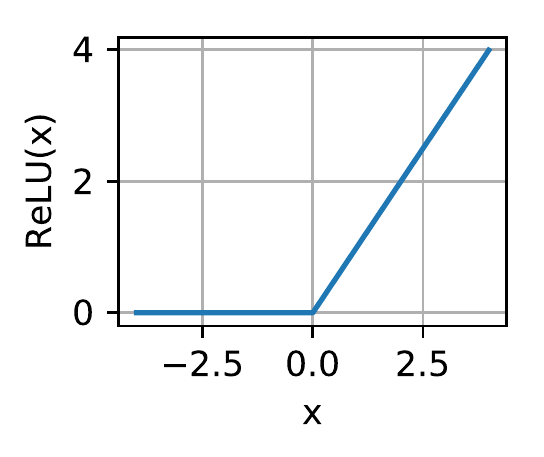}
        \end{minipage}
    \end{figure}
    \vspace*{-0.4cm}
    
    There are some variants of the ReLU function, such as \textit{Leaky-ReLU} or \textit{Exponential Linear Unit (ELU)}. The former has a negative slope for values $x<0$; the latter uses a log curve when $x<0$. These variants have been introduced to solve the \textit{dying ReLU} problem, i.e., since the slope of the ReLU for $x<0$ is zero, the neurons in this region are not trained. Moreover, Leaky-ReLU and ELU are more balanced towards zero if compared to ReLU, and this helps to speed up the training.
    
    \item \textit{Sigmoid} function  normalizes the output in the range $(0,1)$. Contrarily to the ReLU function, it is computationally expensive, as it can be seen from its equation:
    \begin{figure}[H]
        \begin{minipage}[c]{0.49\linewidth}
         \begin{equation*}
        y = \frac{1}{1+e^{-x}}
        \end{equation*}
        \end{minipage}
        \hfill
        \begin{minipage}[c]{0.49\linewidth}
        \centering
        \includegraphics[width=0.8\linewidth]{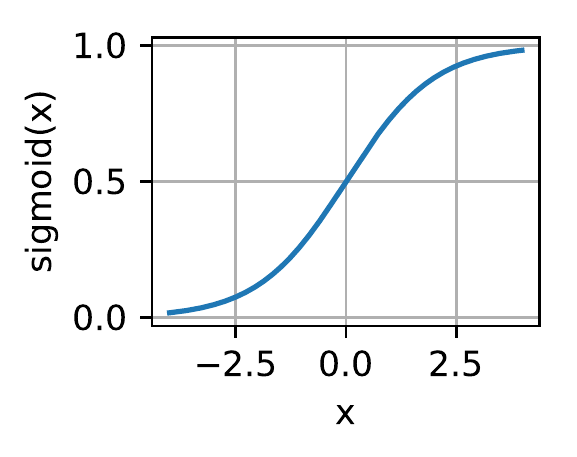}
        \end{minipage}
    \end{figure}
    
    \item \textit{Hyperbolic Tangent} function (TanH) is the equivalent of Sigmoid function to bound activations in the range $(-1,1)$, to model outputs that can assume negative values too. 
    \begin{figure}[h!]
        \begin{minipage}[c]{0.49\linewidth}
        \begin{equation*}
        y = \frac{e^x - e^{-x}}{e^x + e^{-x}}
    \end{equation*}
        \end{minipage}
        \hfill
        \begin{minipage}[c]{0.49\linewidth}
        \centering
        \includegraphics[width=0.8\linewidth]{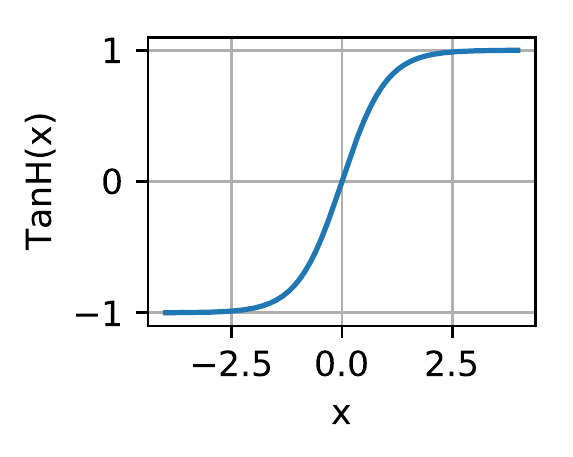}
        \end{minipage}
    \end{figure}
    
    \item \textit{Softmax} function is also know as \textit{normalized exponential function}. It receives a vector of N numbers as an input: each number is normalized in the range $(0,1)$ and the sum of all N numbers is equal to 1. This function is used mainly in output layers if the outputs represent the classification probabilities. 
    \begin{equation*}
        y_i = \frac{e^{x_i}}{\sum_{j=0}^{N-1}e^{x_j}} \hspace{0.5cm} \mbox{for } i=0,1,...,N-1
    \end{equation*}
\end{itemize}

\subsection{Training and Inference} 
\begin{figure}[h]
    \centering
    \includegraphics[width=\linewidth]{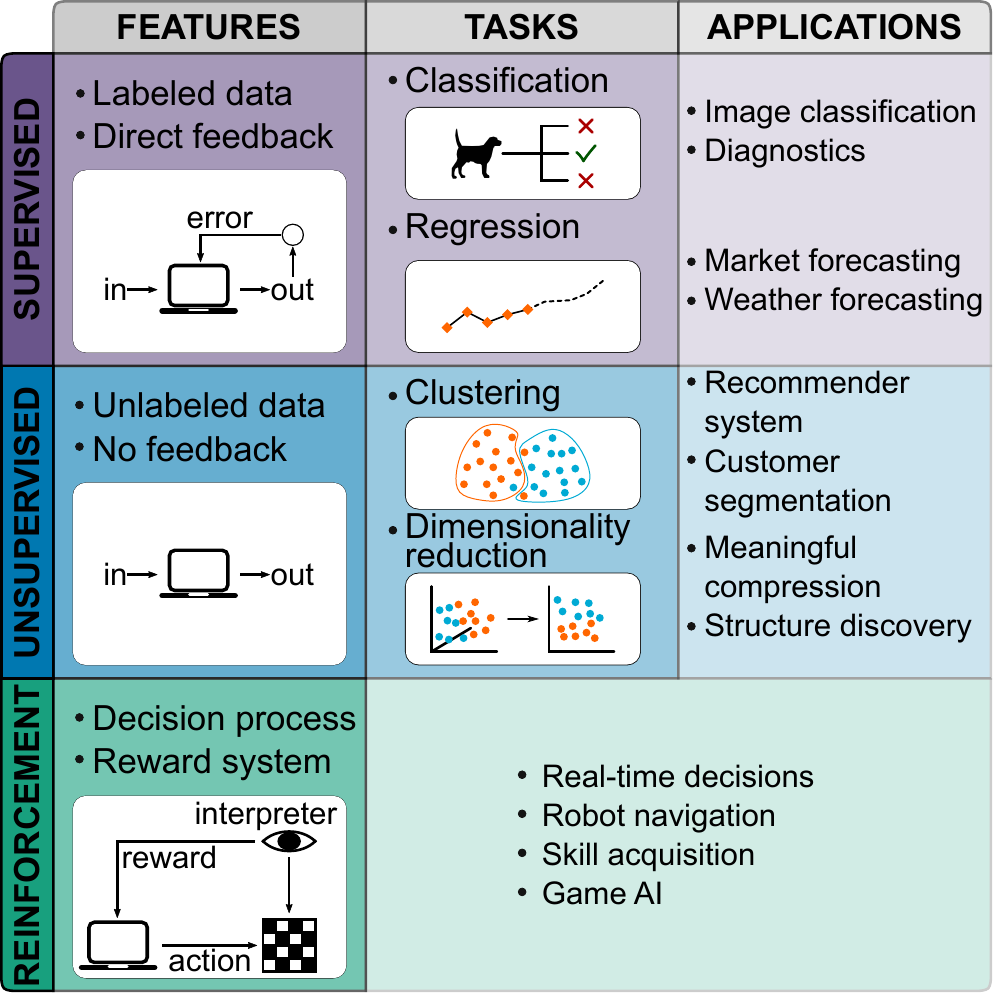}
    \caption{Features, achievable tasks and applications of the three existing ways of training (supervised, unsupervised, and reinforcement)}
    \label{fig:learning}
\end{figure}

A neural network can learn to solve a problem by determining the correct values of the weights and biases of its layers: this process is referred to as \textit{training}. Using a trained NN, with pre-learned weights and biases, is referred to as \textit{inference}. There are different ways of training a NN (see Figure~\ref{fig:learning}): 
\begin{enumerate}[leftmargin=0cm]
    \item[] \textbf{Supervised learning:} It requires a set of labeled input-output pairs, i.e., a set of inputs (\textit{data}) with the corresponding expected output (\textit{labels}). This set of pairs is called a \textit{training set}. During the supervised learning, the model receives a labeled input and updates its parameters based on the discrepancy between the expected output and the actual output. Supervised learning is predominantly used today in a wide range of applications, in the big-data era, thanks to the immense availability of datasets and its good performances. 
    \item[] \textbf{Unsupervised learning:} It is performed when only non-labeled data are available. It lies in finding common patterns in the data. An example of unsupervised learning is \textit{clustering}, that clusters data based on their shared attributes. Neural networks that apply unsupervised learning are, for example, \textit{autoencoders} and \textit{Generative Adversarial Networks} (GANs).
    \item[] \textbf{Reinforcement learning:} Reinforcement learning is the third main type of learning and, similar to the unsupervised learning, it does not need labeled data. The aim of reinforcement learning is the creation of autonomous agents able to make decisions in a given environment. The training scenario is composed of the agent who takes actions in an environment. There is then an interpreter who evaluates the agent's actions in terms of a  reward, which is then fed back to the agent. The goal of the agent is to maximize the reward.  
\end{enumerate}

A supervised-learning algorithm commonly used for the training of DNNs is \textit{gradient backpropagation}, where the input samples are fed into the network, and the outputs are computed using weights $\mathbf{W}$. The network's outputs and expected outputs are compared, and a \textit{loss (L)} is calculated with a \textit{loss function}, such as Euclidean distance or Mean Squared Error (MSE). To perform the learning process, the weights are updated by a quantity proportional to the partial derivative of the loss with respect to the weights themselves, i.e., the gradient. The gradients are efficiently computed with the backpropagation algorithm, which is the \textit{chain rule} of calculus applied to calculate the derivative of the loss starting from the output of the network and going up to the input layer. 


The learning actually takes place by updating the weights and biases of the network, which can be done with different \textit{optimization algorithms}. The simplest optimization algorithm is gradient descent (GD), shown in Eq. \ref{eq:gradient_descent}, where $\theta$ is a parameter of the network and $\eta$ is a scaling factor referred to as \textit{learning rate}. Other algorithms are, e.g., GD with momentum~\cite{momentum}, Nesterov accelerated gradient~\cite{nesterov}, Adagrad~\cite{adagrad}, Adadelta~\cite{adadelta} and Adam~\cite{adam}. 

\begin{equation}
    \label{eq:gradient_descent}
    \theta ^ {t+1} = \theta ^ t - \eta \frac{\partial L}{\partial \theta ^ t}
\end{equation}

During the training of neural networks it is common to encounter the problem of \textit{overfitting}, i.e., if a model is complex and has many parameters, it is possible that it fits the data of the training set too accurately.  The model therefore "memorizes" the correct result for each input rather than learning to generalize, and has a poor performance on the inputs that are never seen before. The solutions to the overfitting problem are either the transition to a simpler model or employing different \textit{regularization techniques}. L1 and L2~\cite{l1l2} are common regularization techniques, both require adding a regularization term to the loss function, which has the effect of reducing the value of the weights. This results in a compressed and simpler model. Another technique that gives good results is \textit{dropout}~\cite{dropout}, i.e., at each iteration some neurons are randomly selected and removed from the model. 

\subsection{DNN Models}
\label{sec:dnnmodels}
\begin{figure}[h]
    \centering
    \includegraphics[width=\linewidth]{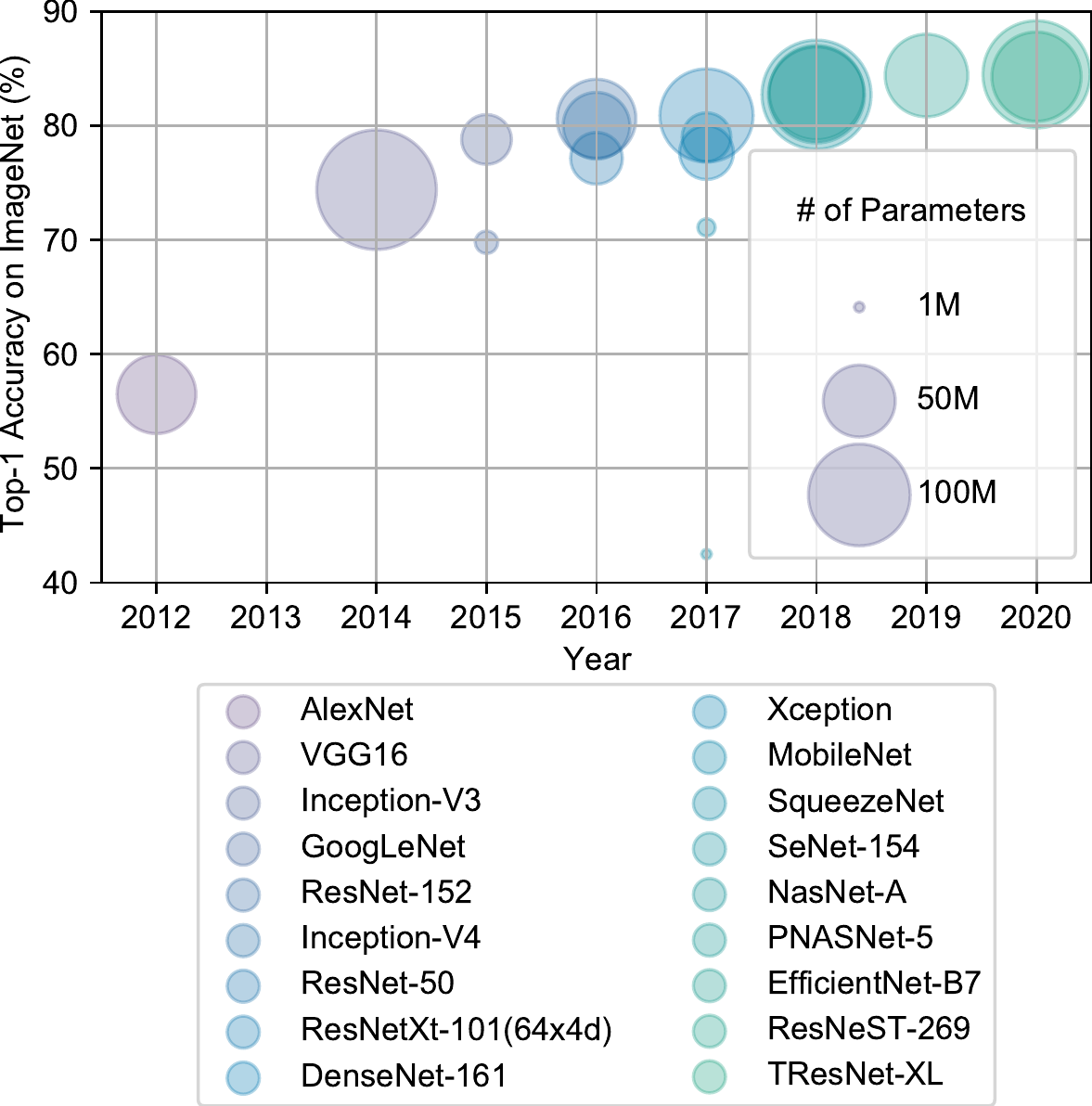}
    \caption{Timeline of significant neural networks models with the accuracies achieved on the ImageNet dataset~\cite{ImageNet} and the number of parameters.}
    \label{fig:modesl}
\end{figure}

Over the years, many CNNs models have been proposed to achieve better to the best-possible performance for a given task. Figure~\ref{fig:modesl} shows a timeline of significant neural network models with their classification accuracy in the image classification task on the ImageNet dataset~\cite{ImageNet} and number of parameters. These models will be discussed in the following paragraphs and compared in Table~\ref{tab:models}. 

\medskip 
\noindent \textbf{LeNet~\cite{MNIST} (1998)}: It has been one of the first neural network trained by backpropagation with a convolutional structure and has been the inspiration for the following research on CNNs. It was designed for the recognition of handwritten digits represented on 32$\times$32 pixels images. LeNet-5 is a version consisting of five layers, of which the first two are convolutional layers, and the last three are FC layers. The Conv layers have 5$\times$5 kernels and are both followed by 2$\times$2 average pooling layers. All the layers use hyperbolic tangent (tanh) as the activation function, except the output layer that applies the softmax function. 

\medskip 
\noindent \textbf{AlexNet~\cite{Krizhevsky2012} (2012)}: It is a CNN built for the ImageNet dataset~\cite{ImageNet}, a database of more than 1.3M of high-resolution 256$\times$256 pixels images divided into 1000 classes. It is the first (deep) CNN architecture to win the ImageNet Large Scale Visual Recognition Challenge 2012 (ILSVRC-2012)~\cite{ILSVRC15}, achieving consistent accuracy improvements compared to the traditional non-convolutional networks winners of the previous editions. AlexNet follows the architecture of LeNet, stacking more Conv layers; it consists of five Conv layers and three FC layers. AlexNet was the first NN to introduce Rectified Linear Units (ReLU) as the activation functions, reducing the training time significantly. Moreover, to overcome the limitations imposed by the memory size of the GPUs, AlexNet adopts a parallel solution, splitting the architecture on two GPUs. To reduce the communication bottleneck,  the GPUs exchange data in two points of the network only. 
\begin{table}[t]
\caption{Comparison of the models presented in Section \ref{sec:dnnmodels}}
\label{tab:models}
\centering 
\resizebox{\linewidth}{!}{ 
\begin{tabular}{|l|c|l|}
\hline
\textbf{Model}                                                 & \textbf{Year}         & \textbf{Contribution}\\ \hline
LeNet \cite{MNIST}   & 1998  & -First popular CNN \\ \hline
AlexNet \cite{Krizhevsky2012} & 2012  & \begin{tabular}[c]{@{}l@{}}-First CNN winner of ILSVRC\\ -Introduction of ReLU\end{tabular} \\ \hline
VGG16 \cite{Simonyan2014}  & 2014 & -Smaller kernels \\ \hline
GoogLeNet \cite{Szegedy2015}  & 2015                  & -Inception block \\ \hline
ResNet \cite{He2015} & 2016 & \begin{tabular}[c]{@{}l@{}}-Skip connections\\ -Residual learning\end{tabular} \\ \hline 
\begin{tabular}[c]{@{}c@{}}ResNetXt-\\ 101\_64x4d\end{tabular} \cite{resnetxt} & 2017                  & -Grouped convolution  \\ \hline
DenseNet161 \cite{Huang2016}& 2017                  & -Regular structure\\ \hline
CapsNet~\cite{capsnet} & 2017 & \begin{tabular}[c]{@{}l@{}}- Capsules \\ - Dynamic Routing\end{tabular}\\ \hline 
SeNet154  \cite{Hu2018} & 2018                  & \begin{tabular}[c]{@{}l@{}}-Dependencies between feature \\ maps are exploted\end{tabular}  \\ \hline
NasNet-A    \cite{nasnet} & 2018                  & \begin{tabular}[c]{@{}l@{}}- Neural Architecture Search\\ - Transfer learning\end{tabular}  \\ \hline
\end{tabular}}
\end{table}

\medskip 
\noindent \textbf{VGG~\cite{Simonyan2014} (2014)}: Thanks to the availability of hardware resources supporting heavier computations, the initial trend in NN research has been the design of deeper and deeper architectures. VGG is a model that takes the structure of AlexNet and furtherly increases the number of Conv layers. In particular, VGG-16 has 13 Conv layers and three FC layers, while VGG-19, with a total of 19 layers, was the winner of ILSVRC-2014.

\begin{figure}[h]
     \centering
     \includegraphics[scale=0.53]{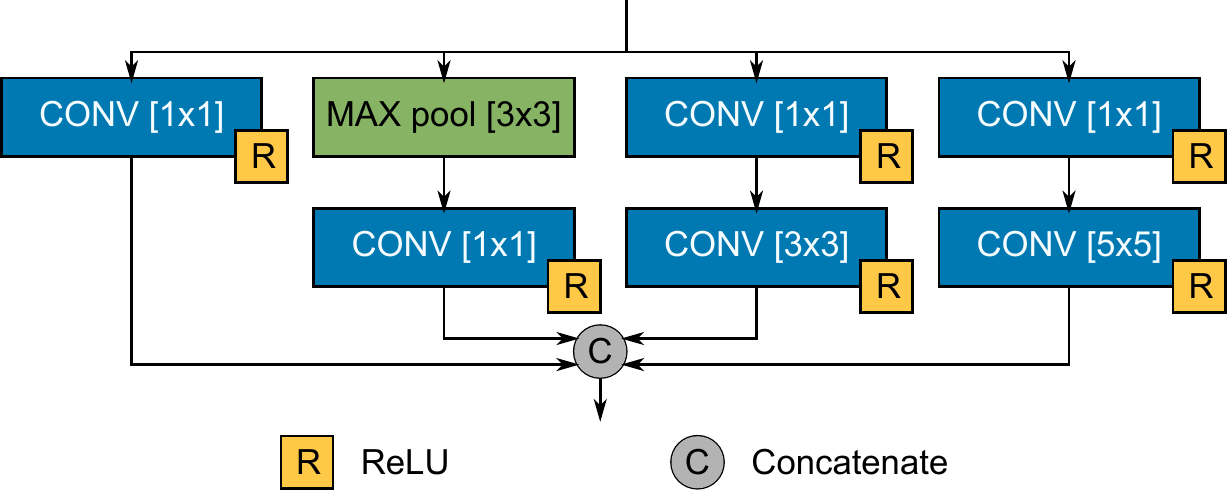}
     \caption{Inception Module used in GoogLeNet~\cite{Szegedy2015}. Convolutions are performed in four parallel branches and the four outputs are concatenated to produce the single output of the Inception module.}
     \label{fig:inceptionmodule}
 \end{figure}

\medskip 
\noindent \textbf{GoogLeNet~\cite{Szegedy2015} (2015)}: It is based on the intuition of finding a dense structure, i.e., an inception module, and then building the network by stacking these modules. An inception module (see Figure~\ref{fig:inceptionmodule}) captures features at various scales and concatenates them at the output, passing to the next layer different levels of information. The increase of the depth of the NNs has allowed to improve their accuracy but has however led to the appearance of the \textit{vanishing gradient problem}. Since during backpropagation the gradients are computed with the chain rule and the values are often in the range $[0,1]$ or $[-1,1]$, the magnitude of the gradients decreases exponentially with the depth of the network. In the earlier layers, the gradients can become so small that they prevent the correct training. To overcome this problem, GoogLeNet has two additional classifiers used for training only that take the activations at earlier stages of the network, and therefore increase the magnitude of their gradients. GoogLeNet successors are Inception-v3~\cite{inceptionv3} and Inception-v4~\cite{inceptionv4}.

\begin{figure}[h]
    \centering
    \includegraphics[scale=0.53]{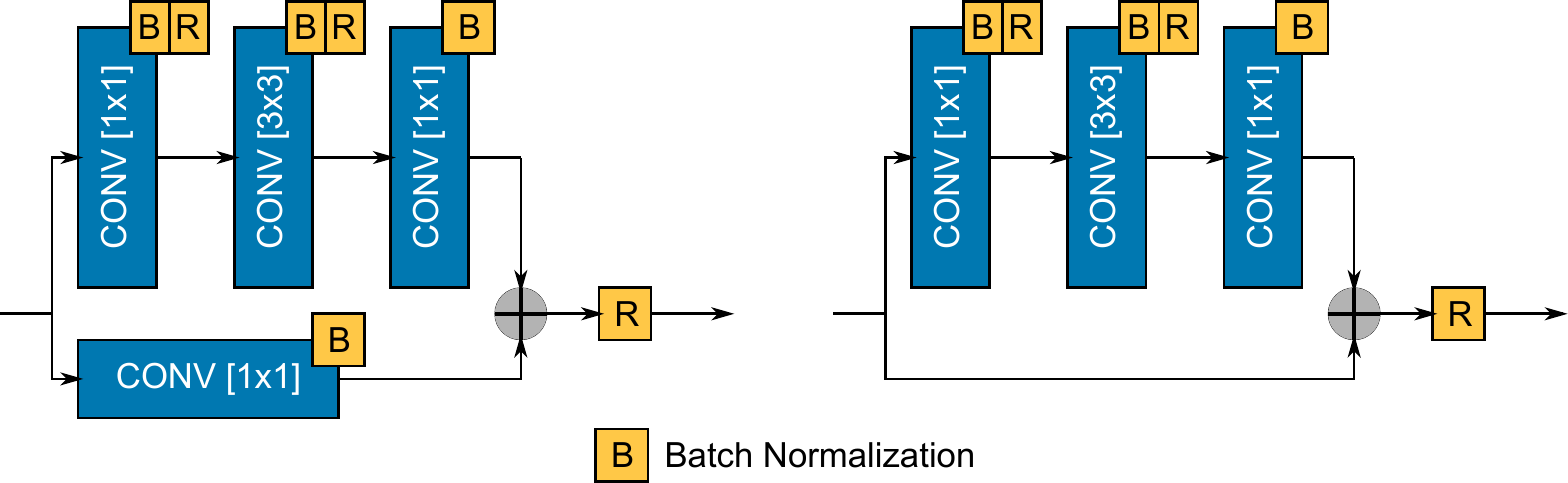}
    \caption{Skip connection modules used in Residual Networks~\cite{He2015}. Three convolution are performed in series and a parallel connection is added. In the parallel connection, it is possible to choose between a 1$\times$1 convolution (\textbf{left}) or the identity function, i.e., no operation (\textbf{right}). The results of the two branches are summed. }
    \label{fig:resnet}
\end{figure}

\medskip 
\noindent \textbf{ResNet~\cite{He2015} (2015).} To work around the vanishing gradients problem, Residual Networks (ResNets) have adopted and made popular \textit{skip connections}, shown in Figure~\ref{fig:resnet}, that run in parallel to a series of Conv layers and avoid excessive degradation of the gradients during backpropagation. Moreover, ResNets are the first architectures to use batch normalization layers. Based on ResNet architecture, different models with higher accuracies have been proposed over the years, such as ResNetXt~\cite{resnetxt}, ResNeST~\cite{resnest}, or TResNet~\cite{tresnet}. 

\begin{figure}[h]
    \centering
    \includegraphics[scale=0.53]{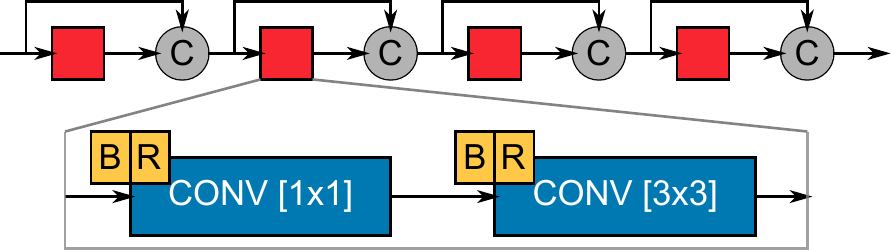}
    \caption{Dense blocks used in DenseNets~\cite{Huang2016}. The output and input of each blue block, that is the series of two convolutions, are concatenated.}
    \label{fig:dense}
\end{figure}

\medskip 
\noindent \textbf{DenseNet~\cite{Huang2016} (2016).} Given the success of skip connections, DenseNets adopt a regular and therefore simpler connection pattern. As shown in Figure~\ref{fig:dense}, in a Dense Block, every layer receives in input a concatenation of the activations of all the preceding layers. A DenseNet is then built by stacking Dense Blocks of different depth, interleaved by Conv and Pooling layers for dimensionality reduction. 

\begin{figure}[ht]
    \centering
    \includegraphics[scale=0.53]{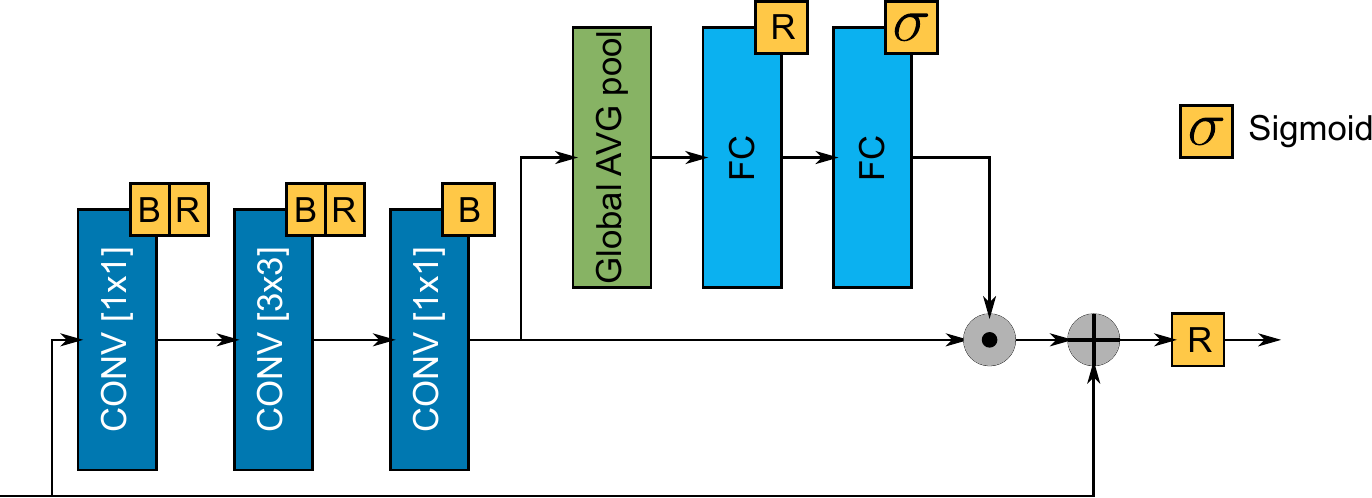}
    \caption{Residual module as modified in Squeeze-and-Excitation Networks~\cite{Hu2018}. A skip connection is inserted in parallel to a pooling and two FC layers, and the output of the two branches are multiplied. As in traditional residual modules, a skip connection runs in parallel to the whole block. }
    \label{fig:seresnet}
\end{figure}

\medskip 
\noindent \textbf{SENet~\cite{Hu2018} (2017).} Squeeze-and-Excitation Networks (SENets) modify traditional layers, e.g., Conv layers, or blocks, e.g., inception or residual modules, to model the relationship between the different channels of the feature maps. Figure~\ref{fig:seresnet} shows how a residual module is modified following the SE approach. SENet-154 is the NN winner of ILSVRC-2017, which is built integrating SE blocks in a version of ResNetXt~\cite{resnetxt}. 

\medskip 
\noindent \textbf{Capsule Network (2017).} The Capsule Networks were created in a try to solve some of the problems of CNNs, such as the loss of data caused by pooling layers or the high sensitivity to input shifts or rotations. The idea of \textit{capsules} was introduced in~\cite{trans_autoenc} and the first network model was proposed in~\cite{capsnet}. In~\cite{capsnet}, the neurons are replaced by capsules, i.e., a vector of neurons. Each element of the vector encodes an instantiation parameter of an entity, e.g., the width or the rotation, and the length of the vector represents the instantiation probability of the entity. Since the length of the vector represents a probability, its value must be in the range $[0,1]$. For this reason, the \textit{squash} function (Eq. \ref{eq:squash}) is used as non-linear activation function in the capsule layers.

\begin{equation}
    \label{eq:squash}
    \vec{y} = \frac{|\vec{x}|^2}{1+|\vec{x}|^2} \frac{\vec{x}}{|\vec{x}|}
\end{equation}

\begin{figure}[h]
    \centering
    \includegraphics[scale=0.6]{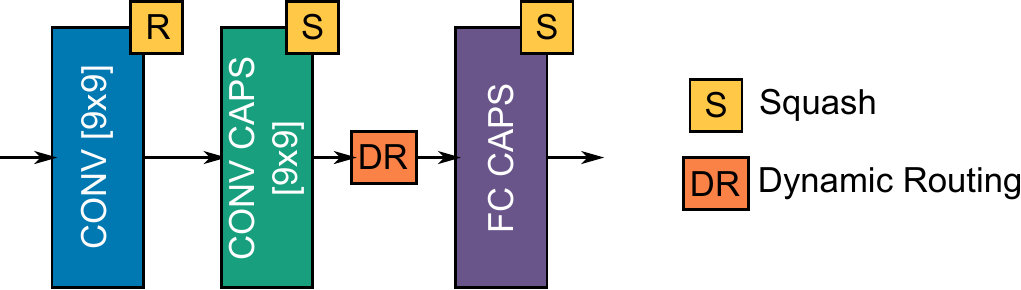}
    \caption{Capsule Network~\cite{capsnet}.}
    \label{fig:capsnet}
\end{figure}

Moreover, in Capsule Networks, the pooling layers are substituted by a \textit{dynamic routing algorithm} that strengthens the connections between capsules of adjacent layers if relevant entities are detected. Figure~\ref{fig:capsnet} shows the Capsule Network model as proposed in~\cite{capsnet}. The work in~\cite{emrouting} proposes instead a model in which the values of the capsules are arranged in matrices, and the dynamic routing is substituted by the \textit{EM routing}.

\medskip 
\noindent \textbf{NASNet~\cite{nasnet} (2018).} NASNet is the first popular neural network model obtained with neural architecture search. The approach of NasNet is the search of a cell for a simple dataset in a small search space. The cells can then be stacked to work on more complex datasets. Other models resulting from neural architecture search are PNASNet-5~\cite{pnasnet} and EfficientNet~\cite{efficientnet}. 

\begin{figure*}[!ht]
    \centering
    \includegraphics[width=\linewidth]{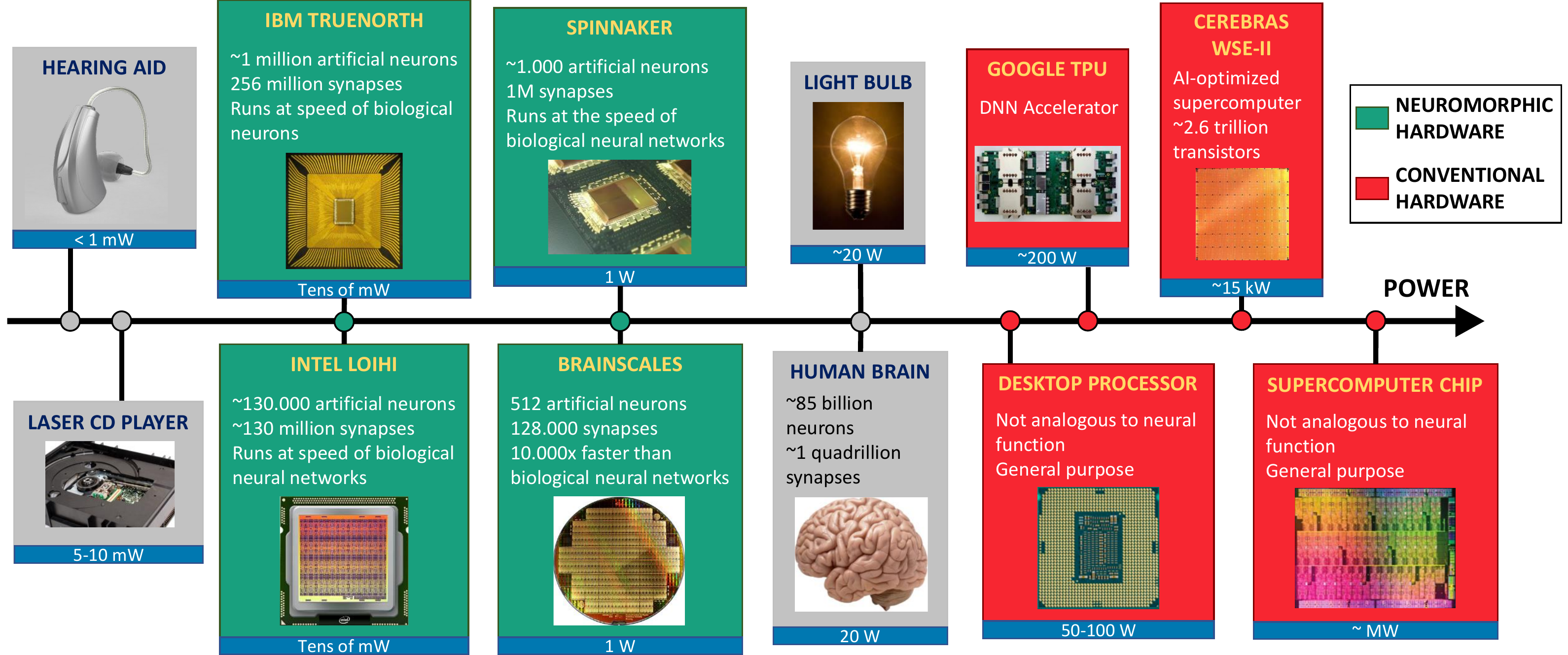}
    \caption{Comparison of power consumption between conventional hardware architectures and neuromorphic architectures.}
    \label{fig:infographic_introduction}
\end{figure*}

\subsection{Spiking Neural Networks (SNNs)}

Recently, Spiking Neural Networks (SNNs), considered as the third generation of neural networks~\cite{Maas1997ThirdGenerationSNN}, have received an increasing interest in the fields of deep learning and neuroscience, because of their extremely energy-efficient nature. SNNs, in contrast to the traditional DNNs, base their computational models much closer to that of the biological neurons, with a spike-based communication mechanism~\cite{Kasabov2019TimeSpaceSNN}. Due to their bio-inspired computations, SNNs bear a high potential to be the most promising solution for bridging the energy efficiency gap between the artificial machines and the human brain. A custom SNN hardware support is provided by neuromorphic computing, a relatively novel branch of computer architecture. The underlying goal is to reproduce in hardware the same computations that are executed in the human brain. Some examples of state-of-the-art neuromorphic designs, like IBM TrueNorth~\cite{Merolla2014Truenorth}, SpiNNaker~\cite{Furber2014Spinnaker}, BrainScale~\cite{Schmitt2017Brainscales} and Intel Loihi~\cite{Davies2018Loihi}, will be discussed in Section~\ref{SNN_acc}. Figure~\ref{fig:infographic_introduction} compares several hardware architectures, showing how efficient in terms of power consumption are neuromorphic solutions, compared to conventional designs~\cite{infograph_intro}. Moreover, another energy efficiency benefit in the neuromorphic research comes from the new sensor data formats. For instance, the event-based sensors such as the dynamic vision sensor (DVS) cameras~\cite{Lichtsteiner2006DVScamera} resemble the behavior of the human retina, in such a way that spikes are generated only when movements of the recorded subjects are detected.

\subsubsection{Spiking Neuron Models}
Modeling a spiking neuron is a challenging task. These models must be at the same time biologically accurate and computationally simple. When an input spike arrives to the neuron, the associated synaptic weight $w_i$ is integrated on the membrane and, consequently, the neuron membrane potential $V_m$ is increased. When the membrane potential overcomes a threshold $V_t$, the neuron fires, emitting a spike at the output, and its membrane potential is reset to a value $V_R$. Moreover, the membrane potential decreases continuously through time due to a leakage, according to a leak rate $\tau_m$ between spikes.

\begin{figure}[!h]
    \centering
    \includegraphics[width=0.9\linewidth]{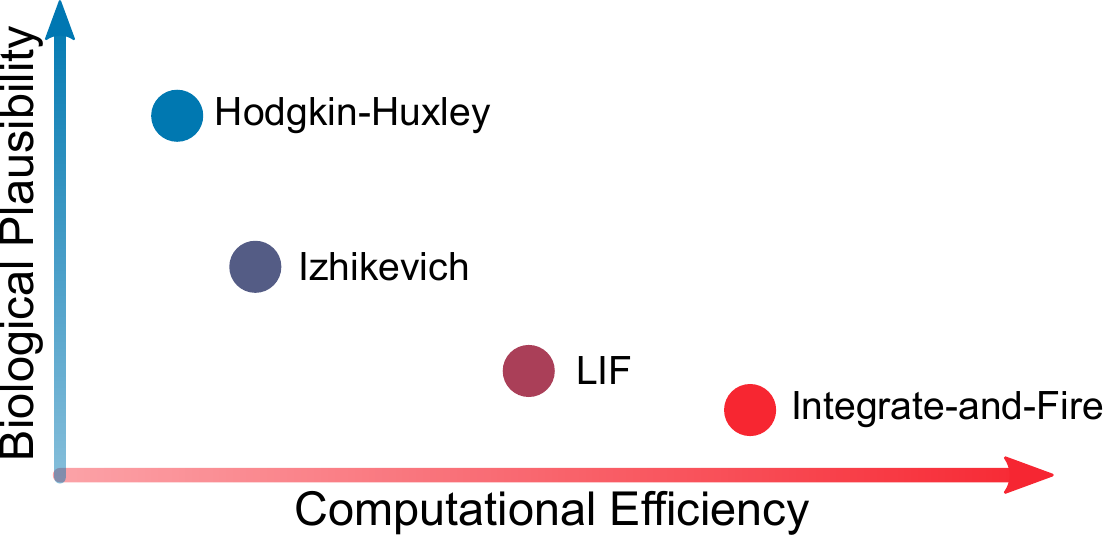}
    \caption{Comparison of different spiking neuron models.}
    \label{fig:neuron_models}
\end{figure}

Different spiking neuron models have been proposed in the literature. Figure~\ref{fig:neuron_models} shows the trade-off between biological plausibility and complexity of these models. The Hodgkin-Huxley model~\cite{Beeman2013HHneuronmodel} is very biologically-plausible, but extremely computational intensive. The Izhikevich model~\cite{Izhikevich2003IzhikevichModel} is slightly less complex, but still very computational intensive. On the other end, the Integrate-and-Fire is too simple and not very accurate in terms of biological plausibility. The most commonly adopted model is the Leaky-Integrate-and-Fire (LIF)~\cite{Wang2014LIF}, which is relatively simple and also takes into account the membrane leakage.

\subsubsection{Spike Encoding}
\begin{figure}[!h]
    \centering
    \includegraphics[width=0.8\linewidth]{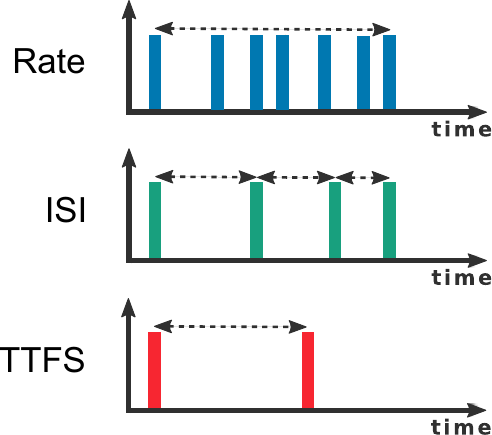}
    \caption{Comparison between Rate, Inter-spike interval (ISI), and Time to first spike (TTFS) encoding techniques for SNNs.}
    \label{fig:spike_codings}
\end{figure}

In order to provide input spikes and to collect the resulting output spikes of the SNN, the information has to be properly coded using spikes. Different approaches used to obtain such a conversion~\cite{Kasinski2011IntroSNN} are shown in Figure~\ref{fig:spike_codings}:
\begin{itemize}
    \item \textbf{Rate coding}: the information is coded as the mean firing rate of the generated spikes in a defined observation period. 
    \item  \textbf{Inter-spike interval (ISI)}: the intensity of the activation is coded as the precise delay between consecutive spikes.
    \item \textbf{Time to first spike (TTFS)}: the information is encoded in the latency that goes from the beginning of the stimulus to the time of the first output spike. This solution enables a very fast information processing, carrying enough information.
\end{itemize}

\subsubsection{SNN Training}
\begin{figure}[!h]
    \centering
    \includegraphics[width=\linewidth]{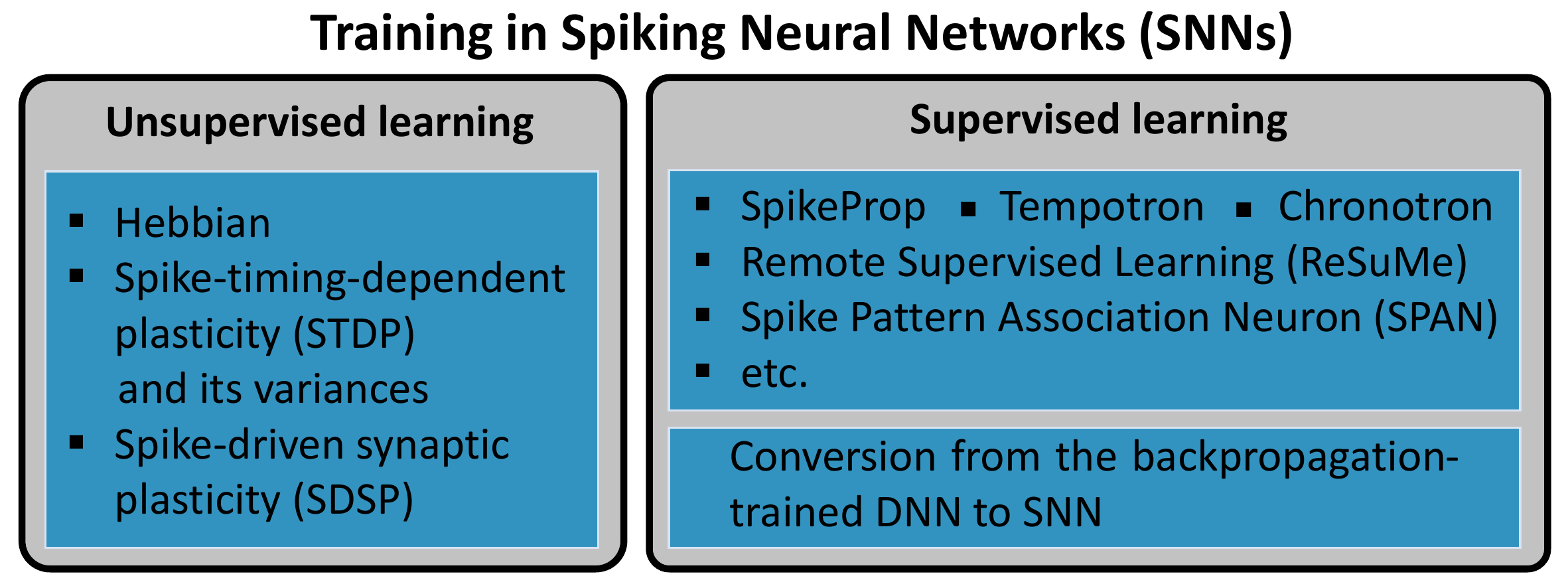}
    \caption{Training techniques for SNNs.}
    \label{fig:SNNtraining}
\end{figure}

Regarding the SNN training algorithm, the different possibilities have been explored are summarized in Figure~\ref{fig:SNNtraining}. For unsupervised learning, the possible algorithms are Hebbian Learning~\cite{Ruf1997hebbian}, the \textbf{Spike-Time-Dependent Plasticity (STDP)}~\cite{Bi1999STDP}\cite{Srinivasan2018STDP}, and the Spike-Driven Synaptic Plasticity (SDSP)~\cite{Fusi2000SDSP}. The most widely adopted method is the STDP, which is based on temporal relations between the \textit{presynaptic} spikes (at the input of the neuron) and the \textit{postsynaptic} spikes (at the output of the neuron). Basically, the synaptic weight is tuned accordingly to the temporal correlation between the presynaptic and postsynaptic spikes. 
The STDP algorithm can be optimized through the FSpiNN framework~\cite{Putra2020FSpiNN}, for executing energy-efficient SNNs on edge devices.

\normalcolor

For supervised learning, a fundamental challenge arises, because the traditional learning method, i.e., the backpropagation, cannot be applied due to the non-differentiability of the SNN loss function~\cite{Bodo2019ClosingTheAccuracyGap}. Hence, two different procedures can be adopted to solve or bypass the problem, thereby achieving supervised learning for SNNs:

\begin{enumerate}[leftmargin=*]
    \item \textit{Approximate the derivative of the spike trains.} This solution has been extensively studied in the works of~\cite{Boht2000SpikeProp}\cite{Gtig2006Tempotron}\cite{Florian2012Chronotron}\cite{Ponulak2005ReSuMe}\cite{Mohemmed2012SPAN}\cite{snnbackp}\cite{Lee2016TrainingSNN}\cite{Shrestha2018SLAYER}\cite{Neftci2019SurrogateGL}\cite{Thiele2020SpikeGrad}, which provide different types of approximations. The advantage is that the network can learn based on the temporal information of the spikes. For example, DECOLLE~\cite{Kaiser2018DECOLLE} introduces a local learning rule for continuous SNN learning. On the other hand, with this approach it is challenging to match the consolidated state-of-the-art high accuracy results of the DNNs.
    \item \textit{Train a DNN offline and convert it to SNN.} This approach~\cite{Rueckauer2017ConversionSNN} allows to use the most advanced training policies and techniques for DNNs. 
    An efficient conversion~\cite{SNNDVSLoihi_shafique} requires a comprehensive study of different conversion parameters to adapt the DNN-to-SNN conversion process to the neuromorphic hardware platform. 
    The main drawback is that a certain accuracy drop is encountered during the conversion. To overcome this, the recent work of~\cite{Rathi2020EnablingSNNbackprop} proposed a hybrid approach consisting of converting the DNN to SNN ad then incrementally training the SNN with an approximated backpropagation. Moreover, the max pooling operations cannot be implemented with spike rates~\cite{Pfeiffer2018DLSNN}. Therefore, max pooling layers are replaced by average pooling, which is easy to implement but shows an accuracy drop.
\end{enumerate}

\section{Hardware solutions and co-design}\label{HW_co-design}\label{sec:hwsolutions}
\subsection{Temporal vs Spatial architectures} \label{temp_vs_spatial}
Neural networks are a class of algorithms with an inherent parallelism. Two types of parallelism can be identified~\cite{parallelism_nn}. The neuron and consequently the FC and Conv layers have a \textit{topological parallelism} since the Multiply-and-Accumulate (MAC) operations that they perform have no data dependencies and can be executed in parallel. Moreover, the training sets consist of a large number of samples, that rather than being processed one at a time can be fed into the network in batches (\textit{operational parallelism}). 

\begin{figure}[h]
    \centering
    \includegraphics[width=\linewidth]{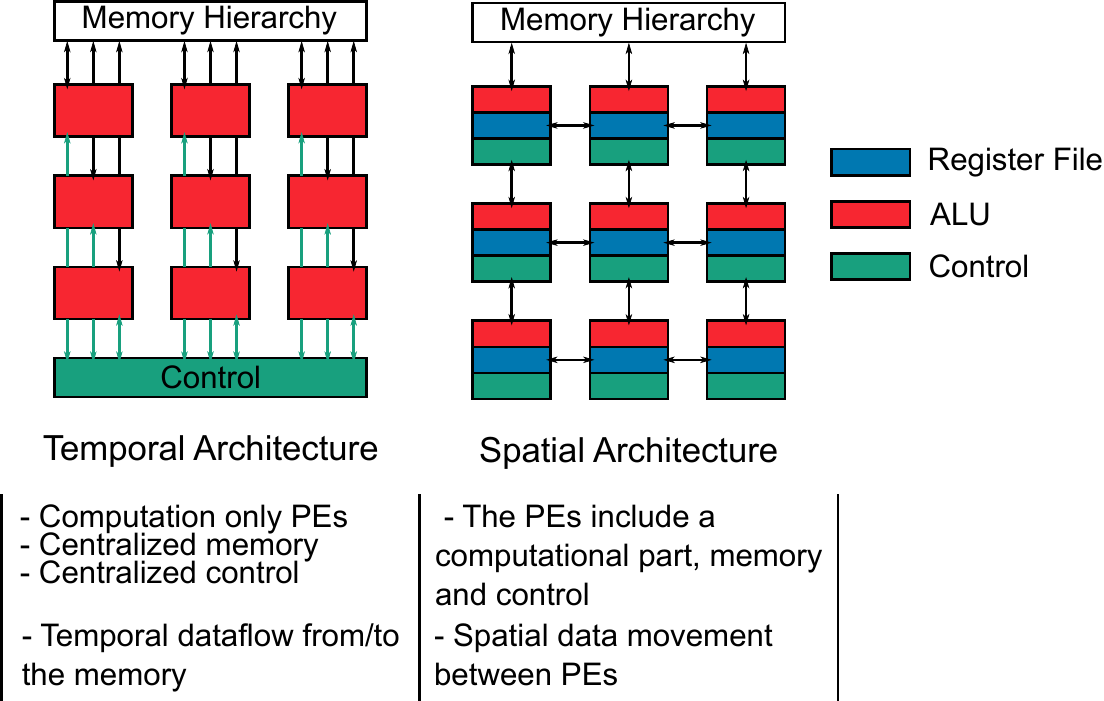}
    \caption{Basic models of temporal \textbf{(left)} and spatial \textbf{(right)} architectures.}
    \label{fig:spatial_temporal}
\end{figure}

The intrinsic parallelism of the layers can be exploited using parallel computing paradigms to increase the performance of the hardware implementations of NNs. Among the various solutions for parallel computation, temporal and spatial architectures~\cite{sze} are distinguished. Both the architectures consist of a large number of Processing Elements (PEs) that perform operations in parallel on the same or different data. In temporal architectures, the PEs can only access data from the central memory, the control is centralized, and there are no inter-PEs connections. In spatial architectures, on the contrary, each PE can also have its control logic and one or more local memory locations. Most importantly, in spatial architectures, the PEs are interconnected to exchange data with each other, creating a processing array. Figure~\ref{fig:spatial_temporal} shows the differences between temporal and spatial architectures. 

In the following, subsections \ref{subsec:temporal} and \ref{subsec:spatial} describe temporal and spatial architectures in detail respectively, and how to efficiently deploy neural networks on them.  

\subsection{Temporal architectures and software optimizations}
\label{subsec:temporal}

\begin{figure}[h]
    \centering
    \includegraphics[width=\linewidth]{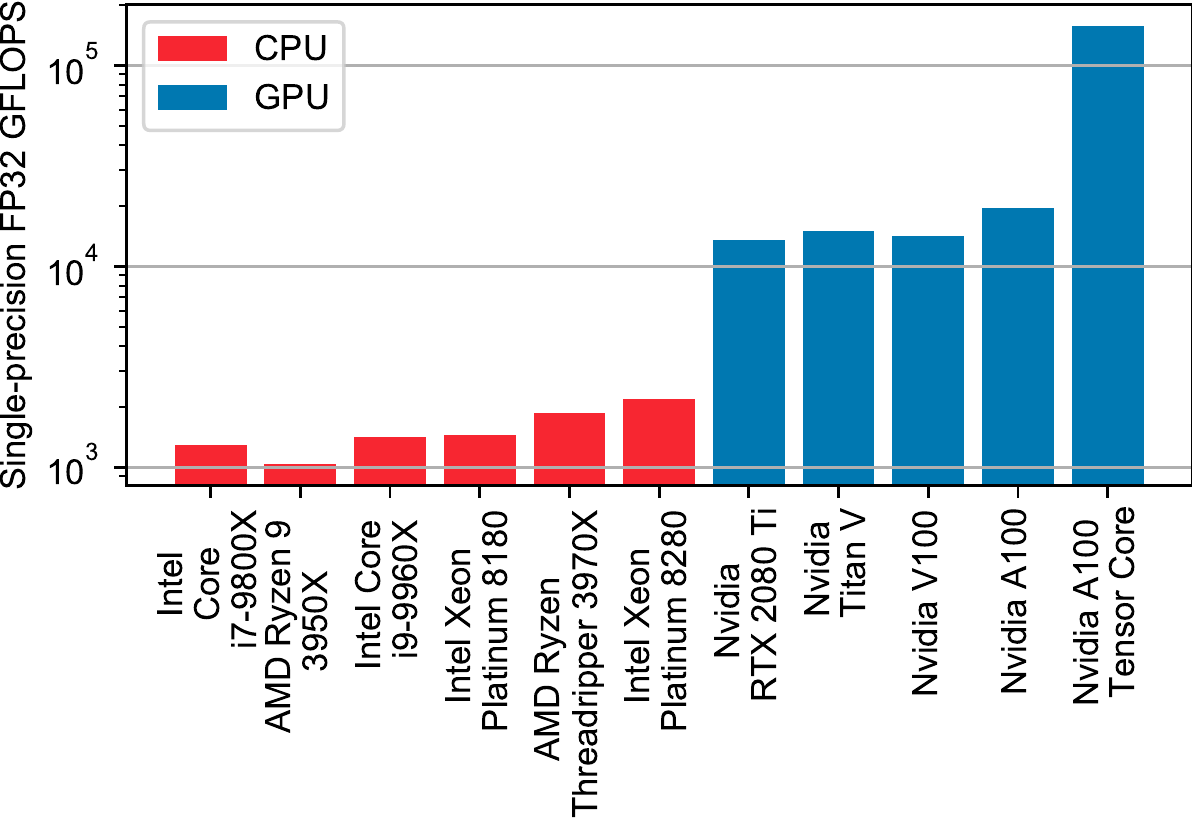}
    \caption{GFLOPS comparison between different CPUs and GPUs.}
    \label{fig:gflops}
\end{figure}

\begin{figure}[h]
    \centering
    \includegraphics[width=\linewidth]{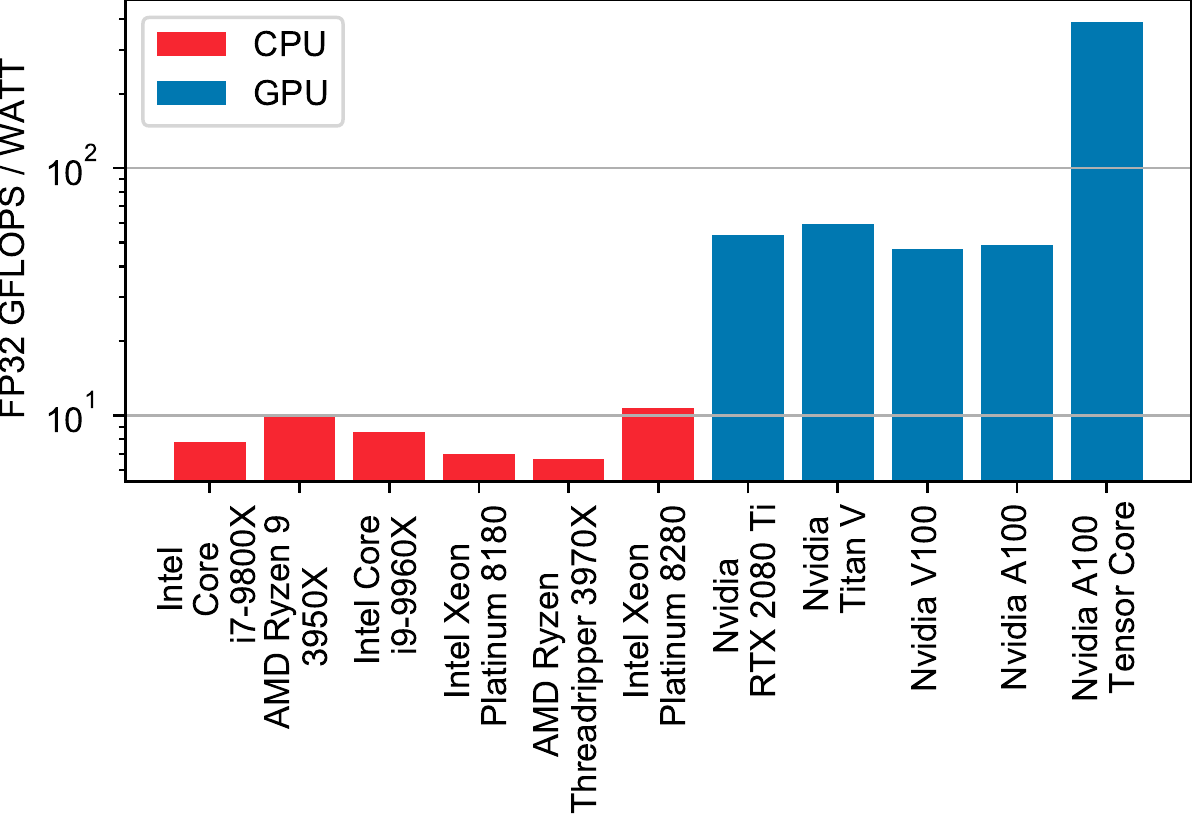}
    \caption{GFLOPS/WATT comparison between different CPUs and GPUs.}
    \label{fig:gflopswatt}
\end{figure}

Temporal architectures are commonly adopted in general-purpose platforms, such as CPUs and GPUs. CPUs can nowadays be realized as \textit{vector processors} (e.g.,  Intel's Xeon Phi x200 and Skylake-X CPUs) with an ability of working with multiple data elements simultaneously rather than with a single data at a time. Vector processors have multiple Arithmetic Logic Units (ALUs) that work synchronously and perform an instruction on a vector of data. Therefore, vector processors use the Single-Instruction-Multiple-Data (SIMD) technique. Among the available hardware platforms, CPUs are often the least used for DNNs inference or training, as they provide lower FLOPS and FLOPS/WATT performance (see Figures~\ref{fig:gflops} and \ref{fig:gflopswatt}). However, manufacturers have recently undertaken measures to accelerate the deployment of NNs on CPUs. For example, at the instruction level, Intel has added the AVX-512 Vector Neural Network Instructions (AVX-512 VNNI) to the AVX-512 Instruction Set~\cite{intelavx} to accelerate CNNs. In addition, Intel announced that the next generation of Cooper Lake and Sky Lake processors will support Brain Floating Point (\textit{bfloat16}) operations~\cite{bfloat16}. \textit{bfloat16} is a floating-radix-point format on 16 bits with a dynamic range comparable with the dynamic range of the 32-bit IEEE 754 floating-point format. \textit{bfloat16} is also supported by ARMv8.6-A and AMD's ROCm library. Intel has also created BigDL~\cite{bigdl}, an ML library for the distributed acceleration of DNN algorithms on CPU clusters. 

GPUs are manycore architectures with up to thousands of cores that are specifically designed for parallel computation (e.g., 5120 cores in Nvidia V100 GPU~\cite{teslav100}). Similarly to vector CPUs, GPUs adopt the Single-Instruction-Multiple- Thread (SIMT) execution model, first introduced by Nvidia. The SIMT model executes a single instruction simultaneously on multiple cores. Each core receives a different data that belongs to multiple threads running in parallel. GPUs are the real workhorses for DNNs training in particular, and in certain cases for inference as well. Among the various GPU manufacturers, Nvidia has put a lot of emphasis on GPU hardware and software optimization for DL. Most DL frameworks support the execution on Nvidia GPUs, e.g., Pytorch~\cite{pytorch}, Tensorflow~\cite{tensorflow}, or Caffe~\cite{caffe}. One of the great advantages of Nvidia GPUs is cuDNN~\cite{chetlur2014cudnn}, a highly optimized library of primitives for DNNs. cuDNN is not the only library for DL, rather all Nvidia libraries for DNN/ML are collected in CUDA-X AI~\cite{cudaxai}. In the latest high-end GPUs, Nvidia has combined traditional CUDA Cores with Tensor Cores~\cite{teslav100}, which are optimized for large matrix operations. Tensor Cores can also support mixed-precision operations. In the new Nvidia A100, the Tensor Cores support a new format, the Tensor Format (TF32), with which performance is 10x higher when compared to the performance of the FP32 format on the V100 architecture~\cite{ampere100}. In addition, Nvidia A100's Tensor Cores can also take advantage of the sparsity of tensors, very common in DNNs, to achieve up to 2x higher performance.

At the software level, several libraries have been created to optimize \textit{Basic Linear Algebra Subroutines} (BLAS) on both CPUs (e.g., AMD Core Math Library (ACML), Intel Math Kernel Library (Intel MKL) or OpenBLAS) and GPUs (e.g., Nvidia cuBLAS or Intel cIBLAS). Among the numerous subroutines implemented, the BLAS also include element-wise matrix multiplication, matrix-vector multiplication and matrix-matrix multiplication, also called General Matrix Multiplication (GeMM). For what concerns neural networks the BLAS come in hand for the FC layer that, as explained in Section~\ref{sec:dnnlayers}, can be seen as a vector-matrix multiplication or as a matrix-matrix multiplication in case of batched computation. 

\begin{figure}[h]
    \centering
    \includegraphics[width=0.9\linewidth]{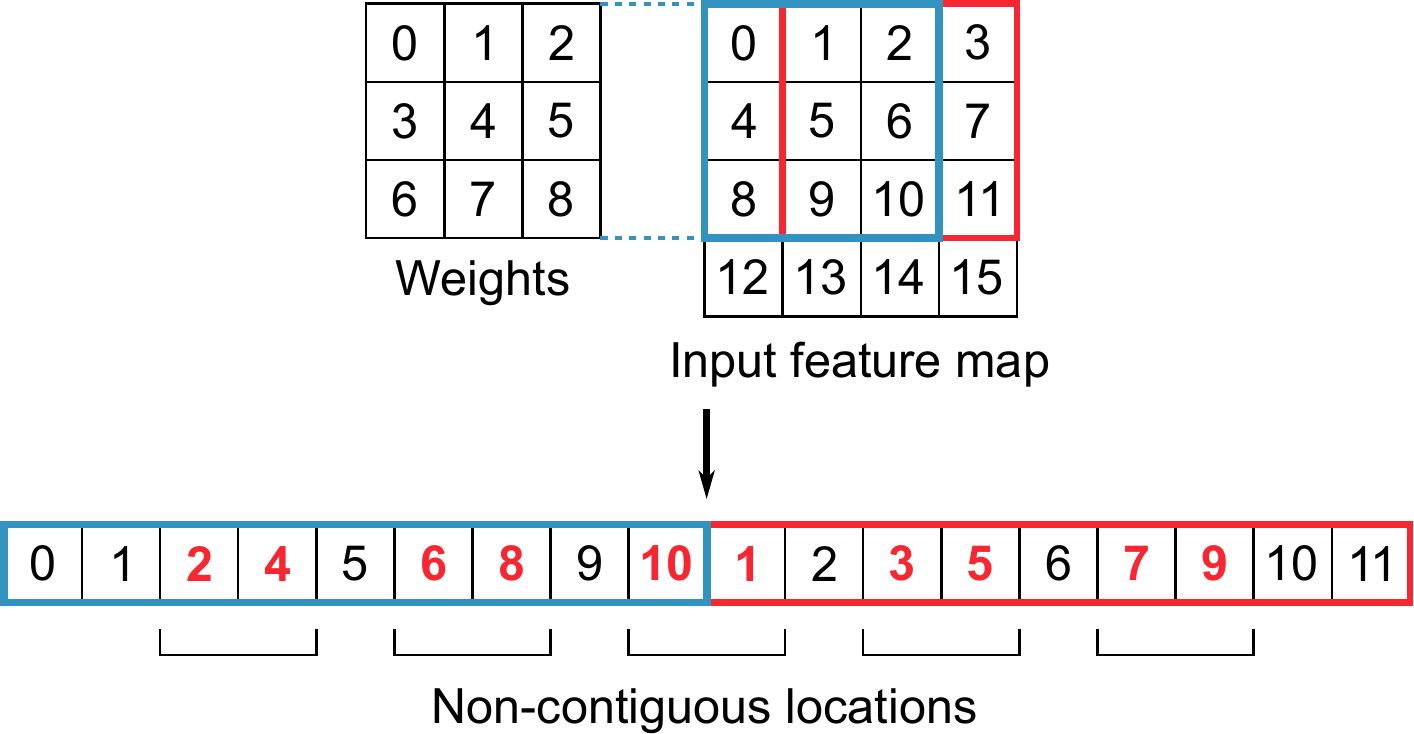}
    \caption{Discontinuity in memory accesses when performing convolution on a matrix stored by rows.}
    \label{fig:mem_discont}
\end{figure}

Optimizing the computation of the Conv layers is a more challenging task. The operations between a weight kernel and the subsets of the input feature maps are simple point-wise multiplications of matrices, but the memory access pattern is complex. Figure~\ref{fig:mem_discont} shows how, if an input feature map is stored by rows, it is necessary to perform accesses to non-contiguous locations of memory.

\begin{figure}[h]
    \centering
    \includegraphics[width=\linewidth]{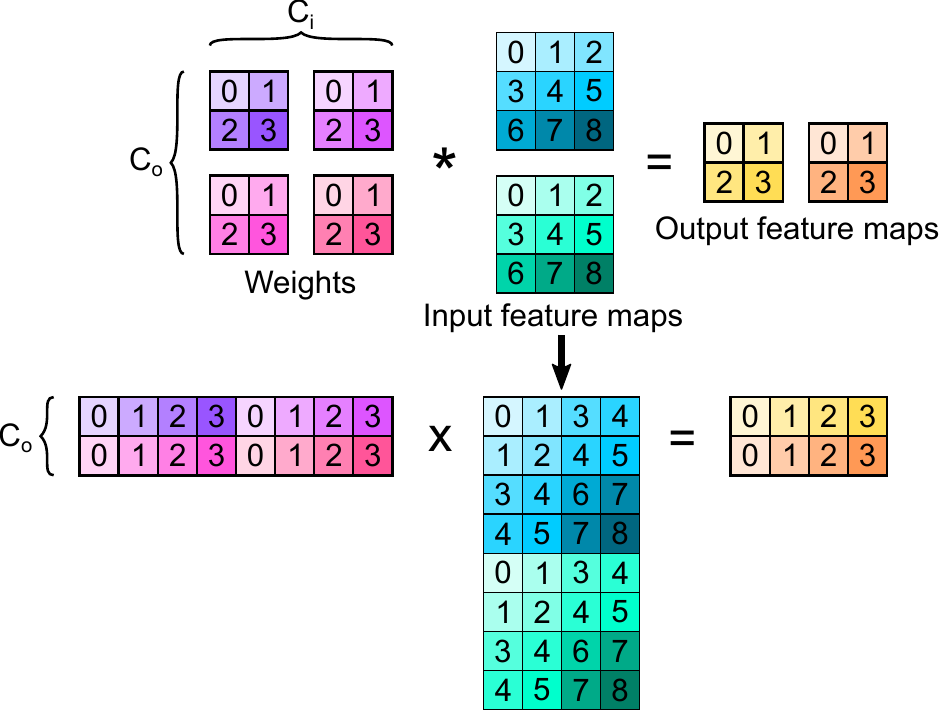}
    \caption{Convolution lowering: mapping a convolution to a matrix-matrix multiplication by rearranging the matrices.}
    \label{fig:conv_lowering}
\end{figure}

Several computational transforms have been proposed to apply the optimized BLAS to Conv layers. Many of the software libraries mentioned above lower the convolution into a GeMM as proposed in~\cite{chellapilla}~\cite{vasudevan} and shown in Figure~\ref{fig:conv_lowering}. A 4D-tensor of weights is flattened to a 2D matrix, while the data in the input feature maps are duplicated and rearranged following a pattern that leads to the correct result of a convolution by performing a matrix multiplication. This method is very efficient since the GeMM routine is highly optimized. However, it requires data to be duplicated up to $H_k\times W_k$ times, with the dimension of the input feature maps moving from $C_i \times H_i \times W_i$ to  $C_iH_kW_k \times H_oW_o$. This approach, therefore, requires a large memory for temporary allocation. 

The GeMM method for convolution can further be optimized by applying the Strassen algorithm~\cite{strassen}\cite{strassen2} that reduces the number of necessary multiplications by partitioning the matrices. The number of multiplications is reduced of 1/8 at each partition, at the cost of a higher number of additions.

A different approach consists of transforming both the input feature maps and the weights from the space domain to the frequency domain with the FFT algorithm~\cite{fft}. In the frequency domain, the convolution operation becomes an element-wise multiplication of matrices. However, the FFT algorithm introduces a high computational overhead for the domain change, and its efficiency has only been proven valid for large weight kernels and unitary strides. Another approach often used is based on the Winograd algorithm~\cite{winograd}\cite{winograd2}, which, unlike the FFT algorithm, is particularly efficient for small kernels.

Direct convolution can also be performed exploiting the parallel hardware solutions offered by modern CPUs and GPUs. In~\cite{blockedconv} and~\cite{blockedconv2} it is shown how to rearrange the tensors to have more efficient memory accesses, and how to perform operations to take full advantage of Intel AVX-512~\cite{intelavx} vector instructions.

\subsection{Spatial architectures and dataflow processing} 
\label{subsec:spatial}

Spatial architectures are commonly implemented on FPGAs and ASICs, that allow for a design tailored on specific applications at the price of less flexibility. Neural networks are particularly suitable for this kind of hardware implementation since the type and order of operations of each layer is fixed and known a priori. Therefore, it is possible to develop specialized and highly optimized circuits.

\begin{figure}[h]
    \centering
    \includegraphics{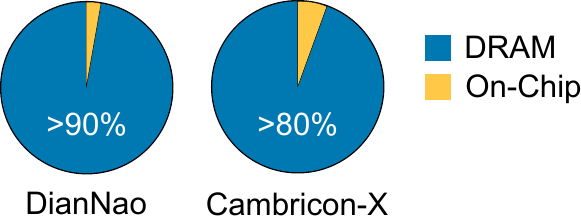}
    \caption{Energy breakdown of two state-of-the-art DNNs accelerators.}
    \label{fig:energy_breakdown}
\end{figure}

The operations carried out in the neural networks are simple, mostly multiply-and-accumulate (MACs), but they must be performed on a large set of data. Therefore, the bottleneck is not caused by computation but by the memory accesses that are necessary to fetch and store the inputs and the results, respectively. Every MAC requires three data elements to be read from memory (input pixel, weight and partial sum) and one data element to be written (updated partial sum). It has been demonstrated that a DRAM access has an energy cost of $\sim 2$ orders of magnitude higher than a MAC operation~\cite{horowitz}. The enormous DRAM access energy cost compared to the computational energy has been observed in many state-of-the-art DNNs accelerators such as DianNao~\cite{diannao} or Cambricon-X~\cite{cambricon-x} (Figure~\ref{fig:energy_breakdown}). 

\begin{figure}[h]
    \centering
    \includegraphics[width=\linewidth]{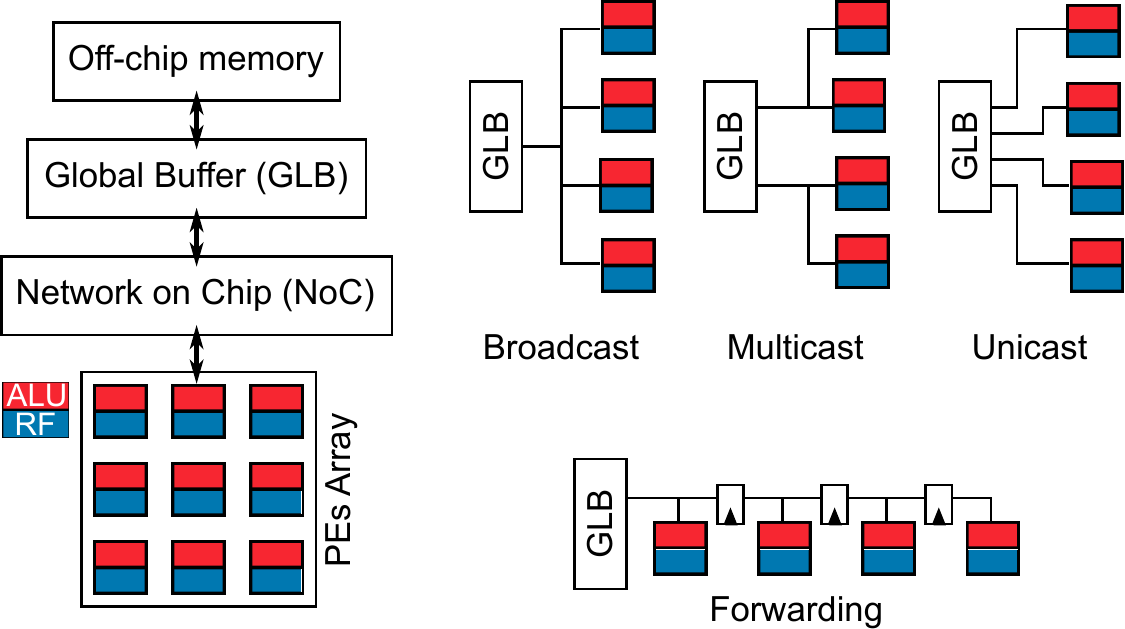}
    \caption{\textbf{(left)} General structure of a hardware accelerator for DNNs and \textbf{(right)} interconnection schemes}
    \label{fig:hw_acc}
\end{figure}

A typical hardware architecture of a DNNs accelerator (Figure~\ref{fig:hw_acc} left) consists of: 
\begin{itemize}[leftmargin=*]
     \item An \textbf{off-chip memory} (usually DRAM), to store the weights and the activations of the whole network. This level of memory can typically contain several GBs of data.
    \item An \textbf{on-chip global buffer (GLB)}, large enough to hold the weights and inputs necessary to feed all the PEs. The energy needed to access the GLB can be two orders of magnitude lower than that of the DRAM~\cite{chen_dataflows}.
    \item An array of hundreds of \textbf{PEs}, each containing an ALU to perform MACs operations in parallel. The PEs usually also include one or more Register Files (RFs) to locally store data with an energy cost-per-access lower than that of the GLB.
    \item The PEs are connected with each other and to the GLB by a \textbf{Network-on-Chip (NoC)}. 
    The data must be moved coordinately through the PEs to obtain the correct result, depending on how the operations are temporally scheduled and spatially distributed on the PEs. The NoC can then assume different configurations to implement various communication patterns, represented in Figure~\ref{fig:hw_acc} right, depending on how data must be delivered.
\end{itemize}

Given the energy cost required by a DRAM access, the design of state-of-the-art DNNs accelerators focuses on the exploitation of \textit{data reuse}, i.e., optimizing the architecture, the mapping of data on the PEs and the temporal scheduling of operations to maximize the reuse of data when they are stored in the lower-level memories such as the RFs or the GLB. 


The different layers in an NN allow for taking advantage of various opportunities of data reuse, as explained in the following. 

\begin{figure}[h]
     \centering
     \includegraphics[width=\linewidth]{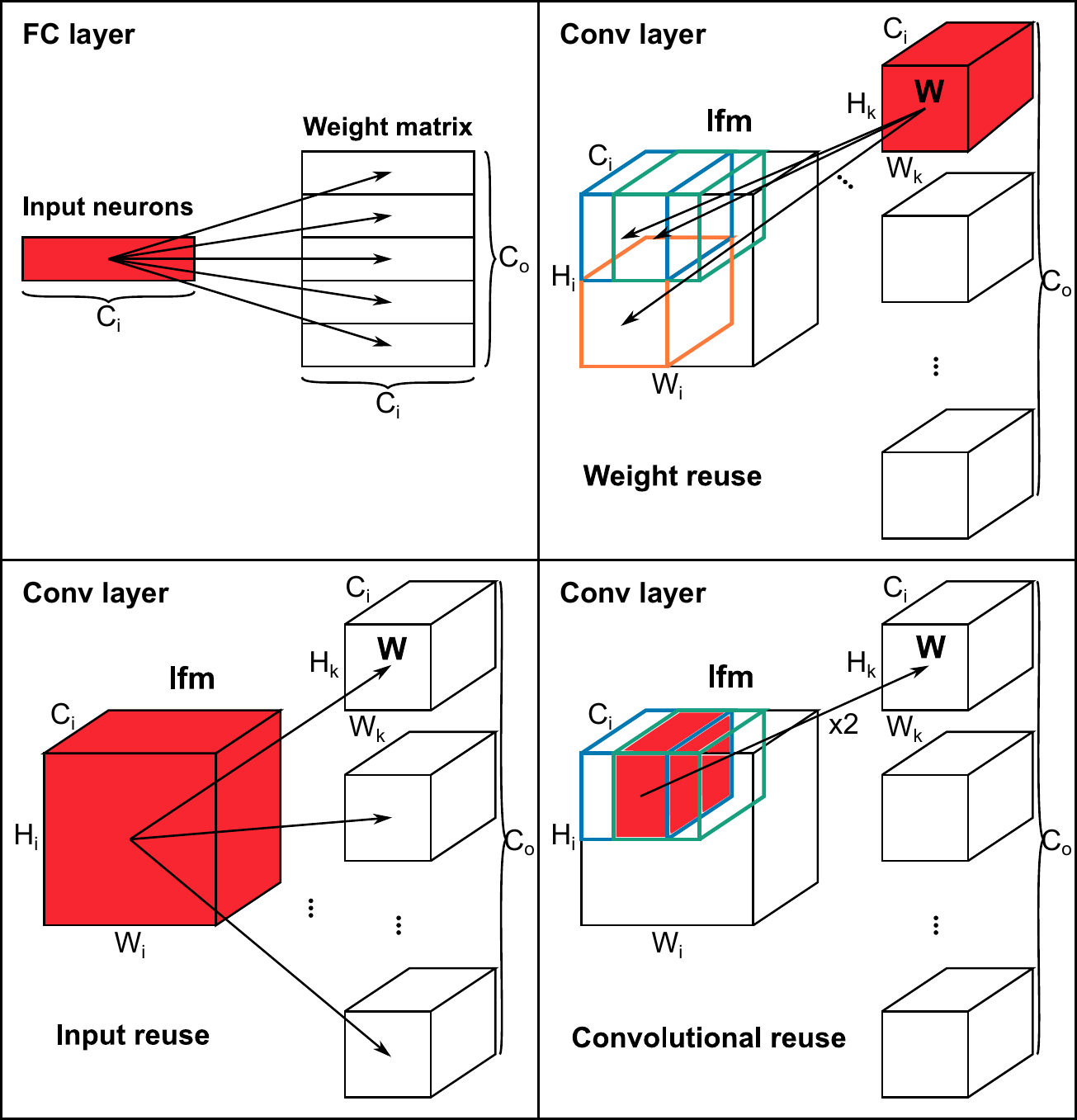}
     \caption{Data reuse in an FC layer and in a Conv layer.}
     \label{fig:FC_reuse}
 \end{figure}

\textbf{FC layer.} A FC layer can be described as a matrix-vector multiplication and it therefore presents an opportunity for \textit{input reuse}, since the vector of the input neurons is dot-multiplied with each row of the matrix of weights (see Figure~\ref{fig:FC_reuse}).  

\textbf{Conv layer.} The Conv layer has three different opportunities for data reuse (see Figure~\ref{fig:FC_reuse}). To perform the convolution operation, a weight kernel is slid over the whole input feature map. There is an opportunity for \textit{weight reuse} since the same weight kernel is multiplied for multiple subsets of the input feature maps. In particular, each of the $C_o$ kernels is reused $H_o \times W_o$ times. 


There is an \textit{input reuse} opportunity too, since the input feature maps are used $C_o$ times to generate all the output feature maps. The last reuse opportunity is defined as \textit{convolutional reuse}~\cite{chen_dataflows}, and it exploits the sliding window mechanism, i.e., when computing two adjacent output pixels, there is usually an intersection between the two subsets of pixels of the input feature map used, as shown in Figure~\ref{fig:FC_reuse}. The width and height of the intersection depends on the dimensions of the kernels ($H_k \times W_k$) and the horizontal and vertical strides ($s_x, s_y$). The convolutional reuse combines both the weight reuse and the input reuse.

\textbf{Pooling layer.} Pooling layers do not demand the use of weights. Therefore there are no opportunities of weight reuse. The stride parameter is usually set to have non-overlapping receptive fields, so it is not possible to exploit the sliding window mechanism for input reuse. These layers do not allow for any data reuse.

\begin{figure}[h]
    \centering
    \includegraphics[width=\linewidth]{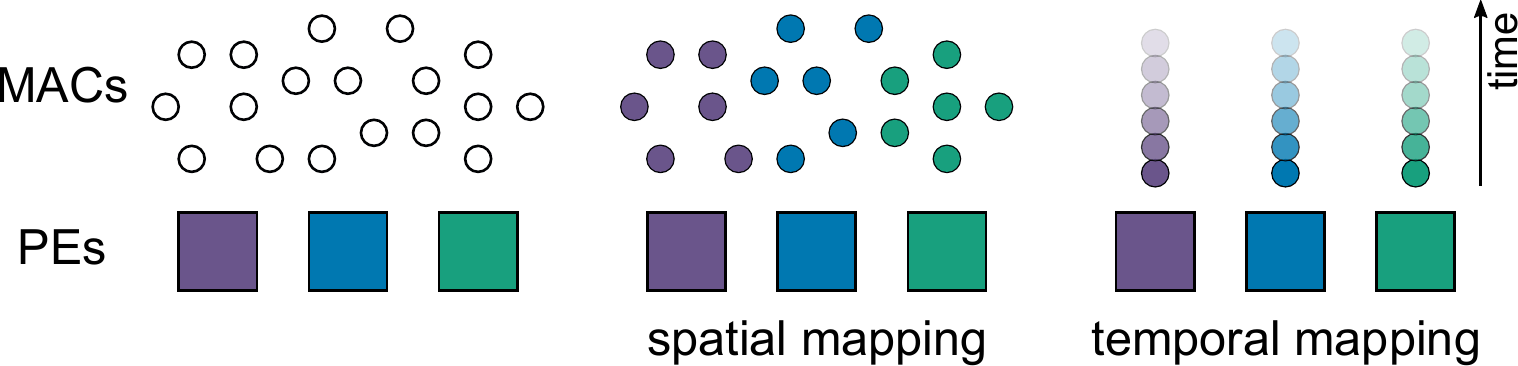}
    \caption{Spatial and temporal mapping of the Multiply-and-Accumulate (MACs) operations to the Processing Elements (PEs).}
    \label{fig:dataflow_mapping}
\end{figure}

\medskip 
Given an array of PEs and all the MACs between weights and input feature maps that must be performed to calculate the output feature maps, each PE will execute a subset of MACs, and a number of MACs equal to the number of PEs will be executed in parallel. The MACs must, therefore, be \textit{spatially} and \textit{temporally mapped} to the PEs array (Figure~\ref{fig:dataflow_mapping}). The mapping consequently defines how data must be loaded and stored from/to the memory hierarchy of the accelerator and how the NoC must be designed to correctly and efficiently deliver and collect the inputs, the weights and the partial sums. The spatial and temporal mapping of the operations is defined as \textit{dataflow}~\cite{chen_dataflows}. 

Considering the high dimensions of the PEs array and the vast number of MACs to be computed, the space of possible mappings on a generic HW accelerator is enormous. Given the considerations on the energy consumption of the memory hierarchy, dataflows usually try to maximally exploit the opportunities of data reuse provided by the different layers of the NNs to minimize the accesses to the off-chip memory and the global buffer, and to use the data stored in the RFs as much as possible. Chen et al.~\cite{chen_dataflows} introduced a taxonomy to classify existing accelerators based on their dataflow and on how they exploit data reuse, that will be explained briefly in the following. 

\begin{figure}[h]
    \centering
    \includegraphics[width=\linewidth]{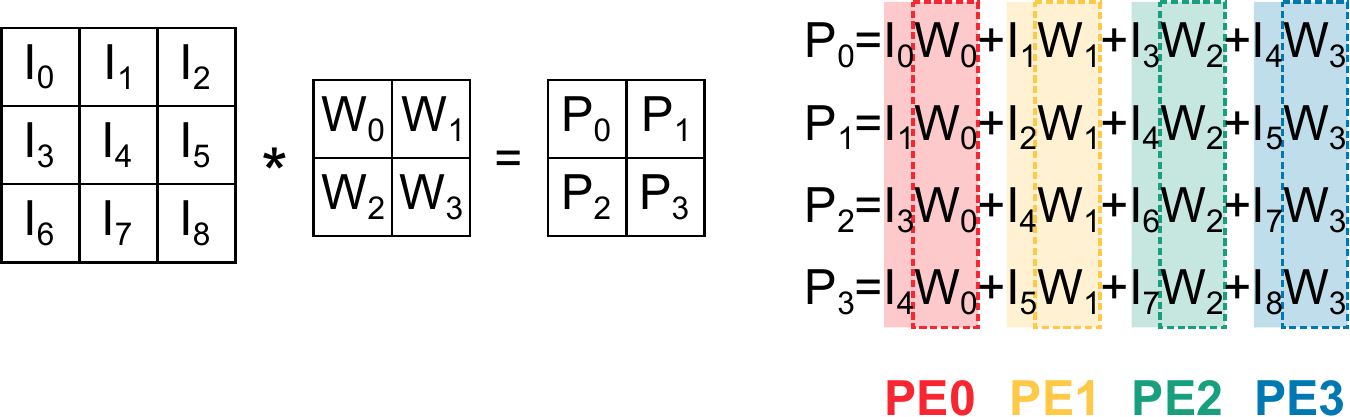}
    \caption{Spatial mapping of the operations in a weight stationary dataflow.}
    \label{fig:weight_stat}
\end{figure}

\textbf{Weight Stationary:} The weight stationary dataflow aims at exploiting mainly the weight reuse to minimize the energy cost necessary to fetch the weights from the DRAM and the GLB. A subset of weights is read from the DRAM/GLB and stored in the RFs of the PEs. All the operations that involve a certain weight are then mapped to the PE where it is stored. Figure~\ref{fig:weight_stat} shows how operations are mapped to an array of 4 PEs to perform a 2$\times$2 convolution on a 3$\times$3 input feature map. 

\begin{figure}[h]
    \centering
    \includegraphics{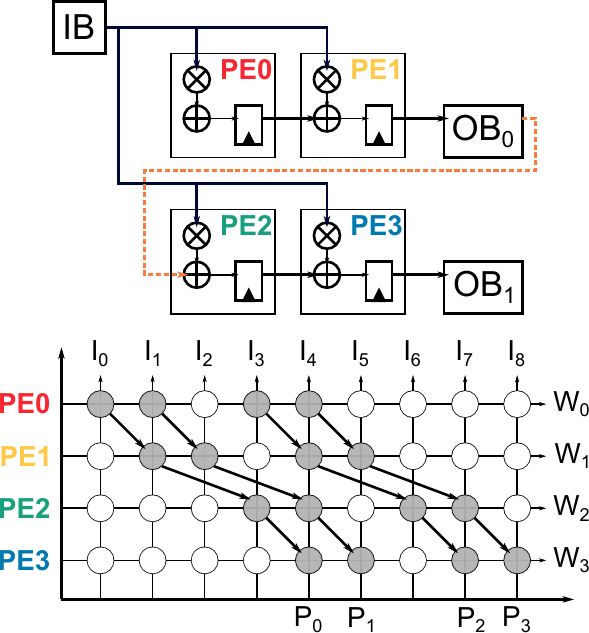}
    \caption{\textbf{(top)} PEs array and \textbf{(bottom)} temporal-spatial mapping of the operations in a weight stationary accelerator with inputs broadcasting and outputs forwarding.}
    \label{fig:weight_stat_broad}
\end{figure}

Since the weights are kept stationary in the PEs, the inputs and the partial sums need to be coordinately moved through the array to optimize the data movement on the NoC too. A possibility consists of broadcasting a single input pixel of the input feature map to all the PEs and in storing each partial sum in a register then to pass it to the adjacent PE on the right. As shown in Figure~\ref{fig:weight_stat_broad}, there are time steps in which some of the PEs perform operations that are not useful for the result (denoted in white). Moreover, the partial sums at the end of each row of processing elements needs to be stalled for $W_i - W_k$ time steps before being passed to the next row of PEs. Therefore, all of these partial sums must be stored in the GLB. The nn-X accelerator~\cite{nnx} allocates instead $H_k$ FIFOs at the end of each row, each of dimension $W_i - W_k$, to introduce the proper delay.  

\begin{figure}[h]
    \centering
    \includegraphics{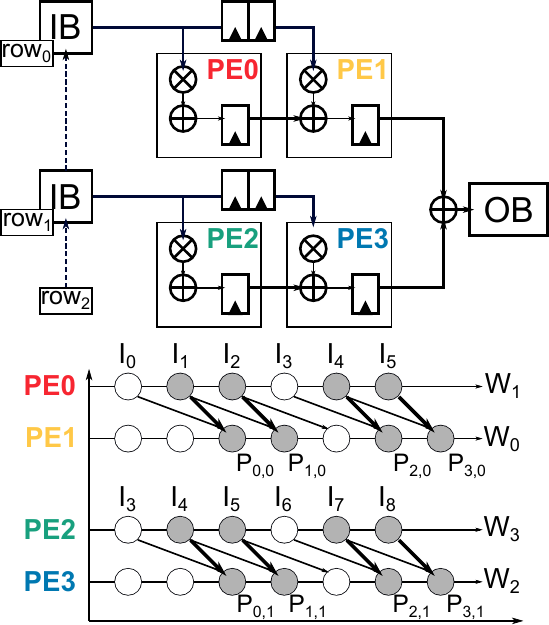}
    \caption{\textbf{(top)} PEs array and \textbf{(bottom)} temporal-spatial mapping of the operations in a weight stationary accelerator with inputs and outputs forwarding.}
    \label{fig:weight_stat_forward}
\end{figure}

The input pixels can be moved with the forwarding scheme to take advantage of the convolutional reuse in addition to the weight reuse. The forwarding scheme consists of placing additional registers in the PEs to store the input pixel that they receive, and to then pass it to the neighboring PEs on the right (horizontally-sliding window). Figure~\ref{fig:weight_stat_forward} shows a dataflow with stationary weights and input forwarding. Both in~\cite{sriram2010} and~\cite{Sankaradas2009}, $H_k$ rows of the input feature map are processed in parallel, and the partial sums of each row are then accumulated. The inputs are therefore stored in $H_k$ buffers, and the pixels of the input feature map are moved from the $(K-1)$ buffer to the $0$ buffer (vertically-sliding window).  

\begin{figure}[ht]
    \centering
    \includegraphics[width=0.9\linewidth]{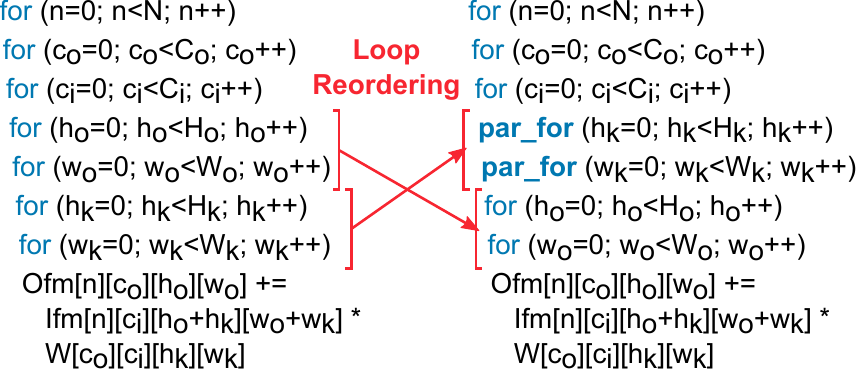}
    \caption{Loop reordering of the 7-nested loops representation of the Conv layer.}
    \label{fig:loop_reordering}
\end{figure}

What characterizes the above-discussed dataflows is that all the operations along dimensions $H_k$ and $W_k$ are mapped to the 2D PE array and executed in parallel. This mapping operation is defined as \textit{spatial unrolling} in~\cite{Yang2018DNNDC}. From a software perspective this is equivalent to replacing the \textbf{for} loops in the 7-nested loop representation with parallel for loops (\textbf{par\_for}) as in Figure~\ref{fig:loop_reordering}. In~\cite{Yang2018DNNDC}, the $H_k$|$W_k$ syntax is adopted to denote which loops are parallelized. The stationarity of the weights is instead equivalent, from the software perspective, to a \textit{loop reordering} operation of the \textbf{for} loops, as shown in Figure~\ref{fig:loop_reordering}. Other architectures that adopt a $H_k$|$W_k$ weight stationary approach are~\cite{Chakradhar2010},~\cite{park2015} and~\cite{Qiu2016}.

\begin{figure}[h]
    \centering
    \includegraphics[width=\linewidth]{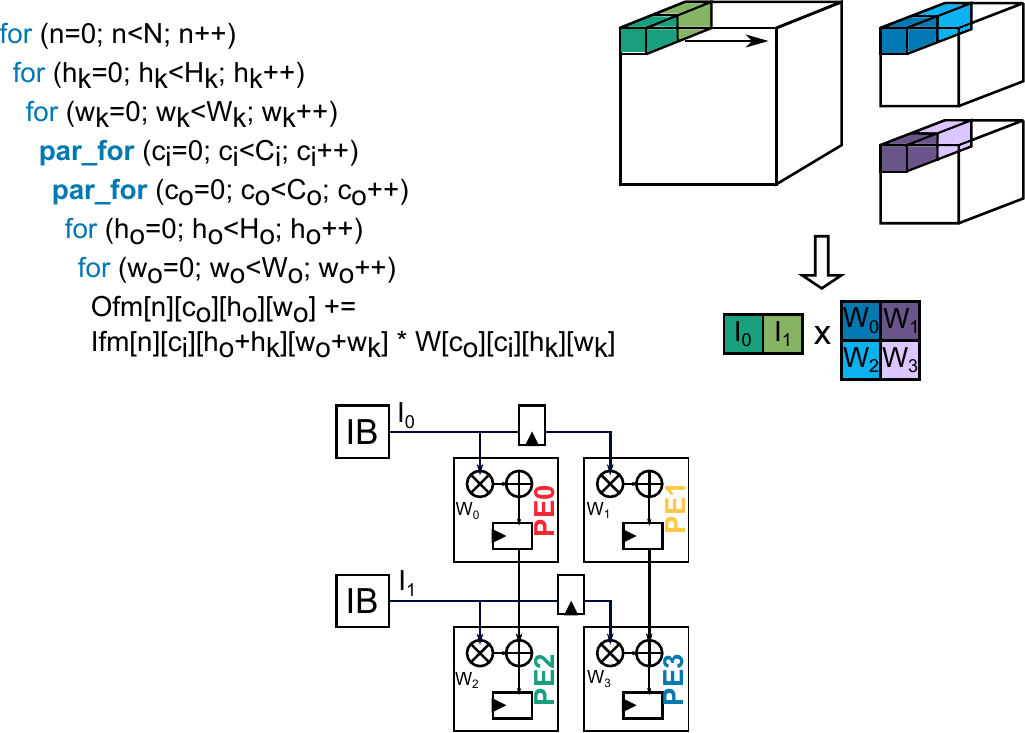}
    \caption{\textbf{(top left)} Reordered loops for a dataflow with weights stationary along dimensions $C_i$ and $C_o$. \textbf{(top right)} Mapping the operations of a Conv layer to a matrix multiplication. \textbf{(bottom)} PEs array in a $C_i$|$C_o$ weight stationary accelerator with input and output forwarding.}
    \label{fig:weight_stat_ck}
\end{figure}

A different dataflow, but in which the weights are still stationary, is obtained by spatially unrolling the dimensions $C_o$ and $C_i$ ($C_o$|$C_i$). As shown in Figure~\ref{fig:weight_stat_ck}, the operations that must be performed are equivalent to a vector-matrix multiplication. It can be realized in hardware with a 2D systolic array. In essence, the weights are internally stored in the PEs, the inputs are horizontally forwarded, and the partial sums are accumulated along the vertical dimension. An example of $C_o$|$C_i$-weight stationary dataflow can be found in in the Tensor Processing Unit (TPU)~\cite{googletpu} developed at Google. TPUs are deployed in datacenters, and it has therefore been possible to obtain statistics and metrics on real-life applications. It has been observed that CNNs, on which the development of HW accelerators is focused, actually represent the 5\% of all applications used in datacenters~\cite{googletpu}. For this reason, Google designers decided to focus on the acceleration of FC layers, which are inherently vector-matrix operations and can, therefore, be directly mapped to the matrix-multiply unit that is the heart of the TPUs. 

Because of its flexibility, the systolic array is often used in configurable architectures that must support various layer types~\cite{scalesim}\cite{mpna_shafique}\cite{rna}. This solution is also adopted for the acceleration of Capsule Networks~\cite{capsacc_shafique}\cite{capstore_shafique}\cite{descnet_shafique}, that consist of Conv layers, Conv layers of capsules and FC layers of capsules. 

\begin{figure}[h]
    \centering
    \includegraphics[width=\linewidth]{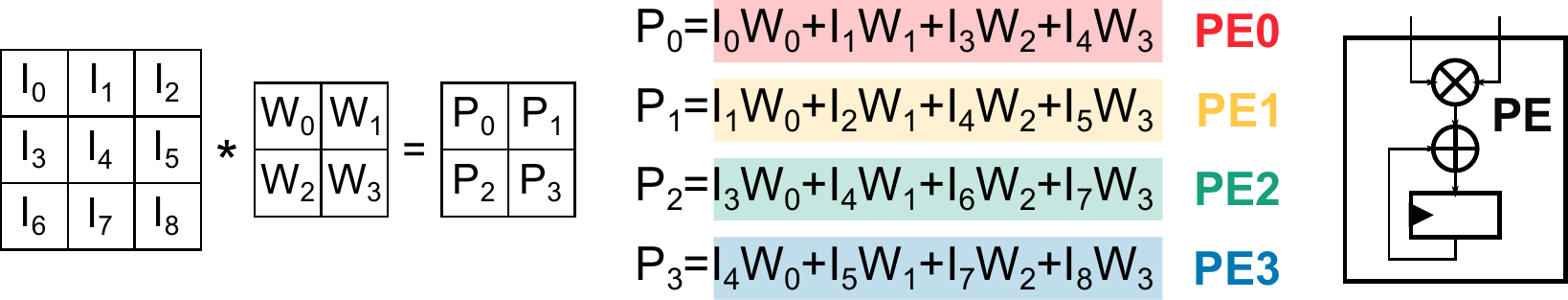}
    \caption{\textbf{(left)} Spatial mapping of the operations in an output stationary dataflow. \textbf{(right)} Modified PE for an output stationary accelerator.}
    \label{fig:output_stat}
\end{figure}

\textbf{Output Stationary:} The output stationary dataflow has the purpose of minimizing the data movement necessary to store and load the partial sums in the GLB. With the weight stationary dataflow of Figure~\ref{fig:weight_stat_broad}, for example, the partial sum of a single output pixel must be stored and reloaded to/from the GLB $(H_k-1)\times C_i$ times. In the output stationary dataflow, the PEs are modified to have the possibility of locally accumulating the results of the MACs that they perform (Figure~\ref{fig:output_stat}). Each PE is therefore responsible for the computations necessary to obtain an output pixel, whose partial sums are accumulated in a single RF.  

\begin{figure}[ht]
    \centering
    \includegraphics[width=\linewidth]{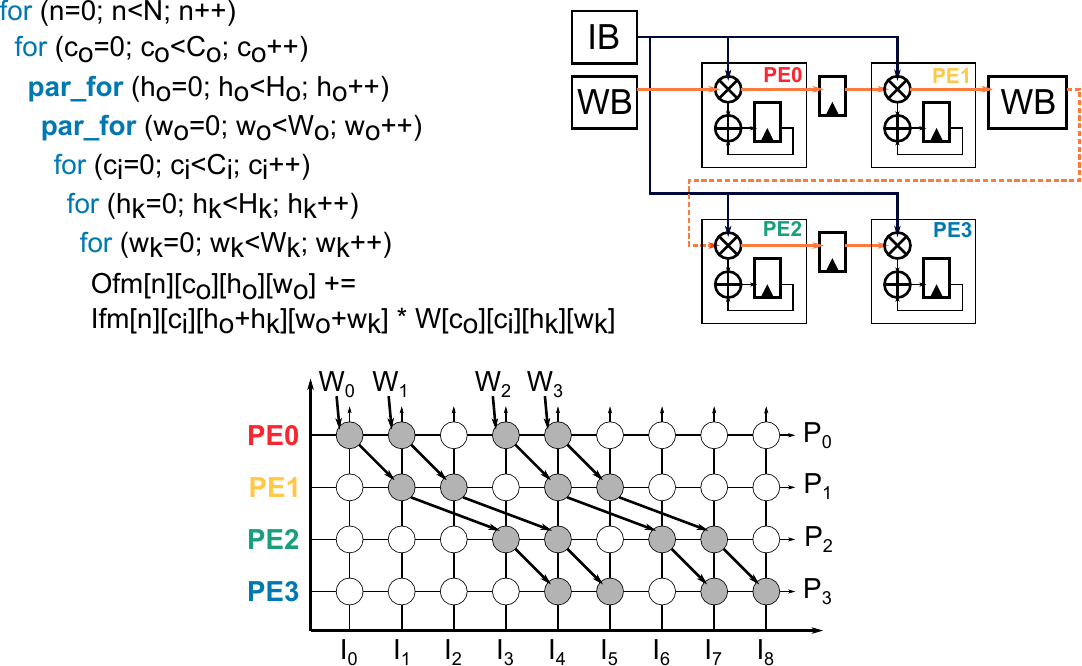}
    \caption{\textbf{(top left)} Reordered loops for an output stationary dataflow . \textbf{(top right)} PEs array and \textbf{(bottom)}  temporal-spatial mapping of the operations in an output stationary accelerator with input broadcasting and weight forwarding.}
    \label{fig:output_stat_broadin}
\end{figure}

Similarly to the weight stationary dataflow, it is possible to spatially unroll the $H_o$ and $W_o$ loops to get an output stationary dataflow. The input pixels and the weights can then be read from the GLB and moved to the PEs array in different ways. It is, for example, possible to broadcast the input pixels to all the PEs and to forward the weights, as shown in Figure~\ref{fig:output_stat_broadin}.

\begin{figure}[h]
    \centering
    \includegraphics[width=0.35\linewidth]{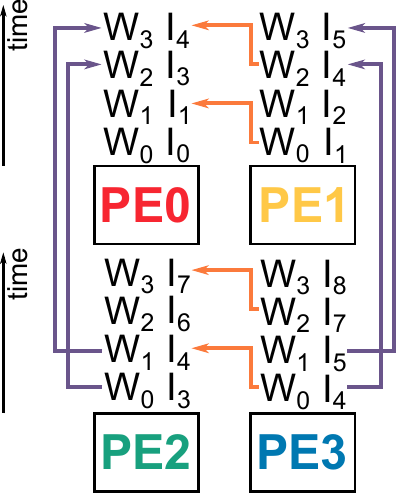}
    \caption{Data movements in ShiDianNao accelerator~\cite{7284058}.}
    \label{fig:shidiannao}
\end{figure}

A popular accelerator that adopts an output stationary dataflow with $H_o$|$W_o$ spatial unrolling is ShiDianNao~\cite{7284058}. Being an output stationary dataflow, each PE in the 2D grid of ShiDianNao processes a pixel of an output feature map, and all the results are then collected and stored in the global buffer. A single weight is broadcasted to all the PEs at every operation cycle. The PEs can read the input pixel either from the GLB, from their right neighbor or their lower neighbor. The PEs have a RF for the partial sum accumulation and two FIFOs to store input pixels for inter-PEs communication. Figure~\ref{fig:shidiannao} schematizes data movement in ShiDianNao for a 2$\times$2 array of PEs. 

\begin{figure}[h]
    \centering
    \includegraphics[width=0.75\linewidth]{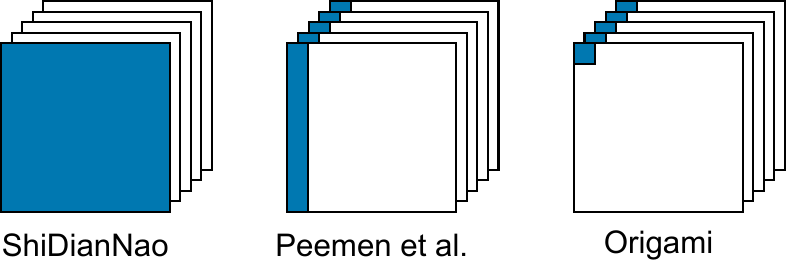}
    \caption{Different solutions~\cite{7284058}\cite{peemen2013}\cite{origami} to spatially unroll the computations in an output stationary dataflow.}
    \label{fig:out_stat_compare}
\end{figure}

Computing the output pixels in parallel along the dimensions $H_o$ and $W_o$ is not the only possible solution to get an output stationary dataflow. Origami accelerator~\cite{origami}, for example, spatially unrolls three loops ($H_k$, $W_k$ and $C_o$) and computes all the pixels along the output channel $C_o$ in parallel, dedicating an accumulator to each one. In a compromise between~\cite{7284058} and~\cite{origami}, in~\cite{peemen2013} the output pixels along dimensions $H_o$ and $C_o$ are computed in parallel. Figure~\ref{fig:out_stat_compare} graphically shows how~\cite{7284058},~\cite{peemen2013} and~\cite{origami} spatially unroll the computation of the output pixels. 

It is important to notice that spatially unrolling the dimensions $C_i$ and $C_o$ can either lead to a weight stationary or an output stationary dataflow. Beyond what data is kept stationary, $C_i$|$C_o$ dataflow is very common because it performs a vector-matrix or matrix-matrix multiplication, and therefore, it allows to easily map both a convolutional and a fully-connected layer to the same array of PEs.

\textbf{Row Stationary:} The row stationary dataflow is introduced in~\cite{chen_dataflows} and used by the Eyeriss accelerator~\cite{eyeriss}. It has the purpose of maximizing the reuse of inputs, weights and partial sums all together, in contrast to weight and output stationary dataflows that focus on a single type of data reuse.  

\begin{figure}[h]
    \centering
    \includegraphics[width=\linewidth]{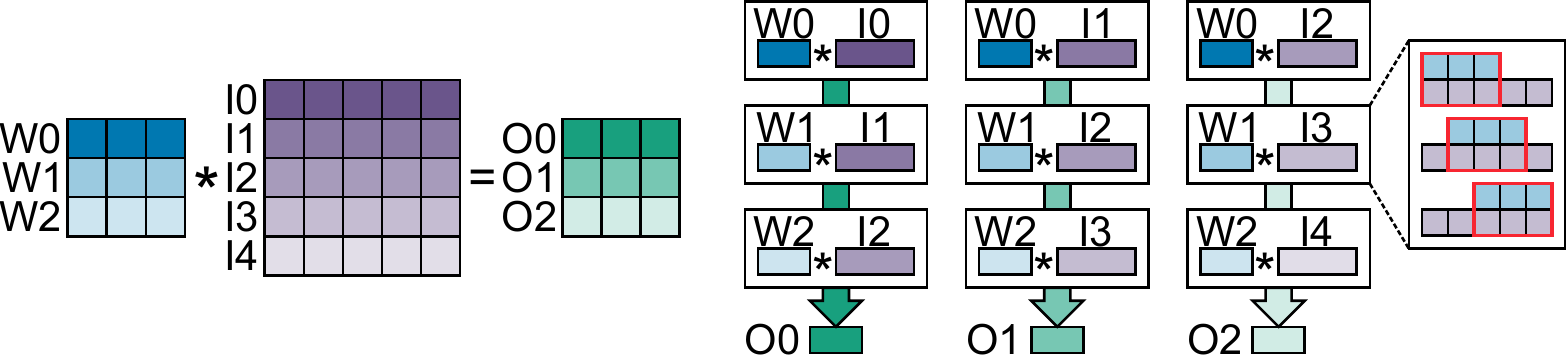}
    \caption{Operations mapping in a row stationary dataflow.}
    \label{fig:row_stat}
\end{figure}

In the row stationary dataflow all the MACs necessary to perform a row of the convolution (1D convolution) are mapped to a single PE. A PE has a RF to keep stationary a row of the weight kernel while the inputs are streamed in the PE exploiting the sliding window mechanism. To perform a whole 2D convolution, it is necessary to have a 2D array of $H_k \times H_o$ PEs. Each column of the array computes the $H_k \times W_o$ partial sums that contribute to a row of the output feature map, that are therefore accumulated. Figure~\ref{fig:row_stat} shows how a 2D convolution with a 3$\times$3 weight kernel is mapped to a row stationary dataflow, and how the partial sums are accumulated along the columns of the PEs array. 

From Figure~\ref{fig:row_stat} it is also possible to see the different types of reuse obtained by the row stationary dataflow. The optimization of data reuse is multi-objective, i.e., a row of PEs shares the same weights, the input pixels are diagonally reused, and the partial sums are vertically accumulated. 

\textbf{No Local Reuse.} The memory elements with higher energy efficiency are those with a low storage capacity, but they are less efficient in terms of area occupation (area/bit). Therefore, a RF has a higher area/bit compared to a scratchpad memory or a SRAM. The no local reuse dataflow maximizes the area dedicated to storage by removing register files from the PEs and allocating all the on-chip memory in the global buffer. Having no local reuse in the PEs, the traffic from and to the GLB on the NoC will be higher. 

\begin{figure}[h]
    \centering
    \includegraphics[scale=0.9]{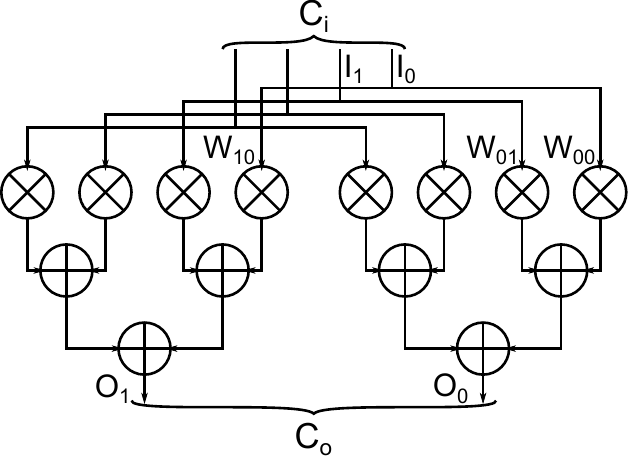}
    \caption{No local reuse dataflow.}
    \label{fig:nlr}
\end{figure}

Which dimension is spatially unrolled on the PEs is not relevant for the no local reuse dataflow. Two accelerators that adopt this dataflow are~\cite{zhang2015},~\cite{diannao} and~\cite{7011421}, which execute the loops along the dimensions $C_i$ and $C_o$ in parallel. In~\cite{zhang2015}, $C_i \times C_o$ multipliers are allocated to multiply the inputs and the weights, and the $C_o$ outputs are then computed with adder trees. An input pixel is multicasted to $C_o$ multipliers (see Figure~\ref{fig:nlr}), while each multiplier reads a different weight from the global buffer. 

\bigskip 

\begin{figure}[ht]
    \centering
    \includegraphics[width=\linewidth]{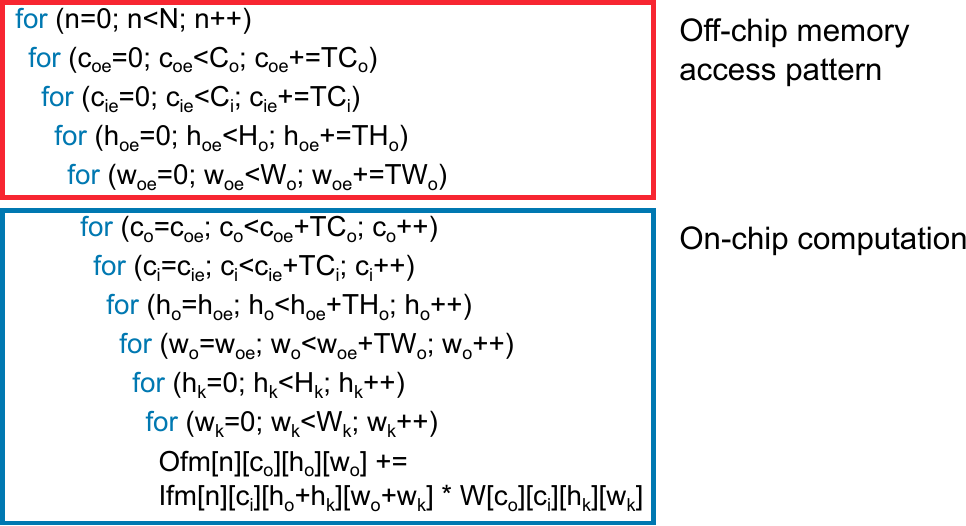}
    \caption{Loop tiling technique applied to the 7-nested loops representation of the Conv layer. The outermost loops describe off-chip memory accesses, and the innermost loops determine the dataflow.}
    \label{fig:tiling}
\end{figure}

A critical aspect of the dataflow definition and accelerator design has not yet been mentioned. Usually, the global buffer size is not sufficient to fully contain the input feature maps, kernel weights and output feature maps. For this reason, it is necessary to apply the \textit{loop tiling} technique, which consists of partitioning the larger tensors into smaller tensors that can be contained in the buffer. The \textbf{for} loops of the 7-nested loops representation of the convolutional layer are therefore split into multiple loops, as shown in Figure~\ref{fig:tiling}. The tiling factors ($TC_o$, $TC_i$, $TH_o$, $TW_o$) define the size of the innermost loops and consequently of the global buffer size. In contrast, the permutations of the outermost loops determine the off-chip memory accesses and how the data are reused. 

\medskip 
Due to the wide variety of layer types and sizes in DNNs models, recently the reconfigurable accelerators that allow to efficiently map different types of layers on the same hardware have gained importance. For example, in~\cite{hybridnn}, there are two 16x16 arrays, whose PEs are divided into general PEs and super PEs. The former are used to map the Conv and FC layers, while the latter are used for the activations functions, Pooling layers, and RNN layers. The arrays can also be partitioned to process multiple layers in parallel and the accelerator supports 8- or 16-bit operations. Another example of a reconfigurable accelerator is the NPU that is at the heart of Project Brainwave~\cite{project_brainwave}, the real-time AI FPGA used in Microsoft's servers. The Project Brainwave NPU is a spatially distributed architecture with efficient matrix-vectors multipliers for the operations between tensors and multifunction units that implement a wide variety of functions. MAERI~\cite{maeri} obtains reconfigurability through the interconnections. The multipliers are arranged in a 1D structure and the inputs are delivered with a flexible distribution network that can be set to implement different dataflows. Similarly, the outputs of the multipliers are collected with an Augmented Reduction Tree of adders. A similar approach is adopted in SIGMA~\cite{sigma}, in which the flexible distribution and reduction networks allow to perform vector dot-products of different sizes simultaneously. Cerebras Wafer Scale Engine is the largest chip ever built, and it is optimized for DL applications. The engine consists of a large amount of flexible cores that target tensor operations but support general operations too. The memory has a high capacity, in the order of gigabytes, and is distributed on-chip. Huawei has released the DaVinci AI core \cite{davinci}, which is completely high-level programmable and consists of a vector engine and a 3D Cube
engine for matrix computations. Two or more DaVinci cores can be combined to work in parallel, as in the Huawei Ascend 910 and 310 AI processors.

\subsection{Tools for design space exploration (DSE)}
From the analysis of possible architectures and dataflows discussed in the previous section it can be understood that many aspects have to be considered during the design of an accelerator, such as, architectural parameters, memory hierarchy, spatial and temporal mapping, and tiling factors. Exploring the whole space of possible designs is a tough task (even an NP-hard problem considering a wide range of design points), especially if the target platform of the accelerator is an ASIC, whose development cost is high in terms of cost and time. For this reason, many researchers have been focusing on the development of tools and frameworks for efficient design space exploration (see Table~\ref{tab:dse}).

\begin{table*}[]
\caption{Comparison of the tools for Design Space Exploration}
\label{tab:dse}
\resizebox{\linewidth}{!}{
\begin{tabular}{|l|l|l|l|l|}
\hline
                     \textbf{Name}                          & \textbf{Year} & \textbf{Motivation}                                                                                                    & \textbf{Target}                                     & \textbf{Characteristics}                                                                                                                                         \\ \hline
\multicolumn{1}{|l|}{Peemen et al.~\cite{peemen2013}}  & 2013          & Minimize off-chip memory accesses                                                                                      & MEM                                                 & DSE of all schedules, given on-chip buffers sizes                                                                                                                \\ \hline
\multicolumn{1}{|l|}{Pouchet et al.~\cite{pochet2013}} & 2013          & Minimize off-chip memory accesses                                                                                      & MEM                                                 & Local memory promotion                                                                                                                                           \\ \hline
\multicolumn{1}{|l|}{Yang et al.~\cite{yang2016}}    & 2016          & Optimize scheduling given multi-level memory hierarchy                                                                 & MEM                                                 & Systematic approach (i.e., iterative optimization) to loop blocking                                                                                              \\ \hline
\multicolumn{1}{|l|}{SmartShuttle~\cite{smartshuttle}}   & 2018          & Minimize off-chip memory accesses                                                                                      & MEM                                                 & Adapt tiling factors and data reuse to the size of the convolution                                                                                               \\ \hline
\multicolumn{1}{|l|}{NNest~\cite{nnest}}          & 2018          & \begin{tabular}[c]{@{}l@{}}Optimize memory hierarchy, memory accesses and \\ computational resources (CR)\end{tabular} & \begin{tabular}[c]{@{}l@{}}MEM + \\ CR\end{tabular} & \begin{tabular}[c]{@{}l@{}}Given the tiling factors, optimization of the computational resources, \\ of the on-chip buffers and of the schedule\end{tabular}     \\ \hline
\multicolumn{1}{|l|}{RomaNet~\cite{romanet_shafique}}        & 2019          & Minimize off-chip memory accesses                                                                                      & MEM                                                 & Given a NN layer, optimization of tiling and partitioning                                                                                                        \\ \hline
\multicolumn{1}{|l|}{MAESTRO~\cite{maestro}}        & 2019          & Analytical cost model used for DSE                                                                                     & CR                                                  & \begin{tabular}[c]{@{}l@{}}Given a DNN, a HW configurations and a schedule, estimation of \\ energy, execution time and NoC\end{tabular}                         \\ \hline
\multicolumn{1}{|l|}{mRNA~\cite{mrna}}           & 2019          & Adapt scheduling to computational resources                                                                            & CR                                                  & Optimize the scheduling targeting MAERI accelerator and its resources                                                                                            \\ \hline
\multicolumn{1}{|l|}{Timeloop~\cite{timeloop}}       & 2019          & Data mapping exploration and evaluation                                                                                & \begin{tabular}[c]{@{}l@{}}MEM + \\ CR\end{tabular} & \begin{tabular}[c]{@{}l@{}}Given an hardware accelerator, explore all the possible data mappings\\ and evaluate them\end{tabular}                                \\ \hline
\multicolumn{1}{|l|}{MAGNet~\cite{magnet}}         & 2019          & Hardware and mapping generation and optimization                                                                       & \begin{tabular}[c]{@{}l@{}}MEM +\\ CR\end{tabular}  & \begin{tabular}[c]{@{}l@{}}Given HW and performances constraint, generate a HW accelerator and \\ a valid mapping; use DSE to co-optimize HW and SW\end{tabular} \\ \hline
\multicolumn{1}{|l|}{XploreDL~\cite{Colucci2020XploreDL}}       & 2020          & DSE of hardware configurations                                                                                         & \begin{tabular}[c]{@{}l@{}}MEM + \\ CR\end{tabular} & Accurate HW simulation and evaluation to perform the DSE                                                                                                         \\ \hline
\multicolumn{1}{|l|}{SuperSlash~\cite{SuperSlash_shafique}}     & 2020          & Minimize off-chip memory accesses                                                                                       & MEM                                                 & Integration of pruning with existing DSE techniques                                                                                                              \\ \hline
\end{tabular}}
\end{table*}

Peemen et al.~\cite{peemen2013} proposed a design flow that selects the best computation schedules to maximize data reuse for a determined on-chip buffer size, exploiting loop reordering and tiling. The whole design space is explored to find the optimized schedule that minimized off-chip memory accesses, discarding the configurations that do not satisfy the memory size requirement.

In~\cite{zhang2015}, loop tiling is realized so that the innermost loops represent on-chip computation and the outermost loops the off-chip memory accesses, as in Figure~\ref{fig:tiling}. Local memory promotion~\cite{pochet2013} is then used to eliminate redundant memory accesses. If one among the input feature maps, output feature maps or weights is not addressed by the index of the innermost off-chip loop ($w_{oe}$ in Figure~\ref{fig:tiling}), then it is reused for all its iterations. Hence, there is no need for continuously loading and storing back the reused tensor. The operations of load and store can consequently be moved out of the innermost loop. A polyhedral-based optimization is used to identify all the possible combinations of loop schedules and tiling factors, and local memory promotion is applied whenever possible. The computational roof and the computation to communication ratio is calculated for each solution to identify the optimal one.   

In~\cite{yang2016}, Yang et al. showed a systematic approach to loop blocking. Given a memory hierarchy, the systematic approach consists of applying a \textit{loop blocking} (i.e., loop tiling and loop reordering) for each level of the memory hierarchy. Exploring the design space for a multi-level memory hierarchy with the proposed methodology is computationally expensive. Therefore, Yang et al. proposed an iterative optimization where the loop blocking is applied to two levels of memory at a time, starting from the lowest level to the highest and re-adjusting the lower levels parameters at each iteration. 

SmartShuttle~\cite{smartshuttle} is a framework that focuses on optimizing off-chip memory accesses exploring the possible loop schedules, that influence the data reuse, and the tiling factors. In~\cite{smartshuttle}, it is noted that convolutional layers with different shapes may benefit from different types of data reuse and various tiling factors. SmartShuttle therefore adaptively varies the ordering and tiling of the loops to match different convolutional layers dynamically. 

NNest~\cite{nnest} is a design space exploration tool for inference accelerators that focuses on the optimization of the memory hierarchy, of the memory accesses and the computational resources too, covering all the main aspects of an accelerator design. In~\cite{nnest} it is proposed a spatial accelerator architecture template that is parametrized, with the possibility of setting the tiling factors, that directly define the size of the on-chip buffers, the size of the PEs array and the possibility of implementing different dataflows and reuse schemes. NNest explores the whole design space and finds the Pareto-optimal solutions for a NN layer. It also allows for a multi-layer fitting.

In ROMANet~\cite{romanet_shafique}, a systematic design space exploration methodology is proposed for reducing the number of memory accesses required for DNN inference. For each layer, an efficient data partitioning and scheduling is designed, based on the available on-chip memory and data reuse factors. Moreover, the proposed DRAM data mapping reduces the number of DRAM row buffer conflicts, while improving the system throughput, compared to a conventional DRAM design.

MAESTRO (Modeling Accelerator Efficiency via Spatio-Temporal Reuse and Occupancy)~\cite{maestro} is an analytical cost model that estimates the execution time, energy and NoC of a hardware configuration applied to a DNN model with a specific dataflow. MAESTRO is a cost model precise and efficient enough to be used for design space exploration, and can be used to determine Pareto-optimal architectural parameters given area, energy or throughput constraints.   

mRNA~\cite{mrna} is a mapper that performs design space exploration to find the optimal mapping targeting the re-configurable DNN accelerator MAERI~\cite{maeri}. Similarly to other design space exploration tools, it explores all the possible permutations of the \textbf{for} loops of the Conv layer 7-nested loop representation and all the possible combinations of tiling factors. Given the high dimensionality of the design space, mRNA reduces it by applying constraints based on domain knowledge, for example setting the tile sizes as multiples of the number of multipliers contained in MAERI to maximize resource utilization. mRNA experiments confirm that dataflows that maximize the usage of available PEs have a shorter runtime and that exploiting data reuse and broadcast/multicast reduces the energy cost. 

Timeloop~\cite{timeloop} is a framework for the exploration of the design space of DNN hardware accelerators and for the evaluation of their performance and energy consumption to make the design more systematic. The users can describe an architectural model following a configurable template and, given a workload, a mapper within Timeloop systematically constructs the map space to be explored and evaluates every possible mapping with its performance, area and cost models. 

MAGNet~\cite{magnet} is a Modular Accelerator Generator for Neural Networks that consists of the following three modules. (1) A MAGNet Designer, that, given a neural network model, hardware constraints and performance goals, generates an accelerator based on a parametric template. (2) A MAGNet mapper that generates a valid mapping of the operations on the accelerator, defining the tiling factors, the spatial and temporal mapping. (3) A MAGNet tuner that uses Bayesian optimization to efficiently explore the design space for hardware-software co-optimizations. 

XploreDL~\cite{Colucci2020XploreDL} is a design space exploration tool for both training and inference accelerators. The tool can be employed in an early stage of the design, because it estimates in a fast yet fidelitous way the Pareto-optimal solutions for specialized accelerators executing CNNs and Capsule Networks, given as optimization objectives the energy-efficiency and the performance-per-area.

Since different level of DNN compression show different on-chip memory accesses, depending upon the pruning strategy, SuperSlash~\cite{SuperSlash_shafique} integrates the pruning techniques with existing design space exploration methodologies, evaluating multiple data reuse strategies for each layer. For instance, the off-chip memory access volume can be reduced by directly using a layer's output as the input for the processing of the subsequent layer.

\subsection{Hardware-Aware Neural Architecture Search}
\begin{figure}[h]
    \centering
    \includegraphics[width=\linewidth]{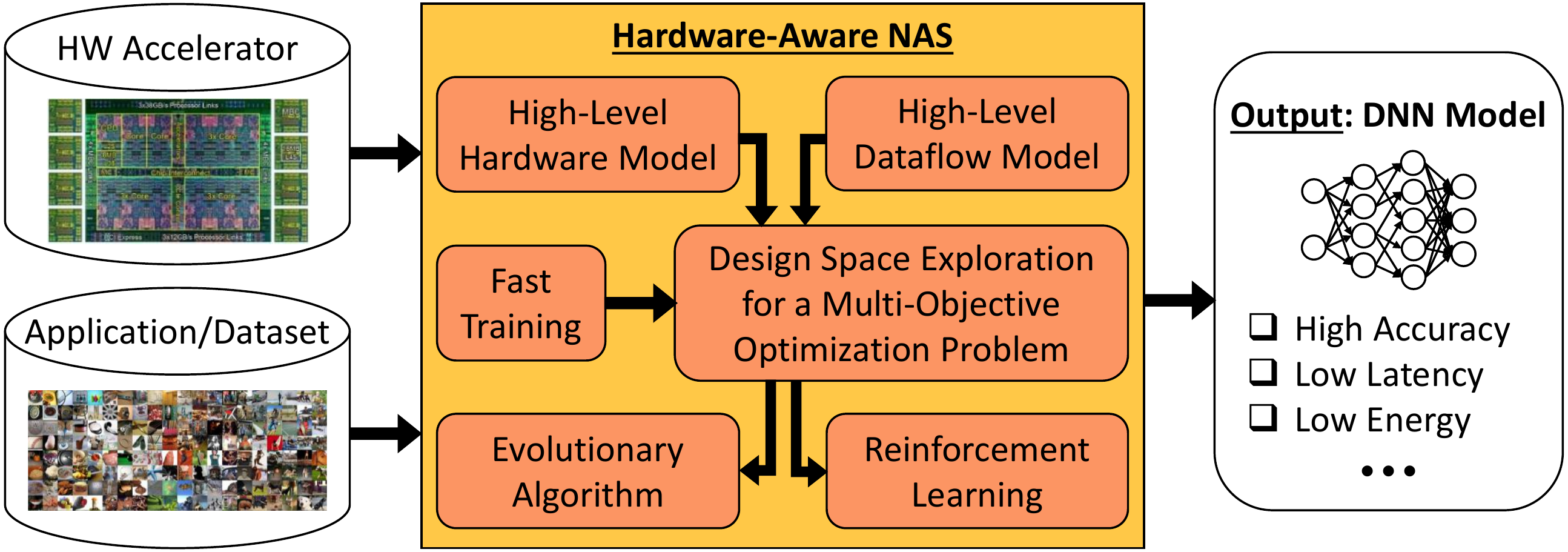}
    \caption{An overview of the hardware-aware NAS problem.}
    \label{fig:hardware_aware_NAS}
\end{figure}

Another interesting design strategy is to customize the DNN based on the underlying hardware. The optimization goal is then to jointly optimize the accuracy and the energy-efficiency, given the underlying hardware (e.g., an accelerator) and the dataset for the target application, as shown in Figure~\ref{fig:hardware_aware_NAS}.

One of the biggest challenges is caused by the explosion of the exploration time and space, when all the hyper-parameters of the DNN are considered. To overcome such a problem, a fast yet accurate evaluation of the energy consumption and performance of the hardware is key. Therefore, a high-level modeling of the scheduling and dataflow, as discussed in the previous sections, is required. Moreover, a smart search is typically employed to speedup the exploration convergence. In the literature, there exist mainly three types of heuristic search algorithms for the hardware-aware neural architecture search, which are (1) evolutionary algorithms, (2) reinforcement learning, and (3) differentiable NAS.

The ProxylessNAS~\cite{Cai2019ProxylessNAS} can reduce the computational demand of the search by executing partial tasks, such as training on a smaller dataset, or learning with only a few blocks, or training just for a few epochs. Afterwards, the framework can directly learn the architectures for the complete tasks and the target hardware platforms. 
The MnasNet approach~\cite{Tan2019MnasNetPN} directly implements and measures the inference latency by executing the model on mobile phones, and incorporates the model latency into the main objective of the search, along with the accuracy. 
In~\cite{Zeng2020BlackBoxNAS}, the authors proposed a black-box profiling-based search in the first stage of the accelerator-aware NAS pipeline using an ISA-based DNN accelerator on FPGA, with a particular focus on the accurate latency evaluation. 
The NASCaps~\cite{Marchisio2020NASCaps} is a framework integrating capsule layers in the search space. With a multi-objective evolutionary algorithm, it jointly optimizes the accuracy and the hardware efficiency of capsule-based DNNs. 
In~\cite{Lu2019ResourceConstrainedNAS}, the authors developed a NAS framework which integrates the quantization and hardware implementation in the design flow. 

The APNAS~\cite{Achararit2020APNAS} is a reinforcement learning-based exploration methodology, searching for high accurate DNNs that also offer high execution performance. To speed-up the search, instead of running millions of DNN configurations on real hardware, the cycle count is estimated by analytical models. 
The FNAS framework~\cite{Jiang2019FPGA-NAS} performs a hardware-aware NAS targeting FPGA acceleration. In particular, it employs an abstraction model to estimate the latency for meeting the specifications. Moreover, a specialized scheduling mechanism is proposed to execute the DNN inference on multiple FPGAs. 
The HotNAS~\cite{Jiang2020HotNAS} is a hardware and neural architecture co-search methodology, which starts the exploration from a set of existing pre-trained models to reduce the training time. In addition, it supports hardware for compressed DNNs and it integrates the compression in the co-search to improve the energy-efficiency. 
With the ENAS approach~\cite{Pham2018ENAS}, the authors proposed to share the parameters between the child DNN models. It allows not only to speedup the search, but also to achieve high accuracy, with similar benefits as for the transfer learning~\cite{Zoph2016TransferLearning}. 

The Single-Path NAS~\cite{Stamoulis2019SinglePathNAS} is a method searching for the optimal building block for the convolutional layers, called superkernel, and then sharing the convolutional kernel weights with a specialized encoding. 
The DNAS~\cite{Wu2019FBNet} is a differentiable NAS framework, where the search space is represented by a stochastic super net. It explores a layer-wise space where each layer of the CNN can choose a different block, and the learning process is done by training the stochastic super net. The SPOS~\cite{Guo2019SPOS} uses in a similar way the supernet concept to perform NAS, where the constraints such as latency and number of FLOPs are applied. 
The HURRICANE framework~\cite{Zhang2020FastHANAS} performs a two-stage search algorithm for the automatic hardware-aware NAS. It can generate different models for different types of hardware platforms for executing the inference. 
In~\cite{Li2019RandomSearchNAS}, the authors demonstrate that competitive results for the NAS can be achieved by using random search. This approach significantly reduces the complexity, compared to other search methods.

\subsection{Full precision vs quantized implementations} 
As discussed in the previous sections, one of the main obstacles to the deployment of DNNs on edge devices is their large memory footprint, the high energy cost of memory accesses, and the energy required for computations. 

One of the most popular methods for reducing memory and computation requirements is quantization. Quantization is the process of mapping values from a continuous or large set to a discrete and smaller set by applying a function that can be either linear or non-linear. The difference between the quantized value $x_q$ and the original value $x$ is the quantization error $e_q$ (Eq. \ref{eq:quant_err}). 

\begin{equation}
    \label{eq:quant_err}
    e_q = x_q - x
\end{equation}

\begin{figure}[h]
    \centering
    \includegraphics[width=\linewidth]{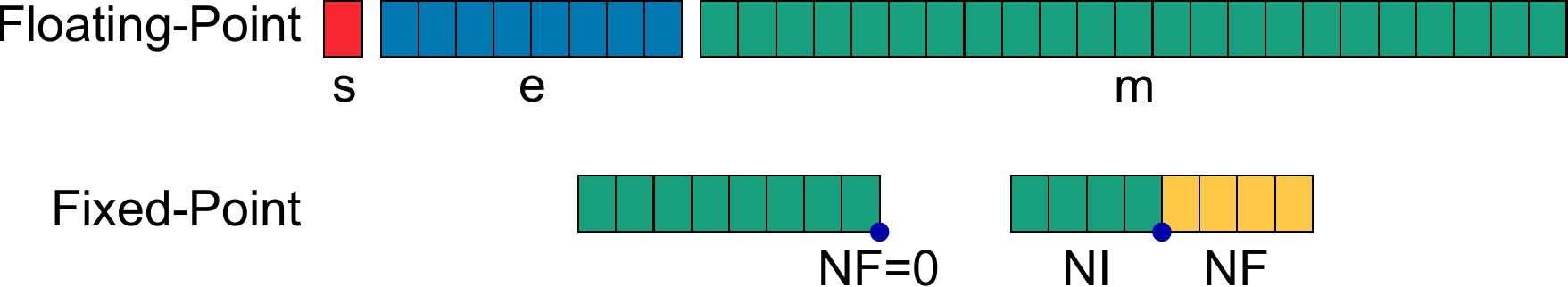}
    \caption{\textbf{(top)} 32-bit floating-point representation. \textbf{(bottom)} 8-bit dynamic fixed-point representation.}
    \label{fig:floating_fixed}
\end{figure}

From a hardware perspective, quantization reduces the precision of the values, and consequently, the number of bits necessary to represent them. It is, therefore, possible to move from the floating-point representation to a shorter fixed-point representation (see Figure~\ref{fig:floating_fixed}). According with the IEEE 754 standard, in 32-bit floating-point representation, one bit expresses the sign $s$ of the number, 8 bits represent the exponent $e$ and 23 bits the mantissa $m$. The value of the number is $(-1)^s \cdot m \cdot 2^{e-127}$ and can be in the range of $10^{-38}$ to $10^{38}$.  An N-bit fixed-point number in two's complement has an integer part of $NI$ bits and a fractional part of $NF$ bits. The width $NF$ of the fractional part can be seen as a \textit{scale factor} that determines the position of the decimal point. Numbers can be in the range $[-2^{NI-1}, 2^{NI-1}-2^{-NF}]$ and the quantization step is $2^{-NF}$.

The scale factor $NF$ can be varied to have different ranges and different precision, making the fixed-point representation \textit{dynamic}. This is particularly useful for neural networks, as weights and activations fall in very different numerical ranges depending on the layer. 

The fixed-point representation allows for memory and energy saving, e.g., a MAC performed on an 8-bit fixed-point number consumes 20x lower energy than a MAC on a 32-bit floating-point number~\cite{horowitz2014}. Moreover, a number expressed on 8 bits has a memory footprint of 4x smaller than one on 32 bits. This allows us to understand the large potential of gain, in terms of energy and memory, that can be achieved through quantization of data. 

The purpose of quantizing the neural networks is to reduce the size of the models, obtaining a lower memory footprint and at the same time, a lower energy cost for both the computations and memory accesses. However, quantization carefully must be applied without reducing the accuracy of the models. 

In NNs, there are three sets of values that can be quantized: the weights, the activations and the gradients. 
Earlier works on quantized NNs focused on the weights only since they directly affect the memory requirements~\cite{binaryconnect}\cite{ternaryweightnet}. While the activations must be quantized at each execution of the algorithm, the weights can only be quantized once off-line after the training. This has two advantages: 
\begin{itemize}[leftmargin=*]
    \item The quantized weights can be further fine-tuned to recover a possible accuracy degradation following the precision reduction. 
    \item Since the weights are quantized offline, it is possible to apply complex quantization functions or stochastic functions, without affecting the computational resource required on-chip. 
\end{itemize}

Recently researchers have started to focus on the quantization of activations too~\cite{jacob2018}\cite{vanhouke2011}\cite{quantizedneuralnetworks}, that affect the memory footprint and bandwidth depending on how the dataflow is implemented, as well as directly affecting the required computational resources. 

The study of gradient quantization is limited, mainly for two reasons: 
\begin{itemize}[leftmargin=*]
    \item The training of a NN is very sensitive to even small variations in weight values, and there is a risk of not achieving convergence. Therefore quantizing the weights is complex.
    \item Usually the training is done only once offline on a GPU or a CPU, and not on the edge devices, so there is no reason to devote too much effort to reduce the size of the model and to reduce the energy consumption.
\end{itemize}

\begin{figure}[ht]
    \centering
    \includegraphics[width=\linewidth]{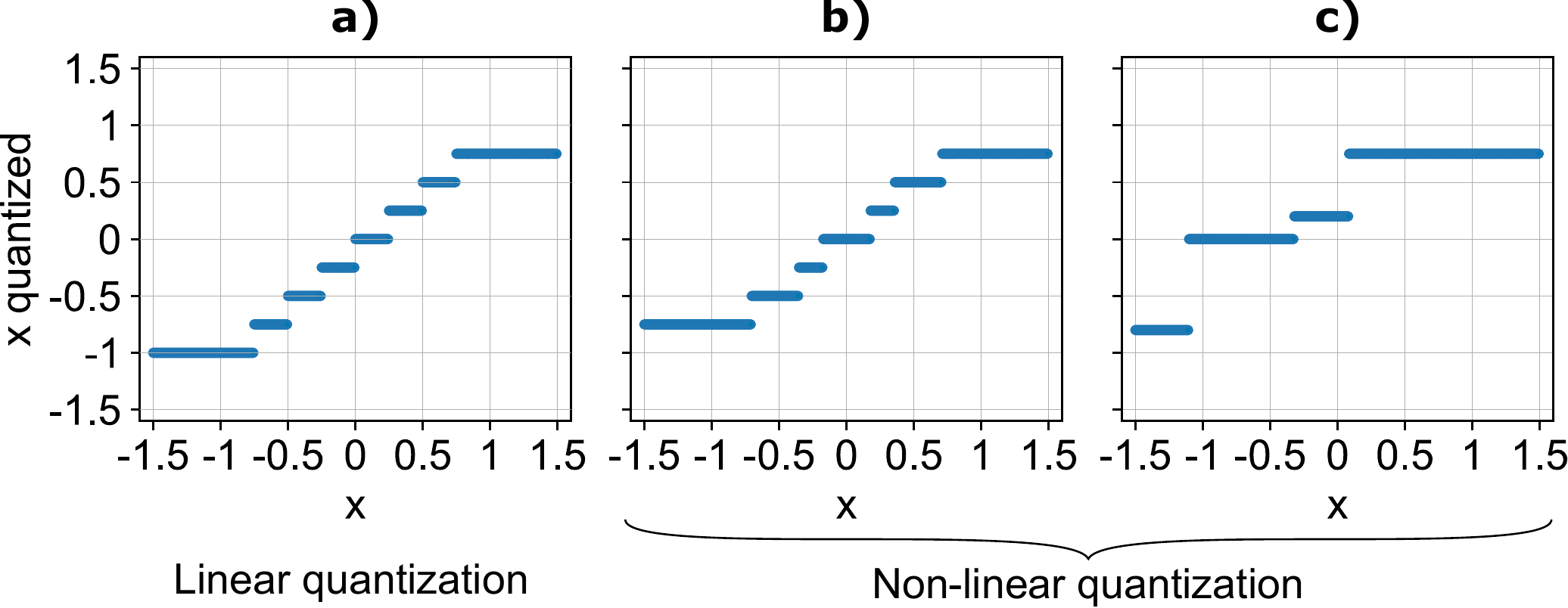}
    \caption{Linear, logarithmic and vector quantization techniques.}
    \label{fig:quantization}
\end{figure}

Several studies have been made on quantization methods~\cite{Guo2018}. In the following, we will provide an overview of hardware-friendly quantization methods, distinguishing between linear and non-linear methods (see Figure~\ref{fig:quantization}). 

\textbf{Linear Quantization:} It is characterized by evenly-spaced quantization intervals, as shown in Figure~\ref{fig:quantization}.a. An example of linear quantization is the above-discussed fixed-point coding, which has been widely studied and applied to NNs because its hardware implementation is well known. Moreover, most CPUs support fixed-point arithmetic on 8, 16, or 32 bits. Given the strong diffusion that the quantization of NNs is having, the Nvidia Tesla GPU supports 8-bit fixed-point operations, and so do the Tensor Processing Units (TPUs)~\cite{googletpu} used in Google datacenters. 

It has been demonstrated by several works that both the weights and activations can be quantized to 8-bit dynamic fixed-point numbers for inference without significantly affecting the accuracy~\cite{gysel2016}~\cite{jacob2018}. The Ristretto framework~\cite{gysel2016} identifies the quantization parameters (bitwidth and scale factor) by running a statistical analysis on the weights and activations. The weights are furtherly fine-tuned with a re-training step. With the Ristretto framework, complex models such as AlexNet~\cite{Krizhevsky2012}, SqueezeNet~\cite{squeezenet} or GoogleNet~\cite{Szegedy2015} are inferred on 8 bits with less than 1\% accuracy loss. In~\cite{vanhouke2011}, an NN for speech recognition is implemented on 8-bit fixed-point numbers exploiting the Intel SSSE3 instruction set for SIMD execution. A speed-up of 7.6x is achieved compared to the floating-point baseline. 

Given the great diversity between the various layers of an NN, it may be useful to use a different precision across the model, i.e., a variable bitwidth depending on the layer. Works in~\cite{Sankaradas2009} and~\cite{sakr2019} show that the bitwidth can be set depending on the position in the model, making a per-layer optimization of the number of bits of weights and activations. In particular, in~\cite{sakr2019}, it is stated that the bitwidth used for the weights can decrease approaching the last layers of the NN, while the bitwidth of the activations remains more or less constant. Following these ideas, Q-CapsNets~\cite{qcapsnets_shafique} analyzes the layer-wise quantization capabilities of weights and activations of CapsNets, with a cross-layer optimization of the bitwidth and a fine-grained tuning for the dynamic routing operations. Finding the optimal bitwidth for each layer of a DNN is a complex task. For this purpose, HAQ, a hardware-aware quantization framework, is introduced in~\cite{haq}. It applies reinforcement learning to determine the optimal bitwidths for weights and activations, using as feedback the results of a hardware simulator. 

The research on fine-grained bitwidth optimization is also backed by the parallel development of hardware accelerators that support flexible bitwidth arithmetic operations. BISMO~\cite{bismo} is a matrix-matrix multiplication core with variable parallelism and precision to adapt to the requirements of different applications. It supports precision from 1 to 8 bits exploiting bit-serial computation.  Stripes~\cite{stripes} is an accelerator for DNNs with flexible bitwidth for the activations that uses bit-serial operations. UNPU~\cite{unpu} has a similar approach, but the bits of the activations are kept constant to 16-bits and the weights have variable bitwidth. Loom~\cite{loom} adopts bit-serial multiplicators and both weights and activations have fully variable bitwidth, from 1-bit to 16-bit. Bit Fusion~\cite{bitfusion} instead implements variable precision operations for DNNs with a spatial approach, using an array of bit-level PEs combined together according to the required bitwidth. BitBlade~\cite{bitblade} is an optimization of Bitfusion, in which bit-wise summations substitutes the shift-add logic. On the industrial front, in 2018 Apple released the A12 Bionic chip with a Neural Processing Unit (NPU) that supports variable precision; Nvidia Turing Tensor Cores, available in the Nvidia Turing architecture~\cite{nvidiaturing}, support operations from 32/16-bit floating-point down to 8/4-bit fixed-point; the Imagination PowerVR Series2NX architecture has adjustable bitwidth from 16 to 4 bits. The above-discussed platforms that provide variable bitwidths are compared in Table~\ref{tab:vb}. 

\begin{table}[h]
\centering
\caption{Comparison of different variable-bitwidth HW platforms.}
\label{tab:vb}
\resizebox{\linewidth}{!}{
\begin{tabular}{|l|c|c|c|c|}
\hline
 \multicolumn{1}{|c|}{\textbf{Name}} & \textbf{Weights} & \textbf{Activations} & \textbf{Features} & \textbf{Target} \\ \hline
BISMO \cite{bismo} & & & & \\ \hline 
Stripes\cite{stripes}        &          16-bit            &            1-bit to 16-bit              & Serial &  ASIC           \\ \hline
UNPU \cite{unpu}       &           1-bit to 16-bit              &           16-bit              & Serial &        ASIC          \\ \hline
Loom   \cite{loom}        &            1-bit to 16-bit          &              1-bit to 16-bit            &        Serial         &      ASIC    \\ \hline
Bit Fusion \cite{bitfusion}         &       1,2,4,8,16-bit               &                1,2,4,8,16-bit          &       Spatial            &        ASIC   \\ \hline
BitBlade  \cite{bitblade}        &            1,2,4,8,16-bit          &            1,2,4,8,16-bit              &          Spatial        &        ASIC       \\ \hline
Turing TC \cite{nvidiaturing} & 64,32,16,8,4-bit & 64,32,16,8,4-bit & & GPU \\ \hline 
PowerVR S2NX & 16,8,4-bit & 16,8,4-bit & & SoC \\ \hline 
\end{tabular}}
\end{table}

Both weights and activations can be quantized to very low bitwidths. BinaryConnect (BC)~\cite{binaryconnect} introduced the idea of binary weights, included in $\{-1, 1\}$. Binary Weight Nets (BWN)~\cite{xnornet} approximate a filter $W$ as $\alpha B$, where $B$ is a filter whose values are binary, and $\alpha$ is a scale factor. The operations are performed between full-precision inputs and binary weights, and the output is then multiplied by $\alpha$. In Ternary Weight Nets (TWN)~\cite{ternaryweightnet} the same approach is adopted but the weights are ternary, i.e., in the set $\{-1, 0, 1\}$.

Quantized Neural Networks~\cite{quantizedneuralnetworks} and DoReFa-Net~\cite{dorefa} have an even more aggressive approach using binary weights and 2-bit activations. Finally, Binarized Neural Networks (BNN)~\cite{binarizedneuralnetworks} and XNOR-Nets~\cite{xnornet} use both binary weights and activations. XNOR-Nets use the same approach as BWN and TWN to limit accuracy reduction by multiplying the outputs with a scaling factor. 

Several hardware accelerators have been proposed to support efficient inference of binary NNs: YodaNN~\cite{yodann} and Hyperdrive~\cite{hyperdrive} are accelerators for binary weights only NNs; BRein~\cite{brein}, XNOR Neural Engine~\cite{xnorneuralengine} and XNORBIN~\cite{xnorbin} accelerate NNs with binary weights and activations, while BRein supports ternary weights too. 

\medskip 

\textbf{Non-Linear Quantization:} Weights and activations in an NN usually have non-uniform distributions, so they can benefit from the non-linear quantization, where the quantization intervals are unevenly distributed, as shown in Figure~\ref{fig:quantization}.b and \ref{fig:quantization}.c. 

An example of a non-linear quantization scheme is the logarithmic quantization, first applied to NNs in~\cite{Miyashita2016}. The dot product between a vector of weights $\textbf{w}$ and activations $\textbf{x}$ can be approximated as follows adopting the logarithmic quantization: 

\begin{equation}
\label{eq:dotlog}
    \textbf{w} \cdot \textbf{x} = \sum_{i=0}^{N} w_i x_i \simeq \sum_{i=0}^{N} w_i 2^{\tilde{x_i}} = \sum_{i=0}^{N} w_i \ll \tilde{x_i}
\end{equation}
\begin{equation}
\label{eq:log}
    \tilde{x_i} = Int(log_2(x_i))
\end{equation}

From Eq. \ref{eq:dotlog} and Eq. \ref{eq:log}, we can notice that the multiplications can be substituted with shift operations. With the same bitwidth used, logarithmic quantization reduces the accuracy loss compared to linear quantization. With respect to a floating-point baseline, the accuracy loss of VGG16 with a 3-bit linear quantization is 6.2\%, while with logarithmic quantization it is only 0.6\%. 

In~\cite{lee2017},~\cite{aqss} and~\cite{vogel2018}, NN accelerators with logarithmic numerical representation are presented. They are characterized by processing elements whose multipliers used for MACs are replaced by barrel-shifters. 

Another type of non-linear quantization is \textit{vector quantization}. It consists of applying clustering algorithms to the weights of an NN. The centroids of the clusters to which the weights belong are used as quantization values. For the first time, this method was applied to the quantization of NNs in~\cite{gong2014}. The vector quantization can be applied offline before inference, so it does not need accelerators with specialized architectures to support it. 

\subsection{Methods for model compression}\label{pruning}
As seen in Section~\ref{sec:dnnmodels}, the trend to achieve greater accuracy has been the development of deeper and deeper NNs with a higher number of layers and parameters. This evolution is hardly compatible with the recent desire to deploy NNs on mobile and edge devices. During the last few years, therefore, there has been a big push towards the research of methods to compress the models of NNs without affecting the achieved accuracy~\cite{surveycompression}. The most prominent works are in the domains of network pruning, architectural choices and knowledge distillation, as described in the following paragraphs. 

\medskip
\noindent \textbf{Network Pruning:} Given the redundancy of the parameters in NNs, \textit{network pruning} consists of removing, i.e., set to zero, those parameters that do not affect the performance (i.e., network accuracy) of the model. Pruning was first explored in Optimal Brain Damage~\cite{braindamage}, where the weight with lower influence on the loss function during the training were pruned. A simpler method~\cite{pruning1} consists of pruning the weights with small magnitude after the training and then in performing a fine-tuning of the remaining weights to recover possible accuracy losses. This method, straightforward and linear, allows to reduce the number of parameters in AlexNet, for example, by 10x~\cite{pruning1}. 

Subsequent works have proposed variations of the pruning method in an attempt to obtain a high yet accuracy-wise effective compression of the models. In~\cite{pruning2}, instead of removing individual weights, entire neurons are pruned. In~\cite{pruning3}, full channels are pruned from feature maps by applying a two-step algorithm based on LOSSA regression for channel selection and least square reconstruction.  In~\cite{han2015deep}, Deep Compression is proposed, a three-stage pipeline that applies, in order, pruning, quantization and Huffman coding. PruNet~\cite{prunet_shafique} iteratively applies a magnitude-based Class-Blind pruning followed by weight retraining. In~\cite{pruning4}, the pruning is guided by an estimate of CNN's energy consumption, to optimize the model's energy performance and not just minimize the number of parameters. A similar approach based on energy constraints is adopted in ECC~\cite{ecc}. In~\cite{clipq}, quantization and pruning are performed jointly, and fine-tuning is run in parallel. In AMC~\cite{amc} and~\cite{automl}, learning-based approaches are adopted to prune and quantize the models for algorithm-hardware co-design. In APQ~\cite{apq}, pruning and quantization are optimized jointly with the NN model avoiding any accuracy loss.

\begin{figure}[h]
    \centering
    \includegraphics[width=\linewidth]{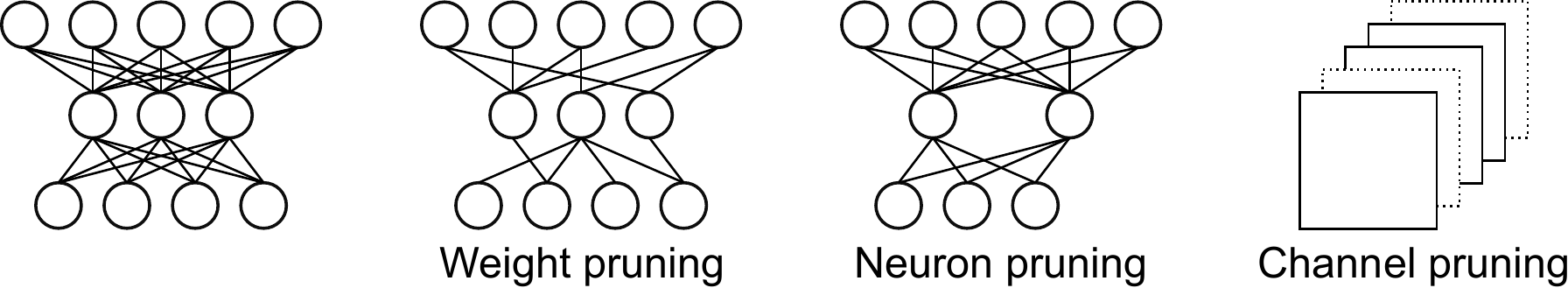}
    \caption{Various pruning techniques.}
    \label{fig:pruning}
\end{figure}

The pruning has the advantage of making the matrices of the weights sparse. Section~\ref{sec:sparsity} explains in detail how it is possible to take advantage of \textit{sparsity} in neural networks.

\begin{figure}[h]
    \centering
    \includegraphics[width=0.8\linewidth]{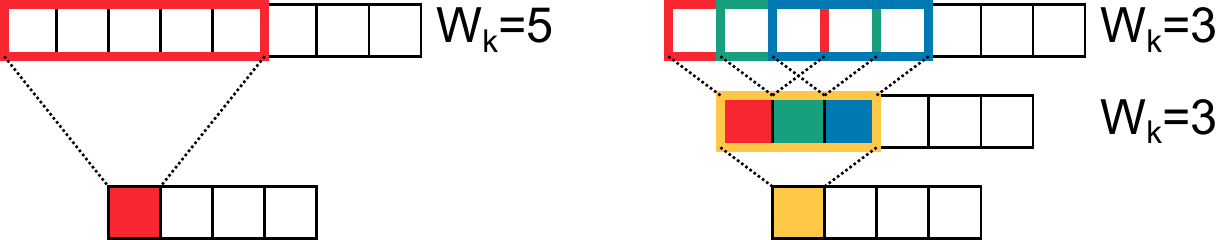}
    \caption{Reduction of the number of parameters by splitting a large kernel in a series of two smaller kernels while maintaining the equivalent receptive field equal.}
    \label{fig:kernel_reduction}
\end{figure}

\medskip
\noindent \textbf{Architectural Choices:} Some researches have explored new architectures with a lower number of parameters by construction. The basic idea is to replace a large kernel with a series of two or more smaller kernels. In this way, an equivalent receptive field is obtained but with fewer parameters. For example, a 5x5 kernel can be replaced by a series of two 3x3 kernels, reducing the number of weights from 25 to 18 (see Figure \ref{fig:kernel_reduction}). In SqueezeNet~\cite{squeezenet}, most of the 3x3 kernels are substituted with 1x1 kernels that have 9x fewer parameters, and the input channels to the 3x3 convolutions are reduced. SqueezeNet achieves the same accuracy of AlexNet with 50x fewer parameters. In MobileNet~\cite{MobileNets}, a standard convolution is divided in a depthwise convolution and a point-wise convolution. The depthwise convolution applies a different kernel to each input channel, while the point-wise convolution uses 1x1 kernels to combine together the output channels of the depthwise convolution. This factorization reduces the number of parameters. Xception~\cite{Xception} adopts this same approach. 

It is also possible to obtain smaller tensors from large tensors after training by applying the Tensor Decomposition, which is a low-rank factorization technique.  The kernels of the convolutional layers are 4D tensors, while the weights of the fully-connected layers are organized in a 2D matrix. With tensor decomposition, these can be broken down into tensors of lower dimensionality by Canonical Polyadic (CP) decomposition~\cite{canonical}. Since CP is not numerically stable for tensors with dimension higher than two, it is possible to adopt Tucker decomposition~\cite{tucker}. 

\medskip
\noindent \textbf{Knowledge Distillation:} Higher accuracies are obtained with very deep models or with \textit{ensembles} of models, whose results are then averaged. Using a deep model or even several models at once requires considerable computational effort. However, it is possible to transfer the knowledge of one or more large models (teachers) into a smaller model (student).  This process is commonly known as \textit{knowledge distillation} and has been introduced in~\cite{kd1} and~\cite{kd2}, for shallow and deep teacher models respectively. In~\cite{kd1} and~\cite{kd2}, the (trained) teacher models receive a dataset of unlabeled data and classify them, producing a synthetically-labelled dataset. This dataset is then used to train the shallow student model, that, therefore, learns to mimic the classifying function of the teachers. The knowledge distillation method has shown promising results and several variations have been proposed in subsequent work~\cite{hinton2015distilling}\cite{romero2014fitnets}\cite{kd3}\cite{kd4}.

\subsection{Activations and Weights Sparsity: Strategies and Encoding}
\label{sec:sparsity}
Recent studies have shown that most DNNs are subject to redundancy concerning the weights. Consequently, it is possible to prune them without affecting the accuracy as demonstrated in~\cite{pruning1} and~\cite{pruning2}. Both works show that the synapses can be reduced to percentages ranging from 20\% to 80\%, depending on the considered layers. As explained in Section~\ref{pruning}, pruning weights results in zero values that make the matrices sparse.
On the other hand, during inference, the ReLU clamps negative activations to zero. The null activations range from about 50\% to 70\%. Hence, these represent the two primary sources for sparse matrices for both activations and weights.
Sparsity represents an excellent opportunity to optimize the inference for two main reasons:
\begin{itemize}[leftmargin=*]
    \item The basic operations of the DL is the multiplication between a weight and an activation. However, whenever one of the two is zero, the operation has no reason to be performed, as the result, will be null too. Therefore, it is possible to skip such operations to speed up execution and save the energy.
    \item By using compression techniques, it is possible to save only the non-null elements and their relative positions in the matrices. This reduces the storage requirements with the possibility of fitting more data into the on-chip SRAM, thereby cutting off the accesses to the off-chip DRAM significantly.
\end{itemize}

\begin{figure*}[ht]
    \centering
    \includegraphics [width=1\linewidth] {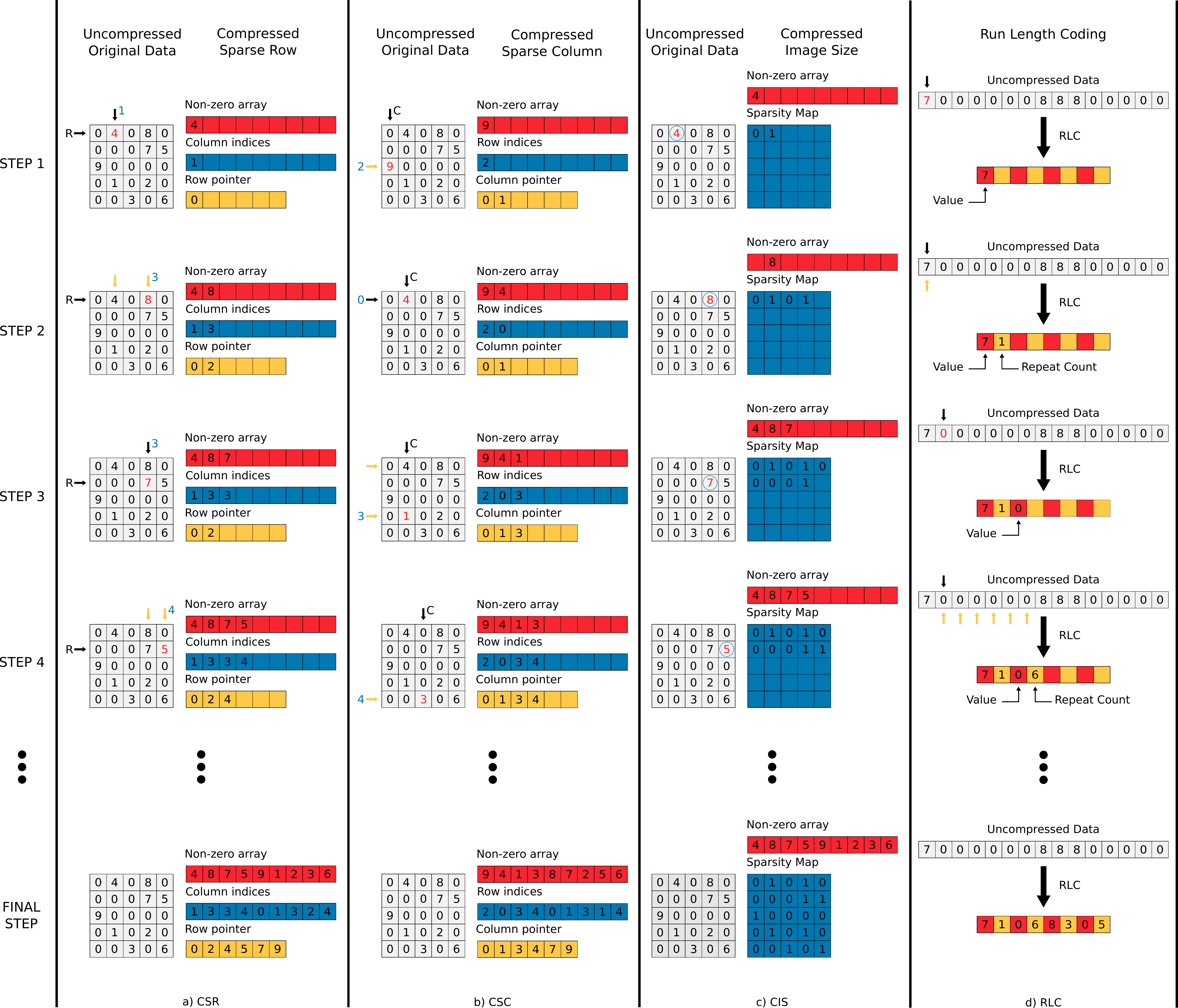}
    \caption{Step-by-step compression formats comparison: (a) Compressed Spare Row (CSR), (b) Compressed Sparse Column (CSC), (c) Compressed Image Size (CIS), (d) Run Length Coding (RLC).}
    \label{fig:comp_form}
\end{figure*}

Although numerous compression techniques can be found in literature, DNNs and CNNs rely mainly on three hardware-friendly methods. Such methods consist of two sets of data: one represents all the non-zero values, while the other represents the metadata or the indices necessary to reconstruct the original pattern. Compressed Sparse Row (CSR) and Compressed Sparse Column (CSC) are two formats belonging to the class of compressed stripe storage~\cite{sparsity_thesis}. Both CSR and CSC can be seen as a collection of scattered vectors, which allows random access to entire rows or columns respectively, equipped with an efficient count of non-zeros within each row or column, as detailed in the following.

\medskip

\noindent\textbf{Compressed Sparse Row (CSR)}: As shown in Figure~\ref{fig:comp_form}(a), a single array (Non-zero array) stores all the non-zero values of the sparse rows in order, and an integer array (Column indices) stores the corresponding column indices. A third array (Row pointer) stores the offsets within the previous two vectors, indicating the number of non-null elements per row in an incremental fashion. Such a structure allows fetching any row thanks to an efficient element enumeration. The number of bits required for such a representation is given as:
\begin{equation}
    I\cdot(Nb\cdot(1-Sp)+Ni\cdot(1-Sp))+(H+1)\cdot No 
\end{equation}
where $I$ is the input size, $Sp$ the sparsity percentage, $H$ the height of the input matrix, and $Nb$, $Ni$ and $No$ are the bitwidths of data, indices and offset, respectively.

\medskip

\noindent\textbf{Compressed Sparse Column (CSC)}: CSC works like CSR, but this time data are organized by columns. A single array (Non-zero array) stores all the non-zero values of the sparse columns in order, and an integer array (Row indices) stores the corresponding row indices. A third array (Column pointer) stores the offset within the previous two vectors, indicating the number of non-null elements per column in an incremental fashion. Figure~\ref{fig:comp_form}(b) shows an example of the CSC coding.  The number of bits required for such a representation is quite similar to the previous one, i.e.,:
\begin{equation}
    I\cdot(Nb\cdot(1-Sp)+Ni\cdot(1-Sp))+(W+1)\cdot No 
\end{equation}
where $W$ is the width of the input matrix.

\medskip

\noindent\textbf{Compressed Image Size (CIS)}: This data format consists of a sparsity map and a non-zero value list, as depicted in Figure~\ref{fig:comp_form}(c). The former is a mask with the same shape of the original data (1D vector, 2D Matrix or 3D matrices array) having one bit per entry. The bit is 0 if the corresponding value is null, 1 otherwise. The latter is an array composed of all non-zero values. With respect to CSR and CSC this technique allows an easier representation with no need for decompression. In this case, the number of required bits has a simpler equation, as given below:
\begin{equation}
    I\cdot(Nb\cdot(1-Sp)+n)
\end{equation}
where $n$ is typically 1 bit.

\medskip

\begin{figure*}[ht]
    \centering
    \includegraphics [width=0.6\textwidth] {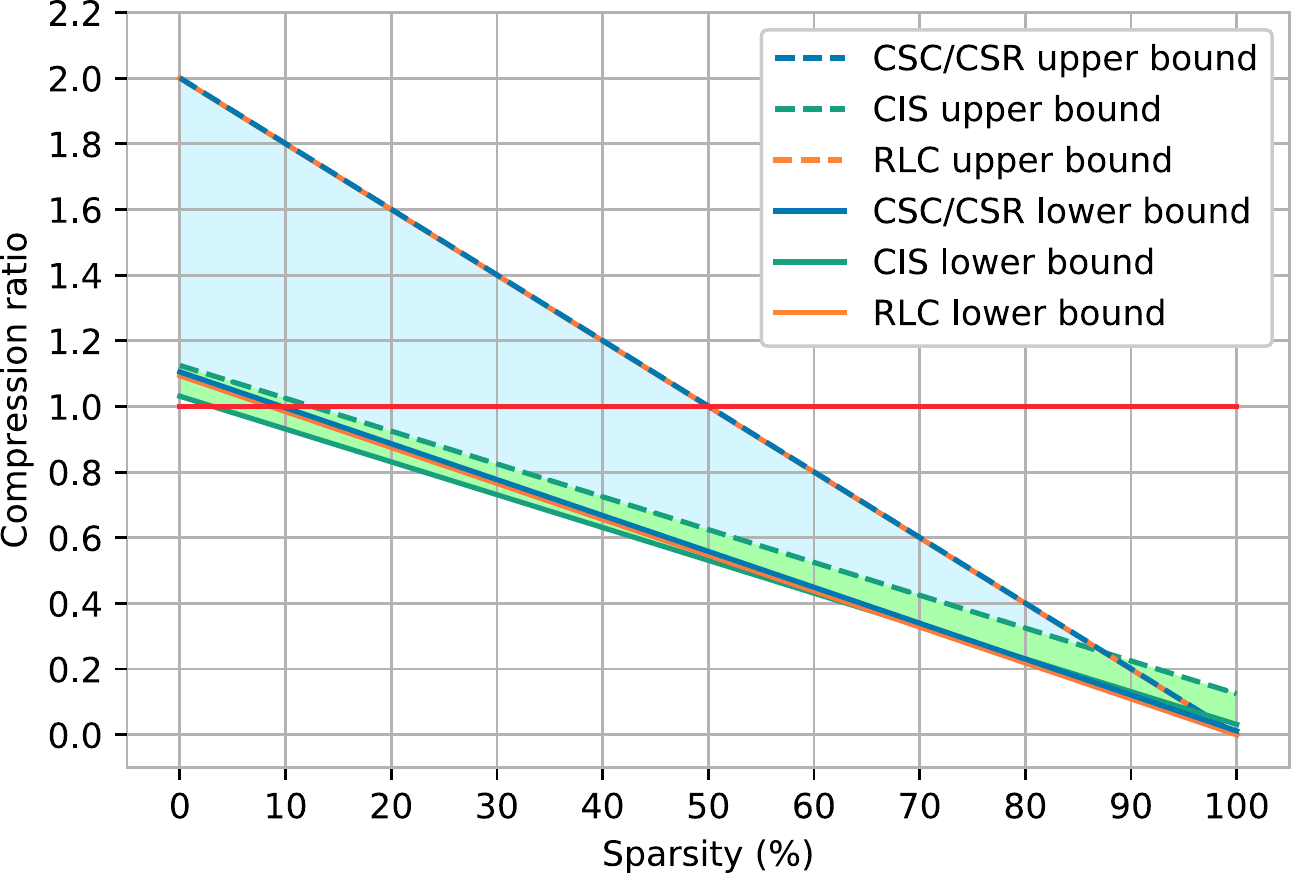}
    \caption{Compressed ratio with variation in sparsity of CSC, CSR, CIS and RLC. The picture includes two boundaries: the upper bound refers to Conv 4 of AlexNet ($3\times3\times384$) with data parallelism of 8 bits, while the lower bound is based on Conv 1 ($11\times11\times3$) represented on 32 bits.}
    \label{fig:comp}
\end{figure*}

Figure~\ref{fig:comp} compares the compression ratio of CSR, CSC and CIS  methods. This is the ratio between the compressed bit size and the original model. The picture includes two boundaries for the three formats. The upper case is based on the filter size of Conv 4 for AlexNet ($3\times3\times384$) with data parallelism of 8 bits, while the lower bound is based on Conv 1 for the same neural network ($11\times11\times3$) represented on 32 bits. As it is possible to notice, the CIS format performs better than the CSR or CSC formats in almost all the sparsity range. However, the coding choice often depends on how the data will be handled by the hardware.

For the sake of completeness, a fourth method should be introduced, namely Run Length Coding (RLC).

\medskip

\noindent\textbf{Run Length Coding (RLC)}. It is a simple data format that is able to compress the consecutive repetition of the same value as depicted in Figure~\ref{fig:comp_form}(d). In the case of sparsity, it is mainly used to compress consecutive zeros in a single zero and the count of them.  Although it is very easy to implement, it is only effective when zeros manifest in a compact and consecutive manner (high percentage of sparsity). In addition, the RLC is designed for data arrays, so it is not optimal when operating on matrices.


The following, and last proposed method represents a milestone in the history of compression. However, for reasons of complexity it is difficult to use in hardware architectures.

\medskip

\noindent\textbf{Huffman coding}. As is well known, Huffam coding is the most efficient method to encode scattered data thanks to its optimal compression rate. However, its complexity makes it difficult to employ since it would require computation-hungry compressor/decompressor schemes (large silicon area). Moreover, the continuous data manipulation would introduce a power overhead, which can hardly be compensated by saved computations. Thus, such a non-friendly-hardware coding approach is only used in software-level implementations.

\medskip

Even though it is proved that the above-mentioned techniques bring benefits, compression introduces irregular data patterns that are reflected in irregular memory accesses. Moreover, ad-hoc hardware support is required to identify useful operations. In this scenario, general-purpose platforms like CPUs and GPUs are not very prone to use sparsity as an advantage, but rather, random memory accesses represent for those a source of inefficiency. 

Thus, many FPGA and ASIC architectures leverage sparse matrices to accelerate the inference stage thanks to custom hardware. For example, Cnvlutin~\cite{cnvlutin} relies on the ReLU function to compress activations with a CSR approach, but it does not consider weight sparsity. On the other hand, Cambricon-X~\cite{cambricon-x} employs the weight sparsity, having a PE-based implementation, where each PE stores compressed synapses for asynchronous computation. SCNN~\cite{SCNN}, instead, is able to take care of both sparsity simultaneously in CNNs by means of an input stationary dataflow. Activations and weights in a CSC scheme are provided to a multiplier array that generates scattered partial products, subsequently added together using a dedicated interconnection mesh. Despite the fact that it reaches an excellent PEs utilization efficiency over convolutional layers, fully connected ones represent a bottleneck since it is impossible to reuse values. EIE~\cite{EIE} encodes the sparse weights using the CSC format as well, avoiding the use of the DRAM for 120x energy saving.
Moreover, the ability to skip zero activations makes its matrix-vector multiplication inference engine extremely efficient. NullHop~\cite{NullHop} is a CNN FPGA-based hardware accelerator which embodies both a zero-skipping ability over null activations and a CIS compression over the synapses. The first comes without any clock cycle waste, while the second allows acting directly on compressed data thanks to its hardware-friendly representation. SqueezeFlow~\cite{squeezeflow} exploits a different approach by introducing concise convolutional rules. Such rules reduce the computation by avoiding part of the useless operations (null values).  The hardware implementation enables the acceleration of dense DNNs without intrusive PE variations.\\
Eyeriss~\cite{eyeriss} simply exploits sparsity by clock-gating the PEs with zero value, i.e. not performing the multiplication. Although this reduces the power consumption, highly sparse DNN models could cause a poor PE array utilization. ZeNA~\cite{ZeNA} was the first zero-aware architecture targeting the CNNs, able to skip ineffective computation induced by both weights and activations. Moreover, it addresses the unbalanced workload among PEs due to the zero-skip operation by introducing a novel load distribution method.  Huan et al.~\cite{huan2017lowpower}, instead, proposed an approximate architecture that skips near-zero multiplications, providing a further reduction of computation (1.92x over LeNet5) with negligible accuracy loss.\\
Table~\ref{tab:sparsity} summarizes what has been discussed in previous paragraphs about the sparse architectures analyzed in this section. It reports the data compression format and which data among activations (A) or weighs (W) is subject to the compression. Besides, the last column reports the type of data for which unnecessary operations are skipped.

\begin{table}[h]
\centering
\caption{Summary table of sparse architectures.}
\label{tab:sparsity}
\resizebox{\linewidth}{!}{%
\begin{tabular}{|c|c|c|c|c|}
\hline
\textbf{Architecture} & \begin{tabular}[c]{@{}c@{}}\textbf{Target} \\ \textbf{device}\end{tabular} & \begin{tabular}[c]{@{}c@{}}\textbf{Compression} \\ \textbf{Format}\end{tabular} & \begin{tabular}[c]{@{}c@{}}\textbf{Compressed}\\ \textbf{Data}\end{tabular} & \begin{tabular}[c]{@{}c@{}}\textbf{Zero} \\ \textbf{Skip}\end{tabular} \\ \hline
Eyeriss ~\cite{eyeriss}    & ASIC                                                     & RLC                                                           & A                                                         & A                                                    \\ \hline
Cnvlutin ~\cite{cnvlutin}   & ASIC                                                     & CSR                                                           & A                                                         & A                                                    \\ \hline
Cambricon-X~\cite{cambricon-x}  & ASIC                                                     & CIS-alike                                                     & W                                                         & W                                                    \\ \hline
SCNN    ~\cite{SCNN}    & ASIC                                                     & CSC                                                           & A+W                                                       & A+W                                                  \\ \hline
EIE     ~\cite{EIE}    & ASIC                                                     & CSC                                                           & W                                                         & A+W                                                  \\ \hline
NullHop   ~\cite{NullHop}  & FPGA                                                     & CIS                                                           & W                                                         & A+W                                                  \\ \hline
SqueezeFlow ~\cite{squeezeflow} & ASIC                                                     & none                                                          & none                                                      & W                                                    \\ \hline
ZeNA  ~\cite{ZeNA}      & ASIC                                                     & none                                                          & none                                                      & A+W                                                  \\ \hline
Huan et al.~\cite{huan2017lowpower} & ASIC                                                     & none                                                          & none                                                      & A+W                                                  \\ \hline
\end{tabular}}
\end{table}

\subsection{Approximate Computing for Deep Learning and their Resilience}
\begin{figure}[h]
    \centering
    \includegraphics[width=\linewidth]{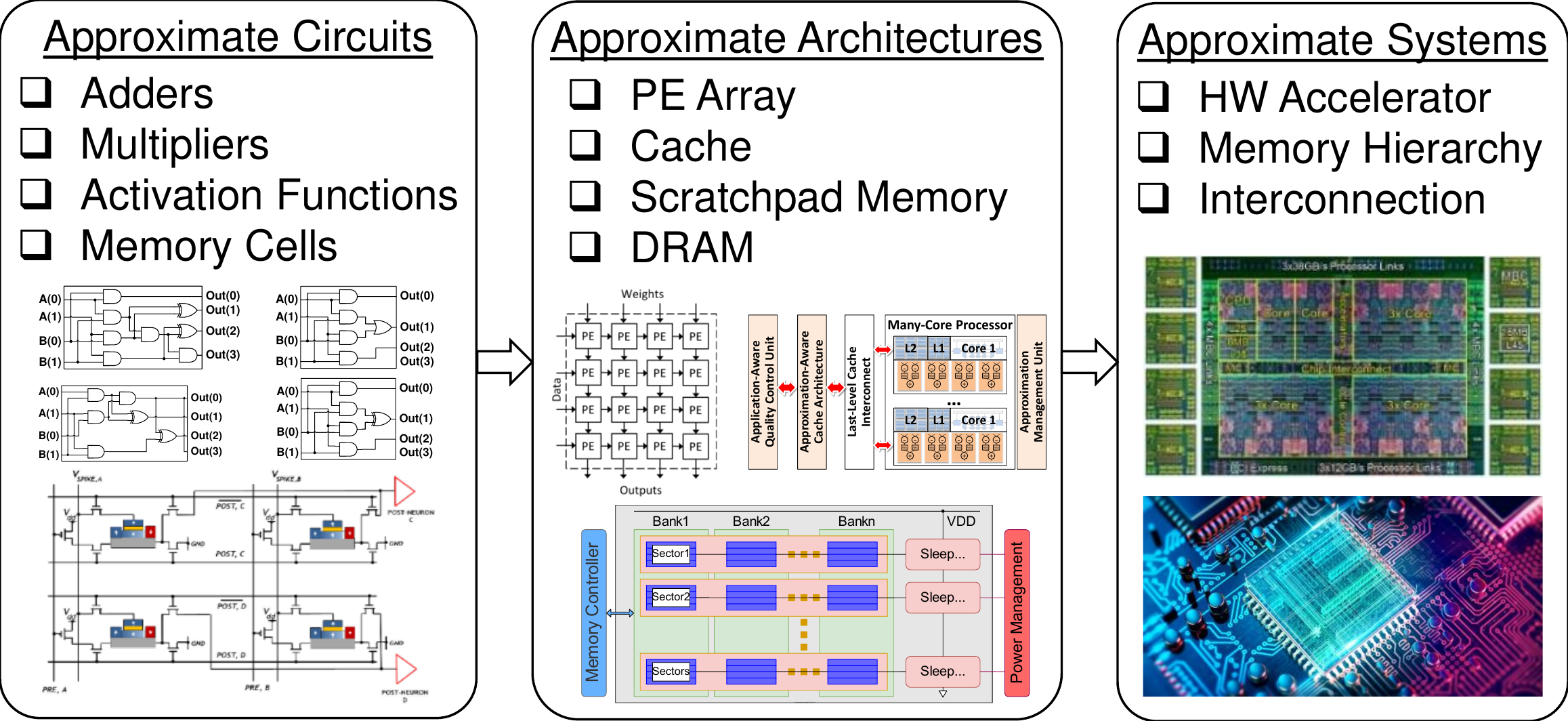}
    \caption{Opportunities for employing approximate computing in deep learning.}
    \label{fig:approx_computing}
\end{figure}

Approximate computing is a well known paradigm, whose basic idea is to trade quality for efficiency, at different abstraction levels~\cite{crosslayerac_shafique}. 
Therefore, it is desirable for non-safety-critical tasks, or for applications that are resilient to approximation errors~\cite{Mittal2016SurveyAC}. 
Many studies have analyzed its applicability on DL-based applications~\cite{Kumar2019AC4ML}. An overview of the possibilities of employing approximate computing is shown in Figure~\ref{fig:approx_computing}.

The most compute-intensive operations that are performed in the inference are the multiplications. Approximate multipliers can be employed in DNN accelerators to reduce the power consumption~\cite{Mrazek2016AM4ANN}.

At the architecture-level, systematic resilience analyses are needed for applying approximate computing in CNNs~\cite{errorresilience_shafique} and CapsNets~\cite{redcane_shafique}. The error generated by approximate MAC units in systolic array-based DNN accelerators can be mitigated by employing curable approximations~\cite{cann_shafique}. This is extremely useful for reducing the critical path and energy consumption of the DNN accelerators, without sacrificing the classification accuracy. 
AxTrain~\cite{He2018AxTrain} is a framework for DNN training that enables approximate inference. Otherwise, a layer-wise approximation of DNN accelerators at the inference stage can be done automatically~\cite{alwann_shafique}. CAxCNN~\cite{CAxCNN_shafique} is a methodology for approximating the filter weights of DNNs without retraining and executing DNN inference with low-complexity multipliers.

Approximate memories~\cite{Deng2015ReducedPrecisionMV}\cite{Koppula2019EDEN} can further reduce the energy consumption of DNN accelerators and systems. 
The work in~\cite{Chen2018AC4DL} optimized the communication network for reducing the computational cost of DL training and inference.

A cross-layer approach~\cite{xdnns_shafique} leads to integrate approximate computing in a compression framework for further reducing the energy consumption of DNN accelerators.

\subsection{Embedded vs Cloud Computing} 
So far, the focus has been mainly on the development of embedded architectures for deep learning. However, it is also necessary to mention the other solution that is gaining ground, namely \textit{cloud computing}. Cloud computing is a paradigm of service delivery, especially storage of data and computational resources, offered by a provider to a client through the Internet.  

Cloud computing offers some advantages when applied to deep learning. As demonstrated in the previous paragraphs, deep learning is based on the availability of a large amount of data, especially during the training phase. For the latest models of neural networks, many computational resources are also required. These resources may not be available and accessible to everyone, so services provided by third parties may be a valid solution. Besides, cloud services have a very flexible availability of resources, which can be scaled during the development of a project to better adapt to different needs. Finally, many cloud computing services for AI and machine learning offer solutions and resources that do not require in-depth technical knowledge, allowing even inexperienced people to approach this field and exploit its potential. There are currently several providers offering cloud computing solutions, among them Alibaba Cloud~\cite{alibaba}, Amazon Web Services (AWS)~\cite{aws}, IBM Cloud~\cite{ibmcloud}, Google Cloud~\cite{gcloud}, Google Colab~\cite{gcolab} and Microsoft Azure~\cite{mazure}. 

However, cloud computing has some disadvantages that do not make it suitable for all applications. First of all, cloud computing is based on the availability of an Internet connection, and it is therefore not ideal for applications that do not permit interruptions of service, such as self-driving vehicles. Data transmission between client and server is also subject to security issues as it is more vulnerable to breaches. Cloud computing is, therefore, not suitable for applications where there are strict regulations on data security, such as government or defence services. Finally, in latency-constrained applications, such as again self-driven vehicles or virtual reality applications, it is preferable to have near-sensor computing rather than relying on the transmission of data via the Internet. However, the advent of 5G could potentially mitigate this problem. 

\subsection{SNNs Hardware Accelerators}\label{SNN_acc}
Modern computing systems based on the Von Neumann architecture are not efficient for the SNNs implementation, because of the physically separated computational and memory units~\cite{Rajendran2019NeuromorphicHardware}. Therefore, novel computational architectures are necessary to implement SNNs with high performance and low energy. Several accelerators for SNNs have been proposed in the literature. The most popular ones are adopting the \textit{neuromorphic} architecture.

SpiNNaker~\cite{Furber2014Spinnaker} is a system designed to implement large SNNs in real-time. Using ARM9 cores as building blocks, it implements event-driven computation and communication, interfacing with Python libraries such as PyNN. Its second version, SpiNNaker~2~\cite{Liu2018SpiKKaker2}, increases the number of cores for implementing deep learning with sparse connectivity.

IBM TrueNorth~\cite{Merolla2014Truenorth} is designed with a 28-nm CMOS technology, with 4,096 neurosynaptic cores. Each core has 12.75 KB of local SRAM and can support up to 256 neurons. The scaling and integration of multiple chips is allowed by the spike-based nature of the communication and routing infrastructure in an asynchronous-based NoC.

Intel Loihi~\cite{Davies2018Loihi} provides highly parallel and power efficient asynchronous computations. The chip implements in a 14-nm CMOS technology a mesh of 128 neurocores, each of them having  1,024 spiking neurons and 2 Mb of SRAM. Scaling is possible through a hierarchical connectivity between chips. Moreover, several neuromorphic learning rules are supported.

BrainScaleS~\cite{Schmitt2017Brainscales} is a mixed analog-digital system, with analog neurons and digital communication. Like SpiNNaker, it also allows PyNN interface.

\section{Memory Hierarchy}\label{memory_hier}


While optimizing the algorithms and accelerating the implementation of computational primitives is of fundamental importance to achieve the best performance, inefficient memory management could undermine all efforts made to achieve high throughput and energy efficiency claimed by the accelerator design~\cite{DRAM_shafique}. Typically, memory accesses dominate the energy consumption of a system~\cite{Inefficiency}. As depicted in Deep Compression~\cite{han2015deep}\cite{cai2019onceforall}, using a \SI{45}{\nano\meter} technology, a 32-bit adder consumes \SI{0.9}{\pico\joule}, while SRAM and DRAM access require respectively 5.5$\times$ and 711$\times$ more energy. 
In such a situation where the storage elements constitute a clear efficiency bottleneck, memory must be taken into account from the earliest design steps as a first-order concern.

Conversely to processors, where the general-purpose structure prevents adaptation to the workload, for other platforms it is possible to make a tightly tailored design on the specific algorithm in order to reduce at minimum the memory transfer. These considerations are crucial, especially in the field of machine learning, where the enormous number of MACs to be performed requires an enormous and continuous data movement towards the processing units. Considering, for example, a 1G fully connected layer running at a typical video recording frame rate (30 fps), using the above DRAM technology, its computation would require (30 fps)(1G)(\SI{640}{\pico\joule})= \SI{19.2}{\watt} that is a considerable amount of power, unaffordable for mobile devices.

\bigskip
\noindent\textbf{Inference vs. Training}: From a memory perspective, training is way more intensive than inference. While in the latter the NN is crossed only once, in the former the backpropagation mechanism imposes to cross it backward, reloading both activations and weights. Thus, the training has an almost double cost.
Generally speaking, in most of the industrial, medical and commonly used applications, there is no reason for on-line training. Usually, neural networks are trained on a dataset off-line and then delivered to the end-users. As the dataset is periodically improved by adding corner cases, networks can be realigned through a new off-line training session.
Since such an operation is performed off-line, there is no need for highly optimized hardware platforms, but rather for high-speed general-purpose architectures, such as GPUs, able to scale down the training time with no constrains over the power envelope. Moreover, even though ML researchers are very interested in speeding up learning, from a business perspective, this represents a small market.
According to the above, and knowing that the nature of the computations required to carry out the backpropagation and the inference is almost identical, from now on we mainly focus on the inference stage that offers more case studies.

\bigskip
\noindent As mentioned before, accelerating large size ML algorithms involves high memory traffic. Focusing on the currently most known and used networks, DNN and CNN, we analyze in detail their main fundamental layers from a typical processor memory organization perspective, providing an idea of the number of operations to be carried out and the possible optimizations.

\bigskip
\noindent\textbf{Fully Connected Layer}: Among NN layers, FC ones are those which require the highest memory transfer due to their topology. Considering a layer composed of $C_i$ input neurons and $C_o$ output neurons, the synapses (i.e. weights) are represented by a $C_i \times C_o$ matrix. Thus, the execution of the entire layer can be summarized in a matrix-vector multiplication that needs a total number of memory transfers equal to $C_i \times C_o + C_i \times C_o + C_o$ where each addendum represents respectively the inputs loaded, the weights loaded and the output written back to the main storage.

The matrix-vector multiplication is a critical operation, especially in case the weight matrix is larger than the lowest cache level capacity. In such a case, it is impossible to reuse the matrix values, and new memory accesses are performed every cycle.
In the case of CPUs and GPUs, this problem can be overcome by batching~\cite{batching}. This technique allows to group multiple input vectors into a single matrix and reuses the weight parameters. However, real-time applications cannot use this optimization because a certain latency is introduced. Therefore batching can only be used during offline training, where the dataset is provided a priori.
As far as inference is concerned, we prefer techniques capable of overcoming the bottleneck represented by memory by spatially and temporally distributing the workload as seen in Section~\ref{temp_vs_spatial}, or compressing the network by reducing the number of parameters as shown in Section~\ref{pruning}. Nevertheless, compression generates irregular patterns that make CPUs and GPUs ineffective.

Theoretically, input activations can be reused for each output one, but unfortunately, their size ranges from few thousands to hundreds of thousands, making them unfeasible to be stored on an L1 cache. Tiling could be used to subdivide the loop over the input neurons. However, it is not possible to perform loop tiling over a factor without affecting the rest of the execution. Indeed, increasing the reuse of the input neurons, the amount of the partial output sums to be stored back to the main storage increases as well. Thus, the input memory bandwidth saved from the tiling is partially compromised by the partial sums write back, but still advantageous.
As far as weights are concerned, they are unique for each input, so it is not possible to reuse them. Moreover, in the DNNs these vary from tens of millions to some billions, making it impossible to store them even on higher cache levels.

\bigskip
\noindent\textbf{Convolutional Layer}: With respect to FC layers, the convolutional ones are built on a 2D scheme (3D considering the channel direction), exhibiting an input and an output feature map. The input feature map can be reused as many times as the number of kernels. More precisely, since the convolutional windows tend to overlap, the single input feature map windows, with size equal to the kernel one, can be reused $\frac{H_k \times W_k}{S_x \times S_y} $ times as shown in Figure~\ref{fig:FC_reuse}, where $H_k$ and $W_k$ are the sizes of the kernel and $S_x$ and $S_y$ are the stride over x and y directions. Consequently, as described above for FC schemes,  it is possible to perform a tiling loop over the two dimensions of the IFM with a reuse strategy to reduce the accesses to the main storage. In this case, tiling the input does not affect the output. Consequently, in GPUs and CPUs, no tiling is performed since it is possible to fit an entire kernel volume in an L1 cache; thus, an entire OFM can be produced without the need to break down the loop. Indeed, typical kernels size is $H_k \times W_k \times C_i$, where $H_k$ and $W_k$ are in the order of ten, while the number of input channels ($C_i$) can reach the hundreds.
For FPGA and ASIC approaches, the reuse strategy can be way more aggressive thanks to ad hoc designs as explained in the following.
Kernels are usually shared, reducing considerably the number of DNN parameters, and consequently the required bandwidth. Nonetheless, the number of output channels can make the synapses unfeasible to be stored in an L1 cache. Indeed, the weight volume expressed as $H_k \times W_k \times N_i \times C_o$, where $C_o$ is the number of OFMs, can easily exceed the lower level of the memory hierarchy. Also, in this case, it is suggested to use a tiling to break the loop over the output feature map, resizing the total capacity in $H_k \times W_k \times C_i \times TC_o$ sets, where the $TC_o$ is the tile size.
In the rare case of not shared weights, as discussed for FC layers, not even the L2 cache could fit them, making reuse impossible.

\bigskip
\noindent\textbf{Pooling Layers}: Unlike the previous layers, pooling has no weights, and the number of OFM is equal to IFM, thus the opportunities to perform data reuse are fewer. The sliding windows, over which the pooling is performed, generally do not overlap, consequently, the bandwidth for input neurons is higher than the convolutional approach. Even with the introduction of the IFM tiling, performance would improve marginally.

\begin{figure*}[ht]
    \centering
    \includegraphics [width=0.75\linewidth] {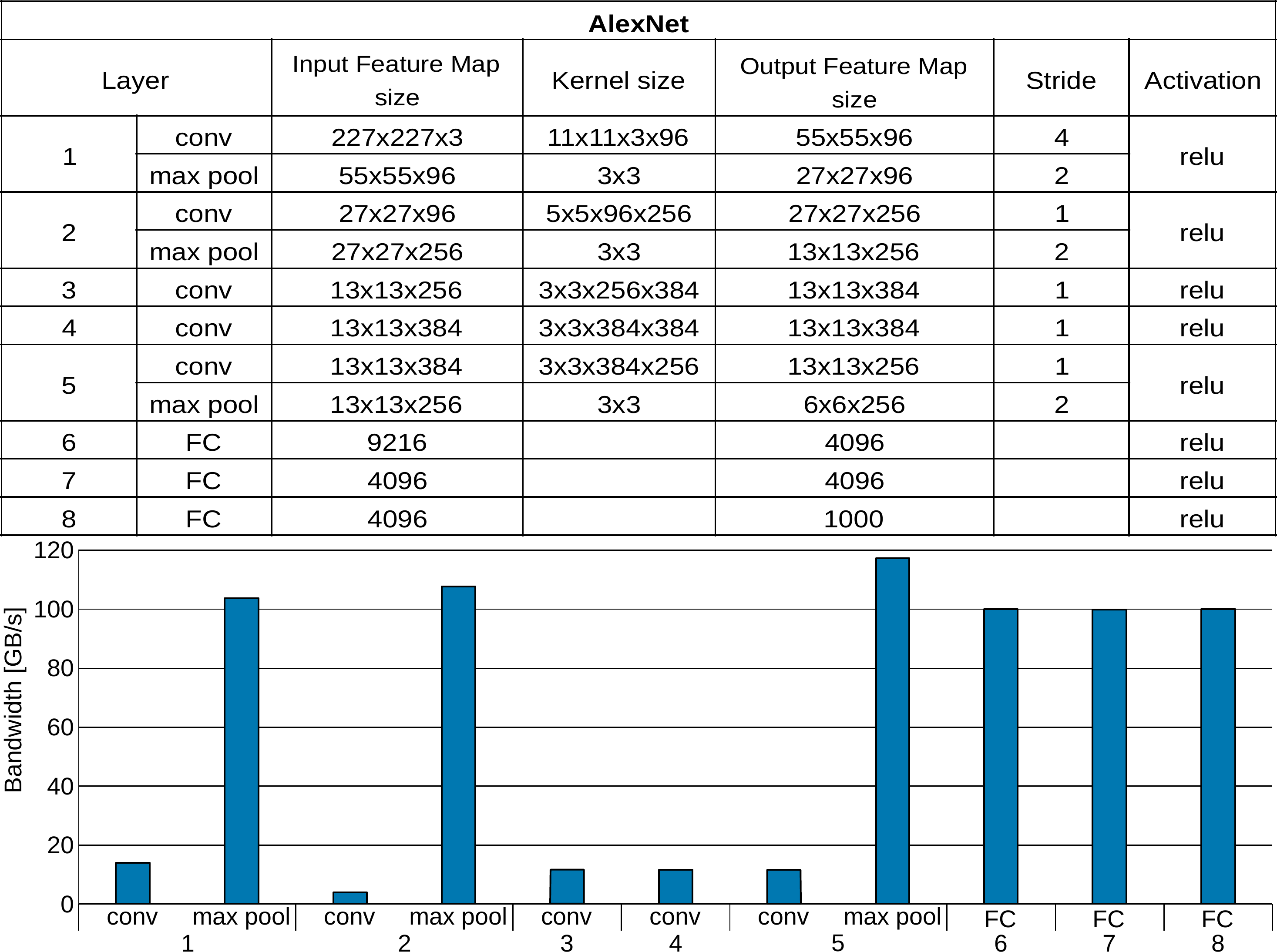}
    \caption{Example of required bandwidth per layer executing Alexnet on a device able to perform 100 Gops/s with 100\% efficiency.}
    \label{fig:bw}
\end{figure*}

Taking into account the three types of layers described above, it is clear that the required bandwidth is profoundly different from each other. Figure~\ref{fig:bw}, for example, shows the bandwidth needed for the execution of AlexNet on a device able to perform 100 Gops/s with 100\% efficiency, i.e. no stall and data dependencies, with no memory constraints. Despite the bitwidth for both activations and weights is just 16 bits, the bandwidth for some layers, especially FC and max pooling, where data reuse is practically impossible, is unattainable from any commercially available memory. This once again highlights how difficult it is to execute ML algorithms efficiently, in particular on devices with rather modest hardware, as may be the case with IoT nodes, which are mainly CPU-based. The architecture of these nodes must guarantee flexibility and speed of execution for a wide range of algorithms, therefore it cannot be optimized for the DL. Researchers over the years have tried to improve the libraries and kernels of basic operations carried out on their processors to maximize the management of storage elements such as the Intel MKL-DNN~\cite{mkl-dnn} and ARM CMSIS-NN~\cite{lai2018cmsisnn}. The former library works on the data format mapping multidimensional arrays into linear memory address spaces. Moreover, it enables lower numerical precision primitives, accelerating the execution of multiple operations, i.e., increasing the number of operations per second, and enhancing the performance of the cache at bandwidth parity. The latter intends to reduce memory overhead and maximize NN execution on Cortex-M processors for low-power applications oriented to IoT devices. Another example is represented by Garofalo et al.~\cite{Garofalo_2019}, who proposed PULP-NN, a library designed for a parallel cluster of tightly-coupled RISC-V processors. Its set of software kernels targets the inference of quantized NN, being capable of exploiting sub-byte bitwidth data.

What just said for CPUs is also valid for GPUs, with the big difference that their capability to parallelize large workloads makes these devices ideal for DNN applications, although expensive from the power point of view. NVIDIA developed the CUDA Deep Neural Network library (cuDNN)~\cite{chetlur2014cudnn}, a special library that also includes the possibility to use a fixed point format at 16 and 32 bits, moreover, transforms convolution operations into multiplications between matrices, which are extensively optimized. This property is reflected in a reduction in the demand for RAM and a consequent increase in the number of supported operations. However, for large DNN models, it is essential to tune the memory usage to fit them into the DRAM. vDNN~\cite{7783721} virtualizes the memory of the CPU and GPU so that it can be simultaneously used for training in a hybrid fashion.
Kim et al.~\cite{8599530} extended the concept of vDNN to a multi-GPU environment employing PCIe-bus. Furthermore, thanks to a prefetching algorithm, they can increase the mini-batch size of 60\%.

Conversely to GPUs, FPGA and ASIC accelerators have a limited amount of memory. In CNP~\cite{5272559} for example, in order to accommodate a large number of DSPs on a Virtex4 SX35 FPGA platform, the authors designed an interface with an external memory capable of performing 8 read/write operations. However, their flexibility and the possibility to design their memory hierarchy tailored to the specific problem can lead to a lower energy envelope. The sizing of on-chip memory buffers is not trivial and depends on many factors such as layer size, layer type, frequency of buffer usage. Wei et al.~\cite{8806999} have proposed an FPGA-based layer conscious framework to allocate on-chip buffers efficiently. Such a paradigm combined with buffer sharing saves resources and enhances their usage.
Since DRAM has an access cost about 130$\times$ higher than SRAM, in some cases, it has been thought to directly remove this storage device as in the case of Park et al.~\cite{7471828}. Exploiting fixed-point data format and the capability of NN to work even in case of reduced precision~\cite{han2015deep}, Park et al. were able to fit the entire DNN model into the on-chip memory, reducing the power consumption drastically. Following the same approach Du et al. have proposed ShiDianNao~\cite{7284058}, an ASIC designed to be integrated into a commercial image chip typical of smartphones. Being in close contact with the sensor, the data it processes is taken directly from the local SRAM, minimizing the power needed. ShiDianNao is the last accelerator of the series started with DianNao~\cite{diannao}, a small-footprint memory-wall aware accelerator for large NN models, and continued with DaDianNao~\cite{7011421}. The latter instead proposes a multi-chip ML architecture with 64 cores in supercomputer style able to achieve a speedup of about 450x over a typical GPU.
While the above architectures distribute the on-chip memory among the PEs, i.e., near computation, there have also been efforts to do the opposite, namely to bring the computation into the storage elements. This is the case of the logic-in-memory (LIM), where easy computational tasks are executed directly inside the memory like in~\cite{8310401},~\cite{8824826},~\cite{8839490},~\cite{8867863}.

\section{Deep Learning Security}\label{security}
Despite the great success and popularity of deep learning in recent years, recent researches showed that DNNs have intrinsic weaknesses that can threaten the security~\cite{Barreno2010SecurityML}\cite{Szegedy2013IntriguingPO}\cite{RobustML_shafique}. Starting from the work of Goodfellow et al.~\cite{Goodfellow2015explainingadvexamples}, many researches have been conducted, with the purpose of identifying weaknesses (\textit{Adversarial Attacks}) and their countermeasures (\textit{Adversarial Defenses})~\cite{SecurityML_shafique}\cite{RobustML2_shafique}. Moreover, machine learning models can be stolen~\cite{Tramer2016modelstealing} or inverted~\cite{Song2017modelinversion}.

\subsection{Adversarial Attacks}

\begin{figure*}[ht]
    \centering
    \includegraphics[width=.7\linewidth]{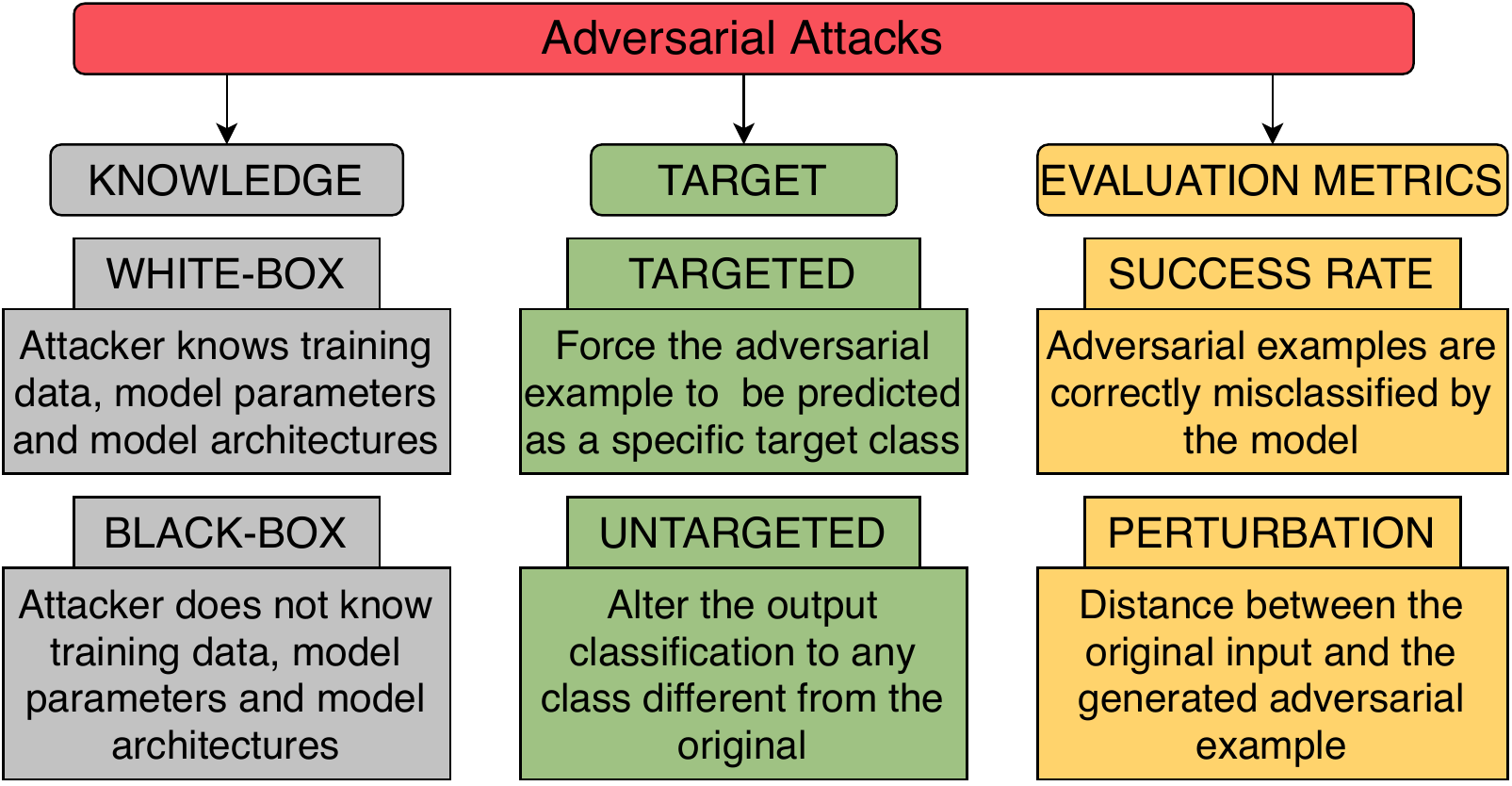}
    \caption{Taxonomy of Adversarial Attacks.}
    \label{fig:adversarial_taxonomy}
\end{figure*}

The basic idea behind an adversarial attack is to make a machine learning model classify a malicious sample wrongly. In case of image classification, the adversarial attack introduces a noise in the input image to create the adversarial example, which is classified wrongly by the DNN. Adversarial attacks can be categorized according to different attributes, e.g., the choice of the class, the kind of the perturbation and the knowledge of the network under attack~\cite{Papernot2017practicalBBattacks}\cite{Yuan2019attacksanddefenses}. We summarize these properties in Figure~\ref{fig:adversarial_taxonomy}.

The goal of an adversarial attack is to be at the same time imperceptible and robust~\cite{Luo2018imperceptibleandrobust}. A successful adversarial example should not have obvious variations perceived from an human eye, compared to the original image. Moreover, an attack is robust if the gap between the probabilities of the adversarial class and the correct class is so large that, after a transformation (e.g., noise filtering, compression or resizing), the misclassification still holds. These kind of attacks have been evaluated also on CapsNets~\cite{CapsAttacks_shafique} and SNNs~\cite{SNNAttack_shafique}. Moreover, if applied on a different domain~\cite{ProbabilisticAA_shafique}, the imperceptibility of the adversarial examples can be improved.

Several types of adversarial attacks have been proposed.

Poisoning attacks~\cite{Biggio2013DataClusteringAdversarial}\cite{Gonzalez2017PoisoningDL}\cite{Shafahi2018PoisoningAttacks} contaminate the training data in such a way that the decision boundaries of the classifier are pushed to incorrect zones, thus reducing its classification accuracy on clean inputs. More specifically, backdoor attacks~\cite{Chen2017BackdoorDL} train a network in a way that, when exposed to a specific noise pattern that plays the role of a trigger, it is fooled. Triggered by an adversarial noise pattern, the NeuroAttack~\cite{NeuroAttack_shafique} introduces a backdoor Trojan to fool DNNs and SNNs with bit-flips. 

Gradient-based attacks like FGSM~\cite{Goodfellow2015explainingadvexamples} and its variants~\cite{Biggio2013EvasionAA}\cite{Kurakin2016AdvExamplesPhysicalWorld}\cite{Dong2018BoostingAA}\cite{Madry2017TowardsDL}\cite{TrISec_shafique} are white-box adversarial attacks that perturb the inputs based on the gradient of the output probabilities with respect to the inputs. They only introduce perturbations at the inference stage, without modifying the training data.

The Carlini \& Wagner attack~\cite{Carlini2016CWAttack} aims at minimizing at the same time (i) the distance between the original image and the adversarial image and (ii) the distance between the maximum output activation and the confidence of the target class.

Decision-based attacks~\cite{Brendel2017DecisionBasedAA}\cite{Brunner2018GuessingSmartAA}\cite{Chen2019BoundaryAttack++}\cite{REDAttack_shafique} are black-box adversarial attacks which estimate the decision boundary and aim at crossing it to obtain a misclassification. The quality of such attacks is measured in terms of number of queries, i.e., the inference passes with different inputs.

Universal perturbations~\cite{MoosaviDezfooli2016UniversalAP} aim at identifying a noise pattern, specific for a given dataset, which, when added to the input, significantly reduces the test accuracy of any deep learning model.

\subsection{Adversarial Defenses}

Several defense methods have been studied and proposed. They aim at increasing the generalization of DNNs, while they perform better against different types of attacks. However, one of the main drawbacks of applying the defenses on DNNs is that the classification accuracy on clean images decreases.

Data protection defenses~\cite{Chakarov2016DebuggingML}\cite{Barcaldo2017MitigatingPoisonA} analyze the impact of the input in order to identify the noise, thus effectively working against poisoning attacks.

Standard DNN compression techniques has been adapted to successfully defend against adversarial attacks. Fine-Pruning~\cite{Liu2018FinePruningDA} removes the redundant connections in DNNs which do not significantly contribute for the accuracy of the clean data in order to remove the effect of the backdoor. A quantization-based defense~\cite{QuSecNets_shafique} reduces the success rate of the attack by quantizing the input pixel intensities.

Adversarial training~\cite{Madry2017TowardsDL} is the de-facto standard defense method against adversarial attacks. Since the adversarial examples are added to the training set, the classifier is able to learn these perturbations. As a drawback, the adversarial training adds a prohibitive overhead in the training process. Further variants of such defense~\cite{Zhang2019YOPO}\cite{Shafahi2019FreeAdversarialT}\cite{Wong2020FastAT} aim at reducing its computational cost and training time overhead.

Different pre-processing techniques can elude the effectiveness of adversarial attacks. Simple pre-processing filters~\cite{FadeML_shafique} completely alter the functionality of the attack. Randomized smoothing~\cite{Cohen2019RandomizedSmoothing} produces a Gaussian noise at the input to mitigate the effect of the adversarial perturbations on the inputs. It has been demonstrated to be effective also on large perturbations on large and complex datasets.

Detectors~\cite{10.1145/3133956.3134057} add a sub-network model to detect whether an input is an adversarial example or not. This algorithm can be successfully executed in specialized hardware and integrated with DNN accelerators~\cite{10.1145/3373376.3378532}.

\section{Benchmarking}\label{sec:benchmarking}

Since Deep Learning is a critical topic in the research community, over the years, big companies have made available a massive series of tools to help the development of new models. These frameworks, in addition to the most recent and updated datasets, are crucial for both software and accelerator development. The possibility to explore new models and evaluate them in terms of workload, the trade-off between complexity and accuracy, access to memory, numerical representation (floating-point vs fixed-point) is a fundamental step in the hardware design phase. All these steps are of great importance to understand what the performance of the accelerator will be.
In this section, we present the main frameworks for the DL and the datasets used to determine the performance of the algorithms. Finally, parameters and metrics for the comparison of hardware platforms are discussed.

\subsection{Frameworks}\label{sec:framework}
Frameworks are working environments that provide the developers with all the basics and support to build new ready-to-use models, as depicted in Table \ref{tab:framework-table}. Having such tools, able to compose a DNN using high-level programming language like Python and then test the performance of the algorithm, speeds up the research work enormously. Moreover, profiling the code execution and understanding where the critical load is located, it is possible to define which parts will have to be translated into hardware and, consequently, what needs to be accelerated.

In the case of CPUs and GPUs instead, libraries and frameworks are essential to parallelize and distribute the effort among the cores.

To be noted that many frameworks transform DNN models into optimal graphs. This is only an effective method for visualizing the operations to be performed sequentially.

\medskip

\noindent \textbf{Tensorflow}~\cite{tensorflow}. Google's Tensorflow is one of the most popular DL framework. It supports many different languages such as JavaScript and Java, C++, Go, C\#, Julia, even though the most convenient client remains Python. It is characterized by a static computation graph, which means that it first defines the graph, and then processes it. Since the model is static, it is not possible to make changes in the structure at run time, but it is necessary to do the training of a new structure. The efficiency of its primitives compensates this lack of flexibility. It is optimized for Tensor Processing Unit (TPU) architectures.

\medskip

\noindent \textbf{PyTorch}~\cite{pytorch}. Created by Facebook, it is the principal competitor of Tensorflow. Unlike the previous one, PyTorch takes advantage of a dynamically updated graph. This means that it is possible to make changes to the DNN  architecture on the fly. Generally speaking, PyTorch is often used in projects in which new training paradigms are exploited. In fact, the dynamic graph property is exploited during the backpropagation task, where the normal graph execution needs to be altered.
Moreover, it supports different data parallelism (suitable for hardware solution exploration) and distributed learning models.

\medskip

\noindent \textbf{Caffe}~\cite{caffe}. This DL-based framework supports C, C++, Python, and MATLAB. It is mainly used to model CNNs. In its repository called Caffe Model Zoo, it is possible to access a wide range of pre-trained models ready-to-use. Thus, whenever there is a problem with image processing, Caffe could be the solution. Since its libraries are mainly written in C++, its strength lies in the speed of execution. However, unlike other frameworks, Caffe does not allow a fine-granularity network layer alteration by the user, which makes it inflexible. Moreover, for recurrent model applications such as the Natural Lenguage Processing, the available resources are poor.

\medskip

\noindent \textbf{MXNet}~\cite{mxnet}. This work environment supports a wide range of languages including C++, Python, R, Go, JavaScript and Julia. The strength of this framework lies in its ability to parallelize execution both on multiple GPUs and on multiple machines, as in the case of Amazon servers.

\medskip

\noindent \textbf{Chainer}~\cite{chainer}. It is the first framework to exploit the dynamic computation graph, allowing for varying length input, a handy feature in problems of natural language processing. Chainer is built on Numpy and Cupy libraries and is completely written in Python.
Since it is faster than other Python-based frameworks, today it is the leading tool for GPU performance in data centres. 

\medskip

\noindent \textbf{Microsoft Cognitive Toolkit}~\cite{microsoftcognitive}. This framework, also known as CNTK, supports Python, C++ and command-line interface. Unlike Caffe, when a new layer model is needed, it can be built thanks to the fine granularity of the base blocks, without the need for low-level code.
Concerning the operation over multiple machines, it presents higher performance compared to Theano and Tensorflow. However, as a result of a lack of support related to ARM architectures, the applications over mobile devices are limited.

\medskip

\noindent \textbf{PaddlePaddle}~\cite{paddle}. This is an industrial-oriented framework equipped with basic libraries and tools for end-to-end product development. It mainly supports CNNs and recurrent neural networks for highly optimized computation and memory recycling. Moreover, it can efficiently scale over heterogeneous architectures to speed up the training process.

\medskip

\noindent \textbf{ONNX}~\cite{onnx}. This is not a framework but a representation format for deep learning models. Microsoft and Facebook collaborated to create such a format in order to make the models portable from a framework to another. In some cases, it is convenient to perform the training on a platform and the inference on another. Moreover, ONNX can also be a valuable resource for developers, researchers and the open-source world, in fact, any pre-trained model can be shared with the community, and every user can choose the most suitable framework.
It is supported by TensorFlow, PyTorch, Caffe2, Chainer, MXNet, Keras, Microsoft Cognitive Toolkit, PaddlePaddle, and many others.

\medskip



\noindent \textbf{Keras}~\cite{chollet2015}. Keras is an Application Programming Interface (API) for ML and DL. It can be used in many of the shells presented above. It is a high-level code abstraction for implementing NNs exploiting the lower level primitives of the corresponding framework. Working at a higher level, it is suitable for fast prototyping and handling wide amounts of data streams thanks to Python generators and serialization/deserialization APIs.

\begin{table*}[h!]
\centering
\caption{Framework summary}
\label{tab:framework-table}
\begin{tabular}{|c|c|c|c|l|}
\hline
\textbf{Name}               & \begin{tabular}[c]{@{}c@{}}\textbf{Year of}\\ \textbf{creation}\end{tabular} & \textbf{Language}                                                                       & \textbf{Creator}                                                     & \multicolumn{1}{c|}{\textbf{Features}}                                                                            \\ \hline
Tensorflow  ~\cite{tensorflow}                & 2015                                                                & \begin{tabular}[c]{@{}c@{}}JavaScript, Java, C++,\\ Go, C\#, Julia, Python\end{tabular} & Google                                                               & \begin{tabular}[c]{@{}l@{}}- Static computation graph\\ - optimized for TPU\end{tabular}                              \\ \hline
Pytorch        ~\cite{pytorch}             & 2016                                                                & \begin{tabular}[c]{@{}c@{}}Python, C++, Java,\\ CUDA\end{tabular}                       & Facebook                                                             & \begin{tabular}[c]{@{}l@{}}- dynamic computation graph\\ - different data parallelism\end{tabular}                    \\ \hline
Caffe        ~\cite{caffe}               & 2016                                                                & \begin{tabular}[c]{@{}c@{}}C, C++, Python,\\ MATLAB\end{tabular}                        & Berkeley                                                             & \begin{tabular}[c]{@{}l@{}}- speed of execution\\ - fine granularity layer not allowed\end{tabular}                   \\ \hline
MXNet        ~\cite{mxnet}               & 2017                                                                & \begin{tabular}[c]{@{}c@{}}C++, Python, R, Go,\\ JavaScript, Julia\end{tabular}         & \begin{tabular}[c]{@{}c@{}}Apache Software\\ Foundation\end{tabular} & \begin{tabular}[c]{@{}l@{}}- parallelized execution on multiple \\ GPUs and machines\end{tabular}                   \\ \hline
Chainer        ~\cite{chainer}             & 2015                                                                & Python                                                                                  & Seiya Tokui                                                          & \begin{tabular}[c]{@{}l@{}}-dynamic computation graph\\ -varying length input support\end{tabular}                  \\ \hline
Microsoft Cognitive Toolkit ~\cite{microsoftcognitive}& 2016                                                                & Python, C++                                                                             & Microsoft                                                            & \begin{tabular}[c]{@{}l@{}}- fine granularity layer basic block\\ management\end{tabular}                                                                                \\ \hline
PaddlePaddle   ~\cite{paddle}            & 2016                                                                & Python, R, Go                                                                           & Baidu                                                                & \begin{tabular}[c]{@{}l@{}}- industrial-oriented\\ - heterogeneous architecture-scale\\ training process\end{tabular} \\ \hline
ONNX           ~\cite{onnx}             & 2017                                                                & C++, Python                                                                             & \begin{tabular}[c]{@{}c@{}}Facebook,\\ Microsoft\end{tabular}        & \begin{tabular}[c]{@{}l@{}}- representation format for DNN\\ - portable\end{tabular}                                  \\ \hline
Keras         ~\cite{chollet2015}              & 2017                                                                & Python                                                                                  & Google                                                               & \begin{tabular}[c]{@{}l@{}}- Application Programming Interface\\ (API)\end{tabular}                                                                                \\ \hline
\end{tabular}
\end{table*}

\subsection{Datasets}

As the frameworks are used to build new DL models, datasets are fundamental to test their performance concerning the designed task. It is essential to underline that there are countless datasets for each specific task (image classification, object detection, etc.). However, datasets of the same task are hardly comparable to each other, the difficulty of each could vary in orders of magnitude as depicted in Table~\ref{tab:dataset}. Considering, for example, MNIST and CIFAR100, both datasets for the image classification, the first is a collection of handwritten digits in grayscale, while the second ranks objects in 100 different classes.
Typically, different datasets reflect different DL models. The more complicated the dataset, the greater the model size in terms of weights and consequently in the number of operations (MAC). 
The metrics used to evaluate the performance of DL models on datasets are mainly two: accuracy in Top-1 and Top-5 mode, and weights size. Top-5 means that if in the 5 classes with the highest score there is the correct one, then it is counted as correct. Top-1 instead needs the highest score class to be the correct one.

\medskip

\noindent \textbf{MNIST}\cite{MNIST}. This dataset is composed of 70,000 images divided into 10 classes representing handwritten digits. 60,000 are for training, while the remaining are the test set. Each image is 28x28 pixel size in grayscale.

\medskip

\noindent \textbf{ImageNet}\cite{ImageNet}. This dataset is composed of 1.3M training images, 100,000 for test and a final 50,000 for validation. All the images are divided into 1000 classes organized according to the WordNet. This latter allows managing synonyms and ambiguities. Each image has a size of 256x256 pixels in color. ImageNet is the core of a famous challenge where DNNs and CNNs try to score the best Top-1 and Top-5 accuracy.

\medskip

\noindent \textbf{CIFAR}\cite{CIFAR}. Under the name of CIFAR two different datasets fall, namely CIFAR-10 and CIFAR-100. Both are composed of 60,000 32x32 pixels coloured images, but while the former ranks them in 10 classes, the latter uses a finer classification in 100 classes. Both datasets have 50,000 images for training and the remaining for test purposes.

\medskip

\noindent \textbf{COCO}\cite{COCO}. This dataset aims to advance the object recognition state-of-the-art by putting together also segmentation and captioning. It is composed of 328,000 images of everyday scenes for a total of 91 stuff categories and 80 object classes.

\medskip

\noindent \textbf{Open Images V6}\cite{OpenImageV4}. This dataset contains 9M images equipped with labels, objects bounding boxes, segmentation, narratives and relationship views. Each image contains 8.3 objects on average. With 16M bounding boxes over 600 categories, it is the largest object localization dataset ever realized.
Moreover, it is able to provide 19,957 classes with an annotation at the image-level.

\medskip

\noindent \textbf{CORe50}\cite{CORe50}. This is the first collection of images designed for continual object recognition, i.e., learning new classes online. It is composed of 11 sessions of 300 RGB-D images that can be classified by objects (50) or by categories (10). The objects are held and moved by the operator who is recording a 15 seconds video at 20 fps (300 frames in total) from a subjective point of view. 

\medskip

\noindent \textbf{ObjectNet}~\cite{objectnet}. ObjectNet is a dataset composed only of a test set of 50.000 images. The aim is to test object recognition applications in a real-world scenario in which images present background, rotation and often the viewpoint are random. ObjectNet showed that applications performing at the top in their respective datasets present a lack of generalization with a 40-45\% drop in the performance. Fine tuning-robust, this dataset represents one of the best challenges for the generalization of object recognition tasks.

\medskip

\begin{table*}[!ht]
\centering
\caption{Dataset summary}
\label{tab:dataset}
\begin{tabular}{|c|c|c|c|c|c|c|}
\hline
\textbf{Dataset} & \textbf{Format}  & \textbf{Instances} & \textbf{Size} & \textbf{Task} & \begin{tabular}[c]{@{}c@{}} \textbf{Year of} \\ \textbf{creation} \end{tabular} & \textbf{Creator} \\ 
\hline
MNIST & \begin{tabular}[c]{@{}c@{}}Images,\\ text\end{tabular} & 60,000 & 50MB  & Classification & 
1998 & \begin{tabular}[c]{@{}c@{}}LeCun et al.~\cite{MNIST}\end{tabular} \\ 
\hline
ImageNet  & \begin{tabular}[c]{@{}c@{}}Images, \\ text\end{tabular} & 14,197,122 & 150GB & \begin{tabular}[c]{@{}c@{}}Classification, \\ Object recognition\end{tabular} & 2009 
& J. Deng et al.~\cite{ImageNet} \\ 
\hline
CIFAR & \begin{tabular}[c]{@{}c@{}}Images,\\ text\end{tabular}  & 60,000 & 170MB & Classification & 
2009 & A. Krizhevsky et al.~\cite{CIFAR}  \\ 
\hline
COCO & \begin{tabular}[c]{@{}c@{}}Images,\\ text\end{tabular} & 2,500,000 & 25GB & Object recognition & 
2015 & T. Lin et al.~\cite{COCO}   \\
\hline
Open Image V6 & \begin{tabular}[c]{@{}c@{}}Images,\\ Segmentation masks,\\ text\end{tabular} & 9,000,000 & 500GB         & \begin{tabular}[c]{@{}c@{}}Classification,\\ Object recognition,\\ Object detection,\\ Object segmentation\end{tabular} & 2020 & Kuznetsova et al.~\cite{OpenImageV4} \\ 
\hline
CORe50 & Images & 3,300 & 25.8GB  & \begin{tabular}[c]{@{}c@{}}Continuous\\ Object recognition\end{tabular} & 
2017 & Lomonaco et al.~\cite{CORe50}  \\ 
\hline
ObjectNet  & \begin{tabular}[c]{@{}c@{}}Images,\\ text\end{tabular} & 50,000 & -  & Object recognition                   & 2019 & \begin{tabular}[c]{@{}c@{}}Barbu et al.~\cite{objectnet}\end{tabular} \\ 
\hline
\end{tabular}%
\end{table*}

\subsection{Neural Networks Model Metrics}
The attributes of DL models can be evaluated, considering a few important metrics. 
\begin{itemize}[leftmargin=*]
    \item \textbf{Accuracy}. The accuracy (Top-5 or Top-1) of a model with respect to a specific Dataset is an important metric. Besides the dataset, the training properties must be reported, e.g., the number of epochs, learning rate, data augmentation.
    \item \textbf{Model Architecture}. The shape of the DL model is the foundation for understanding how it operates. The number and type of layers, the size of feature maps and the number of channels, the number of filters and their size are all fundamental properties to understand how an algorithm elaborates the incoming raw data.
    \item \textbf{Workload}. The size of the input feature map and the number of kernels define the total amount of operations (MACs) to be performed. The MAC count is one of the basic metrics to evaluate a DNN, and it also defines the throughput and the energy effort of the target hardware platform. To be noted that as explained in Section~\ref{sec:sparsity}, only effective MACs should be counted.
    \item \textbf{Memory requirement}. The amount of weights (non-null) determines the storage impact of the model. A large number of weights could represent a limit for the target hardware platform and for the power envelope as well.
    \item \textbf{Training time}. Typically, the higher the complexity of the model, the more accurate it is. 
    On the other hand, complexity results in difficulties in training the model. In fact, the more the weights and the number of layers, the more epochs will be needed. This metric can be expressed as number of training epochs or GPU hours related to a specific dataset to obtain a certain accuracy.
    \item \textbf{Adversarial Robustness}. As explained in Section~\ref{security}, DNNs are vulnerable to adversarial attacks, which is one of the hottest research topics in the development of new deep learning models. Therefore, it is of fundamental importance to provide the model with defense algorithms and perform an exhaustive and correct evaluation of adversarial attacks.

\end{itemize}

\subsection{Hardware Accelerator Metrics}
The fundamental metrics to evaluate the hardware platforms are:
\begin{itemize}[leftmargin=*]
    \item \textbf{Power}. The power consumption of the device determines the final application for which it can be exploited. In addition, an important metric is energy efficiency defined as pJ for MAC. Note that the power consumed must also include that spent on readings from the off-chip memory, as explained in Section~\ref{memory_hier}.
    \item \textbf{Throughput}. Throughput and latency depend directly on the working frequency of the device and the memory bandwidth. Such metrics are crucial to define how often the hardware platform can perform a complete inference or backpropagation of a model. The throughput is often defined as billions of operations per second (Gop/s) or as billions of MAC per second (GMAC/s) where 1GMAC/s $\simeq$ 2Gop/s.
    \item \textbf{Area}. From the area of the device, cost and integration capacity in larger systems are derived. On the other hand, the area depends on the technological node and the amount of memory. Memory, as in the case of power, plays a critical role also for the area. This represents another reason to optimize its use.
\end{itemize}
In addition to the main metrics listed above, there are others that could be defined as application-dependent. For example, the flexibility with which an accelerator can be adapted for more complex models, parallelizing its architecture, or how a device can scale with respect to the required computing accuracy, having a tuning bitwidth.
Although metrics are easily definable, comparisons between different hardware platforms are not always straightforward. There are many factors on which possible comparisons depend, moreover the benchmarks used to evaluate performance are not always impartial. It is, therefore, necessary to assess all the side-factors on a case-by-case basis.

\section{Challenges and the Road Ahead}\label{sec:challenges}
As discussed in Section~\ref{sec:introduction}, AI and DL are adopted in numerous and various fields, and the number of their applications is growing over time. The number of investments that have been made in AI startups over the years is a clear demonstration of this growth. In 2010, investments amounted to \$1.3B, while in 2018, they reached \$40.4B~\cite{aiindex2019}. The average annual growth rate was 48\%, and this trend does not seem to stop. Many technical reports~\cite{aiindex2019}~\cite{ecchipreport} indicate that, in the years to come, AI will be a driving force to the economy. 

\begin{figure}[h]
    \centering
    \includegraphics[width=\linewidth]{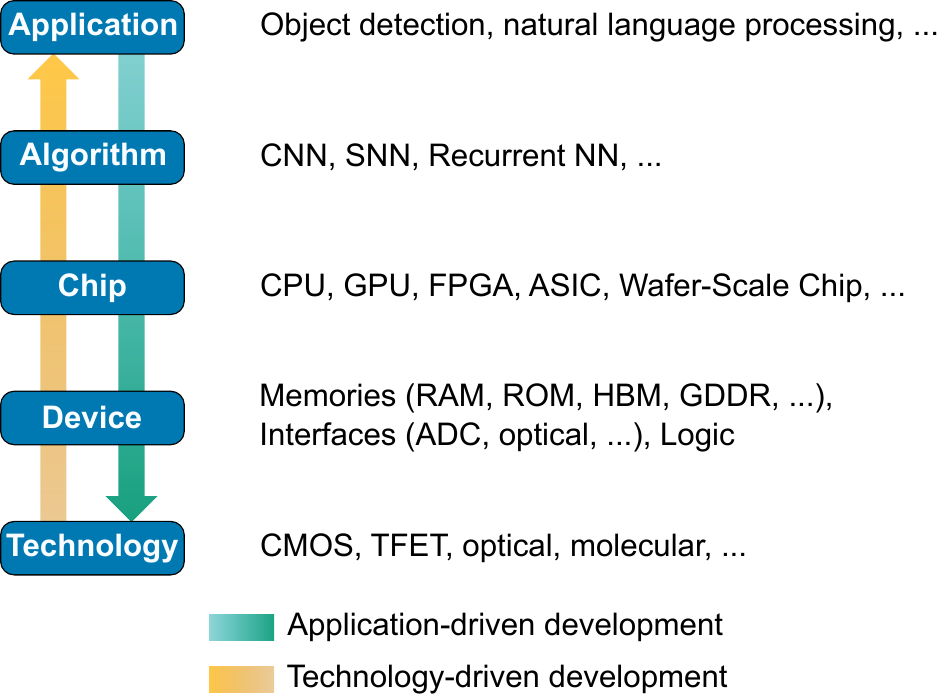}
    \caption{Development flows that run from the application to the technology and vice-versa.}
    \label{fig:flows}
\end{figure}

The development and diffusion of AI applications are closely related to technological advancements, i.e., the chips. There is a flow that runs from the application, that is expressed as an algorithm. The algorithm is deployed on a chip, which consists of some devices realized in a technology, e.g., CMOS technology (see Figure~\ref{fig:flows}). The growth of AI applications in number and complexity has required more and more performance from the hardware (\textit{application-driven development}). Before 2012, the chip compute doubled about every two years. After 2012, the year of DL boom, AI chips started to double the compute every 3.4 months~\cite{aiindex2019}. On the other hand, the development of new technologies and hardware improvements allowed to develop more complex and therefore accurate applications (\textit{technology-driven development}). The two directions of development continue to feed each other in a virtuous circle. It is estimated that, in 2025, the AI chip market will reach a value of \$29B, while in 2017, it was only \$2B~\cite{ecchipreport}.

To maintain such a high growth rate, the industrial and academic world will face new challenges in the coming years, which we summarize in the following, together with the possible solutions, research trends and future directions. 

\textbf{Von Neuman Bottleneck:} One of the biggest challenges the developers are facing currently is the Von Neuman bottleneck, i.e., the bandwidth that modern memories can provide is not sufficient for the huge amount of data that AI chips need to process. To get around the problem, it is possible to modify the algorithms to reduce the number of data items to be used (e.g., model compression, pruning, or quantization). 

\begin{figure}[h]
    \centering
    \includegraphics[width=\linewidth]{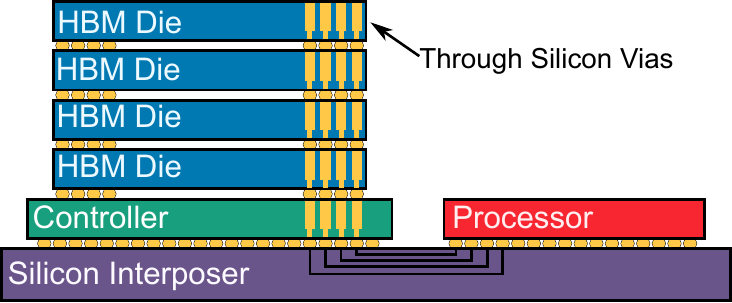}
    \caption{High Bandwidth Memory (HBM) scheme.}
    \label{fig:hbm}
\end{figure}

To solve the problem, it is necessary to act at the memories level. One possible solution is the increase of the memory bandwidth, and this is the purpose of High Bandwidth Memories (HBMs) (see Figure~\ref{fig:hbm}). HBM is a stacked DRAM integrated with the processing elements through a silicon interposer. A single HBM2 block has a bandwidth of 256~GB/s, lower than the 616~GB/s bandwidth of a more traditional Graphics Double Data Rate 6 (GDDR6) memory. However, a stack with four HBM blocks reaches a 1~TB/s bandwidth. HBM2 memories are currently used in the Nvidia V100 and P100 GPUs. 

Another possibility is in-memory computing (IMC), which consists of moving the logic inside the memory. IMC is particularly suitable for the DNNs operations since DNNs algorithms are deterministic, and it is possible to know when and where data items will be required in advance. IMC  wants to enhance DNN acceleration by reducing the latency and power needed to access the memory hierarchy in traditional Von Neuman architectures. Moreover, it increases the parallelization by working with all the memory cells simultaneously. Researchers are currently studying the application of IMC to DNNs algorithms and obtaining promising results~\cite{brein}~\cite{8310401}~\cite{xnorsram}~\cite{imc}, and Mythic startup produces IMC accelerators for AI with a 40~nm process. 

\begin{figure}[h]
    \centering
    \includegraphics[width=\linewidth]{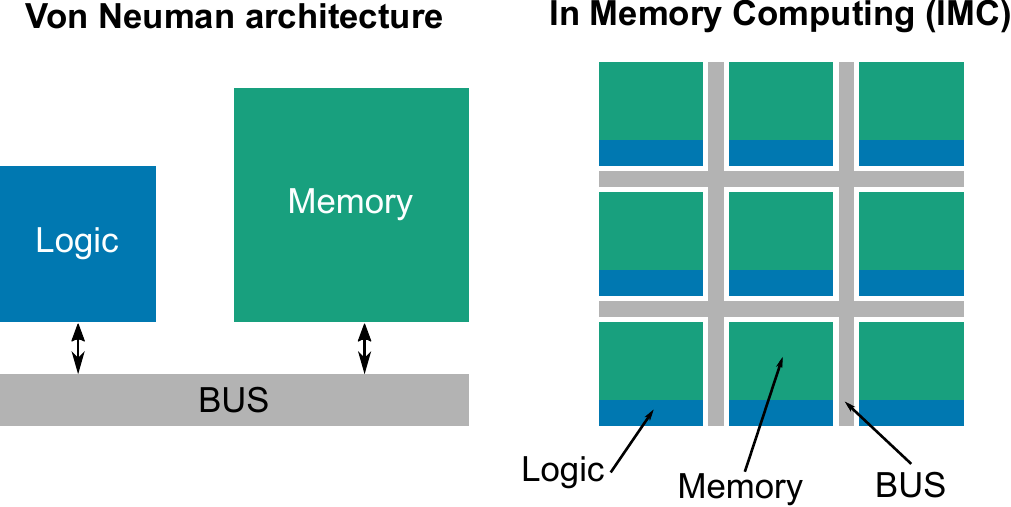}
    \caption{Comparison between the structures of a Von Neuman Architecture (\textbf{left}) and an In Memory Computing (IMC) architecture (\textbf{right}).}
    \label{fig:imc}
\end{figure}

\textbf{CMOS Technology Limitations:} From the '60s onwards, CMOS technology has been scaling following Moore's Law, according to which the number of transistors on a chip doubles every 24 months. However, this pace of scaling is beginning to stop and it will not be sustainable in the future for technological and economic reasons~\cite{endmoore}. Researchers are currently exploring new physical possibilities and a lot of effort is placed in the emerging memories, such as Phase Change Memories (PCMs)~\cite{phasechangememory}~\cite{phasechangememory2}, Spin-Torque-Transfer Magnetoresistive RAM (STT-MRAM)~\cite{sttmram}~\cite{sttmram2}, or Resistive RAM (ReRAM)~\cite{reram}~\cite{sttmram2}. Beyond emerging memories, several new technologies are being studied, such as Tunnel FETs, organic FETs, molecular transistors, and spintronic devices. Despite the possible gains deriving from moving to a beyond-CMOS technology, replacing CMOS technology with emerging ones will not be an immediate procedure since it is considered very reliable and easy to manufacture. Moreover, foundries and production lines have been calibrated to this technology and cannot be dismantled until production has paid for the initial investment.

\textbf{AI Toolchains:} Besides the special-purpose ASICs and the programmable CPUs/GPUs, flexible hyper-scale AI accelerators are gaining importance, e.g., Google TPUs or Cerebras Wafer Scale Engine. As seen in Section \ref{sec:framework}, there are several high-level frameworks, mainly python-based, for the description of DLs algorithms. However, there is not yet a unified method to program AI accelerators from a unified high-level language. So far there are the compiler toolchains for CPUs and GPUs, and there is the synthesis toolchain for the FPGAs. The development of a toolchain for AI accelerators programming will be a huge step forward for their diffusion. 

\textbf{General AI: } Even though DL models can perform various tasks at a better-than-human level, e.g., object detection or language processing, AI is to be considered still at an early development stage. Indeed, scientists are very far from the so-called Artificial General Intelligence (AGI), e.g., an algorithm able to perform multiple tasks and of taking decisions. Even if an AGI algorithm existed, at the moment, probably, hardware systems could not provide enough computational power for its deployment. For a long future, it will be necessary to combine different algorithms to perform complex tasks. For this reason, it will be important to develop hardware platforms able to support multiple algorithms, easily programmable or reconfigurable.

\textbf{AI At The Edge:} It has been listed in the 2020 Top Technological Trends~\cite{toptrends} by the IEEE Computing Society. Thanks to the diffusion of 5G connectivity and IoT sensors, ML algorithms will spread into the edge devices. If compared to AI cloud platforms, the edge devices have completely different requirements. During the development of AI edge devices, the focus must be placed on low power and low latency.  For this purpose, many roads can be taken. At the application level, it is necessary to develop models co-optimized with the hardware for a more efficient resource handling. At the hardware level, new possibilities are being explored beyond the traditional low-power techniques, e.g., moving the computation in the analog part of the circuit to save energy~\cite{analog1}~\cite{analog2}~\cite{analog3}. 

\begin{figure}[h]
    \centering
    \includegraphics[width=0.85\linewidth]{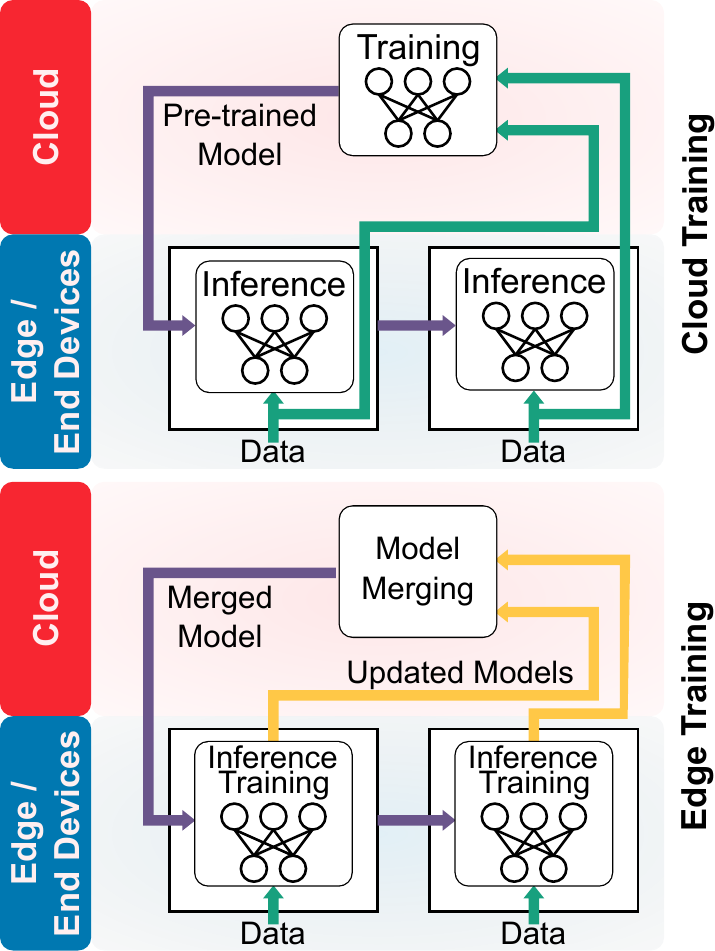}
    \caption{Comparison between training performed in the cloud (\textbf{top}) and on edge devices (\textbf{bottom}).}
    \label{fig:edgetr}
\end{figure}

Presently, most of the AI edge devices perform the inference only. The collected data must be sent to the cloud for model training (see Figure~\ref{fig:edgetr} top). In the future, it will be important to move the learning to the edge device for several reasons. The learning in the sensors guarantees real-time lifelong learning, i.e., the device can immediately learn from every sample received and adapt the model consequently. The connection to the cloud is no more needed continuously and higher data privacy can be guaranteed since it is no more necessary to communicate the data but only the models (see Figure~\ref{fig:edgetr} bottom). 

\section{Distinction from Other Surveys}\label{sec:relatedworks}
Over the years, many works have been proposed to give an overview of the research carried out and the recent state-of-the-art. However, Deep Learning is currently a hot topic, so research is progressing fast with continuous discoveries and improvements. The same applies to hardware architectures. Although the fundamental blocks are fixed, the paradigms with which they can be combined and exploited are many and varied. Therefore, it is essential to have surveys that periodically collect the newest material and the recent advancements to keep researchers up to date. This is the idea behind this work, which wants to inform hardware designers about the latest architectures and techniques employed in the DL field. 

This paper is intended as complementary to the surveys already available in the literature. The authors aim to focus on the hardware architectures for DL that have become available in the last five years, with a cross-cut on the different platforms. Schuman et al.~\cite{neuromorphic} have collected and summarized the previous 35 years of discoveries related to neuromorphic computing, with various examples of hardware for neural networks. However, the paper does not offer much discussion on the collected material, remaining very compact. On the other hand, Chen et al.~\cite{chen_survey} offer a much broader view, but the examples are limited, and topics such as SNN and adversarial attacks are not covered. Deng et al.~\cite{compr_acc} propose a very complete and extensive work that deals thoroughly with the compression of data taking into account the sparsity and quantization. The work is very comprehensive; nevertheless, it remains very biased toward the compression and lacks of considerations on the SNN. Our work stems from the survey proposed by Sze et al.~\cite{sze}, however, updating it with the numerous advances of the last three years and completing it with comprehensive sections on SNNs and adversarial attacks. Table~\ref{tab:comparison} compares a list of state-of-the-art surveys, showing the key aspects that characterize each work.

\begin{table*}[]
\centering
\caption{Comparison among state-of-the-art surveys.}
\label{tab:comparison}
\resizebox{\textwidth}{!}{%
\begin{tabular}{|c|c|c|c|c|c|c|c|c|c|c|c|c|c|c|c|c|}
\hline
\multirow{2}{*}[-4.5em]{\rotatebox[origin=c]{90}{\textbf{Reference}}} & \multirow{2}{*}[-5.5em]{\rotatebox[origin=c]{90}{\textbf{Year}}} & \multirow{2}{*}[-4.5em]{\rotatebox[origin=c]{90}{\textbf{Citation \#}}} & \multicolumn{7}{c|}{\textbf{DNN and general ML}}                                                                                                                                                                                                                                    & \multicolumn{2}{c|}{\textbf{SNN}}                        & \textbf{Security }                                                     & \multicolumn{3}{c|}{\textbf{Benchmarking}}                                        & \multirow{2}{*}[-1.2em]{\rotatebox[origin=c]{90}{\begin{tabular}[c]{@{}c@{}}\textbf{CPU/GPU/FPGA/ASIC} \\ \textbf{Comparison}\end{tabular}}} \\ \cline{4-16}
                           &                       &                              & \multicolumn{1}{l|}{\rotatebox[origin=c]{90}{\textbf{Techniques}}} & \multicolumn{1}{l|}{\rotatebox[origin=c]{90}{\textbf{Dataflow}}} & \begin{tabular}[c]{@{}c@{}}\rotatebox[origin=c]{90}{\textbf{Tools for Design}}\end{tabular} & \rotatebox[origin=c]{90}{\textbf{Quantization}}          & \rotatebox[origin=c]{90}{\textbf{Sparsity}}               & \begin{tabular}[c]{@{}c@{}}\rotatebox[origin=c]{90}{\hspace{0.05cm} \textbf{Approximate  Computing} \hspace{0.05cm}}\end{tabular} & \rotatebox[origin=c]{90}{\textbf{Memory  Hierarchy}}       & \rotatebox[origin=c]{90}{\textbf{Techniques}}             & \rotatebox[origin=c]{90}{\textbf{Architectures} }         & \begin{tabular}[c]{@{}c@{}}\rotatebox[origin=c]{90}{\textbf{Adversarial  Attack}}\end{tabular} & \rotatebox[origin=c]{90}{\textbf{Frameworks}}             & \rotatebox[origin=c]{90}{\textbf{Datasets}}               & \rotatebox[origin=c]{90}{\textbf{Metrics}}                &                                                                                         \\ \hline
                          ~\cite{ref6}& 2016                  & 2                            & $\approx$          & \xmark         & \xmark                                       & \xmark & \xmark  & \xmark                                            & \xmark  & $\approx$ & $\approx$ & \xmark                                         & \xmark  & $\approx$ & \xmark  & \xmark                                                                   \\ \hline
                          ~\cite{neuromorphic} & 2017                  &     -                      & \cmark           & $\approx$         & \xmark                                       & \xmark & \xmark & \xmark                                           & $\approx$ & \cmark  & \cmark  & \xmark                                         & \cmark & \xmark & \xmark & \cmark                                                                  \\ \hline
                          ~\cite{sze} & 2017                  & 247                          & \cmark           & \cmark         & \xmark                                       & \cmark & $\approx$ & \cmark                                           & \cmark & \xmark  & \xmark  & \xmark                                         & \cmark & \cmark & \cmark & \cmark                                                                  \\ \hline
                          ~\cite{ref5} & 2017                  & 8                            & \cmark           & \xmark         & \xmark                                       & \xmark & \xmark  & \xmark                                            & \xmark  & \xmark  & \xmark  & \xmark                                         & \xmark  & \xmark  & $\approx$ & \xmark                                                                   \\ \hline
                          ~\cite{ref8} & 2018                  & 2                           & \xmark           & \xmark         & $\approx$                                      & \xmark & \xmark  & \xmark                                            & \xmark  & \xmark  & \xmark  & \xmark                                         & \xmark  & \xmark  & $\approx$ & \cmark                                                                  \\ \hline
                           ~\cite{ref10} & 2018                  & 3                            & \xmark           & \xmark         & \xmark                                       & \xmark & \xmark  & \xmark                                            & \xmark  & \xmark  & \xmark  & \xmark                                         & \xmark  & \cmark & \cmark & \cmark                                                                  \\ \hline
                         ~\cite{ref9} & 2019                  & 14                            & \cmark           & \xmark         & \xmark                                       & \xmark & \xmark  & \xmark                                            & \cmark & \xmark  & \xmark  & \xmark                                         & \xmark  & \xmark  & \xmark  & \xmark                                                                   \\ \hline
                        ~\cite{ref1}  & 2019                  & 11                            & \cmark           & \xmark         & \xmark                                       & \xmark & \xmark  & \xmark                                            & \xmark  & \xmark  & \xmark  & \xmark                                         & \xmark  & \xmark  & $\approx$ & \xmark                                                                   \\ \hline
                         ~\cite{ref3} & 2019                  & 7                           & \cmark           & \xmark         & \xmark                                       & \xmark & $\approx$ & \xmark                                            & \xmark  & \xmark  & \xmark  & \xmark                                         & \xmark  & \xmark  & \xmark  & \xmark                                                                   \\ \hline
                           
                          ~\cite{ref13} & 2019                  & -                            & \cmark           & \xmark         & $\approx$                                      & \xmark & \xmark  & \xmark                                            & \xmark  & \xmark  & \xmark  & \xmark                                         & \xmark  & \xmark  & $\approx$ & $\approx$                                                                  \\ \hline
                          ~\cite{ref12}& 2019                  & 82                            & \cmark           & \cmark         & \cmark                                       & \xmark & \xmark  & \xmark                                            & \xmark  & \xmark  & \xmark  & \xmark                                         & $\approx$ & $\approx$ & \xmark  & \xmark                                         \\ \hline 
                         ~\cite{chen_survey} & 2020                  & 1                            & \cmark           & \cmark         & \xmark                                       & $\approx$ & \cmark  & \xmark                                           & \cmark & \xmark  & \xmark  & \xmark                                        & \xmark  & \xmark  & $\approx$   &    \xmark                                                              \\ \hline
                          ~\cite{compr_acc} & 2020                  & -                            & \cmark           & \cmark         & \xmark                                       & \cmark & \cmark  & $\approx$                                           & \cmark & \xmark  & \xmark  & $\approx$                                       & \cmark & \xmark  & \xmark  & $\approx$                                                                  \\ \hline

                         ~\cite{ref7} & 2020                  & -                            & $\approx$          & \xmark         & \xmark                                       & \xmark & $\approx$ & \xmark                                            & \xmark  & $\approx$ & \xmark  & \xmark                                         & \xmark  & \xmark  & $\approx$ & \xmark                                                                   \\ \hline

                          ~\cite{ref11}& 2020                  & -                            & \cmark           & \xmark         & \xmark                                       & \xmark & $\approx$ & \xmark                                            & \xmark  & \xmark  & \xmark  & $\approx$                                        & \cmark & \cmark & \xmark  & \xmark                                                                   \\ \hline
                          
                          \textbf{our work} & \textbf{2020}                  & \textbf{-}                            & \textbf{\cmark}           & \textbf{\cmark}         & \textbf{\cmark}                                       & \textbf{\cmark} & \textbf{\cmark}  & \textbf{\cmark}                                            & \textbf{\cmark}  & \textbf{\cmark}  & \textbf{\cmark}  & \textbf{\cmark}                                         & \textbf{\cmark} & \textbf{\cmark} & \textbf{\cmark}  & \textbf{\cmark}                                         \\ \hline 
\end{tabular}
}
\end{table*}

\section{Conclusion}\label{sec:conclusion}
The focus on Deep Learning (DL) has grown exponentially in recent years, as well as the performance of the algorithms and the number of applications that involve it. However, with the increasing complexity of algorithms, the need for hardware devices capable of satisfying the requirements has also increased. The DL has always stood out for its high workload and for being computation-hungry. Moreover, today’s trend is to move towards mobile and possibly wearable devices that are part of the IoT whose architectures are heterogeneous, in which general-purpose processors are coupled with dedicated accelerators. The IoT introduces even tighter power constraints considering that many of its nodes are battery-powered or rely on energy harvest systems.

Therefore, it is essential to take into consideration the critical aspects of the hardware already in the design phase. In this regard, there are a large number of techniques to design hardware architectures with high energy-efficiency and high performance without sacrificing accuracy.

This work surveys most of the known techniques to produce energy-efficient dataflows, handling especially the aspects related to memory. The memory hierarchy is deeply analyzed to understand to which levels it is convenient to intervene and, in the ad-hoc architectures, how it must be modelled in order to reduce to the minimum the power consumption. For example, starting from the memory that is the most power-greedy element, it is possible to define a dataflow with a related memory hierarchy that maximizes the data reuse,  avoiding continuous access to memory.

The article mainly refers to three models: Deep neural Networks (DNNs), convolutional neural Networks (CNNs) and Spiking Neural Networks (SNNs). While the first two cases are important for the performance and accuracy they have managed to achieve, often beyond the human one, the latter is interesting for the low-power profile and paradigm they represent, considered by many as the third generation of NNs.
 
In addition to the techniques for developing accelerator architectures, other factors need to be considered like the cybersecurity. In the DL world, security attacks are often represented by noise injection into the input sample to the NN to ensure its misclassification. Many different types of attacks exist, according to the knowledge of the network under attack, the type of perturbation and the target class.

Finally, this work presents which frameworks to use to create or modify models and any datasets to test them on. Benchmarking is a key step in establishing the properties of the networks, but also the hardware on which they are developed. The most critical metrics to define their goodness and comparison with other platforms are examined.

\bibliographystyle{ieeetr}
\bibliography{biblio.bib} 

\begin{thebibliography}{100}

\bibitem{energy_eff}
T.~Hamada, K.~Benkrid, K.~Nitadori, and M.~Taiji, ``A comparative study on
  asic, fpgas, gpus and general purpose processors in the o(n-2) gravitational
  n-body simulation,'' {\em Proceedings - 2009 NASA/ESA Conference on Adaptive
  Hardware and Systems, AHS 2009}, 07 2009.

\bibitem{SurpassingHuman-Level}
K.~He, X.~Zhang, S.~Ren, and J.~Sun, ``Delving deep into rectifiers: Surpassing
  human-level performance on imagenet classification,'' in {\em Proceedings of
  the 2015 IEEE International Conference on Computer Vision (ICCV)}, ICCV
  ’15, (USA), p.~1026–1034, IEEE Computer Society, 2015.

\bibitem{8633218}
D.~{Zhang} and S.~{Liu}, ``Top-down saliency object localization based on
  deep-learned features,'' in {\em 2018 11th International Congress on Image
  and Signal Processing, BioMedical Engineering and Informatics (CISP-BMEI)},
  pp.~1--9, 2018.

\bibitem{8793320}
T.~{Treebupachatsakul} and S.~{Poomrittigul}, ``Bacteria classification using
  image processing and deep learning,'' in {\em 2019 34th International
  Technical Conference on Circuits/Systems, Computers and Communications
  (ITC-CSCC)}, pp.~1--3, 2019.

\bibitem{minaee2020image}
S.~Minaee, Y.~Boykov, F.~Porikli, A.~Plaza, N.~Kehtarnavaz, and D.~Terzopoulos,
  ``Image segmentation using deep learning: {A} survey,'' {\em CoRR},
  vol.~abs/2001.05566, 2020.

\bibitem{8840035}
M.~{Aibin}, ``Deep learning for cloud resources allocation: Long-short term
  memory in eons,'' in {\em 2019 21st International Conference on Transparent
  Optical Networks (ICTON)}, pp.~1--4, 2019.

\bibitem{8876949}
H.~C. {Kaskavalci} and S.~{Gören}, ``A deep learning based distributed smart
  surveillance architecture using edge and cloud computing,'' in {\em 2019
  International Conference on Deep Learning and Machine Learning in Emerging
  Applications (Deep-ML)}, pp.~1--6, 2019.

\bibitem{8959715}
R.~{Zanc}, T.~{Cioara}, and I.~{Anghel}, ``Forecasting financial markets using
  deep learning,'' in {\em 2019 IEEE 15th International Conference on
  Intelligent Computer Communication and Processing (ICCP)}, pp.~459--466,
  2019.

\bibitem{8258131}
J.~J. {Ying}, P.~{Huang}, C.~{Chang}, and D.~{Yang}, ``A preliminary study on
  deep learning for predicting social insurance payment behavior,'' in {\em
  2017 IEEE International Conference on Big Data (Big Data)}, pp.~1866--1875,
  2017.

\bibitem{8701943}
V.~{Ha}, D.~{Lu}, G.~S. {Choi}, H.~{Nguyen}, and B.~{Yoon}, ``Improving credit
  risk prediction in online peer-to-peer (p2p) lending using feature selection
  with deep learning,'' in {\em 2019 21st International Conference on Advanced
  Communication Technology (ICACT)}, pp.~511--515, 2019.

\bibitem{8875896}
A.~K. {Arslan}, \c{S}. {Ya\c{s}ar}, and C.~{\c{C}olak}, ``An intelligent system
  for the classification of lung cancer based on deep learning strategy,'' in
  {\em 2019 International Artificial Intelligence and Data Processing Symposium
  (IDAP)}, pp.~1--4, 2019.

\bibitem{brain}
H.~Mohsen, E.-S. El-Dahshan, E.-S. El-Horbarty, and A.-B. M.Salem,
  ``Classification using deep learning neural networks for brain tumors,'' {\em
  Future Computing and Informatics Journal}, vol.~3, 12 2017.

\bibitem{8759561}
C.~{Barata} and J.~S. {Marques}, ``Deep learning for skin cancer diagnosis with
  hierarchical architectures,'' in {\em 2019 IEEE 16th International Symposium
  on Biomedical Imaging (ISBI 2019)}, pp.~841--845, 2019.

\bibitem{Grigorescu_2019}
S.~Grigorescu, B.~Trasnea, T.~Cocias, and G.~Macesanu, ``A survey of deep
  learning techniques for autonomous driving,'' {\em Journal of Field
  Robotics}, Nov 2019.

\bibitem{humanoid}
T.~S. {Li}, P.~{Kuo}, C.~{Chang}, H.~{Hsu}, Y.~{Chen}, and C.~{Chang}, ``Deep
  belief network–based learning algorithm for humanoid robot in a pitching
  game,'' {\em IEEE Access}, vol.~7, pp.~165659--165670, 2019.

\bibitem{assistive}
C.~{Wang}, D.~{Freer}, J.~{Liu}, and G.~{Yang}, ``Vision-based automatic
  control of a 5-fingered assistive robotic manipulator for activities of daily
  living,'' in {\em 2019 IEEE/RSJ International Conference on Intelligent
  Robots and Systems (IROS)}, pp.~627--633, 2019.

\bibitem{swarm}
J.~{Guan}, W.~{Zhou}, S.~{Kang}, Y.~{Sun}, and Z.~{Liu}, ``Robot formation
  control based on internet of things technology platform,'' {\em IEEE Access},
  vol.~8, pp.~96767--96776, 2020.

\bibitem{drone}
D.~Palossi, A.~Loquercio, F.~Conti, E.~Flamand, D.~Scaramuzza, and L.~Benini,
  ``Ultra low power deep-learning-powered autonomous nano drones,'' {\em CoRR},
  vol.~abs/1805.01831, 2018.

\bibitem{8792636}
N.~{Tsang}, C.~{Cao}, S.~{Wu}, Z.~{Yan}, A.~{Yousefi}, A.~{Fred-Ojala}, and
  I.~{Sidhu}, ``Autonomous household energy management using deep reinforcement
  learning,'' in {\em 2019 IEEE International Conference on Engineering,
  Technology and Innovation (ICE/ITMC)}, pp.~1--7, 2019.

\bibitem{8403442}
C.~{Heghedus}, A.~{Chakravorty}, and C.~{Rong}, ``Energy load forecasting using
  deep learning,'' in {\em 2018 IEEE International Conference on Energy
  Internet (ICEI)}, pp.~146--151, 2018.

\bibitem{IoT_shafique}
M.~{Shafique}, T.~{Theocharides}, C.~{Bouganis}, M.~A. {Hanif}, F.~{Khalid},
  R.~{Hafız}, and S.~{Rehman}, ``An overview of next-generation architectures
  for machine learning: Roadmap, opportunities and challenges in the iot era,''
  in {\em 2018 Design, Automation Test in Europe Conference Exhibition (DATE)},
  pp.~827--832, 2018.

\bibitem{edge_shafique}
A.~{Marchisio}, M.~A. {Hanif}, F.~{Khalid}, G.~{Plastiras}, C.~{Kyrkou},
  T.~{Theocharides}, and M.~{Shafique}, ``Deep learning for edge computing:
  Current trends, cross-layer optimizations, and open research challenges,'' in
  {\em 2019 IEEE Computer Society Annual Symposium on VLSI (ISVLSI)},
  pp.~553--559, 2019.

\bibitem{edgecomp}
M.~Capra, R.~Peloso, G.~Masera, M.~Ruo~Roch, and M.~Martina, ``Edge computing:
  A survey on the hardware requirements in the internet of things world,'' {\em
  Future Internet}, vol.~11, no.~4, 2019.

\bibitem{diannao}
T.~Chen, Z.~Du, N.~Sun, J.~Wang, C.~Wu, Y.~Chen, and O.~Temam, ``Diannao: A
  small-footprint high-throughput accelerator for ubiquitous
  machine-learning,'' {\em International Conference on Architectural Support
  for Programming Languages and Operating Systems - ASPLOS}, vol.~49,
  pp.~269--284, 02 2014.

\bibitem{Freeman2005biological}
S.~{Freeman} and H.~{Hamilton} {\em Biological Science}, 2005.

\bibitem{mcculloch1943}
W.~Mcculloch and W.~Pitts, ``A logical calculus of ideas immanent in nervous
  activity,'' {\em Bulletin of Mathematical Biophysics}, vol.~5, pp.~127--147,
  1943.

\bibitem{rosenblatt}
F.~Rosenblatt, {\em The Perceptron: A Perceiving and Recognizing Automaton
  (Project PARA). Report No. 85-460-1}.
\newblock Cornell Aeronautical Laboratory, 1957.

\bibitem{Bengio2009}
Y.~Bengio, ``Learning deep architectures for {AI},'' {\em Foundations and
  Trends in Machine Learning}, vol.~2, no.~1, pp.~1--127, 2009.

\bibitem{MNIST}
Y.~Lecun, L.~Bottou, Y.~Bengio, and P.~Haffner, ``Gradient-based learning
  applied to document recognition,'' {\em Proceedings of the IEEE}, vol.~86,
  pp.~2278 -- 2324, 12 1998.

\bibitem{hubel1959}
D.~H. Hubel and T.~N. Wiesel, ``Receptive fields of single neurons in the cat's
  striate cortex,'' {\em Journal of Physiology}, vol.~148, pp.~574--591, 1959.

\bibitem{batchnorm}
S.~Ioffe and C.~Szegedy, ``Batch normalization: Accelerating deep network
  training by reducing internal covariate shift,'' in {\em Proceedings of the
  32nd International Conference on Machine Learning} (F.~Bach and D.~Blei,
  eds.), vol.~37 of {\em Proceedings of Machine Learning Research}, (Lille,
  France), pp.~448--456, PMLR, 07--09 Jul 2015.

\bibitem{momentum}
N.~Qian, ``On the momentum term in gradient descent learning algorithms,'' {\em
  Neural Networks}, vol.~12, no.~1, pp.~145--151, 1999.

\bibitem{nesterov}
Y.~Nesterov, ``A method for unconstrained convex minimization problem with the
  rate of convergence o(1/k$^2$),'' {\em Doklady AN USSR}, vol.~269,
  pp.~543--547, 1983.

\bibitem{adagrad}
J.~Duchi, E.~Hazan, and Y.~Singer, ``Adaptive subgradient methods for online
  learning and stochastic optimization,'' {\em J. Mach. Learn. Res.}, vol.~12,
  p.~2121–2159, july 2011.

\bibitem{adadelta}
M.~D. Zeiler, ``{ADADELTA:} an adaptive learning rate method,'' {\em CoRR},
  vol.~abs/1212.5701, 2012.

\bibitem{adam}
D.~Kingma and J.~Ba, ``Adam: A method for stochastic optimization,'' {\em
  International Conference on Learning Representations}, 12 2014.

\bibitem{l1l2}
A.~Y. Ng, ``Feature selection, l1 vs. l2 regularization, and rotational
  invariance,'' in {\em Proceedings of the Twenty-First International
  Conference on Machine Learning}, ICML ’04, (New York, NY, USA), p.~78,
  Association for Computing Machinery, 2004.

\bibitem{dropout}
N.~Srivastava, G.~Hinton, A.~Krizhevsky, I.~Sutskever, and R.~Salakhutdinov,
  ``Dropout: A simple way to prevent neural networks from overfitting,'' {\em
  Journal of Machine Learning Research}, vol.~15, no.~56, pp.~1929--1958, 2014.

\bibitem{ImageNet}
J.~Deng, W.~Dong, R.~Socher, L.-J. Li, K.~Li, and F.~F. Li, ``Imagenet: a
  large-scale hierarchical image database,'' {\em IEEE Conference on Computer
  Vision and Pattern Recognition}, pp.~248--255, 06 2009.

\bibitem{Krizhevsky2012}
A.~Krizhevsky, I.~Sutskever, and G.~E. Hinton, ``Imagenet classification with
  deep convolutional neural networks,'' in {\em Proceedings of the 25th
  International Conference on Neural Information Processing Systems - Volume
  1}, NIPS’12, (Red Hook, NY, USA), p.~1097–1105, Curran Associates Inc.,
  2012.

\bibitem{ILSVRC15}
O.~Russakovsky, J.~Deng, H.~Su, J.~Krause, S.~Satheesh, S.~Ma, Z.~Huang,
  A.~Karpathy, A.~Khosla, M.~Bernstein, A.~C. Berg, and L.~Fei-Fei, ``{ImageNet
  Large Scale Visual Recognition Challenge},'' {\em International Journal of
  Computer Vision (IJCV)}, vol.~115, no.~3, pp.~211--252, 2015.

\bibitem{Simonyan2014}
K.~Simonyan and A.~Zisserman, ``Very deep convolutional networks for
  large-scale image recognition,'' in {\em 3rd International Conference on
  Learning Representations, {ICLR} 2015, San Diego, CA, USA, May 7-9, 2015,
  Conference Track Proceedings} (Y.~Bengio and Y.~LeCun, eds.), 2015.

\bibitem{Szegedy2015}
C.~{Szegedy}, {Wei Liu}, {Yangqing Jia}, P.~{Sermanet}, S.~{Reed},
  D.~{Anguelov}, D.~{Erhan}, V.~{Vanhoucke}, and A.~{Rabinovich}, ``Going
  deeper with convolutions,'' in {\em 2015 IEEE Conference on Computer Vision
  and Pattern Recognition (CVPR)}, pp.~1--9, 2015.

\bibitem{He2015}
K.~He, X.~Zhang, S.~Ren, and J.~Sun, ``Deep residual learning for image
  recognition,'' {\em 2016 IEEE Conference on Computer Vision and Pattern
  Recognition (CVPR)}, pp.~770--778, 2015.

\bibitem{resnetxt}
S.~{Xie}, R.~{Girshick}, P.~{Dollár}, Z.~{Tu}, and K.~{He}, ``Aggregated
  residual transformations for deep neural networks,'' in {\em 2017 IEEE
  Conference on Computer Vision and Pattern Recognition (CVPR)},
  pp.~5987--5995, July 2017.

\bibitem{Huang2016}
G.~Huang, Z.~Liu, and K.~Q. Weinberger, ``Densely connected convolutional
  networks,'' {\em 2017 IEEE Conference on Computer Vision and Pattern
  Recognition (CVPR)}, pp.~2261--2269, 2016.

\bibitem{capsnet}
S.~Sabour, N.~Frosst, and G.~E. Hinton, ``Dynamic routing between capsules,''
  in {\em Proceedings of the 31st International Conference on Neural
  Information Processing Systems}, NIPS’17, (Red Hook, NY, USA),
  p.~3859–3869, Curran Associates Inc., 2017.

\bibitem{Hu2018}
J.~{Hu}, L.~{Shen}, and G.~{Sun}, ``Squeeze-and-excitation networks,'' in {\em
  2018 IEEE/CVF Conference on Computer Vision and Pattern Recognition},
  pp.~7132--7141, June 2018.

\bibitem{nasnet}
B.~Zoph, V.~Vasudevan, J.~Shlens, and Q.~V. Le, ``Learning transferable
  architectures for scalable image recognition,'' in {\em 2018 {IEEE}
  Conference on Computer Vision and Pattern Recognition, {CVPR} 2018, Salt Lake
  City, UT, USA, June 18-22, 2018}, pp.~8697--8710, {IEEE} Computer Society,
  2018.

\bibitem{inceptionv3}
C.~Szegedy, V.~Vanhoucke, S.~Ioffe, J.~Shlens, and Z.~Wojna, ``Rethinking the
  inception architecture for computer vision,'' in {\em 2016 {IEEE} Conference
  on Computer Vision and Pattern Recognition, {CVPR} 2016, Las Vegas, NV, USA,
  June 27-30, 2016}, pp.~2818--2826, {IEEE} Computer Society, 2016.

\bibitem{inceptionv4}
C.~Szegedy, S.~Ioffe, V.~Vanhoucke, and A.~A. Alemi, ``Inception-v4,
  inception-resnet and the impact of residual connections on learning,'' in
  {\em Proceedings of the Thirty-First {AAAI} Conference on Artificial
  Intelligence, February 4-9, 2017, San Francisco, California, {USA}} (S.~P.
  Singh and S.~Markovitch, eds.), pp.~4278--4284, {AAAI} Press, 2017.

\bibitem{resnest}
H.~Zhang, C.~Wu, Z.~Zhang, Y.~Zhu, Z.~Zhang, H.~Lin, Y.~Sun, T.~He, J.~Mueller,
  R.~Manmatha, M.~Li, and A.~J. Smola, ``Resnest: Split-attention networks,''
  {\em CoRR}, vol.~abs/2004.08955, 2020.

\bibitem{tresnet}
T.~Ridnik, H.~Lawen, A.~Noy, and I.~Friedman, ``Tresnet: High performance
  gpu-dedicated architecture,'' {\em CoRR}, vol.~abs/2003.13630, 2020.

\bibitem{trans_autoenc}
G.~E. Hinton, A.~Krizhevsky, and S.~D. Wang, ``Transforming auto-encoders,'' in
  {\em Proceedings of the 21th International Conference on Artificial Neural
  Networks - Volume Part I}, ICANN’11, (Berlin, Heidelberg), p.~44–51,
  Springer-Verlag, 2011.

\bibitem{emrouting}
G.~E. Hinton, S.~Sabour, and N.~Frosst, ``Matrix capsules with {EM} routing,''
  in {\em 6th International Conference on Learning Representations, {ICLR}
  2018, Vancouver, BC, Canada, April 30 - May 3, 2018, Conference Track
  Proceedings}, OpenReview.net, 2018.

\bibitem{pnasnet}
C.~Liu, B.~Zoph, M.~Neumann, J.~Shlens, W.~Hua, L.~Li, L.~Fei{-}Fei, A.~L.
  Yuille, J.~Huang, and K.~Murphy, ``Progressive neural architecture search,''
  in {\em Computer Vision - {ECCV} 2018 - 15th European Conference, Munich,
  Germany, September 8-14, 2018, Proceedings, Part {I}} (V.~Ferrari, M.~Hebert,
  C.~Sminchisescu, and Y.~Weiss, eds.), vol.~11205 of {\em Lecture Notes in
  Computer Science}, pp.~19--35, Springer, 2018.

\bibitem{efficientnet}
M.~Tan and Q.~V. Le, ``Efficientnet: Rethinking model scaling for convolutional
  neural networks,'' in {\em Proceedings of the 36th International Conference
  on Machine Learning, {ICML} 2019, 9-15 June 2019, Long Beach, California,
  {USA}} (K.~Chaudhuri and R.~Salakhutdinov, eds.), vol.~97 of {\em Proceedings
  of Machine Learning Research}, pp.~6105--6114, {PMLR}, 2019.

\bibitem{Maas1997ThirdGenerationSNN}
W.~Maas, ``Networks of spiking neurons: The third generation of neural network
  models,'' {\em Trans. Soc. Comput. Simul. Int.}, 1997.

\bibitem{Kasabov2019TimeSpaceSNN}
N.~K. Kasabov, {\em Time-Space, Spiking Neural Networks and Brain-Inspired
  Artificial Intelligence}.
\newblock Springer-Verlag Berlin Heidelberg, 2019.

\bibitem{Merolla2014Truenorth}
P.~A. Merolla, J.~V. Arthur, R.~Alvarez-Icaza, A.~S. Cassidy, J.~Sawada,
  F.~Akopyan, B.~L. Jackson, N.~Imam, C.~Guo, Y.~Nakamura, B.~Brezzo, I.~Vo,
  S.~K. Esser, R.~Appuswamy, B.~Taba, A.~Amir, M.~D. Flickner, W.~P. Risk,
  R.~Manohar, and D.~S. Modha, ``A million spiking-neuron integrated circuit
  with a scalable communication network and interface,'' {\em Science}, 2014.

\bibitem{Furber2014Spinnaker}
S.~B. {Furber}, F.~{Galluppi}, S.~{Temple}, and L.~A. {Plana}, ``The spinnaker
  project,'' {\em Proceedings of the IEEE}, vol.~102, no.~5, pp.~652--665,
  2014.

\bibitem{Schmitt2017Brainscales}
S.~Schmitt, J.~Klahn, G.~Bellec, A.~Grubl, M.~Guttler, A.~Hartel, S.~Hartmann,
  D.~Husmann, K.~Husmann, S.~Jeltsch, and et~al., ``Neuromorphic hardware in
  the loop: Training a deep spiking network on the brainscales wafer-scale
  system,'' {\em 2017 International Joint Conference on Neural Networks
  (IJCNN)}, 2017.

\bibitem{Davies2018Loihi}
M.~{Davies}, N.~{Srinivasa}, T.~{Lin}, G.~{Chinya}, Y.~{Cao}, S.~H. {Choday},
  G.~{Dimou}, P.~{Joshi}, N.~{Imam}, S.~{Jain}, Y.~{Liao}, C.~{Lin},
  A.~{Lines}, R.~{Liu}, D.~{Mathaikutty}, S.~{McCoy}, A.~{Paul}, J.~{Tse},
  G.~{Venkataramanan}, Y.~{Weng}, A.~{Wild}, Y.~{Yang}, and H.~{Wang}, ``Loihi:
  A neuromorphic manycore processor with on-chip learning,'' {\em IEEE Micro},
  vol.~38, no.~1, pp.~82--99, 2018.

\bibitem{infograph_intro}
``Building a silicon brain.''
  \url{https://www.the-scientist.com/features/building-a-silicon-brain-65738}.
\newblock Accessed: 2020-04-23.

\bibitem{Lichtsteiner2006DVScamera}
P.~{Lichtsteiner}, C.~{Posch}, and T.~{Delbruck}, ``A 128 x 128 120db 30mw
  asynchronous vision sensor that responds to relative intensity change,'' in
  {\em ISSCC}, 2006.

\bibitem{Beeman2013HHneuronmodel}
D.~Beeman, ``Hodgkin-huxley model,'' in {\em Encyclopedia of Computational
  Neuroscience} (D.~Jaeger and R.~Jung, eds.), pp.~1--13, New York, NY:
  Springer New York, 2013.

\bibitem{Izhikevich2003IzhikevichModel}
E.~Izhikevich, ``Simple model of spiking neurons,'' {\em IEEE transactions on
  neural networks / a publication of the IEEE Neural Networks Council},
  vol.~14, pp.~1569--72, 02 2003.

\bibitem{Wang2014LIF}
Z.~Wang, L.~Guo, and M.~Adjouadi, ``A generalized leaky integrate-and-fire
  neuron model with fast implementation method,'' {\em International journal of
  neural systems}, vol.~24, p.~1440004, 08 2014.

\bibitem{Kasinski2011IntroSNN}
F.~Ponulak and A.~Kasiński, ``Introduction to spiking neural networks:
  Information processing, learning and applications,'' {\em Acta neurobiologiae
  experimentalis}, 2011.

\bibitem{Ruf1997hebbian}
B.~Ruf and M.~Schmitt, ``Hebbian learning in networks of spiking neurons using
  temporal coding,'' in {\em Biological and Artificial Computation: From
  Neuroscience to Technology} (J.~Mira, R.~Moreno-D{\'i}az, and J.~Cabestany,
  eds.), (Berlin, Heidelberg), pp.~380--389, Springer Berlin Heidelberg, 1997.

\bibitem{Bi1999STDP}
G.~Bi and M.-m. Poo, ``Synaptic modifications in cultured hippocampal neurons:
  Dependence on spike timing, synaptic strength, and postsynaptic cell type,''
  {\em The Journal of neuroscience : the official journal of the Society for
  Neuroscience}, vol.~18, pp.~10464--72, 01 1999.

\bibitem{Srinivasan2018STDP}
G.~Srinivasan, P.~Panda, and K.~Roy, ``Stdp-based unsupervised feature learning
  using convolution-over-time in spiking neural networks for energy-efficient
  neuromorphic computing,'' {\em J. Emerg. Technol. Comput. Syst.}, 2018.

\bibitem{Fusi2000SDSP}
S.~Fusi, M.~Annunziato, D.~Badoni, A.~Salamon, and D.~J. Amit, ``Spike-driven
  synaptic plasticity: Theory, simulation, vlsi implementation,'' {\em Neural
  Comput.}, vol.~12, p.~2227–2258, Oct. 2000.

\bibitem{Putra2020FSpiNN}
R.~V.~W. Putra and M.~Shafique, ``Fspinn: An optimization framework for memory-
  and energy-efficient spiking neural networks,'' {\em IEEE Transactions on
  Computer-Aided Design of Integrated Circuits and Systems (TCAD),
  ESWeek-Special Issue}, 2020.

\bibitem{Bodo2019ClosingTheAccuracyGap}
B.~R{\"{u}}ckauer, N.~K{\"{a}}nzig, S.~Liu, T.~Delbr{\"{u}}ck, and
  Y.~Sandamirskaya, ``Closing the accuracy gap in an event-based visual
  recognition task,'' {\em CoRR}, 2019.

\bibitem{Boht2000SpikeProp}
S.~Boht{\'e}, J.~Kok, and H.~L. Poutr{\'e}, ``Spikeprop: backpropagation for
  networks of spiking neurons,'' in {\em ESANN}, 2000.

\bibitem{Gtig2006Tempotron}
R.~G{\"u}tig and H.~Sompolinsky, ``The tempotron: a neuron that learns spike
  timing–based decisions,'' {\em Nature Neuroscience}, vol.~9, pp.~420--428,
  2006.

\bibitem{Florian2012Chronotron}
R.~Florian, ``The chronotron: A neuron that learns to fire temporally precise
  spike patterns,'' {\em PLoS ONE}, vol.~7, 2012.

\bibitem{Ponulak2005ReSuMe}
F.~Ponulak, ``Resume-new supervised learning method for spiking neural
  networks,'' 2005.

\bibitem{Mohemmed2012SPAN}
A.~Mohemmed, S.~Schliebs, S.~Matsuda, and N.~Kasabov, ``Span: Spike pattern
  association neuron for learning spatio-temporal spike patterns,'' {\em
  International journal of neural systems}, vol.~22 4, p.~1250012, 2012.

\bibitem{snnbackp}
S.~Bohte, J.~Kok, and H.~Poutré, ``Error-backpropagation in temporally encoded
  networks of spiking neurons,'' {\em Neurocomputing}, vol.~48, pp.~17--37, 02
  2001.

\bibitem{Lee2016TrainingSNN}
J.~H. Lee, T.~Delbruck, and M.~Pfeiffer, ``Training deep spiking neural
  networks using backpropagation,'' {\em Frontiers in Neuroscience}, vol.~10,
  p.~508, 2016.

\bibitem{Shrestha2018SLAYER}
S.~B. Shrestha and G.~Orchard, ``Slayer: Spike layer error reassignment in
  time,'' in {\em NeurIPS}, 2018.

\bibitem{Neftci2019SurrogateGL}
E.~O. Neftci, H.~Mostafa, and F.~Zenke, ``Surrogate gradient learning in
  spiking neural networks: Bringing the power of gradient-based optimization to
  spiking neural networks,'' {\em IEEE Signal Processing Magazine}, vol.~36,
  pp.~51--63, 2019.

\bibitem{Thiele2020SpikeGrad}
J.~C. Thiele, O.~Bichler, and A.~Dupret, ``Spikegrad: An ann-equivalent
  computation model for implementing backpropagation with spikes,'' in {\em
  International Conference on Learning Representations}, 2020.

\bibitem{Kaiser2018DECOLLE}
J.~Kaiser, H.~Mostafa, and E.~Neftci, ``Synaptic plasticity dynamics for deep
  continuous local learning,'' {\em ArXiv}, vol.~abs/1811.10766, 2018.

\bibitem{Rueckauer2017ConversionSNN}
B.~Rueckauer, I.-A. Lungu, Y.~Hu, M.~Pfeiffer, and S.-C. Liu, ``Conversion of
  continuous-valued deep networks to efficient event-driven networks for image
  classification,'' {\em Frontiers in Neuroscience}, vol.~11, p.~682, 2017.

\bibitem{SNNDVSLoihi_shafique}
R.~Massa, A.~Marchisio, M.~Martina, and M.~Shafique, ``An efficient spiking
  neural network for recognizing gestures with a dvs camera on the loihi
  neuromorphic processor,'' {\em 2020 International Joint Conference on Neural
  Networks (IJCNN)}, 2020.

\bibitem{Rathi2020EnablingSNNbackprop}
N.~Rathi, G.~Srinivasan, P.~Panda, and K.~Roy, ``Enabling deep spiking neural
  networks with hybrid conversion and spike timing dependent backpropagation,''
  in {\em International Conference on Learning Representations}, 2020.

\bibitem{Pfeiffer2018DLSNN}
M.~Pfeiffer and T.~Pfeil, ``Deep learning with spiking neurons: Opportunities
  and challenges,'' {\em Frontiers in Neuroscience}, vol.~12, p.~774, 2018.

\bibitem{parallelism_nn}
D.~Al-Dabass, P.~Vindlacheruvu, and D.~J. Evans, ``Parallelism in neural
  nets,'' {\em Parallel Algorithms and Applications}, vol.~11, no.~3-4,
  pp.~169--185, 1997.

\bibitem{sze}
V.~{Sze}, Y.~{Chen}, T.~{Yang}, and J.~S. {Emer}, ``Efficient processing of
  deep neural networks: A tutorial and survey,'' {\em Proceedings of the IEEE},
  vol.~105, pp.~2295--2329, Dec 2017.

\bibitem{intelavx}
R.~James, ``Intel avx-512 instructions,'' {\em Intel}, 06 2017.

\bibitem{bfloat16}
{\em BFLOAT16 – Hardware Numerics Definition}, 2018.

\bibitem{bigdl}
S.~L. Gogar, ``Bigdl – scale-out deep learning on apache spark* cluster,''
  {\em
  \url{https://software.intel.com/content/www/us/en/develop/articles/bigdl-scale-out-deep-learning-on-apache-spark-cluster.html}},
  2017.

\bibitem{teslav100}
{\em NVIDIA TESLA V100 GPU ARCHITECTURE}, 2017.

\bibitem{pytorch}
A.~Paszke, S.~Gross, S.~Chintala, G.~Chanan, E.~Yang, Z.~DeVito, Z.~Lin,
  A.~Desmaison, L.~Antiga, and A.~Lerer, ``Automatic differentiation in
  pytorch,'' in {\em NIPS 2017 Workshop on Autodiff}, 2017.

\bibitem{tensorflow}
M.~Abadi, A.~Agarwal, P.~Barham, E.~Brevdo, Z.~Chen, C.~Citro, G.~S. Corrado,
  A.~Davis, J.~Dean, M.~Devin, S.~Ghemawat, I.~Goodfellow, A.~Harp, G.~Irving,
  M.~Isard, Y.~Jia, R.~Jozefowicz, L.~Kaiser, M.~Kudlur, J.~Levenberg,
  D.~Man\'{e}, R.~Monga, S.~Moore, D.~Murray, C.~Olah, M.~Schuster, J.~Shlens,
  B.~Steiner, I.~Sutskever, K.~Talwar, P.~Tucker, V.~Vanhoucke, V.~Vasudevan,
  F.~Vi\'{e}gas, O.~Vinyals, P.~Warden, M.~Wattenberg, M.~Wicke, Y.~Yu, and
  X.~Zheng, ``{TensorFlow}: Large-scale machine learning on heterogeneous
  systems,'' 2015.
\newblock Software available from tensorflow.org.

\bibitem{caffe}
Y.~Jia, E.~Shelhamer, J.~Donahue, S.~Karayev, J.~Long, R.~B. Girshick,
  S.~Guadarrama, and T.~Darrell, ``Caffe: Convolutional architecture for fast
  feature embedding,'' in {\em Proceedings of the {ACM} International
  Conference on Multimedia, {MM} '14, Orlando, FL, USA, November 03 - 07, 2014}
  (K.~A. Hua, Y.~Rui, R.~Steinmetz, A.~Hanjalic, A.~Natsev, and W.~Zhu, eds.),
  pp.~675--678, {ACM}, 2014.

\bibitem{chetlur2014cudnn}
S.~Chetlur, C.~Woolley, P.~Vandermersch, J.~Cohen, J.~Tran, B.~Catanzaro, and
  E.~Shelhamer, ``cudnn: Efficient primitives for deep learning,'' {\em CoRR},
  vol.~abs/1410.0759, 2014.

\bibitem{cudaxai}
{\em \url{https://developer.nvidia.com/gpu-accelerated-libraries}}.

\bibitem{ampere100}
{\em NVIDIA A100 Tensor Core GPU Architecture}, 2020.

\bibitem{chellapilla}
K.~Chellapilla, S.~Puri, and P.~Simard, ``{High Performance Convolutional
  Neural Networks for Document Processing},'' in {\em {Tenth International
  Workshop on Frontiers in Handwriting Recognition}} (G.~Lorette, ed.), (La
  Baule (France)), {Universit{\'e} de Rennes 1}, {Suvisoft}, Oct 2006.
\newblock http://www.suvisoft.com.

\bibitem{vasudevan}
A.~{Vasudevan}, A.~{Anderson}, and D.~{Gregg}, ``Parallel multi channel
  convolution using general matrix multiplication,'' in {\em 2017 IEEE 28th
  International Conference on Application-specific Systems, Architectures and
  Processors (ASAP)}, pp.~19--24, 2017.

\bibitem{strassen}
V.~Strassen, ``Gaussian elimination is not optimal,'' {\em Numer. Math.},
  vol.~13, p.~354–356, aug 1969.

\bibitem{strassen2}
J.~Cong and B.~Xiao, ``Minimizing computation in convolutional neural
  networks,'' in {\em Artificial Neural Networks and Machine Learning - {ICANN}
  2014 - 24th International Conference on Artificial Neural Networks, Hamburg,
  Germany, September 15-19, 2014. Proceedings} (S.~Wermter, C.~Weber, W.~Duch,
  T.~Honkela, P.~D. Koprinkova{-}Hristova, S.~Magg, G.~Palm, and A.~E.~P.
  Villa, eds.), vol.~8681 of {\em Lecture Notes in Computer Science},
  pp.~281--290, Springer, 2014.

\bibitem{fft}
M.~Mathieu, M.~Henaff, and Y.~LeCun, ``Fast training of convolutional networks
  through ffts,'' in {\em 2nd International Conference on Learning
  Representations, {ICLR} 2014, Banff, AB, Canada, April 14-16, 2014,
  Conference Track Proceedings} (Y.~Bengio and Y.~LeCun, eds.), 2014.

\bibitem{winograd}
S.~Winograd, {\em {Arithmetic complexity of computations}}, vol.~33 of {\em
  CBMS-NSF Regional Conference Series in Applied Mathematics}.
\newblock Philadelphia: Society for Industrial and Applied Mathematics, 1980.

\bibitem{winograd2}
A.~{Lavin} and S.~{Gray}, ``Fast algorithms for convolutional neural
  networks,'' in {\em 2016 IEEE Conference on Computer Vision and Pattern
  Recognition (CVPR)}, pp.~4013--4021, June 2016.

\bibitem{blockedconv}
Y.~Liu, Y.~Wang, R.~Yu, M.~Li, V.~Sharma, and Y.~Wang, ``Optimizing cnn model
  inference on cpus,'' in {\em Proceedings of the 2019 USENIX Conference on
  Usenix Annual Technical Conference}, USENIX ATC ’19, (USA), p.~1025–1040,
  USENIX Association, 2019.

\bibitem{blockedconv2}
A.~Rodriguez, E.~Segal, E.~Meiri, E.~Fomenko, Y.~Kim, H.~Shen, and B.~Ziv,
  ``Lower numerical precision deep learning inference and training,'' {\em
  Intel White Paper}, 01 2018.

\bibitem{horowitz}
M.~Horowitz, ``Energy table for 45nm process.''

\bibitem{cambricon-x}
S.~{Zhang}, Z.~{Du}, L.~{Zhang}, H.~{Lan}, S.~{Liu}, L.~{Li}, Q.~{Guo},
  T.~{Chen}, and Y.~{Chen}, ``Cambricon-x: An accelerator for sparse neural
  networks,'' in {\em 2016 49th Annual IEEE/ACM International Symposium on
  Microarchitecture (MICRO)}, pp.~1--12, Oct 2016.

\bibitem{chen_dataflows}
Y.-H. Chen, J.~S. Emer, and V.~Sze, ``Using dataflow to optimize energy
  efficiency of deep neural network accelerators,'' {\em IEEE Micro}, vol.~37,
  pp.~12--21, 2017.

\bibitem{nnx}
V.~{Gokhale}, J.~{Jin}, A.~{Dundar}, B.~{Martini}, and E.~{Culurciello}, ``A
  240 g-ops/s mobile coprocessor for deep neural networks,'' in {\em 2014 IEEE
  Conference on Computer Vision and Pattern Recognition Workshops},
  pp.~696--701, June 2014.

\bibitem{sriram2010}
V.~{Sriram}, D.~{Cox}, K.~H. {Tsoi}, and W.~{Luk}, ``Towards an embedded
  biologically-inspired machine vision processor,'' in {\em 2010 International
  Conference on Field-Programmable Technology}, pp.~273--278, Dec 2010.

\bibitem{Sankaradas2009}
M.~{Sankaradas}, V.~{Jakkula}, S.~{Cadambi}, S.~{Chakradhar}, I.~{Durdanovic},
  E.~{Cosatto}, and H.~P. {Graf}, ``A massively parallel coprocessor for
  convolutional neural networks,'' in {\em 2009 20th IEEE International
  Conference on Application-specific Systems, Architectures and Processors},
  pp.~53--60, July 2009.

\bibitem{Yang2018DNNDC}
X.~Yang, M.~Gao, J.~Pu, A.~Nayak, Q.~Liu, S.~Bell, J.~Setter, K.~Cao, H.~Ha,
  C.~Kozyrakis, and M.~Horowitz, ``{DNN} dataflow choice is overrated,'' {\em
  CoRR}, vol.~abs/1809.04070, 2018.

\bibitem{Chakradhar2010}
S.~Chakradhar, M.~Sankaradas, V.~Jakkula, and S.~Cadambi, ``A dynamically
  configurable coprocessor for convolutional neural networks,'' in {\em
  Proceedings of the 37th Annual International Symposium on Computer
  Architecture}, ISCA ’10, (New York, NY, USA), p.~247–257, Association for
  Computing Machinery, 2010.

\bibitem{park2015}
S.~{Park}, K.~{Bong}, D.~{Shin}, J.~{Lee}, S.~{Choi}, and H.~{Yoo}, ``4.6
  a1.93tops/w scalable deep learning/inference processor with tetra-parallel
  mimd architecture for big-data applications,'' in {\em 2015 IEEE
  International Solid-State Circuits Conference - (ISSCC) Digest of Technical
  Papers}, pp.~1--3, Feb 2015.

\bibitem{Qiu2016}
J.~Qiu, J.~Wang, S.~Yao, K.~Guo, B.~Li, E.~Zhou, J.~Yu, T.~Tang, N.~Xu,
  S.~Song, and et~al., ``Going deeper with embedded fpga platform for
  convolutional neural network,'' in {\em Proceedings of the 2016 ACM/SIGDA
  International Symposium on Field-Programmable Gate Arrays}, FPGA ’16, (New
  York, NY, USA), p.~26–35, Association for Computing Machinery, 2016.

\bibitem{googletpu}
N.~P. Jouppi, C.~Young, N.~Patil, D.~Patterson, G.~Agrawal, R.~Bajwa, S.~Bates,
  S.~Bhatia, N.~Boden, A.~Borchers, and et~al., ``In-datacenter performance
  analysis of a tensor processing unit,'' {\em SIGARCH Comput. Archit. News},
  vol.~45, p.~1–12, jun 2017.

\bibitem{scalesim}
A.~Samajdar, Y.~Zhu, P.~N. Whatmough, M.~Mattina, and T.~Krishna, ``Scale-sim:
  Systolic {CNN} accelerator simulator,'' {\em CoRR}, vol.~abs/1811.02883,
  2019.

\bibitem{mpna_shafique}
M.~A. Hanif, R.~V.~W. Putra, M.~Tanvir, R.~Hafiz, S.~Rehman, and M.~Shafique,
  ``{MPNA:} {A} massively-parallel neural array accelerator with dataflow
  optimization for convolutional neural networks,'' {\em CoRR},
  vol.~abs/1810.12910, 2018.

\bibitem{rna}
C.~{Luo}, Y.~{Wang}, W.~{Cao}, P.~H.~W. {Leong}, and L.~{Wang}, ``Rna: An
  accurate residual network accelerator for quantized and reconstructed deep
  neural networks,'' in {\em 2018 28th International Conference on Field
  Programmable Logic and Applications (FPL)}, pp.~60--603, 2018.

\bibitem{capsacc_shafique}
A.~{Marchisio}, M.~A. {Hanif}, and M.~{Shafique}, ``Capsacc: An efficient
  hardware accelerator for capsulenets with data reuse,'' in {\em 2019 Design,
  Automation Test in Europe Conference Exhibition (DATE)}, pp.~964--967, 2019.

\bibitem{capstore_shafique}
A.~Marchisio and M.~Shafique, ``Capstore: Energy-efficient design and
  management of the on-chip memory for capsulenet inference accelerators,''
  {\em CoRR}, vol.~abs/1902.01151, 2019.

\bibitem{descnet_shafique}
A.~Marchisio, V.~Mrazek, M.~Hanif, and M.~Shafique, ``Descnet: Developing
  efficient scratchpad memories for capsule network hardware,'' {\em IEEE
  Transactions on Computer-Aided Design of Integrated Circuits and Systems},
  2020.

\bibitem{7284058}
Z.~{Du}, R.~{Fasthuber}, T.~{Chen}, P.~{Ienne}, L.~{Li}, T.~{Luo}, X.~{Feng},
  Y.~{Chen}, and O.~{Temam}, ``Shidiannao: Shifting vision processing closer to
  the sensor,'' in {\em 2015 ACM/IEEE 42nd Annual International Symposium on
  Computer Architecture (ISCA)}, pp.~92--104, June 2015.

\bibitem{origami}
L.~{Cavigelli} and L.~{Benini}, ``Origami: A 803-gop/s/w convolutional network
  accelerator,'' {\em IEEE Transactions on Circuits and Systems for Video
  Technology}, vol.~27, pp.~2461--2475, Nov 2017.

\bibitem{peemen2013}
M.~{Peemen}, A.~A.~A. {Setio}, B.~{Mesman}, and H.~{Corporaal},
  ``Memory-centric accelerator design for convolutional neural networks,'' in
  {\em 2013 IEEE 31st International Conference on Computer Design (ICCD)},
  pp.~13--19, Oct 2013.

\bibitem{eyeriss}
Y.~{Chen}, T.~{Krishna}, J.~S. {Emer}, and V.~{Sze}, ``Eyeriss: An
  energy-efficient reconfigurable accelerator for deep convolutional neural
  networks,'' {\em IEEE Journal of Solid-State Circuits}, vol.~52,
  pp.~127--138, Jan 2017.

\bibitem{zhang2015}
C.~Zhang, P.~Li, G.~Sun, Y.~Guan, B.~Xiao, and J.~Cong, ``Optimizing fpga-based
  accelerator design for deep convolutional neural networks,'' in {\em
  Proceedings of the 2015 ACM/SIGDA International Symposium on
  Field-Programmable Gate Arrays}, FPGA ’15, (New York, NY, USA),
  p.~161–170, Association for Computing Machinery, 2015.

\bibitem{7011421}
Y.~{Chen}, T.~{Luo}, S.~{Liu}, S.~{Zhang}, L.~{He}, J.~{Wang}, L.~{Li},
  T.~{Chen}, Z.~{Xu}, N.~{Sun}, and O.~{Temam}, ``Dadiannao: A machine-learning
  supercomputer,'' in {\em 2014 47th Annual IEEE/ACM International Symposium on
  Microarchitecture}, pp.~609--622, Dec 2014.

\bibitem{hybridnn}
S.~{Yin}, P.~{Ouyang}, S.~{Tang}, F.~{Tu}, X.~{Li}, L.~{Liu}, and S.~{Wei}, ``A
  1.06-to-5.09 tops/w reconfigurable hybrid-neural-network processor for deep
  learning applications,'' in {\em 2017 Symposium on VLSI Circuits},
  pp.~C26--C27, 2017.

\bibitem{project_brainwave}
J.~{Fowers}, K.~{Ovtcharov}, M.~{Papamichael}, T.~{Massengill}, M.~{Liu},
  D.~{Lo}, S.~{Alkalay}, M.~{Haselman}, L.~{Adams}, M.~{Ghandi}, S.~{Heil},
  P.~{Patel}, A.~{Sapek}, G.~{Weisz}, L.~{Woods}, S.~{Lanka}, S.~K.
  {Reinhardt}, A.~M. {Caulfield}, E.~S. {Chung}, and D.~{Burger}, ``A
  configurable cloud-scale dnn processor for real-time ai,'' in {\em 2018
  ACM/IEEE 45th Annual International Symposium on Computer Architecture
  (ISCA)}, pp.~1--14, 2018.

\bibitem{maeri}
H.~Kwon, A.~Samajdar, and T.~Krishna, ``Maeri: Enabling flexible dataflow
  mapping over dnn accelerators via reconfigurable interconnects,'' in {\em
  Proceedings of the Twenty-Third International Conference on Architectural
  Support for Programming Languages and Operating Systems}, ASPLOS ’18, (New
  York, NY, USA), p.~461–475, Association for Computing Machinery, 2018.

\bibitem{sigma}
E.~{Qin}, A.~{Samajdar}, H.~{Kwon}, V.~{Nadella}, S.~{Srinivasan}, D.~{Das},
  B.~{Kaul}, and T.~{Krishna}, ``Sigma: A sparse and irregular gemm accelerator
  with flexible interconnects for dnn training,'' in {\em 2020 IEEE
  International Symposium on High Performance Computer Architecture (HPCA)},
  pp.~58--70, 2020.

\bibitem{davinci}
H.~{Liao}, J.~{Tu}, J.~{Xia}, and X.~{Zhou}, ``Davinci: A scalable architecture
  for neural network computing,'' in {\em 2019 IEEE Hot Chips 31 Symposium
  (HCS)}, pp.~1--44, 2019.

\bibitem{pochet2013}
L.-N. Pouchet, P.~Zhang, P.~Sadayappan, and J.~Cong, ``Polyhedral-based data
  reuse optimization for configurable computing,'' in {\em Proceedings of the
  ACM/SIGDA International Symposium on Field Programmable Gate Arrays}, FPGA
  ’13, (New York, NY, USA), p.~29–38, Association for Computing Machinery,
  2013.

\bibitem{yang2016}
X.~Yang, J.~Pu, B.~B. Rister, N.~Bhagdikar, S.~Richardson, S.~Kvatinsky,
  J.~Ragan{-}Kelley, A.~Pedram, and M.~Horowitz, ``A systematic approach to
  blocking convolutional neural networks,'' {\em CoRR}, vol.~abs/1606.04209,
  2016.

\bibitem{smartshuttle}
J.~{Li}, G.~{Yan}, W.~{Lu}, S.~{Jiang}, S.~{Gong}, J.~{Wu}, and X.~{Li},
  ``Smartshuttle: Optimizing off-chip memory accesses for deep learning
  accelerators,'' in {\em 2018 Design, Automation Test in Europe Conference
  Exhibition (DATE)}, pp.~343--348, March 2018.

\bibitem{nnest}
L.~Ke, X.~He, and X.~Zhang, ``Nnest: Early-stage design space exploration tool
  for neural network inference accelerators,'' in {\em Proceedings of the
  International Symposium on Low Power Electronics and Design}, ISLPED ’18,
  (New York, NY, USA), Association for Computing Machinery, 2018.

\bibitem{romanet_shafique}
R.~V.~W. Putra, M.~A. Hanif, and M.~Shafique, ``Romanet: Fine-grained
  reuse-driven data organization and off-chip memory access management for deep
  neural network accelerators,'' {\em CoRR}, vol.~abs/1902.10222, 2019.

\bibitem{maestro}
H.~Kwon, P.~Chatarasi, M.~Pellauer, A.~Parashar, V.~Sarkar, and T.~Krishna,
  ``Understanding reuse, performance, and hardware cost of dnn dataflow: A
  data-centric approach,'' in {\em Proceedings of the 52nd Annual IEEE/ACM
  International Symposium on Microarchitecture}, MICRO ’52, (New York, NY,
  USA), p.~754–768, Association for Computing Machinery, 2019.

\bibitem{mrna}
Z.~{Zhao}, H.~{Kwon}, S.~{Kuhar}, W.~{Sheng}, Z.~{Mao}, and T.~{Krishna},
  ``mrna: Enabling efficient mapping space exploration for a reconfiguration
  neural accelerator,'' in {\em 2019 IEEE International Symposium on
  Performance Analysis of Systems and Software (ISPASS)}, pp.~282--292, March
  2019.

\bibitem{timeloop}
A.~{Parashar}, P.~{Raina}, Y.~S. {Shao}, Y.~{Chen}, V.~A. {Ying}, A.~{Mukkara},
  R.~{Venkatesan}, B.~{Khailany}, S.~W. {Keckler}, and J.~{Emer}, ``Timeloop: A
  systematic approach to dnn accelerator evaluation,'' in {\em 2019 IEEE
  International Symposium on Performance Analysis of Systems and Software
  (ISPASS)}, pp.~304--315, March 2019.

\bibitem{magnet}
R.~{Venkatesan}, Y.~S. {Shao}, M.~{Wang}, J.~{Clemons}, S.~{Dai}, M.~{Fojtik},
  B.~{Keller}, A.~{Klinefelter}, N.~{Pinckney}, P.~{Raina}, Y.~{Zhang},
  B.~{Zimmer}, W.~J. {Dally}, J.~{Emer}, S.~W. {Keckler}, and B.~{Khailany},
  ``Magnet: A modular accelerator generator for neural networks,'' in {\em 2019
  IEEE/ACM International Conference on Computer-Aided Design (ICCAD)},
  pp.~1--8, Nov 2019.

\bibitem{Colucci2020XploreDL}
A.~Colucci, A.~Marchisio, B.~Bussolino, V.~Mrazek, M.~Martina, G.~Masera, and
  M.~Shafique, ``A fast design space exploration framework for the deep
  learning accelerators,'' in {\em IEEE International Conference on
  Hardware-Software Codesign and System Synthesis (CODES+ISSS), WiP}, 2020.

\bibitem{SuperSlash_shafique}
H.~Ahmad, T.~Arif, M.~A. Hanif, R.~Hafiz, and M.~Shafique, ``Superslash: A
  unified design space exploration and model compression methodology for design
  of deep learning accelerators with reduced off-chip memory access,'' {\em
  IEEE Transactions on Computer-Aided Design of Integrated Circuits and Systems
  (TCAD), ESWeek-Special Issue}, 2020.

\bibitem{Cai2019ProxylessNAS}
H.~Cai, L.~Zhu, and S.~Han, ``Proxyless{NAS}: Direct neural architecture search
  on target task and hardware,'' in {\em International Conference on Learning
  Representations}, 2019.

\bibitem{Tan2019MnasNetPN}
M.~Tan, B.~Chen, R.~Pang, V.~Vasudevan, and Q.~V. Le, ``Mnasnet: Platform-aware
  neural architecture search for mobile,'' {\em 2019 IEEE/CVF Conference on
  Computer Vision and Pattern Recognition (CVPR)}, pp.~2815--2823, 2019.

\bibitem{Zeng2020BlackBoxNAS}
S.~{Zeng}, H.~{Sun}, Y.~{Xing}, X.~{Ning}, Y.~{Shan}, X.~{Chen}, Y.~{Wang}, and
  H.~{Yang}, ``Black box search space profiling for accelerator-aware neural
  architecture search,'' in {\em 2020 25th Asia and South Pacific Design
  Automation Conference (ASP-DAC)}, pp.~518--523, 2020.

\bibitem{Marchisio2020NASCaps}
A.~Marchisio, A.~Massa, V.~Mrazek, B.~Bussolino, M.~Martina, and M.~Shafique,
  ``Nascaps: A framework for neural architecture search to optimize the
  accuracy and hardware efficiency of convolutional capsule networks,'' in {\em
  2020 IEEE/ACM International Conference on Computer-Aided Design (ICCAD)},
  2020.

\bibitem{Lu2019ResourceConstrainedNAS}
Q.~Lu, W.~Jiang, X.~Xu, Y.~Shi, and J.~Hu, ``On neural architecture search for
  resource-constrained hardware platforms,'' {\em ArXiv}, vol.~abs/1911.00105,
  2019.

\bibitem{Achararit2020APNAS}
P.~{Achararit}, M.~A. {Hanif}, R.~V.~W. {Putra}, M.~{Shafique}, and
  Y.~{Hara-Azumi}, ``Apnas: Accuracy-and-performance-aware neural architecture
  search for neural hardware accelerators,'' {\em IEEE Access}, pp.~1--1, 2020.

\bibitem{Jiang2019FPGA-NAS}
W.~Jiang, X.~Zhang, E.~H.-M. Sha, L.~Yang, Q.~Zhuge, Y.~Shi, and J.~Hu,
  ``Accuracy vs. efficiency: Achieving both through fpga-implementation aware
  neural architecture search,'' in {\em Proceedings of the 56th Annual Design
  Automation Conference 2019}, DAC ’19, (New York, NY, USA), Association for
  Computing Machinery, 2019.

\bibitem{Jiang2020HotNAS}
W.~Jiang, L.~Yang, S.~Dasgupta, J.~Hu, and Y.~Shi, ``Standing on the shoulders
  of giants: Hardware and neural architecture co-search with hot start,'' {\em
  ArXiv}, vol.~abs/2007.09087, 2020.

\bibitem{Pham2018ENAS}
H.~Pham, M.~Y. Guan, B.~Zoph, Q.~V. Le, and J.~Dean, ``Efficient neural
  architecture search via parameter sharing,'' in {\em ICML}, 2018.

\bibitem{Zoph2016TransferLearning}
B.~Zoph, D.~Yuret, J.~May, and K.~Knight, ``Transfer learning for low-resource
  neural machine translation,'' in {\em Proceedings of the 2016 Conference on
  Empirical Methods in Natural Language Processing}, (Austin, Texas),
  pp.~1568--1575, Association for Computational Linguistics, Nov. 2016.

\bibitem{Stamoulis2019SinglePathNAS}
D.~Stamoulis, R.~Ding, D.~Wang, D.~Lymberopoulos, B.~Priyantha, J.~Liu, and
  D.~Marculescu, ``Single-path nas: Designing hardware-efficient convnets in
  less than 4 hours,'' in {\em ECML/PKDD}, 2019.

\bibitem{Wu2019FBNet}
B.~Wu, X.~Dai, P.~Zhang, Y.~Wang, F.~Sun, Y.~Wu, Y.~Tian, P.~Vajda, Y.~Jia, and
  K.~Keutzer, ``Fbnet: Hardware-aware efficient convnet design via
  differentiable neural architecture search,'' {\em 2019 IEEE/CVF Conference on
  Computer Vision and Pattern Recognition (CVPR)}, pp.~10726--10734, 2019.

\bibitem{Guo2019SPOS}
Z.~Guo, X.~Zhang, H.~Mu, W.~Heng, Z.~Liu, Y.~Wei, and J.~Sun, ``Single path
  one-shot neural architecture search with uniform sampling,'' {\em ArXiv},
  vol.~abs/1904.00420, 2019.

\bibitem{Zhang2020FastHANAS}
L.~Zhang, Y.~Yang, Y.~Jiang, W.~Zhu, and Y.~Liu, ``Fast hardware-aware neural
  architecture search,'' {\em 2020 IEEE/CVF Conference on Computer Vision and
  Pattern Recognition Workshops (CVPRW)}, pp.~2959--2967, 2020.

\bibitem{Li2019RandomSearchNAS}
L.~Li and A.~Talwalkar, ``Random search and reproducibility for neural
  architecture search,'' in {\em UAI}, 2019.

\bibitem{horowitz2014}
M.~{Horowitz}, ``1.1 computing's energy problem (and what we can do about
  it),'' in {\em 2014 IEEE International Solid-State Circuits Conference Digest
  of Technical Papers (ISSCC)}, pp.~10--14, Feb 2014.

\bibitem{binaryconnect}
M.~Courbariaux, Y.~Bengio, and J.-P. David, ``Binaryconnect: Training deep
  neural networks with binary weights during propagations,'' in {\em NIPS},
  2015.

\bibitem{ternaryweightnet}
F.~Li and B.~Liu, ``Ternary weight networks,'' {\em CoRR}, vol.~abs/1605.04711,
  2016.

\bibitem{jacob2018}
B.~{Jacob}, S.~{Kligys}, B.~{Chen}, M.~{Zhu}, M.~{Tang}, A.~{Howard},
  H.~{Adam}, and D.~{Kalenichenko}, ``Quantization and training of neural
  networks for efficient integer-arithmetic-only inference,'' in {\em 2018
  IEEE/CVF Conference on Computer Vision and Pattern Recognition},
  pp.~2704--2713, June 2018.

\bibitem{vanhouke2011}
V.~Vanhoucke, A.~Senior, and M.~Z. Mao, ``Improving the speed of neural
  networks on cpus,'' in {\em Deep Learning and Unsupervised Feature Learning
  Workshop, NIPS 2011}, 2011.

\bibitem{quantizedneuralnetworks}
I.~Hubara, M.~Courbariaux, D.~Soudry, R.~El-Yaniv, and Y.~Bengio, ``Quantized
  neural networks: Training neural networks with low precision weights and
  activations,'' {\em Journal of Machine Learning Research}, vol.~18, no.~187,
  pp.~1--30, 2018.

\bibitem{Guo2018}
Y.~Guo, ``A survey on methods and theories of quantized neural networks,'' {\em
  CoRR}, vol.~abs/1808.04752, 2018.

\bibitem{gysel2016}
P.~{Gysel}, J.~{Pimentel}, M.~{Motamedi}, and S.~{Ghiasi}, ``Ristretto: A
  framework for empirical study of resource-efficient inference in
  convolutional neural networks,'' {\em IEEE Transactions on Neural Networks
  and Learning Systems}, vol.~29, no.~11, pp.~5784--5789, 2018.

\bibitem{squeezenet}
F.~N. Iandola, M.~W. Moskewicz, K.~Ashraf, S.~Han, W.~J. Dally, and K.~Keutzer,
  ``Squeezenet: Alexnet-level accuracy with 50x fewer parameters and
  {\textless}1mb model size,'' {\em CoRR}, vol.~abs/1602.07360, 2016.

\bibitem{sakr2019}
C.~Sakr and N.~Shanbhag, ``Per-tensor fixed-point quantization of the
  back-propagation algorithm,'' in {\em ICLR}, 2019.

\bibitem{qcapsnets_shafique}
A.~Marchisio, B.~Bussolino, A.~Colucci, M.~Martina, G.~Masera, and M.~Shafique,
  ``Q-capsnets: {A} specialized framework for quantizing capsule networks,'' in
  {\em Proceedings of the 57th Annual Design Automation Conference 2020}.

\bibitem{haq}
K.~{Wang}, Z.~{Liu}, Y.~{Lin}, J.~{Lin}, and S.~{Han}, ``Haq: Hardware-aware
  automated quantization with mixed precision,'' in {\em 2019 IEEE/CVF
  Conference on Computer Vision and Pattern Recognition (CVPR)},
  pp.~8604--8612, 2019.

\bibitem{bismo}
Y.~{Umuroglu}, L.~{Rasnayake}, and M.~{Själander}, ``Bismo: A scalable
  bit-serial matrix multiplication overlay for reconfigurable computing,'' in
  {\em 2018 28th International Conference on Field Programmable Logic and
  Applications (FPL)}, pp.~307--3077, 2018.

\bibitem{stripes}
P.~{Judd}, J.~{Albericio}, and A.~{Moshovos}, ``Stripes: Bit-serial deep neural
  network computing,'' {\em IEEE Computer Architecture Letters}, vol.~16,
  no.~1, pp.~80--83, 2017.

\bibitem{unpu}
J.~{Lee}, C.~{Kim}, S.~{Kang}, D.~{Shin}, S.~{Kim}, and H.~{Yoo}, ``Unpu: An
  energy-efficient deep neural network accelerator with fully variable weight
  bit precision,'' {\em IEEE Journal of Solid-State Circuits}, vol.~54, no.~1,
  pp.~173--185, 2019.

\bibitem{loom}
S.~{Sharify}, A.~D. {Lascorz}, K.~{Siu}, P.~{Judd}, and A.~{Moshovos}, ``Loom:
  Exploiting weight and activation precisions to accelerate convolutional
  neural networks,'' in {\em 2018 55th ACM/ESDA/IEEE Design Automation
  Conference (DAC)}, pp.~1--6, 2018.

\bibitem{bitfusion}
H.~{Sharma}, J.~{Park}, N.~{Suda}, L.~{Lai}, B.~{Chau}, V.~{Chandra}, and
  H.~{Esmaeilzadeh}, ``Bit fusion: Bit-level dynamically composable
  architecture for accelerating deep neural network,'' in {\em 2018 ACM/IEEE
  45th Annual International Symposium on Computer Architecture (ISCA)},
  pp.~764--775, 2018.

\bibitem{bitblade}
S.~{Ryu}, H.~{Kim}, W.~{Yi}, and J.~{Kim}, ``Bitblade: Area and
  energy-efficient precision-scalable neural network accelerator with bitwise
  summation,'' in {\em 2019 56th ACM/IEEE Design Automation Conference (DAC)},
  pp.~1--6, 2019.

\bibitem{nvidiaturing}
Nvidia, {\em NVIDIA TURING GPU ARCHITECTURE. Graphics Reinvented}, 2018.

\bibitem{xnornet}
M.~Rastegari, V.~Ordonez, J.~Redmon, and A.~Farhadi, ``Xnor-net: Imagenet
  classification using binary convolutional neural networks,'' in {\em Computer
  Vision -- ECCV 2016} (B.~Leibe, J.~Matas, N.~Sebe, and M.~Welling, eds.),
  (Cham), pp.~525--542, Springer International Publishing, 2016.

\bibitem{dorefa}
S.~Zhou, Z.~Ni, X.~Zhou, H.~Wen, Y.~Wu, and Y.~Zou, ``Dorefa-net: Training low
  bitwidth convolutional neural networks with low bitwidth gradients,'' {\em
  CoRR}, vol.~abs/1606.06160, 2016.

\bibitem{binarizedneuralnetworks}
I.~Hubara, M.~Courbariaux, D.~Soudry, R.~El-Yaniv, and Y.~Bengio, ``Binarized
  neural networks,'' in {\em Advances in Neural Information Processing Systems
  29} (D.~D. Lee, M.~Sugiyama, U.~V. Luxburg, I.~Guyon, and R.~Garnett, eds.),
  pp.~4107--4115, Curran Associates, Inc., 2016.

\bibitem{yodann}
R.~{Andri}, L.~{Cavigelli}, D.~{Rossi}, and L.~{Benini}, ``Yodann: An ultra-low
  power convolutional neural network accelerator based on binary weights,'' in
  {\em 2016 IEEE Computer Society Annual Symposium on VLSI (ISVLSI)},
  pp.~236--241, 2016.

\bibitem{hyperdrive}
R.~{Andri}, L.~{Cavigelli}, D.~{Rossi}, and L.~{Benini}, ``Hyperdrive: A
  multi-chip systolically scalable binary-weight cnn inference engine,'' {\em
  IEEE Journal on Emerging and Selected Topics in Circuits and Systems},
  vol.~9, no.~2, pp.~309--322, 2019.

\bibitem{brein}
K.~{Ando}, K.~{Ueyoshi}, K.~{Orimo}, H.~{Yonekawa}, S.~{Sato}, H.~{Nakahara},
  S.~{Takamaeda-Yamazaki}, M.~{Ikebe}, T.~{Asai}, T.~{Kuroda}, and
  M.~{Motomura}, ``Brein memory: A single-chip binary/ternary reconfigurable
  in-memory deep neural network accelerator achieving 1.4 tops at 0.6 w,'' {\em
  IEEE Journal of Solid-State Circuits}, vol.~53, no.~4, pp.~983--994, 2018.

\bibitem{xnorneuralengine}
F.~{Conti}, P.~D. {Schiavone}, and L.~{Benini}, ``Xnor neural engine: A
  hardware accelerator ip for 21.6-fj/op binary neural network inference,''
  {\em IEEE Transactions on Computer-Aided Design of Integrated Circuits and
  Systems}, vol.~37, no.~11, pp.~2940--2951, 2018.

\bibitem{xnorbin}
A.~{Al Bahou}, G.~{Karunaratne}, R.~{Andri}, L.~{Cavigelli}, and L.~{Benini},
  ``Xnorbin: A 95 top/s/w hardware accelerator for binary convolutional neural
  networks,'' in {\em 2018 IEEE Symposium in Low-Power and High-Speed Chips
  (COOL CHIPS)}, pp.~1--3, 2018.

\bibitem{Miyashita2016}
D.~Miyashita, E.~H. Lee, and B.~Murmann, ``Convolutional neural networks using
  logarithmic data representation,'' {\em CoRR}, vol.~abs/1603.01025, 2016.

\bibitem{lee2017}
E.~H. {Lee}, D.~{Miyashita}, E.~{Chai}, B.~{Murmann}, and S.~S. {Wong},
  ``Lognet: Energy-efficient neural networks using logarithmic computation,''
  in {\em 2017 IEEE International Conference on Acoustics, Speech and Signal
  Processing (ICASSP)}, pp.~5900--5904, 2017.

\bibitem{aqss}
T.~{Ueki}, I.~{Keisuke}, T.~{Matsubara}, and T.~{Kurokawa}, ``Aqss: Accelerator
  of quantization neural networks with stochastic approach,'' in {\em 2018
  Sixth International Symposium on Computing and Networking Workshops
  (CANDARW)}, pp.~138--144, 2018.

\bibitem{vogel2018}
S.~{Vogel}, M.~{Liang}, A.~{Guntoro}, W.~{Stechele}, and G.~{Ascheid},
  ``Efficient hardware acceleration of cnns using logarithmic data
  representation with arbitrary log-base,'' in {\em 2018 IEEE/ACM International
  Conference on Computer-Aided Design (ICCAD)}, pp.~1--8, 2018.

\bibitem{gong2014}
Y.~Gong, L.~Liu, M.~Yang, and L.~D. Bourdev, ``Compressing deep convolutional
  networks using vector quantization,'' {\em CoRR}, vol.~abs/1412.6115, 2014.

\bibitem{surveycompression}
Y.~{Cheng}, D.~{Wang}, P.~{Zhou}, and T.~{Zhang}, ``Model compression and
  acceleration for deep neural networks: The principles, progress, and
  challenges,'' {\em IEEE Signal Processing Magazine}, vol.~35, no.~1,
  pp.~126--136, 2018.

\bibitem{braindamage}
Y.~LeCun, J.~S. Denker, and S.~A. Solla, ``Optimal brain damage,'' in {\em
  Advances in Neural Information Processing Systems 2} (D.~S. Touretzky, ed.),
  pp.~598--605, Morgan-Kaufmann, 1990.

\bibitem{pruning1}
S.~Han, J.~Pool, J.~Tran, and W.~J. Dally, ``Learning both weights and
  connections for efficient neural networks,'' in {\em Proceedings of the 28th
  International Conference on Neural Information Processing Systems - Volume
  1}, NIPS’15, (Cambridge, MA, USA), p.~1135–1143, MIT Press, 2015.

\bibitem{pruning2}
S.~Srinivas and R.~V. Babu, ``Data-free parameter pruning for deep neural
  networks,'' in {\em Proceedings of the British Machine Vision Conference
  2015, {BMVC} 2015, Swansea, UK, September 7-10, 2015} (X.~Xie, M.~W. Jones,
  and G.~K.~L. Tam, eds.), pp.~31.1--31.12, {BMVA} Press, 2015.

\bibitem{pruning3}
Y.~{He}, X.~{Zhang}, and J.~{Sun}, ``Channel pruning for accelerating very deep
  neural networks,'' in {\em 2017 IEEE International Conference on Computer
  Vision (ICCV)}, pp.~1398--1406, 2017.

\bibitem{han2015deep}
S.~Han, H.~Mao, and W.~J. Dally, ``Deep compression: Compressing deep neural
  network with pruning, trained quantization and huffman coding,'' in {\em 4th
  International Conference on Learning Representations, {ICLR} 2016, San Juan,
  Puerto Rico, May 2-4, 2016, Conference Track Proceedings} (Y.~Bengio and
  Y.~LeCun, eds.), 2016.

\bibitem{prunet_shafique}
A.~{Marchisio}, M.~A. {Hanif}, M.~{Martina}, and M.~{Shafique}, ``Prunet:
  Class-blind pruning method for deep neural networks,'' in {\em 2018
  International Joint Conference on Neural Networks (IJCNN)}, pp.~1--8, 2018.

\bibitem{pruning4}
T.-J. {Yang}, Y.-H. {Chen}, and V.~{Sze}, ``Designing energy-efficient
  convolutional neural networks using energy-aware pruning,'' in {\em 2017 IEEE
  Conference on Computer Vision and Pattern Recognition (CVPR)},
  pp.~6071--6079, 2017.

\bibitem{ecc}
H.~Yang, Y.~Zhu, and J.~Liu, ``{ECC:} platform-independent energy-constrained
  deep neural network compression via a bilinear regression model,'' in {\em
  {IEEE} Conference on Computer Vision and Pattern Recognition, {CVPR} 2019,
  Long Beach, CA, USA, June 16-20, 2019}, pp.~11206--11215, Computer Vision
  Foundation / {IEEE}, 2019.

\bibitem{clipq}
F.~{Tung} and G.~{Mori}, ``Clip-q: Deep network compression learning by
  in-parallel pruning-quantization,'' in {\em 2018 IEEE/CVF Conference on
  Computer Vision and Pattern Recognition}, pp.~7873--7882, 2018.

\bibitem{amc}
Y.~He, J.~Lin, Z.~Liu, H.~Wang, L.-J. Li, and S.~Han, ``Amc: Automl for model
  compression and acceleration on mobile devices,'' in {\em Computer Vision --
  ECCV 2018} (V.~Ferrari, M.~Hebert, C.~Sminchisescu, and Y.~Weiss, eds.),
  (Cham), pp.~815--832, Springer International Publishing, 2018.

\bibitem{automl}
H.~{Cai}, J.~{Lin}, Y.~{Lin}, Z.~{Liu}, K.~{Wang}, T.~{Wang}, L.~{Zhu}, and
  S.~{Han}, ``Automl for architecting efficient and specialized neural
  networks,'' {\em IEEE Micro}, vol.~40, no.~1, pp.~75--82, 2020.

\bibitem{apq}
T.~{Wang}, K.~{Wang}, H.~{Cai}, J.~{Lin}, Z.~{Liu}, H.~{Wang}, Y.~{Lin}, and
  S.~{Han}, ``Apq: Joint search for network architecture, pruning and
  quantization policy,'' in {\em 2020 IEEE/CVF Conference on Computer Vision
  and Pattern Recognition (CVPR)}, pp.~2075--2084, 2020.

\bibitem{MobileNets}
A.~G. Howard, M.~Zhu, B.~Chen, D.~Kalenichenko, W.~Wang, T.~Weyand,
  M.~Andreetto, and H.~Adam, ``Mobilenets: Efficient convolutional neural
  networks for mobile vision applications,'' {\em CoRR}, vol.~abs/1704.04861,
  2017.

\bibitem{Xception}
F.~{Chollet}, ``Xception: Deep learning with depthwise separable
  convolutions,'' in {\em 2017 IEEE Conference on Computer Vision and Pattern
  Recognition (CVPR)}, pp.~1800--1807, 2017.

\bibitem{canonical}
V.~Lebedev, Y.~Ganin, M.~Rakhuba, I.~V. Oseledets, and V.~S. Lempitsky,
  ``Speeding-up convolutional neural networks using fine-tuned
  cp-decomposition,'' in {\em 3rd International Conference on Learning
  Representations, {ICLR} 2015, San Diego, CA, USA, May 7-9, 2015, Conference
  Track Proceedings} (Y.~Bengio and Y.~LeCun, eds.), 2015.

\bibitem{tucker}
Y.~Kim, E.~Park, S.~Yoo, T.~Choi, L.~Yang, and D.~Shin, ``Compression of deep
  convolutional neural networks for fast and low power mobile applications,''
  in {\em 4th International Conference on Learning Representations, {ICLR}
  2016, San Juan, Puerto Rico, May 2-4, 2016, Conference Track Proceedings}
  (Y.~Bengio and Y.~LeCun, eds.), 2016.

\bibitem{kd1}
C.~Bucilu\v{a}, R.~Caruana, and A.~Niculescu-Mizil, ``Model compression,'' in
  {\em Proceedings of the 12th ACM SIGKDD International Conference on Knowledge
  Discovery and Data Mining}, KDD ’06, (New York, NY, USA), p.~535–541,
  Association for Computing Machinery, 2006.

\bibitem{kd2}
L.~Ba and R.~Caruana, ``Do deep nets really need to be deep?,'' {\em Advances
  in Neural Information Processing Systems}, vol.~3, pp.~2654--2662, 01 2014.

\bibitem{hinton2015distilling}
G.~E. Hinton, O.~Vinyals, and J.~Dean, ``Distilling the knowledge in a neural
  network,'' {\em CoRR}, vol.~abs/1503.02531, 2015.

\bibitem{romero2014fitnets}
A.~Romero, N.~Ballas, S.~E. Kahou, A.~Chassang, C.~Gatta, and Y.~Bengio,
  ``Fitnets: Hints for thin deep nets,'' in {\em 3rd International Conference
  on Learning Representations, {ICLR} 2015, San Diego, CA, USA, May 7-9, 2015,
  Conference Track Proceedings} (Y.~Bengio and Y.~LeCun, eds.), 2015.

\bibitem{kd3}
J.~{Yim}, D.~{Joo}, J.~{Bae}, and J.~{Kim}, ``A gift from knowledge
  distillation: Fast optimization, network minimization and transfer
  learning,'' in {\em 2017 IEEE Conference on Computer Vision and Pattern
  Recognition (CVPR)}, pp.~7130--7138, 2017.

\bibitem{kd4}
Y.~{Zhang}, T.~{Xiang}, T.~M. {Hospedales}, and H.~{Lu}, ``Deep mutual
  learning,'' in {\em 2018 IEEE/CVF Conference on Computer Vision and Pattern
  Recognition}, pp.~4320--4328, 2018.

\bibitem{sparsity_thesis}
R.~W. Vuduc and J.~W. Demmel, {\em Automatic Performance Tuning of Sparse
  Matrix Kernels}.
\newblock PhD thesis, University of California, Berkeley, 2003.
\newblock AAI3121741.

\bibitem{cnvlutin}
J.~{Albericio}, P.~{Judd}, T.~{Hetherington}, T.~{Aamodt}, N.~E. {Jerger}, and
  A.~{Moshovos}, ``Cnvlutin: Ineffectual-neuron-free deep neural network
  computing,'' in {\em 2016 ACM/IEEE 43rd Annual International Symposium on
  Computer Architecture (ISCA)}, pp.~1--13, June 2016.

\bibitem{SCNN}
A.~{Parashar}, M.~{Rhu}, A.~{Mukkara}, A.~{Puglielli}, R.~{Venkatesan},
  B.~{Khailany}, J.~{Emer}, S.~W. {Keckler}, and W.~J. {Dally}, ``Scnn: An
  accelerator for compressed-sparse convolutional neural networks,'' in {\em
  2017 ACM/IEEE 44th Annual International Symposium on Computer Architecture
  (ISCA)}, pp.~27--40, June 2017.

\bibitem{EIE}
S.~Han, X.~Liu, H.~Mao, J.~Pu, A.~Pedram, M.~A. Horowitz, and W.~J. Dally,
  ``{EIE:} efficient inference engine on compressed deep neural network,'' in
  {\em 43rd {ACM/IEEE} Annual International Symposium on Computer Architecture,
  {ISCA} 2016, Seoul, South Korea, June 18-22, 2016}, pp.~243--254, {IEEE}
  Computer Society, 2016.

\bibitem{NullHop}
A.~{Aimar}, H.~{Mostafa}, E.~{Calabrese}, A.~{Rios-Navarro},
  R.~{Tapiador-Morales}, I.~{Lungu}, M.~B. {Milde}, F.~{Corradi},
  A.~{Linares-Barranco}, S.~{Liu}, and T.~{Delbruck}, ``Nullhop: A flexible
  convolutional neural network accelerator based on sparse representations of
  feature maps,'' {\em IEEE Transactions on Neural Networks and Learning
  Systems}, vol.~30, pp.~644--656, March 2019.

\bibitem{squeezeflow}
J.~{Li}, S.~{Jiang}, S.~{Gong}, J.~{Wu}, J.~{Yan}, G.~{Yan}, and X.~{Li},
  ``Squeezeflow: A sparse cnn accelerator exploiting concise convolution
  rules,'' {\em IEEE Transactions on Computers}, vol.~68, pp.~1663--1677, Nov
  2019.

\bibitem{ZeNA}
D.~{Kim}, J.~{Ahn}, and S.~{Yoo}, ``Zena: Zero-aware neural network
  accelerator,'' {\em IEEE Design Test}, vol.~35, pp.~39--46, Feb 2018.

\bibitem{huan2017lowpower}
Y.~Huan, Y.~Qin, Y.~You, L.~Zheng, and Z.~Zou, ``A low-power accelerator for
  deep neural networks with enlarged near-zero sparsity,'' {\em CoRR},
  vol.~abs/1705.08009, 2017.

\bibitem{crosslayerac_shafique}
M.~{Shafique}, R.~{Hafiz}, S.~{Rehman}, W.~{El-Harouni}, and J.~{Henkel},
  ``Invited: Cross-layer approximate computing: From logic to architectures,''
  in {\em 2016 53rd ACM/EDAC/IEEE Design Automation Conference (DAC)},
  pp.~1--6, 2016.

\bibitem{Mittal2016SurveyAC}
S.~Mittal, ``A survey of techniques for approximate computing,'' {\em ACM
  Comput. Surv.}, vol.~48, Mar. 2016.

\bibitem{Kumar2019AC4ML}
V.~Kumar and R.~Kant, ``Approximate computing for machine learning,'' in {\em
  Proceedings of 2nd International Conference on Communication, Computing and
  Networking} (C.~R. Krishna, M.~Dutta, and R.~Kumar, eds.), (Singapore),
  pp.~607--613, Springer Singapore, 2019.

\bibitem{Mrazek2016AM4ANN}
V.~Mrazek, S.~S. Sarwar, L.~Sekanina, Z.~Vasicek, and K.~Roy, ``Design of
  power-efficient approximate multipliers for approximate artificial neural
  networks,'' in {\em Proceedings of the 35th International Conference on
  Computer-Aided Design}, ICCAD ’16, (New York, NY, USA), Association for
  Computing Machinery, 2016.

\bibitem{errorresilience_shafique}
M.~A. {Hanif}, R.~{Hafiz}, and M.~{Shafique}, ``Error resilience analysis for
  systematically employing approximate computing in convolutional neural
  networks,'' in {\em 2018 Design, Automation and Test in Europe Conference
  (DATE)}, pp.~913--916, 2018.

\bibitem{redcane_shafique}
A.~Marchisio, V.~Mrazek, M.~A. Hanif, and M.~Shafique, ``Red-cane: A systematic
  methodology for resilience analysis and design of capsule networks under
  approximations,'' in {\em 2020 Design, Automation and Test in Europe
  Conference (DATE)}, 2020.

\bibitem{cann_shafique}
M.~A. Hanif, F.~Khalid, and M.~Shafique, ``Cann: Curable approximations for
  high-performance deep neural network accelerators,'' in {\em Proceedings of
  the 56th Annual Design Automation Conference 2019}, DAC ’19, (New York, NY,
  USA), Association for Computing Machinery, 2019.

\bibitem{He2018AxTrain}
X.~He, L.~Ke, W.~Lu, G.~Yan, and X.~Zhang, ``Axtrain: Hardware-oriented neural
  network training for approximate inference,'' in {\em Proceedings of the
  International Symposium on Low Power Electronics and Design}, ISLPED ’18,
  (New York, NY, USA), Association for Computing Machinery, 2018.

\bibitem{alwann_shafique}
V.~{Mrazek}, Z.~{Vasicek}, L.~{Sekanina}, M.~A. {Hanif}, and M.~{Shafique},
  ``Alwann: Automatic layer-wise approximation of deep neural network
  accelerators without retraining,'' in {\em 2019 IEEE/ACM International
  Conference on Computer-Aided Design (ICCAD)}, pp.~1--8, 2019.

\bibitem{CAxCNN_shafique}
M.~{Riaz}, R.~{Hafiz}, S.~A. {Khaliq}, M.~{Faisal}, H.~T. {Iqbal}, M.~{Ali},
  and M.~{Shafique}, ``Caxcnn: Towards the use of canonic sign digit based
  approximation for hardware-friendly convolutional neural networks,'' {\em
  IEEE Access}, vol.~8, pp.~127014--127021, 2020.

\bibitem{Deng2015ReducedPrecisionMV}
Z.~Deng, C.~Xu, Q.~Cai, P.~Faraboschi, and H.~Packard, ``Reduced-precision
  memory value approximation for deep learning,'' 2015.

\bibitem{Koppula2019EDEN}
S.~Koppula, L.~Orosa, A.~G. Ya\u{g}lundefinedk\c{c}undefined, R.~Azizi,
  T.~Shahroodi, K.~Kanellopoulos, and O.~Mutlu, ``Eden: Enabling
  energy-efficient, high-performance deep neural network inference using
  approximate dram,'' in {\em Proceedings of the 52nd Annual IEEE/ACM
  International Symposium on Microarchitecture}, MICRO ’52, (New York, NY,
  USA), p.~166–181, Association for Computing Machinery, 2019.

\bibitem{Chen2018AC4DL}
C.~{Chen}, J.~{Choi}, K.~{Gopalakrishnan}, V.~{Srinivasan}, and
  S.~{Venkataramani}, ``Exploiting approximate computing for deep learning
  acceleration,'' in {\em 2018 Design, Automation Test in Europe Conference
  Exhibition (DATE)}, pp.~821--826, 2018.

\bibitem{xdnns_shafique}
M.~A. Hanif, A.~Marchisio, T.~Arif, R.~Hafiz, S.~Rehman, and M.~Shafique,
  ``X-dnns: Systematic cross-layer approximations for energy-efficient deep
  neural networks,'' {\em J. Low Power Electronics}, vol.~14, no.~4,
  pp.~520--534, 2018.

\bibitem{alibaba}
``\url{https://alibabacloud.com}.''

\bibitem{aws}
``\url{https://aws.amazon.com}.''

\bibitem{ibmcloud}
``\url{https://www.ibm.com/cloud}.''

\bibitem{gcloud}
``\url{https://cloud.google.com}.''

\bibitem{gcolab}
``\url{https://colab.research.google.com}.''

\bibitem{mazure}
``\url{https://azure.microsoft.com}.''

\bibitem{Rajendran2019NeuromorphicHardware}
B.~{Rajendran}, A.~{Sebastian}, M.~{Schmuker}, N.~{Srinivasa}, and
  E.~{Eleftheriou}, ``Low-power neuromorphic hardware for signal processing
  applications: A review of architectural and system-level design approaches,''
  {\em IEEE Signal Processing Magazine}, vol.~36, no.~6, pp.~97--110, 2019.

\bibitem{Liu2018SpiKKaker2}
C.~Liu, G.~Bellec, B.~Vogginger, D.~Kappel, J.~Partzsch, F.~Neumärker,
  S.~Höppner, W.~Maass, S.~B. Furber, R.~Legenstein, and C.~G. Mayr,
  ``Memory-efficient deep learning on a spinnaker 2 prototype,'' {\em Frontiers
  in Neuroscience}, vol.~12, p.~840, 2018.

\bibitem{DRAM_shafique}
R.~V.~W. Putra, M.~A. Hanif, and M.~Shafique, ``Drmap: {A} generic {DRAM} data
  mapping policy for energy-efficient processing of convolutional neural
  networks,'' in {\em Proceedings of the 57th Annual Design Automation
  Conference 2020}, 2020.

\bibitem{Inefficiency}
R.~Hameed, W.~Qadeer, M.~Wachs, O.~Azizi, A.~Solomatnikov, B.~C. Lee,
  S.~Richardson, C.~Kozyrakis, and M.~Horowitz, ``Understanding sources of
  inefficiency in general-purpose chips,'' in {\em Proceedings of the 37th
  Annual International Symposium on Computer Architecture}, ISCA ’10, (New
  York, NY, USA), p.~37–47, Association for Computing Machinery, 2010.

\bibitem{cai2019onceforall}
H.~Cai, C.~Gan, T.~Wang, Z.~Zhang, and S.~Han, ``Once-for-all: Train one
  network and specialize it for efficient deployment,'' 2019.

\bibitem{batching}
E.~Hoffer, T.~Ben{-}Nun, I.~Hubara, N.~Giladi, T.~Hoefler, and D.~Soudry,
  ``Augment your batch: better training with larger batches,'' {\em CoRR},
  vol.~abs/1901.09335, 2019.

\bibitem{mkl-dnn}
Y.~Liu, Y.~Wang, R.~Yu, M.~Li, V.~Sharma, and Y.~Wang, ``Optimizing {CNN} model
  inference on cpus,'' in {\em 2019 {USENIX} Annual Technical Conference,
  {USENIX} {ATC} 2019, Renton, WA, USA, July 10-12, 2019} (D.~Malkhi and
  D.~Tsafrir, eds.), pp.~1025--1040, {USENIX} Association, 2019.

\bibitem{lai2018cmsisnn}
L.~Lai, N.~Suda, and V.~Chandra, ``{CMSIS-NN:} efficient neural network kernels
  for arm cortex-m cpus,'' {\em CoRR}, vol.~abs/1801.06601, 2018.

\bibitem{Garofalo_2019}
A.~Garofalo, M.~Rusci, F.~Conti, D.~Rossi, and L.~Benini, ``Pulp-nn:
  accelerating quantized neural networks on parallel ultra-low-power risc-v
  processors,'' {\em Philosophical Transactions of the Royal Society A:
  Mathematical, Physical and Engineering Sciences}, vol.~378, p.~20190155, Dec
  2019.

\bibitem{7783721}
M.~{Rhu}, N.~{Gimelshein}, J.~{Clemons}, A.~{Zulfiqar}, and S.~W. {Keckler},
  ``vdnn: Virtualized deep neural networks for scalable, memory-efficient
  neural network design,'' in {\em 2016 49th Annual IEEE/ACM International
  Symposium on Microarchitecture (MICRO)}, pp.~1--13, Oct 2016.

\bibitem{8599530}
Y.~{Kim}, J.~{Lee}, J.~{Kim}, H.~{Jei}, and H.~{Roh}, ``Efficient multi-gpu
  memory management for deep learning acceleration,'' in {\em 2018 IEEE 3rd
  International Workshops on Foundations and Applications of Self* Systems
  (FAS*W)}, pp.~37--43, Sep. 2018.

\bibitem{5272559}
C.~{Farabet}, C.~{Poulet}, J.~Y. {Han}, and Y.~{LeCun}, ``Cnp: An fpga-based
  processor for convolutional networks,'' in {\em 2009 International Conference
  on Field Programmable Logic and Applications}, pp.~32--37, Aug 2009.

\bibitem{8806999}
X.~{Wei}, Y.~{Liang}, and J.~{Cong}, ``Overcoming data transfer bottlenecks in
  fpga-based dnn accelerators via layer conscious memory management,'' in {\em
  2019 56th ACM/IEEE Design Automation Conference (DAC)}, pp.~1--6, June 2019.

\bibitem{7471828}
J.~{Park} and W.~{Sung}, ``Fpga based implementation of deep neural networks
  using on-chip memory only,'' in {\em 2016 IEEE International Conference on
  Acoustics, Speech and Signal Processing (ICASSP)}, pp.~1011--1015, March
  2016.

\bibitem{8310401}
W.~{Khwa}, J.~{Chen}, J.~{Li}, X.~{Si}, E.~{Yang}, X.~{Sun}, R.~{Liu},
  P.~{Chen}, Q.~{Li}, S.~{Yu}, and M.~{Chang}, ``A 65nm 4kb algorithm-dependent
  computing-in-memory sram unit-macro with 2.3ns and 55.8tops/w fully parallel
  product-sum operation for binary dnn edge processors,'' in {\em 2018 IEEE
  International Solid - State Circuits Conference - (ISSCC)}, pp.~496--498, Feb
  2018.

\bibitem{8824826}
T.~{Yoo}, H.~{Kim}, Q.~{Chen}, T.~T. {Kim}, and B.~{Kim}, ``A logic compatible
  4t dual embedded dram array for in-memory computation of deep neural
  networks,'' in {\em 2019 IEEE/ACM International Symposium on Low Power
  Electronics and Design (ISLPED)}, pp.~1--6, July 2019.

\bibitem{8839490}
S.~{Angizi}, Z.~{He}, D.~{Reis}, X.~S. {Hu}, W.~{Tsai}, S.~J. {Lin}, and
  D.~{Fan}, ``Accelerating deep neural networks in processing-in-memory
  platforms: Analog or digital approach?,'' in {\em 2019 IEEE Computer Society
  Annual Symposium on VLSI (ISVLSI)}, pp.~197--202, July 2019.

\bibitem{8867863}
S.~{Yin}, Z.~{Jiang}, M.~{Kim}, T.~{Gupta}, M.~{Seok}, and J.~{Seo}, ``Vesti:
  Energy-efficient in-memory computing accelerator for deep neural networks,''
  {\em IEEE Transactions on Very Large Scale Integration (VLSI) Systems},
  vol.~28, pp.~48--61, Jan 2020.

\bibitem{Barreno2010SecurityML}
M.~Barreno, B.~Nelson, A.~D. Joseph, and J.~D. Tygar, ``The security of machine
  learning,'' {\em Machine Learning}, vol.~81, pp.~121--148, 2010.

\bibitem{Szegedy2013IntriguingPO}
C.~Szegedy, W.~Zaremba, I.~Sutskever, J.~Bruna, D.~Erhan, I.~J. Goodfellow, and
  R.~Fergus, ``Intriguing properties of neural networks,'' {\em ICLR}, 2014.

\bibitem{RobustML_shafique}
M.~{Shafique}, M.~{Naseer}, T.~{Theocharides}, C.~{Kyrkou}, O.~{Mutlu},
  L.~{Orosa}, and J.~{Choi}, ``Robust machine learning systems:
  Challenges,current trends, perspectives, and the road ahead,'' {\em IEEE
  Design Test}, vol.~37, no.~2, pp.~30--57, 2020.

\bibitem{Goodfellow2015explainingadvexamples}
I.~Goodfellow, J.~Shlens, and C.~Szegedy, ``Explaining and harnessing
  adversarial examples,'' in {\em ICLR}, 2015.

\bibitem{SecurityML_shafique}
F.~{Khalid}, M.~A. {Hanif}, S.~{Rehman}, and M.~{Shafique}, ``Security for
  machine learning-based systems: Attacks and challenges during training and
  inference,'' in {\em 2018 International Conference on Frontiers of
  Information Technology (FIT)}, pp.~327--332, 2018.

\bibitem{RobustML2_shafique}
J.~J. Zhang, K.~Liu, F.~Khalid, M.~A. Hanif, S.~Rehman, T.~Theocharides,
  A.~Artussi, M.~Shafique, and S.~Garg, ``Building robust machine learning
  systems: Current progress, research challenges, and opportunities,'' in {\em
  Proceedings of the 56th Annual Design Automation Conference 2019}, DAC ’19,
  (New York, NY, USA), Association for Computing Machinery, 2019.

\bibitem{Tramer2016modelstealing}
F.~Tram\`{e}r, F.~Zhang, A.~Juels, M.~K. Reiter, and T.~Ristenpart, ``Stealing
  machine learning models via prediction apis,'' in {\em Proceedings of the
  25th USENIX Conference on Security Symposium}, SEC’16, (USA), p.~601–618,
  USENIX Association, 2016.

\bibitem{Song2017modelinversion}
C.~Song, T.~Ristenpart, and V.~Shmatikov, ``Machine learning models that
  remember too much,'' in {\em Proceedings of the 2017 ACM SIGSAC Conference on
  Computer and Communications Security}, CCS ’17, (New York, NY, USA),
  p.~587–601, Association for Computing Machinery, 2017.

\bibitem{Papernot2017practicalBBattacks}
N.~Papernot, P.~McDaniel, I.~Goodfellow, S.~Jha, Z.~B. Celik, and A.~Swami,
  ``Practical black-box attacks against machine learning,'' in {\em ASIA CCS},
  2017.

\bibitem{Yuan2019attacksanddefenses}
X.~{Yuan}, P.~{He}, Q.~{Zhu}, and X.~{Li}, ``Adversarial examples: Attacks and
  defenses for deep learning,'' {\em IEEE Trans. Neural Netw. Learn. Syst.},
  2019.

\bibitem{Luo2018imperceptibleandrobust}
B.~Luo, Y.~Liu, L.~Wei, and Q.~Xu, ``Towards imperceptible and robust
  adversarial example attacks against neural networks,'' in {\em AAAI}, 2018.

\bibitem{CapsAttacks_shafique}
A.~Marchisio, G.~Nanfa, F.~Khalid, M.~A. Hanif, M.~Martina, and M.~Shafique,
  ``Capsattacks: Robust and imperceptible adversarial attacks on capsule
  networks,'' {\em CoRR}, vol.~abs/1901.09878, 2019.

\bibitem{SNNAttack_shafique}
A.~Marchisio, G.~Nanfa, F.~Khalid, M.~A. Hanif, M.~Martina, and M.~Shafique,
  ``Is spiking secure? a comparative study on the security vulnerabilities of
  spiking and deep neural networks,'' {\em 2020 International Joint Conference
  on Neural Networks (IJCNN)}, 2020.

\bibitem{ProbabilisticAA_shafique}
Z.~{Yahya}, M.~{Hassan}, S.~{Younis}, and M.~{Shafique}, ``Probabilistic
  analysis of targeted attacks using transform-domain adversarial examples,''
  {\em IEEE Access}, vol.~8, pp.~33855--33869, 2020.

\bibitem{Biggio2013DataClusteringAdversarial}
B.~Biggio, I.~Pillai, S.~Rota~Bul\`{o}, D.~Ariu, M.~Pelillo, and F.~Roli, ``Is
  data clustering in adversarial settings secure?,'' in {\em Proceedings of the
  2013 ACM Workshop on Artificial Intelligence and Security}, AISec ’13, (New
  York, NY, USA), p.~87–98, Association for Computing Machinery, 2013.

\bibitem{Gonzalez2017PoisoningDL}
L.~Mu{\~n}oz-Gonz{\'a}lez, B.~Biggio, A.~Demontis, A.~Paudice, V.~Wongrassamee,
  E.~C. Lupu, and F.~Roli, ``Towards poisoning of deep learning algorithms with
  back-gradient optimization,'' in {\em 10th ACM Workshop on Artificial
  Intelligence and Security}, pp.~27--38, ACM, ACM, 2017.

\bibitem{Shafahi2018PoisoningAttacks}
A.~Shafahi, W.~R. Huang, M.~Najibi, O.~Suciu, C.~Studer, T.~Dumitras, and
  T.~Goldstein, ``Poison frogs! targeted clean-label poisoning attacks on
  neural networks,'' in {\em Proceedings of the 32nd International Conference
  on Neural Information Processing Systems}, NIPS’18, (Red Hook, NY, USA),
  p.~6106–6116, Curran Associates Inc., 2018.

\bibitem{Chen2017BackdoorDL}
X.~Chen, C.~Liu, B.~Li, K.~Lu, and D.~Song, ``Targeted backdoor attacks on deep
  learning systems using data poisoning,'' {\em CoRR}, vol.~abs/1712.05526,
  2017.

\bibitem{NeuroAttack_shafique}
V.~Venceslai, A.~Marchisio, I.~Alouani, M.~Martina, and M.~Shafique,
  ``Neuroattack: Undermining spiking neural networks security through
  externally triggered bit-flips,'' {\em 2020 International Joint Conference on
  Neural Networks (IJCNN)}, 2020.

\bibitem{Biggio2013EvasionAA}
B.~Biggio, I.~Corona, D.~Maiorca, B.~Nelson, N.~Srndic, P.~Laskov, G.~Giacinto,
  and F.~Roli, ``Evasion attacks against machine learning at test time,'' {\em
  ArXiv}, vol.~abs/1708.06131, 2013.

\bibitem{Kurakin2016AdvExamplesPhysicalWorld}
A.~Kurakin, I.~J. Goodfellow, and S.~Bengio, ``Adversarial examples in the
  physical world,'' {\em CoRR}, vol.~abs/1607.02533, 2016.

\bibitem{Dong2018BoostingAA}
Y.~Dong, F.~Liao, T.~Pang, H.~Su, J.~Zhu, X.~Hu, and J.~Li, ``Boosting
  adversarial attacks with momentum,'' {\em 2018 IEEE/CVF Conference on
  Computer Vision and Pattern Recognition}, pp.~9185--9193, 2018.

\bibitem{Madry2017TowardsDL}
A.~Madry, A.~Makelov, L.~Schmidt, D.~Tsipras, and A.~Vladu, ``Towards deep
  learning models resistant to adversarial attacks,'' in {\em 6th International
  Conference on Learning Representations, {ICLR} 2018, Vancouver, BC, Canada,
  April 30 - May 3, 2018, Conference Track Proceedings}, OpenReview.net, 2018.

\bibitem{TrISec_shafique}
F.~Khalid, M.~A. Hanif, S.~Rehman, R.~Ahmed, and M.~Shafique, ``Trisec:
  Training data-unaware imperceptible security attacks on deep neural
  networks,'' {\em 2019 IEEE 25th International Symposium on On-Line Testing
  and Robust System Design (IOLTS)}, pp.~188--193, 2018.

\bibitem{Carlini2016CWAttack}
N.~Carlini and D.~A. Wagner, ``Towards evaluating the robustness of neural
  networks,'' {\em 2017 IEEE Symposium on Security and Privacy (SP)},
  pp.~39--57, 2016.

\bibitem{Brendel2017DecisionBasedAA}
W.~Brendel, J.~Rauber, and M.~Bethge, ``Decision-based adversarial attacks:
  Reliable attacks against black-box machine learning models,'' in {\em 6th
  International Conference on Learning Representations, {ICLR} 2018, Vancouver,
  BC, Canada, April 30 - May 3, 2018, Conference Track Proceedings},
  OpenReview.net, 2018.

\bibitem{Brunner2018GuessingSmartAA}
T.~Brunner, F.~Diehl, M.~Truong-Le, and A.~Knoll, ``Guessing smart: Biased
  sampling for efficient black-box adversarial attacks,'' {\em 2019 IEEE/CVF
  International Conference on Computer Vision (ICCV)}, pp.~4957--4965, 2018.

\bibitem{Chen2019BoundaryAttack++}
J.~Chen and M.~I. Jordan, ``Boundary attack++: Query-efficient decision-based
  adversarial attack,'' {\em CoRR}, vol.~abs/1904.02144, 2019.

\bibitem{REDAttack_shafique}
F.~Khalid, H.~Ali, M.~A. Hanif, S.~Rehman, R.~Ahmed, and M.~Shafique,
  ``Red-attack: Resource efficient decision based attack for machine
  learning,'' {\em CoRR}, vol.~abs/1901.10258, 2019.

\bibitem{MoosaviDezfooli2016UniversalAP}
S.-M. Moosavi-Dezfooli, A.~Fawzi, O.~Fawzi, and P.~Frossard, ``Universal
  adversarial perturbations,'' {\em 2017 IEEE Conference on Computer Vision and
  Pattern Recognition (CVPR)}, pp.~86--94, 2016.

\bibitem{Chakarov2016DebuggingML}
A.~Chakarov, A.~V. Nori, S.~K. Rajamani, S.~Sen, and D.~Vijaykeerthy,
  ``Debugging machine learning tasks,'' {\em CoRR}, vol.~abs/1603.07292, 2016.

\bibitem{Barcaldo2017MitigatingPoisonA}
N.~Baracaldo, B.~Chen, H.~Ludwig, and J.~A. Safavi, ``Mitigating poisoning
  attacks on machine learning models: A data provenance based approach,'' in
  {\em Proceedings of the 10th ACM Workshop on Artificial Intelligence and
  Security}, AISec ’17, (New York, NY, USA), p.~103–110, Association for
  Computing Machinery, 2017.

\bibitem{Liu2018FinePruningDA}
K.~Liu, B.~Dolan{-}Gavitt, and S.~Garg, ``Fine-pruning: Defending against
  backdooring attacks on deep neural networks,'' in {\em Research in Attacks,
  Intrusions, and Defenses - 21st International Symposium, {RAID} 2018,
  Heraklion, Crete, Greece, September 10-12, 2018, Proceedings} (M.~Bailey,
  T.~Holz, M.~Stamatogiannakis, and S.~Ioannidis, eds.), vol.~11050 of {\em
  Lecture Notes in Computer Science}, pp.~273--294, Springer, 2018.

\bibitem{QuSecNets_shafique}
F.~{Khalid}, H.~{Ali}, H.~{Tariq}, M.~A. {Hanif}, S.~{Rehman}, R.~{Ahmed}, and
  M.~{Shafique}, ``Qusecnets: Quantization-based defense mechanism for securing
  deep neural network against adversarial attacks,'' in {\em 2019 IEEE 25th
  International Symposium on On-Line Testing and Robust System Design (IOLTS)},
  pp.~182--187, 2019.

\bibitem{Zhang2019YOPO}
D.~Zhang, T.~Zhang, Y.~Lu, Z.~Zhu, and B.~Dong, ``You only propagate once:
  Accelerating adversarial training via maximal principle,'' in {\em NeurIPS},
  2019.

\bibitem{Shafahi2019FreeAdversarialT}
A.~Shafahi, M.~Najibi, A.~Ghiasi, Z.~Xu, J.~P. Dickerson, C.~Studer, L.~S.
  Davis, G.~Taylor, and T.~Goldstein, ``Adversarial training for free!,'' in
  {\em NeurIPS}, 2019.

\bibitem{Wong2020FastAT}
E.~Wong, L.~Rice, and J.~Z. Kolter, ``Fast is better than free: Revisiting
  adversarial training,'' in {\em ICLR}, 2020.

\bibitem{FadeML_shafique}
F.~{Khalid}, M.~A. {Hanif}, S.~{Rehman}, J.~{Qadir}, and M.~{Shafique},
  ``Fademl: Understanding the impact of pre-processing noise filtering on
  adversarial machine learning,'' in {\em 2019 Design, Automation Test in
  Europe Conference Exhibition (DATE)}, pp.~902--907, 2019.

\bibitem{Cohen2019RandomizedSmoothing}
J.~Cohen, E.~Rosenfeld, and Z.~Kolter, ``Certified adversarial robustness via
  randomized smoothing,'' in {\em Proceedings of the 36th International
  Conference on Machine Learning} (K.~Chaudhuri and R.~Salakhutdinov, eds.),
  vol.~97 of {\em Proceedings of Machine Learning Research}, (Long Beach,
  California, USA), pp.~1310--1320, PMLR, 09--15 Jun 2019.

\bibitem{10.1145/3133956.3134057}
D.~Meng and H.~Chen, ``Magnet: A two-pronged defense against adversarial
  examples,'' in {\em Proceedings of the 2017 ACM SIGSAC Conference on Computer
  and Communications Security}, CCS ’17, (New York, NY, USA), p.~135–147,
  Association for Computing Machinery, 2017.

\bibitem{10.1145/3373376.3378532}
X.~Wang, R.~Hou, B.~Zhao, F.~Yuan, J.~Zhang, D.~Meng, and X.~Qian, ``Dnnguard:
  An elastic heterogeneous dnn accelerator architecture against adversarial
  attacks,'' in {\em Proceedings of the Twenty-Fifth International Conference
  on Architectural Support for Programming Languages and Operating Systems},
  ASPLOS ’20, (New York, NY, USA), p.~19–34, Association for Computing
  Machinery, 2020.

\bibitem{mxnet}
T.~Chen, M.~Li, Y.~Li, M.~Lin, N.~Wang, M.~Wang, T.~Xiao, B.~Xu, C.~Zhang, and
  Z.~Zhang, ``Mxnet: {A} flexible and efficient machine learning library for
  heterogeneous distributed systems,'' {\em CoRR}, vol.~abs/1512.01274, 2015.

\bibitem{chainer}
S.~Tokui, R.~Okuta, T.~Akiba, Y.~Niitani, T.~Ogawa, S.~Saito, S.~Suzuki,
  K.~Uenishi, B.~Vogel, and H.~Yamazaki~Vincent, ``Chainer: A deep learning
  framework for accelerating the research cycle,'' in {\em Proceedings of the
  25th ACM SIGKDD International Conference on Knowledge Discovery \& Data
  Mining}, pp.~2002--2011, ACM, 2019.

\bibitem{microsoftcognitive}
F.~Seide and A.~Agarwal, ``Cntk: Microsoft’s open-source deep-learning
  toolkit,'' in {\em Proceedings of the 22nd ACM SIGKDD International
  Conference on Knowledge Discovery and Data Mining}, KDD ’16, (New York, NY,
  USA), p.~2135, Association for Computing Machinery, 2016.

\bibitem{paddle}
Y.~Ma, D.~Yu, T.~Wu, and H.~Wang, ``Paddlepaddle: An open-source deep learning
  platform from industrial practice,'' {\em Frontiers of Data and Domputing},
  vol.~1, no.~1, p.~105, 2019.

\bibitem{onnx}
J.~Bai, F.~Lu, K.~Zhang, {\em et~al.}, ``Onnx: Open neural network exchange.''
  \url{https://github.com/onnx/onnx}, 2019.

\bibitem{chollet2015}
F.~Chollet, ``keras.'' \url{https://github.com/fchollet/keras}, 2015.

\bibitem{CIFAR}
A.~Krizhevsky, ``Learning multiple layers of features from tiny images,'' tech.
  rep., Department of Computer Science, University of Toronto, 2009.

\bibitem{COCO}
T.~Lin, M.~Maire, S.~J. Belongie, J.~Hays, P.~Perona, D.~Ramanan,
  P.~Doll{\'{a}}r, and C.~L. Zitnick, ``Microsoft {COCO:} common objects in
  context,'' in {\em Computer Vision - {ECCV} 2014 - 13th European Conference,
  Zurich, Switzerland, September 6-12, 2014, Proceedings, Part {V}} (D.~J.
  Fleet, T.~Pajdla, B.~Schiele, and T.~Tuytelaars, eds.), vol.~8693 of {\em
  Lecture Notes in Computer Science}, pp.~740--755, Springer, 2014.

\bibitem{OpenImageV4}
A.~Kuznetsova, H.~Rom, N.~Alldrin, J.~Uijlings, I.~Krasin, J.~Pont-Tuset,
  S.~Kamali, S.~Popov, M.~Malloci, A.~Kolesnikov, and et~al., ``The open images
  dataset v4,'' {\em International Journal of Computer Vision}, Mar 2020.

\bibitem{CORe50}
V.~Lomonaco and D.~Maltoni, ``Core50: a new dataset and benchmark for
  continuous object recognition,'' in {\em Proceedings of the 1st Annual
  Conference on Robot Learning} (S.~Levine, V.~Vanhoucke, and K.~Goldberg,
  eds.), vol.~78 of {\em Proceedings of Machine Learning Research}, pp.~17--26,
  PMLR, 13--15 Nov 2017.

\bibitem{objectnet}
A.~Barbu, D.~Mayo, J.~Alverio, W.~Luo, C.~Wang, D.~Gutfreund, J.~Tenenbaum, and
  B.~Katz, ``Objectnet: A large-scale bias-controlled dataset for pushing the
  limits of object recognition models,'' in {\em NeurIPS}, 2019.

\bibitem{aiindex2019}
R.~Perrault, Y.~Shoham, E.~Brynjolfsson, J.~Clark, J.~Etchemendy, B.~Grosz,
  T.~Lyons, J.~Manyika, S.~Mishra, and J.~C. Niebles, ``The ai index 2019
  annual report,'' tech. rep., AI Index Steering Committee, Human-Centered AI
  Institute, Stanford University, 12 2019.

\bibitem{ecchipreport}
A.~Jadhav and P.~Kakade, ``Deep learning chip market by chip type (gpu, asic,
  fpga, cpu, and others), technology (system-on-chip, system-in-package,
  multi-chip module, and others), and industry vertical (media \& advertising,
  bfsi, it \& telecom, retail, healthcare, automotive \& transportation, and
  others) - global opportunity analysis and industry forecast, 2018-2025,''
  tech. rep., Allied Market Research, 07 2018.

\bibitem{xnorsram}
Z.~{Jiang}, S.~{Yin}, M.~{Seok}, and J.~{Seo}, ``Xnor-sram: In-memory computing
  sram macro for binary/ternary deep neural networks,'' in {\em 2018 IEEE
  Symposium on VLSI Technology}, pp.~173--174, 2018.

\bibitem{imc}
S.~{Okumura}, M.~{Yabuuchi}, K.~{Hijioka}, and K.~{Nose}, ``A ternary based bit
  scalable, 8.80 tops/w cnn accelerator with many-core processing-in-memory
  architecture with 896k synapses/mm2,'' in {\em 2019 Symposium on VLSI
  Technology}, pp.~C248--C249, 2019.

\bibitem{endmoore}
T.~N. {Theis} and H.~.~P. {Wong}, ``The end of moore's law: A new beginning for
  information technology,'' {\em Computing in Science Engineering}, vol.~19,
  no.~2, pp.~41--50, 2017.

\bibitem{phasechangememory}
V.~Joshi, M.~Le~Gallo, S.~Haefeli, I.~Boybat, S.~R. Nandakumar, C.~Piveteau,
  M.~Dazzi, B.~Rajendran, A.~Sebastian, and E.~Eleftheriou, ``Accurate deep
  neural network inference using computational phase-change memory,'' {\em
  Nature Communications}, vol.~11, p.~2473, May 2020.

\bibitem{phasechangememory2}
S.~R. {Nandakumar}, I.~{Boybat}, V.~{Joshi}, C.~{Piveteau}, M.~{Le Gallo},
  B.~{Rajendran}, A.~{Sebastian}, and E.~{Eleftheriou}, ``Phase-change memory
  models for deep learning training and inference,'' in {\em 2019 26th IEEE
  International Conference on Electronics, Circuits and Systems (ICECS)},
  pp.~727--730, 2019.

\bibitem{sttmram}
H.~{Yan}, H.~R. {Cherian}, E.~C. {Ahn}, X.~{Qian}, and L.~{Duan}, ``icelia: A
  full-stack framework for stt-mram-based deep learning acceleration,'' {\em
  IEEE Transactions on Parallel and Distributed Systems}, vol.~31, no.~2,
  pp.~408--422, 2020.

\bibitem{sttmram2}
S.~{Chattopadhyay}, K.~{Brahma}, A.~{Ray}, and M.~{Sharad}, ``Stt-mram for low
  power access for read-intensive parallel deep-learning architectures,'' in
  {\em 2017 IEEE International Symposium on Nanoelectronic and Information
  Systems (iNIS)}, pp.~61--65, 2017.

\bibitem{reram}
L.~{Song}, X.~{Qian}, H.~{Li}, and Y.~{Chen}, ``Pipelayer: A pipelined
  reram-based accelerator for deep learning,'' in {\em 2017 IEEE International
  Symposium on High Performance Computer Architecture (HPCA)}, pp.~541--552,
  2017.

\bibitem{toptrends}
A.~Mutiara, ``Ieee computer society's top 12 technology trends for 2020,'' 12
  2019.

\bibitem{analog1}
A.~{Shafiee}, A.~{Nag}, N.~{Muralimanohar}, R.~{Balasubramonian}, J.~P.
  {Strachan}, M.~{Hu}, R.~S. {Williams}, and V.~{Srikumar}, ``Isaac: A
  convolutional neural network accelerator with in-situ analog arithmetic in
  crossbars,'' in {\em 2016 ACM/IEEE 43rd Annual International Symposium on
  Computer Architecture (ISCA)}, pp.~14--26, 2016.

\bibitem{analog2}
Y.~Kim, H.~Kim, D.~Ahn, and J.-J. Kim, ``Input-splitting of large neural
  networks for power-efficient accelerator with resistive crossbar memory
  array,'' ISLPED '18, (New York, NY, USA), Association for Computing
  Machinery, 2018.

\bibitem{analog3}
D.~{Bankman}, L.~{Yang}, B.~{Moons}, M.~{Verhelst}, and B.~{Murmann}, ``An
  always-on 3.8$\mu$j/86\% cifar-10 mixed-signal binary cnn processor with all
  memory on chip in 28nm cmos,'' in {\em 2018 IEEE International Solid - State
  Circuits Conference - (ISSCC)}, pp.~222--224, 2018.

\bibitem{neuromorphic}
C.~D. Schuman, T.~E. Potok, R.~M. Patton, J.~D. Birdwell, M.~E. Dean, G.~S.
  Rose, and J.~S. Plank, ``A survey of neuromorphic computing and neural
  networks in hardware,'' {\em CoRR}, vol.~abs/1705.06963, 2017.

\bibitem{chen_survey}
Y.~Chen, Y.~Xie, L.~Song, F.~Chen, and T.~Tang, ``A survey of accelerator
  architectures for deep neural networks,'' {\em Engineering}, 2020.

\bibitem{compr_acc}
B.~L. {Deng}, G.~{Li}, S.~{Han}, L.~{Shi}, and Y.~{Xie}, ``Model compression
  and hardware acceleration for neural networks: A comprehensive survey,'' {\em
  Proceedings of the IEEE}, vol.~108, no.~4, pp.~485--532, 2020.

\bibitem{ref6}
A.~A. {Ratnaparkhi}, E.~{Pilli}, and R.~C. {Joshi}, ``Survey of scaling
  platforms for deep neural networks,'' in {\em 2016 International Conference
  on Emerging Trends in Communication Technologies (ETCT)}, pp.~1--6, 2016.

\bibitem{ref5}
C.~{Zhang} and W.~{Xu}, ``Neural networks: Efficient implementations and
  applications,'' in {\em 2017 IEEE 12th International Conference on ASIC
  (ASICON)}, pp.~1029--1032, 2017.

\bibitem{ref8}
M.~{Kotlar}, D.~{Bojic}, M.~{Punt}, and V.~{Milutinovic}, ``A survey of deep
  neural networks: Deployment location and underlying hardware,'' in {\em 2018
  14th Symposium on Neural Networks and Applications (NEUREL)}, pp.~1--6, 2018.

\bibitem{ref10}
A.~{Erofei}, C.~{Druţa}, and C.~{Daniel Căleanu}, ``Embedded solutions for
  deep neural networks implementation,'' in {\em 2018 IEEE 12th International
  Symposium on Applied Computational Intelligence and Informatics (SACI)},
  pp.~000425--000430, 2018.

\bibitem{ref9}
A.~{Sebastian}, I.~{Boybat}, M.~{Dazzi}, I.~{Giannopoulos}, V.~{Jonnalagadda},
  V.~{Joshi}, G.~{Karunaratne}, B.~{Kersting}, R.~{Khaddam-Aljameh}, S.~R.
  {Nandakumar}, A.~{Petropoulos}, C.~{Piveteau}, T.~{Antonakopoulos},
  B.~{Rajendran}, M.~L. {Gallo}, and E.~{Eleftheriou}, ``Computational
  memory-based inference and training of deep neural networks,'' in {\em 2019
  Symposium on VLSI Technology}, pp.~T168--T169, 2019.

\bibitem{ref1}
C.~{Ababei} and M.~G. {Moghaddam}, ``A survey of prediction and classification
  techniques in multicore processor systems,'' {\em IEEE Transactions on
  Parallel and Distributed Systems}, vol.~30, no.~5, pp.~1184--1200, 2019.

\bibitem{ref3}
S.~{Sun}, Z.~{Cao}, H.~{Zhu}, and J.~{Zhao}, ``A survey of optimization methods
  from a machine learning perspective,'' {\em IEEE Transactions on
  Cybernetics}, pp.~1--14, 2019.

\bibitem{ref13}
M.~Alam, M.~D. Samad, L.~Vidyaratne, A.~Glandon, and K.~M. Iftekharuddin,
  ``Survey on deep neural networks in speech and vision systems,'' 2019.

\bibitem{ref12}
A.~Khan, A.~Sohail, U.~Zahoora, and A.~S. Qureshi, ``A survey of the recent
  architectures of deep convolutional neural networks,'' {\em Artificial
  Intelligence Review}, Apr 2020.

\bibitem{ref7}
H.~{Yingge}, I.~{Ali}, and K.~{Lee}, ``Deep neural networks on chip - a
  survey,'' in {\em 2020 IEEE International Conference on Big Data and Smart
  Computing (BigComp)}, pp.~589--592, 2020.

\bibitem{ref11}
Z.~Li, W.~Yang, S.~Peng, and F.~Liu, ``A survey of convolutional neural
  networks: Analysis, applications, and prospects,'' 2020.

\end{thebibliography}

\clearpage
 \begin{IEEEbiography}[{\includegraphics[width=1in,height=1.25in,clip,keepaspectratio]{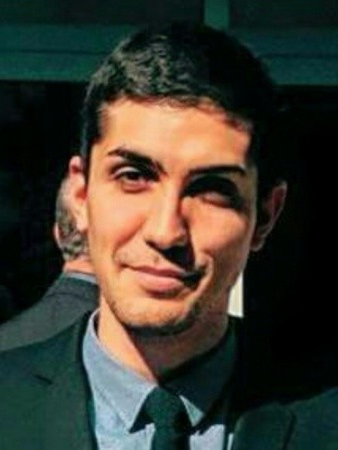}}]{Maurizio Capra} received the Bachelor's degree in Electronic Engineering, in October 2015, and the Master's degree in Electronic Engineering, with career in Electronic Systems, in April 2018 from Politecnico di Torino, Turin, Italy. Now he is pursuing a PhD in Electronics and Communication Engineering under the supervision of Prof. Maurizio Martina at Politecnico di Torino. His research activity is focused on newly dedicated architectures for machine learning with emphasis on on-chip learning based on a vertical approach: starting from the algorithm (top) going to the physical implementation (bottom). He is an IEEE student member and part of the board of the IEEE Student Branch at Politecnico di Torino.
\end{IEEEbiography}
\vspace*{-20pt}

 \begin{IEEEbiography}[{\includegraphics[width=1in,height=1.25in,clip,keepaspectratio]{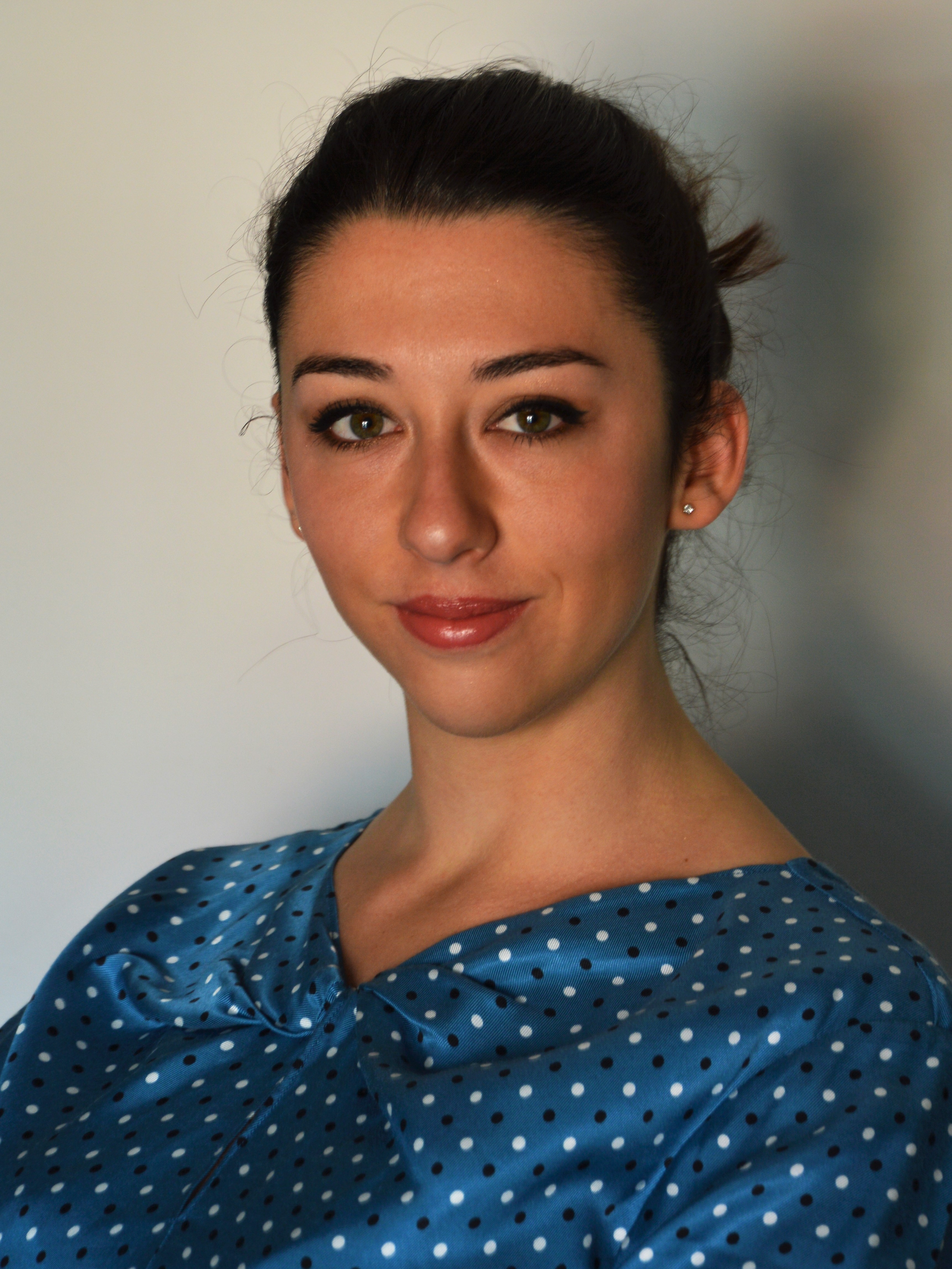}}]{Beatrice Bussolino} received the B.Sc. and M.Sc. degrees in Electronic Engineering, in October 2017 and October 2019 respectively, from Politecnico di Torino, Turin, Italy.  She is now pursuing the Ph.D. degree in Electrical, Electronics and Communications Engineering at Politecnico di Torino under the supervision of Prof. Maurizio Martina. She is an IEEE student member. Her current research interests are in the field of Machine Learning and Deep Neural Networks (DNNs) in particular. The focus of her research activity is the development of on-chip architectures for the edge deployment of DNNs. In 2020, she received the Richard Newton Young Fellow Award and won the DAC Young Fellow Poster Presentation Award.
\end{IEEEbiography}
\vspace*{-23pt}

 \begin{IEEEbiography}[{\includegraphics[width=1in,height=1.25in,clip,keepaspectratio]{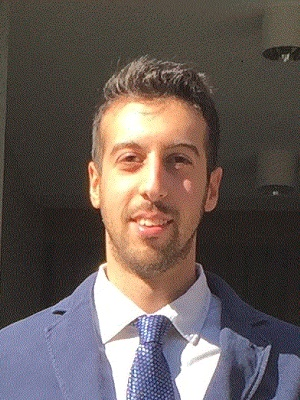}}]{Alberto Marchisio} (S'18) received his B.Sc. degree in Electronic Engineering from Politecnico di Torino, Turin, Italy, in October 2015. He received his M.Sc. degree in Electronic Engineering (Electronic Systems) from Politecnico di Torino, Turin, Italy, in April 2018. Currently, he is Ph.D. Student at Computer Architecture and Robust Energy-Efficient Technologies (CARE-Tech.) lab, Institute of Computer Engineering, Technische Universit{\"a}t Wien (TU Wien), Vienna, Austria, under the supervision of Prof. Dr. Muhammad Shafique. He is also a student IEEE member. His main research interests include hardware and software optimizations for machine learning, brain-inspired computing, VLSI architecture design, emerging computing technologies, robust design, and approximate computing for energy efficiency. He received the honorable mention at the Italian National Finals of Maths Olympic Games in 2012, and the Richard Newton Young Fellow Award in 2019.
\end{IEEEbiography}

 \begin{IEEEbiography}[{\includegraphics[width=1in,height=1.25in,clip,keepaspectratio]{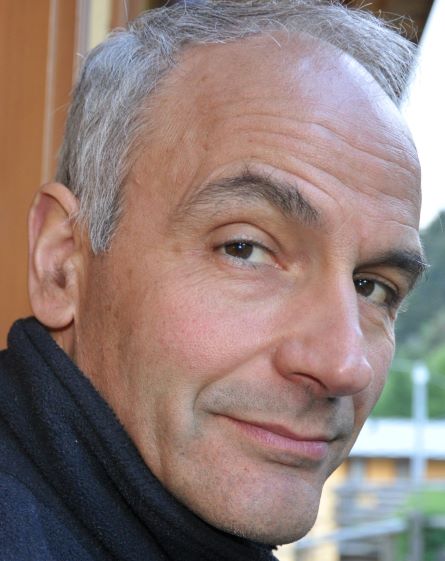}}]{Guido Masera} (SM’07)   received   the   Dr.-Ing. (summa cum laude) and Ph.D. degrees in 
Electronic Engineering  from Politecnico  di Torino, Italy, in 1986 and 1992, respectively.
He   is a  Professor  with  the  Electronic  Department,  Politecnico  di  Torino,  since  1992.
His research  interests  include  several  aspects  in  the design of digital integrated circuits 
and systems, with a special emphasis on high-performance architectures for communications, 
forward error correction, image  and  video  coding,  cryptography and hardware accelerators 
for machine learning. He has more than 200 publications in the fields of VLSI design and 
communications. Dr. Masera  is an Associate  Editor  of the IEEE TRANSACTIONS ON CIRCUITS 
AND SYSTEMS I and MDPI Electronics and a former Associate Editor of IEEE TRANSACTIONS ON 
CIRCUITS AND SYSTEMS II and the IET Circuits, Devices \& Systems.
\end{IEEEbiography}

 \begin{IEEEbiography}[{\includegraphics[width=1in,height=1.25in,clip,keepaspectratio]{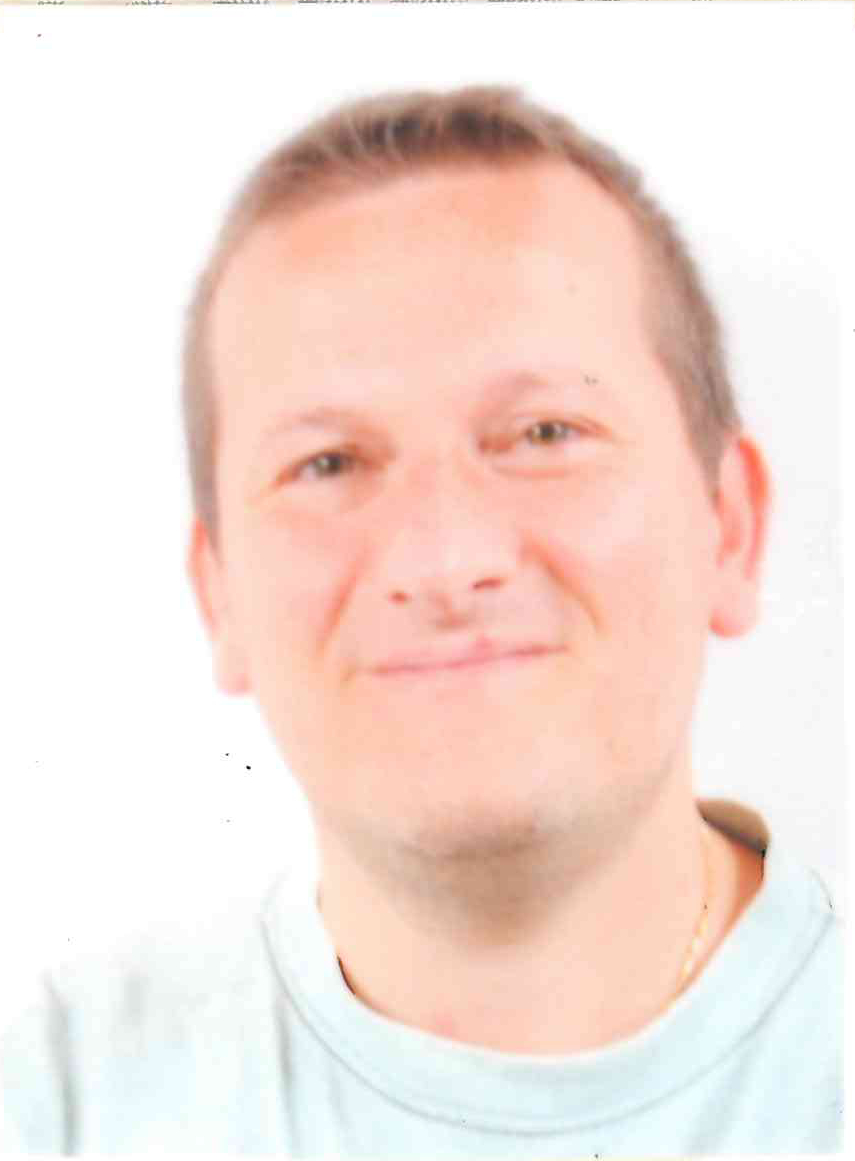}}]{Maurizio Martina} (S’98-M’04-SM’15) received
the M.S. and Ph.D. in electrical engineering from
Politecnico di Torino, Italy, in 2000 and 2004, respectively.
He is currently an Associate Professor
of the VLSI-Lab group, Politecnico di Torino. His
research interests include VLSI design and implementation
of architectures for digital signal processing,
video coding, communications, artificial intelligence,
machine learning and event-based processing.
He edited one book and published 3 book chapters on VLSI architectures
and digital circuits for video coding, wireless communications and
error correcting codes. He has more than 100 scientific publications and is
author of 2 patents. He is now an Associate Editor of IEEE Transactions
on Circuits and Systems - I. 
He had been part of the organizing and technical committee of several 
international conferences, including BioCAS 2017, ICECS 2019, AICAS 2020.
Currently, he is the counselor of the IEEE Student Branch at Politecnico di
Torino and a professional member of IEEE HKN.
\end{IEEEbiography}

\begin{IEEEbiography}[{\includegraphics[width=1in,height=1.25in,clip,keepaspectratio]{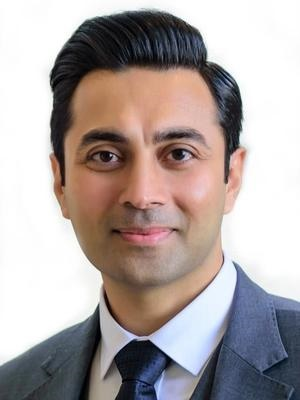}}]{Muhammad Shafique} (M'11-SM'16) received the Ph.D. degree in computer science from the Karlsruhe Institute of Technology (KIT), Germany, in 2011. Afterwards, he established and led a highly recognized research group at KIT for several years as well as conducted impactful R\&D activities in Pakistan. In Oct.2016, he joined the Institute of Computer Engineering at the Faculty of Informatics, Technische Universit{\"a}t Wien (TU Wien), Vienna, Austria as a Full Professor of Computer Architecture and Robust, Energy-Efficient Technologies. Since Sep.2020, he is with the Division of Engineering, New York University Abu Dhabi (NYU AD), United Arab Emirates, and is a Global Network faculty at the NYU Tandon School of Engineering, USA.

\noindent
His research interests are in brain-inspired computing, AI \& machine learning hardware and system-level design, energy-efficient systems, robust computing, hardware security, emerging technologies, FPGAs, MPSoCs, and embedded systems. His research has a special focus on cross-layer analysis, modeling, design, and optimization of computing and memory systems. The researched technologies and tools are deployed in application use cases from Internet-of-Things (IoT), smart Cyber-Physical Systems (CPS), and ICT for Development (ICT4D) domains.

\noindent
Dr. Shafique has given several Keynotes, Invited Talks, and Tutorials, as well as organized many special sessions at premier venues. He has served as the PC Chair, General Chair, Track Chair, and PC member for several prestigious IEEE/ACM conferences. Dr. Shafique holds one U.S. patent has (co-)authored 6 Books, 10+ Book Chapters, and over 250 papers in premier journals and conferences. He received the 2015 ACM/SIGDA Outstanding New Faculty Award, AI 2000 Chip Technology Most Influential Scholar Award in 2020, six gold medals, and several best paper awards and nominations at prestigious conferences.
\end{IEEEbiography}

\EOD

\end{document}